\documentclass[rmp,aps,preprint,amssymb,nofootinbib]{revtex4}
\usepackage{graphics}
\usepackage{epsfig}

\def\bfgamma{\mbox{\boldmath $\gamma$}}
\def\bfnabla{\mbox{\boldmath $\nabla$}}
\def\bfsigma{\mbox{\boldmath $\sigma$}}
\def\bfxi{\mbox{\boldmath $\xi$}}

\def\bfPi{\mbox{\boldmath $\Pi$}}
\def\lQ{\Lambda_{\rm QCD}}
\newcommand{\one}{1\!\!{\rm l}}

\newcommand{\nn}{\nonumber}
\newcommand{\be}{\begin{equation}}
\newcommand{\ee}{\end{equation}}
\newcommand{\bea}{\begin{eqnarray}}
\newcommand{\eea}{\end{eqnarray}}
\def\al{\alpha}
\def\als{\alpha_{\rm s}}
\def\siml{{\
\lower-1.2pt\vbox{\hbox{\rlap{$<$}\lower6pt\vbox{\hbox{$\sim$}}}}\ }}  
\def\simg{{\ \lower-1.2pt\vbox{\hbox{\rlap{$>$}\lower6pt\vbox{\hbox{$\sim$}}}}\ }}
\def\vbfD{{\ \lower-8pt\vbox{\hbox{\rlap{$\!\leftrightarrow$}\lower8pt\vbox{\hbox{$\!\bf D$}}}}\ }} 
\def\dsl{\,\raise.15ex\hbox{/}\mkern-13.5mu D}

\def\vac{\hbox{vac}}
\def\lla{\langle\!\langle}
\def\rra{\rangle\!\rangle}

\newcommand{\MS}{\overline{\rm MS}}
\newcommand{\RS}{\rm RS}
\newcommand{\PS}{\rm PS}

\begin{document}
\title{Effective field theories for heavy quarkonium}
\author {Nora Brambilla$^1$, Antonio Pineda$^2$, Joan Soto$^2$ and Antonio Vairo$^1$}
\affiliation{$^1$ INFN and Dipartimento di Fisica dell'Universit\`a di Milano \\
via Celoria 16, 20133 Milan, Italy}
\affiliation{$^2$ Departament d'Estructura i Constituents de la Mat\`eria, 
     Universitat \\ de Barcelona, Diagonal 647, E-08028 Barcelona, Catalonia, Spain}

\begin{abstract}  
We review recent theoretical developments in heavy quarkonium physics from
the point of view of Effective Field Theories of QCD.
We discuss Non-Relativistic QCD and concentrate on potential 
Non-Relativistic QCD. Our main goal will be to derive
QCD Schr\"odinger-like equations that govern heavy quarkonium
physics in the weak and strong coupling regime.
We also discuss a selected set of applications, which include 
spectroscopy, inclusive decays and electromagnetic threshold production.
\end{abstract}                                                                 

\date{22-03-2005}
\maketitle
\tableofcontents

\section*{TABLE OF ACRONYMS}

\noindent
For ease of convenience, we list below the acronyms used in the review.

\noindent
{\bf 2PI}: Two-particle irreducible\\
{\bf 2PR}: Two-particle reducible\\
{\bf DR}: Dimensional Regularization\\
{\bf EFT}: Effective Field Theory\\
{\bf IR}: Infrared\\
{\bf HQET}: Heavy quark effective theory\\
{\bf LO}: Leading order\\
{\bf MS}: Minimal subtraction\\
{\bf NLO}: Next-to-leading order\\
{\bf NNLO}: Next-to-next-to-leading order\\
{\bf NNNLO}: Next-to-next-to-next-to-leading order\\
{\bf LL}: Leading-logarithm order\\
{\bf NLL}: Next-to-leading-logarithm order\\
{\bf NNLL}: Next-to-next-to-leading-logarithm order\\
{\bf NR}: non-relativistic\\
{\bf NRQCD}: Non-relativistic Quantum Chromodynamics\\
{\bf NRQED}: Non-relativistic Quantum Electrodynamics\\
{\bf pNRQCD}: potential Non-relativistic Quantum Chromodynamics\\
{\bf PS}: Potential-subtracted \\
{\bf QFT}: Quantum Field Theory\\
{\bf QCD}: Quantum Chromodynamics\\
{\bf QED}: Quantum Electrodynamics\\
{\bf RG}: Renormalization group \\
{\bf RS}: Renormalon-subtracted\\
{\bf SCET}: Soft-collinear effective theory\\
{\bf US}: Ultrasoft\\
{\bf UV}: Ultraviolet\\
{\bf vNRQCD}: velocity Non-relativistic Quantum Chromodynamics\\

\section{Introduction}
\label{sec:intro}
In order to understand human scale processes, a classical NR
picture of physics based on Galilean symmetry proves sufficient. Until the
beginning of the last century, this picture, supplemented with
electromagnetism, was enough in order to understand the majority of processes
observed in nature. At the start of the quantum age, it is again a NR equation,
the Schr\"odinger equation, which proved to be the most successful in
explaining the atomic and nuclear spectra.

High-energy processes are far away from human scale processes. They are
described in present days by relativistic QFTs. 
Under some circumstances however, high-energy processes develop
a NR regime and produce bound states that behave very much like atoms.

The discovery of the $J/\psi$, a heavy resonance with a very narrow width, in
Brookhaven and SLAC \cite{Aubert:1974js,Augustin:1974xw}, which was later on
identified with a bound state of a new (heavy) quark, charm, and its
antiquark, namely a charmonium ($c {\bar c}$) state, opened up the possibility
to use a NR picture in the realm of QCD, the fundamental QFT of strong
interactions. This possibility was enhanced three years later by the discovery
of the $\Upsilon$, an even heavier and narrower resonance, which was again
interpreted as a bound state of a new (heavier) quark, bottom, and its
antiquark, namely a bottomonium ($b \bar b$) state \cite{Herb:1977ek}. In
fact, the narrow width of these resonances proved to be crucial to establish
QCD as the sector of the Standard Model that describes the strong interaction
\cite{Appelquist:1975zd,DeRujula:1975nx}.  From that moment on, charmonia and
bottomonia have been throughly studied, and still are a subject of intensive
theoretical and experimental research (see for instance
\cite{Skwarnicki:2003wn,Brambilla:2004wf}).  They can indeed be classified in
terms of the quantum numbers of a NR bound state, and the spacing of the
radial excitations and of the fine and hyperfine splittings has a pattern
similar to the ones in positronium, a well studied QED NR bound state.  A
related system, the $b {\bar c}$ bound state ($B_c$) has also been found in
nature \cite{Abe:1998wi}.  The heaviest of the quarks, the top, which has
recently been found at Tevatron \cite{Abe:1994xt}, has a large decay width
(due to weak interactions) and is not expected to form narrow $t$-$\bar t$
resonances. However, the production of $t$-$\bar t$ near threshold, namely in
the NR regime, will be one of the major programs at the Next Linear Collider. 

These systems will be denoted by heavy quarkonium.
They are characterized by, at least, three widely separated scales: the hard
scale (the mass $m$ of the heavy quarks), the soft scale (the relative
momentum of the heavy-quark--antiquark $|{\bf p}| \equiv p \sim mv$, $ v \ll
1$), and the US scale (the typical kinetic energy $E \sim mv^2$ of
the heavy quark and antiquark).  Moreover, by definition of heavy quark, $m$
is large in comparison with the typical hadronic scale $\lQ$.  Hence,
processes that happen at the scale $m$ are expected to be successfully
described in perturbation theory, due to the asymptotic freedom of QCD. This
explains why the narrow heavy quarkonium widths could be qualitatively
understood as a manifestation of asymptotic freedom.  However, lower scales,
like $|{\bf p}|$ and $E$, which are responsible for the binding, may or may
not be accessible to perturbation theory.  The appearance of all these scales
in the dynamics of heavy quarkonium makes its quantitative study extremely
difficult. This is even so in the weak-coupling regime, where the system
becomes Coulombic. Nevertheless, by exploiting the hierarchies $m \gg p \gg E$
and $m \gg \lQ$ the problem can be considerably simplified. This may be done
in any particular calculation for a given observable, or, alternatively, using
EFTs. In the latter, the hierarchies of scales are
exploited at the action level producing {\it universal} results independent of
particular observables, which is far more advantageous. The basic idea behind
EFTs is that to describe observables of a particular (low) energy region, one
can integrate out the degrees of freedom of the other regions.  This produces
an effective action (for the EFT) involving only the degrees of freedom in the
region we are interested in.  Calculating with the effective (EFT) or with the
fundamental (QCD) action gives equivalent physical results as far as that
particular region is concerned, but calculations are much simpler with the
EFT.  In heavy quarkonium, we are interested in physics at the low energy
scale $E$.  Hence EFTs, which have energy scales larger than $E$ integrated
out, can be and have been built. They have led to a major progress in our
understanding of heavy quarkonium in recent years. We will devote this review
to these new developments.  Before that, let us put this progress in a
historical perspective.

The discovery of bottomonium and charmonium triggered the use of NR potential
models (where the physics of the bound state is described by a Schr\"odinger
equation). The main input in this approach is the potential introduced. At
lowest order in the weak-coupling regime ($|{\bf p}| \gg \lQ$), the potential
is Coulombic.  Higher-order corrections to the potential in perturbation
theory were obtained over the years
\cite{Gupta:1981pd,Gupta:1982qc,Buchmuller:1981aj,Pantaleone:1985uf,Titard:1993nn},
even though the computations were difficult due to the several scales
involved. It was also not clear how to systematically incorporate US
effects (for instance, let us mention the infrared sensitivity found in the
static potential \cite{Appelquist:1978es} or in the one-loop calculations of
$P$-wave decays \cite{Barbieri:1980yp}).  In any case, the observed
bottomonium and charmonium spectra turned out not to be Coulombic and
phenomenologically fine-tuned potentials were necessary to reproduce them
(\cite{Eichten:1978tg}, see \cite{Brambilla:1999ja} for more references).
This motivated attempts to derive the heavy-quarkonium potential from QCD
without relying on perturbation theory.  The idea was to find gauge invariant
expressions for the potentials (within an expansion in $1/m$) in terms of the
expectation values of Wilson loops. Several methods have been worked out over
the years and expressions for the spin-dependent and -independent potentials
up to ${\cal O}(1/m^2)$ were obtained
\cite{Wilson:1974sk,Susskind:1976pi,Brown:1979ya,
Eichten:1981mw,Peskin:1983up,Gromes:1984ma,Barchielli:1988zs,Barchielli:1990zp,
Szczepaniak:1997tk}. All the obtained potentials have been investigated on the
lattice (see \cite{Bali:2000gf} for a recent review).  However, these results
had a number of shortcomings.  
\cite{Lucha:1991vn} pointed out that, if
calculated in perturbation theory, the potentials obtained from the Wilson
loop approach missed the hard logarithms $\sim \ln m$ present in the
potentials directly computed from QCD. More recently, \cite{Brambilla:2000gk} also
pointed out that not only hard logarithms, but some of the potentials, 
relevant at relative order $\als^2$ in the spectrum, were missed as well.
Finally, the IR divergences in the perturbative
computation of $P$-wave decays seemed impossible to accommodate in that
framework. Overall, a more systematic and controlled derivation of the
NR dynamics from QCD was required.

Independently of the line above, NRQED, an EFT for NR
leptons, was introduced \cite{Caswell:1986ui}. It turned out to provide the
first and decisive link in the chain of developments that we will review
here. NRQED is obtained from QED by integrating out the hard scale $m$. It is
characterized by an UV cut-off much smaller than the mass $m$
and much larger than any other scale. NRQCD, which also appears in the title
of \cite{Caswell:1986ui}, was born soon afterwards \cite{Lepage:1987gg}.  The
Lagrangian of NRQCD can be organized in powers of $1/m$, thus making explicit
its NR nature. To each power in $1/m$, a set of operators (which encode the
low-energy content of the theory) and matching coefficients (which encode the
effects due to degrees of freedom with energy of ${\cal O}(m)$ that have been
integrated out from QCD) are associated.  Namely, in NRQCD the contributions
coming from the hard scale $m$ are factorized.  NRQCD had two major virtues
that we would like to pont out here: (i) it could be rigorously derived from QCD
in a systematic manner (providing an optimized framework for lattice
simulations \cite{Thacker:1991bm}) and (ii) it solved the problem of the IR
divergences of the $P$-wave decays of heavy quarkonium.  This solution,
however, came to the price of introducing the so-called color-octet matrix
elements, which could not be incorporated in the Schr\"odinger-like
formulations available at that time.  In spite of this, it was noted in
\cite{Chen:1995dg} that if the non-perturbative potentials were calculated
starting from NRQCD instead of from QCD, the problem of the missing hard logarithms
mentioned above disappeared\footnote{These are included in the matching
coefficients of the theory and may be transferred to the potentials by
expanding Green functions in NRQCD instead of in QCD
\cite{Chen:1995dg,Bali:1997am,Brambilla:1998vm}.}.  This raised again some
hope that NR potential models could eventually be regarded as EFTs of QCD. It
also made it evident that the potential models available, even those in which the
potentials were obtained in terms of Wilson loops, were not controlled
derivations from QCD and that first-principles computations of heavy quarkonia
should better be done within the framework of NRQCD.

NRQCD itself was, however, not free of shortcomings.  The main problem was
related to the fact that both soft and US degrees of freedom were
entangled. This had effects on (i) the power counting rules, which were not
homogeneous (the power counting by \cite{Lepage:1992tx}, which assumed that
$\lQ \siml mv^2$, catches the leading order contribution of the matrix
elements but there are also subleading contributions in $v$); and on (ii) the
perturbative calculations, which were dependent on two scales and, therefore,
still difficult to compute.  Another problem was that the first computations
in NRQCD were based on cut-off regularization\footnote{In any case, the
simplifications compared with purely relativistic Bethe-Salpeter-like
\cite{Bethe:1951aa} computations were enormous and led to a plethora of new
results in QED, see for instance \cite{Kinoshita:1995mt,
Labelle:1997uw,Hoang:1997ki,Hill:2000qi,Hill:2000zy}.}, 
whereas the calculations in QCD are often done in
DR. Attempts to perform the matching between QCD
and NRQCD using DR had the drawback that the naive incorporation of the
kinetic term in the quark propagator jeopardized the power counting rules.
 
A solution to the last problem was first proposed by
\cite{Manohar:1997qy}. There, it was argued that the matching between QCD and
NRQCD in the bilinear sector of the theory in DR should be performed just as
in HQET (see \cite{Neubert:1994mb} for a
review), namely treating the kinetic-energy term as a perturbation.  Along the
same lines, the matching of QCD to NRQCD in the 4-fermion sector, where the
Coulomb pole enhancement starts playing a role, was performed soon after by
\cite{Pineda:1998bj,Pineda:1998kj}.  The key point was that, in order to carry
out the matching, it is not so important to know the power counting of each
term in the effective theory, but to know that the remaining dynamical scales
of the effective theory are much lower than the mass: $m \gg |{\bf p}|,\; E,\;
\lQ$.

Coming back to the main problem, the first works addressing the entanglement
of the soft and US scales in NRQCD tried to classify the different momentum
regions existing in a purely perturbative version of NRQCD and/or to reformulate
NRQCD in such a way that some of these regions were explicitly displayed by
introducing new fields in the NRQCD Lagrangian.  In particular, we mention
\cite{Labelle:1998en} where a diagrammatic approach to NRQED was used and the
subsequent work by \cite{Luke:1997hj,Grinstein:1998gv,Luke:1998ys} in
NRQCD. All these early attempts turned out to be missing some relevant
intermediate degrees of freedom.

The first complete solution came in \cite{Pineda:1998bj}.  The idea was to
build an EFT containing only the degrees of freedom relevant for ${\bar
Q}$--$Q$ systems near threshold, i.e. those with $E \sim mv^2$, and as close
as possible to a Schr\"odinger-like formulation (see also
\cite{Lepage:1997cs}). All other degrees of freedom were to be integrated
out. The EFT, which was called pNRQCD had, roughly, the
following structure:
\begin{eqnarray*}
\,\left.
\begin{array}{ll}
{\cal L} = \Phi({\bf r})^\dagger &
\displaystyle{ \Bigg(i\partial_0-{{\bf p}^2 \over 2m}-V^{(0)}(r) } 
\\
& \displaystyle{~~+ \hbox{\rm  corrections to the potential}}
\\
& \displaystyle{~~+\hbox{interactions with other low-energy degrees of freedom}} \Bigg) 
\Phi({\bf r}) 
\end{array} \right\}
{\rm pNRQCD}
\end{eqnarray*}
where $V^{(0)}(r)$ is the static potential ($-C_F\als/r$ in the perturbative
case) and $\Phi({\bf r})$ is the field associated with the ${\bar Q}$--$Q$
state.  This EFT turned out to meet all our expectations: it achieved the
factorization between US and higher energy modes, had a definite power
counting (at least in the perturbative regime), and was very close to a NR
Schr\"odinger-like formulation of the heavy quarkonium dynamics. In the
Lagrangian, there appear potentials.  These are the matching coefficients of
the theory and are calculated by matching with NRQCD amplitudes, either using
Feynman diagrams (see
\cite{Pineda:1998kn,Czarnecki:1999mw,Beneke:1999qg,Kniehl:2001ju} for specific
examples in QCD and QED and sec.~\ref{pNRmatchingI} for further details) or
Wilson-loop amplitudes \cite{Brambilla:1999qa, Brambilla:1999xf}.  In the
perturbative regime, a confirmation that pNRQCD was able to catch all the
relevant dynamical regions came from diagrammatic studies.
\cite{Beneke:1998zp} made the most complete classification of (perturbative)
momentum regions to date by a diagrammatic study called the {\it threshold
expansion}. In the language of the {\it threshold expansion}, the matching
between NRQCD and QCD corresponds to integrating out the hard region and pNRQCD
is obtained from NRQCD by integrating out what are called soft quarks and
gluons and potential gluons.  Finally, we mention two later works, which dealt
with reformulating NRQCD within an effective Lagrangian formalism.  In
\cite{Griesshammer:1998wz} all degrees of freedom of NRQCD were made explicit
in the Lagrangian. In \cite{Luke:1999kz} the question on how to obtain 
RG equations for NR systems was addressed
for the first time. The resulting formalism is now known as vNRQCD (for
a review on this theory see \cite{Hoang:2002ae}). All these formulations should 
be equivalent to pNRQCD, once the same degrees of freedom have been integrated out.

This closed the circle, connecting QCD with a properly modified
Schr\"odinger-like formulation in the weak-coupling regime.  Compared with the
traditional methods, perturbative computations are optimized since only one
scale appears in each of the Feynman integrals. The interaction with US gluons
is treated in a quantum field theory fashion but yet everything can be encoded
in a Schr\"odinger-like formulation. The applications of these ideas to QED
have also been very successful.  We refer to sec.~\ref{pNRweakObservables} for
references.

The natural question then was: what happens in the strong-coupling regime?
The application of EFTs has also led to a well-founded connection with QCD in
this regime. The potentials are now understood as matching coefficients to be
obtained by comparison with NRQCD.  This (along with new computational
techniques) has solved the problems mentioned before allowing for the complete
computation of the potential at ${\cal O}(1/m^2)$
\cite{Brambilla:2000gk,Pineda:2000sz} (as well as identifying new non-analytic
terms in the $1/m$ expansion \cite{Brambilla:2003mu}), and the solution of the
IR sensitivity of the $P$-wave decays in terms of singlet fields and
potentials only \cite{Brambilla:2001xy}.  Again, the use of EFTs has allowed
to close the circle and connect QCD with a properly modified
Schr\"odinger-like formulation in the strong-coupling regime.

Heavy quarkonium lives nowadays in a new golden age. In the early seventies, its
high-energy nature helped to establish asymptotic freedom and QCD as the
fundamental theory of the strong interaction. Later on, its NR nature served
as a playground for many models of the low-energy dynamics of QCD. Since the
nineties, due to the rise of EFTs for heavy quarks, heavy quarkonium observables
can be rigorously derived from QCD, low and high energy modes factorized,
large logarithms systematically resummed. From a conceptual point of view, the origin
and the exact meaning of a QCD Schr\"odinger-like equation has been
clarified. In the weak-coupling regime, this opens up the possibility to have
precision determinations of the Standard Model parameters to which heavy
quarkonium is sensitive: $\als$ and the heavy quark masses.  In the strong-coupling 
regime, heavy quarkonia are, thanks to their wealth of scales, an
ideal laboratory in which to probe the structure of the QCD vacuum.

It is our aim to review here the recent developments in heavy quarkonium
physics mentioned above from the point of view of EFTs. Our main goal will be
to derive the QCD Schr\"odinger-like equation that governs heavy
quarkonium physics in the weak and strong coupling regime. We will not be
exhaustive in most of the derivations but concentrate on the main ideas and
general lines of development with spotted examples to illustrate the
procedure. Then we will discuss a selected set of applications.  The review is
not exhaustive.  In particular, we will not discuss one of the major
phenomenological successes of NRQCD: to provide an explanation of the
heavy quarkonium production rate at the Tevatron
\cite{Braaten:1996pv,Beneke:1997av,Kramer:2001hh,Bodwin:2003kc}.

\medskip

Before moving to the main body of the review, we list here our main notational
choices \cite{Yndurain:1999ui}. The QCD Lagrangian density reads 
\be
{\cal L} = 
\sum_{i=1}^{N_f} \bar q_i (i \dsl  - m_i) q_i  
-\frac{1}{4}G^{\mu\nu \, a}G_{\mu \nu}^a, 
\ee
where $D_{\mu}=\partial_{\mu} +igA_{\mu}$, $i\,g\,G_{\mu\nu} = [D_\mu,D_\nu]$,
$q_i$ are the quark fields and $m_i$ their current masses. $N_f$ is the total number of quark flavors. 
In the review, we will often indicate with the capital letter, $Q_i$, the heavy quark fields and
always set to zero the light quark masses.  In the EFT, the heavy quark masses
will be also indicated by $m_i$, but always understood, if not differently
specified, as pole masses.  The strong-coupling constant, $\als=g^2/(4\pi)$,  in the presence of
$n_f$ light quarks runs, at energies below the heavy quark thresholds, as 
\be 
\nu {d \als \over d
\nu}=-2\als\left\{\beta_0{\als \over 4 \pi}+\beta_1\left({\als \over 4
\pi}\right)^2 + \cdots\right\}, 
\ee 
where
$$
  \beta_0 = \frac{11}{3}C_A-\frac{4}{3} T_Fn_f\,, \qquad
  \beta_1 = \frac{34}{3}C_A^2-\frac{20}{3} C_A T_Fn_f - 4\,C_FT_Fn_f,
  ~~~~~~~\dots,  
$$
and $C_A = N_c =3$, $C_F = (N_c^2-1)/(2\,N_c) = 4/3$ and $T_F =1/2$.

The basic computational techniques for perturbative QCD used through the review can be found in \cite{Pascual:1984zb}.

\section{NRQCD}
\label{sec:NRQCD}

\subsection{Degrees of freedom}
NRQCD is designed to describe the dynamics of a heavy quark and a heavy
antiquark (not necessarily of the same flavor) at energy scales (in the center
of mass frame) much smaller than their masses, which are assumed to be much
larger than $\lQ$, the typical hadronic scale. At these energies, further heavy 
quark-antiquark pairs cannot be created so it is sufficient, and convenient, to
use Pauli spinors for both the heavy quark and the heavy anti-quark degrees of
freedom.  We shall denote by $\psi (x)$ the Pauli spinor field that annihilates a
quark and by $\chi (x)$ the one that creates an antiquark. Both $\psi (x)$ and
$\chi (x)$ transform in the fundamental representation of color SU(3). The
remaining (light) degrees of freedom are the same as in QCD, except for the UV
cut-offs as we shall discuss below. In particular, the gluon fields will appear 
in covariant derivatives $D_{\mu}$ and field strengths $G_{\mu\nu}$. For instance, we shall see 
that the leading-order Lagrangian density for the heavy quark and antiquark fields reads
\be
{\cal L}
=
\psi^{\dagger} \left( i D_0 + {1\over 2 m} {\bf D}^2 \right)\psi  + 
\chi^{\dagger} \left( i D_0 - {1\over 2 m} {\bf D}^2 \right)\chi \, .
\ee
In a NR frame, the energy and three-momentum of the heavy particles scale in a
different way and hence a different UV cut-off may be introduced for each: 
$\nu_s$ and $\nu_p$ respectively.
However, NRQCD is usually considered as having a single UV
cut-off $\nu_{NR}=\{\nu_p,\nu_s\}$ satisfying $E, p, \lQ \ll \nu_{NR} \ll m$;
$\nu_p$ is the UV cut-off of the relative three-momentum of the heavy quark
and antiquark; $\nu_s$ is the UV cut-off of the energy of the heavy quark and
the heavy antiquark, and of the four-momentum of the gluons and light quarks.

From a Wilson RG point of view, NRQCD is obtained from QCD by integrating out
energy fluctuations about the heavy quark (heavy antiquark) mass and
three-momentum fluctuations up to the scale $\nu_{NR}$ for the heavy quark
(heavy antiquark) fields, and four-momentum fluctuations up to the same scale
for the fields of the light degrees of freedom. Since $\nu_{NR} \gg \lQ$, this
can be carried out in practice perturbatively in $\als (\nu_{NR})$.  Within
the {\it threshold expansion} framework \cite{Beneke:1998zp}, this corresponds
to integrating out the {\it hard} modes of QCD.

If the quark and antiquark have the same flavor, they can annihilate into hard
gluons, which have already been integrated out and are not present in the
NRQCD Lagrangian.  This implies that, in this case, the QCD Lagrangian must,
and will, contain imaginary Wilson coefficients. The non-Hermiticity of the
NRQCD Lagrangian, which at first sight may appear rather unpleasant, if not
disastrous, turns out to provide an extremely powerful tool for calculating
inclusive decay widths to light particles.

\subsection{Power counting}
\label{sec:powerNRQCD}
From the discussion above, it follows that the NRQCD Lagrangian can be
organized as a power series in $1/m_Q$ (and $1/m_{\bar Q}$).  
The Wilson (matching) coefficients of each operator depend logarithmically on
$m_Q$  ($m_{\bar Q}$), $\nu_{NR}$ and, as mentioned before, can be calculated
in perturbation theory in $\als (\nu_{NR})$. Hence the importance of a given
operator for a practical calculation not only depends on its size (power
counting), which we will briefly discuss next, but also on the leading power
of $\als$ that its matching coefficient has.

Since several scales ($E$, $|{\bf p}|$, $\lQ$)  remain dynamical, it is not possible to
assign a size to each operator unambiguously without extra
assumptions: no homogeneous power counting exists. As we will see below, the introduction of pNRQCD facilitates this
task.  The original power  counting introduced by \cite{Bodwin:1994jh} assumes
$\lQ\sim E \sim mv^2$, and  hence $|{\bf p}| \sim mv\gg \lQ$, $v\sim \als (mv)\ll
1$. 
We will see that this implies that
the bound state is Coulombic (positronium-like). In this case homogeneous
power counting rules can be given using  pNRQCD in the weak-coupling regime
(ch.~\ref{sec:pNRQCDII}). Nevertheless, it is  unlikely that the whole heavy
quarkonium spectrum can be described by this power counting and 
alternatives need to be explored. 
We only anticipate here that in the strong-coupling 
regime of pNRQCD the following scaling will be considered: $E\ll |{\bf p}| \sim \lQ$. 
The issue of the power counting of NRQCD has also been addressed  
by \cite{Beneke:1997av} and \cite{Fleming:2000ib} (see also the discussion in sec.~\ref{sec:NRapplications}). 
In both cases, the authors allow for some freedom in the possible size of the NRQCD matrix
elements by introducing a parameter $\lambda$ that interpolates
between different power countings.

\subsection{Lagrangian, currents and symmetries}
\label{sec:NRLag}
The allowed
operators in the Lagrangian are constrained by the symmetries of
QCD. However, due to the particular kinematic region on which we are focusing on, Lorentz
invariance is not linearly realized in the heavy quark sector, and it is not
straightforward (though certainly possible, as will be discussed below) to
implement. One has, in a first stage, to content oneself with implementing the
rotational subgroup only. Including $n_f$ light quarks, the NRQCD Lagrangian density for a quark of mass $m_1$ 
and an antiquark of mass $m_2$ ($m_1\; ,\; m_2 \gg \lQ$) reads at ${\cal O}(1/m^2)$ 
\footnote{
We also include the ${\bf D}^4/(8\, m^3)$ terms since they will be necessary
later on.}, $m=m_1 , m_2$ \cite{Caswell:1986ui,Bodwin:1994jh,Manohar:1997qy,Bauer:1998gs}:
\bea
&& 
{\cal L}_{\rm NRQCD}={\cal L}_g+{\cal L}_l+{\cal L}_{\psi}+{\cal L}_{\chi}+{\cal L}_{\psi\chi},
\label{LagNRQCD}
\\
\nn
\\
&&
{\cal L}_g=-\frac{1}{4}G^{\mu\nu \, a}G_{\mu \nu}^a +
{1\over 4}
\left({c_1^{g\,(1)} \over m_1^2} + {c_1^{g\,(2)} \over m_2^2} \right)
g f_{abc} G_{\mu\nu}^a G^{\mu \, b}{}_\alpha G^{\nu\alpha\, c},
\label{Lg}
\\
\nn
\\
&&
{\cal L}_l = \sum_{i=1}^{n_f} \bar q_i \, i \dsl \, q_i 
+ {1\over 8}\left( {c_1^{ll\,(1)}\over m_1^2} + {c_1^{ll\,(2)} \over m_2^2}\right) \, g^2 \,
\sum_{i,j =1}^{n_f} \bar{q_i} T^a \gamma^\mu q_i \ \bar{q}_j T^a \gamma_\mu q_j  
\nn
\\
&& \qquad\qquad
+ {1\over 8}\left( {c_2^{ll\,(1)}\over m_1^2} + {c_2^{ll\,(2)} \over m_2^2}\right) \, g^2 \,
\sum_{i,j=1}^{n_f}\bar{q_i} T^a \gamma^\mu \gamma_5 q_i \ \bar{q}_j T^a \gamma_\mu \gamma_5 q_j 
\nn
\\ 
&& \qquad\qquad
+ {1\over 8}\left( {c_3^{ll\,(1)}\over m_1^2} + {c_3^{ll\,(2)} \over m_2^2}\right) \, g^2 \,
\sum_{i,j=1}^{n_f} \bar{q_i}  \gamma^\mu q_i \ \bar{q}_j \gamma_\mu q_j 
\nn
\\
&& \qquad\qquad
+ {1\over 8}\left( {c_4^{ll\,(1)}\over m_1^2} + {c_4^{ll\,(2)} \over m_2^2}\right) \, g^2 \,
\sum_{i,j=1}^{n_f}\bar{q_i} \gamma^\mu \gamma_5  q_i \ \bar{q}_j \gamma_\mu \gamma_5 q_j,
\label{Ll}
\\
\nn
\\
&&
{\cal L}_{\psi}=
\psi^{\dagger} \Biggl\{ i D_0 + {c_k^{(1)}\over 2 m_1} {\bf D}^2 + {c_4^{(1)} \over 8 m_1^3} {\bf D}^4 
+ {c_F^{(1)} \over 2 m_1} {\bfsigma \cdot g{\bf B}} 
\nn
\\
&& \qquad\qquad
+ { c_D^{(1)} \over 8 m_1^2} \left({\bf D} \cdot g{\bf E} - g{\bf E} \cdot {\bf D} \right) 
+ i \, { c_S^{(1)} \over 8 m_1^2} 
{\bfsigma \cdot \left({\bf D} \times g{\bf E} -g{\bf E} \times {\bf D}\right) }
\Biggr\} \psi
\nn
\\
&& \qquad\qquad
+{c_1^{hl\,(1)} \over 8m^2_1}\, g^2 \,\sum_{i=1}^{n_f}\psi^{\dagger} T^a \psi \ \bar{q}_i\gamma_0 T^a q_i 
+{c_2^{hl\,(1)} \over 8m^2_1}\, g^2 \,\sum_{i=1}^{n_f}\psi^{\dagger}\gamma^\mu\gamma_5
T^a \psi \ \bar{q}_i\gamma_\mu\gamma_5 T^a q_i 
\nn
\\
&& \qquad\qquad
+{c_3^{hl\,(1)}\over 8m^2_1}\, g^2 \,\sum_{i=1}^{n_f}\psi^{\dagger} \psi \ \bar{q}_i\gamma_0 q_i
+{c_4^{hl\,(1)}\over 8m^2_1}\, g^2 \,\sum_{i=1}^{n_f}\psi^{\dagger}\gamma^\mu\gamma_5
\psi \ \bar{q}_i\gamma_\mu\gamma_5 q_i,
\label{Lhl}
\\
\nn
\\
&& {\cal L}_{\chi} = \hbox{c.c. of ${\cal L}_{\psi}(1 \leftrightarrow 2)$},
\nn
\\
\nn
\\
&&
{\cal L}_{\psi\chi} =
  {f_1(^1S_0) \over m_1 m_2}O_1(^1S_0) 
+ {f_1(^3S_1) \over m_1 m_2}O_1(^3S_1) 
+ {f_8(^1S_0) \over m_1 m_2} O_8(^1S_0)
+ {f_8(^3S_1) \over m_1 m_2} O_8(^3S_1),
\label{Lhh}
\eea
where
\bea 
O_1(^1S_0)&=\psi^{\dag} \chi \, \chi^{\dag} \psi &, 
\quad O_1(^3S_1)=\psi^{\dag} {\bfsigma} \chi \, \chi^{\dag} {\bfsigma} \psi , 
\\
O_8(^1S_0)&=\psi^{\dag} {\rm T}^a \chi \, \chi^{\dag} {\rm T}^a \psi &, 
\quad O_8(^3S_1)= \psi^{\dag} {\rm T}^a {\bfsigma} \chi \, \chi^{\dag} {\rm T}^a {\bfsigma} \psi .
\label{def4fops}
\eea
The matching coefficients are symmetric under the exchange 
$m_1 \leftrightarrow m_2$, $\bfsigma$ are the Pauli matrices,
$i D^0=i\partial^0 -gA^0$, $i{\bf D}=i\bfnabla+g{\bf A}$,
${\bf E}^i = G^{i0}$, ${\bf B}^i = -\epsilon_{ijk}G^{jk}/2$, $\epsilon_{ijk}$ being 
the usual three-dimensional antisymmetric tensor\footnote{
In DR several prescriptions are possible for 
the $\epsilon_{ijk}$ tensors and $\bfsigma$.
Therefore, if DR is used, one has to make sure that 
one uses the same prescription as that one used to calculate the matching coefficients.}
($({\bf a} \times {\bf b})^i \equiv \epsilon_{ijk} {\bf a}^j {\bf b}^k$)
with $\epsilon_{123}=1$, and c.c. stands for charge conjugate 
($\psi^c=-i\sigma^2\chi^\ast$, $\chi^c=i\sigma^2\psi^\ast$ and $A_\mu^c=-A_\mu^T$).

The NRQCD Lagrangian is defined up to field redefinitions. In the expression 
adopted here, we have made use of this freedom. Powers larger than one of $iD_0$ 
applied to the quark fields have been eliminated. We have also redefined the 
gluon fields in such a way that the coefficient in front of $- G^{\mu\nu \, a}G_{\mu \nu}^a/4$ 
in  ${\cal L}_g$ is one. This turns out to be equivalent to redefining the coupling constant 
in such a way that it runs with $N_f - 2 = n_f$ flavors (for $m_1 \neq m_2$,  
$N_f - 1 = n_f$ for $m_1 = m_2$), where $N_f$ are the flavors in QCD 
\cite{Pineda:1998kj} (see also \cite{Griesshammer:1998fh} for a calculation of the $\beta$ 
function in NRQCD). A possible term $D^\mu\, G_{\mu \alpha}^a\, D_\nu\, G^{\nu \alpha \, a}$ has been 
eliminated through the identity \cite{Manohar:1997qy}: 
\be
\displaystyle \int d^4x \, \left(
2\, D^\mu\, G_{\mu \alpha}^a\, D_\nu\, G^{\nu \alpha \, a} 
+ 2\ g \, f_{abc} \, G_{\mu\nu}^a G^{\mu \, b}{}_\alpha G^{\nu\alpha\, c}
+ G_{\mu \nu}^a\, D^2\, G^{\mu \nu \, a}\right) = 0 .
\ee 
Finally, a possible term like $c\, G_{\mu\nu}^a D^2 G^{\mu\nu \, a}$
has been eliminated through the field redefinition 
$A_\mu \to A_\mu + 2 \, c \, [D^\alpha,G_{\alpha\mu}]$ \cite{Pineda:2000sz}. 

The Wilson coefficients appearing in the NRQCD Lagrangian will be discussed 
in sec.~\ref{NRmatching}. The ${\cal O}(1/m^3)$ Lagrangian (without 
the light-fermion sector) can be found in \cite{Manohar:1997qy}.
The Feynman rules associated to the first two lines of
Eq.~(\ref{Lhl}) can be found in \cite{Bodwin:1998mn}.

NR currents should also be considered, since they appear 
in inclusive (electromagnetic) decays, NR sum rules or 
$t$-$\bar t$ production near threshold. Similarly to the Lagrangian, they can 
be written as an expansion in $1/m$ times some hard matching coefficients 
times some NR (local) operators. For instance, the electromagnetic
vector and axial-vector currents read (see \cite{Hoang:1999zc})
\bea
\label{NRcurrent}
 j_k^v(x) &=& b^v_{1,\rm NR}
\,\Big({ \psi}^\dagger \sigma_k 
\chi\Big)(x) -
\frac{b^v_{2,\rm NR} }{6 m^2}\,\Big({ \psi}^\dagger \sigma_k
(\mbox{$-\frac{i}{2}$} 
\stackrel{\leftrightarrow}{\mbox{\boldmath $D$}})^2
  \chi\Big)(x) + \ldots
\,, 
\\
j_k^a(x)  &=& \frac{b^a_{1,\rm NR}
}{m}\,\Big({ \psi}^\dagger 
(\mbox{$-\frac{i}{2}$} 
\stackrel{\leftrightarrow}{\mbox{\boldmath $D$}}\!\!
       \times{\mbox{\boldmath $\sigma$}})_k\, 
 \chi\Big)(x) + \ldots\,, 
\label{NRcurrenta}
\eea
where $\psi^{\dagger} \stackrel{\leftrightarrow}{\mbox{\boldmath $D$}} \chi 
\equiv \psi^{\dagger} ({\bf D} \chi)-({\bf D} \psi)^{\dagger}\chi$ and the dots stand for 
corrections, which do not contribute at NNLO order for $S$-waves. 
In practice, most of the physical information can be extracted from the 
imaginary parts of the 4-fermion operators, which are discussed in 
sec.~\ref{NRdecay}. In particular, the matching coefficients $b^v_{1,\rm NR}$ 
and $b^v_{2, \rm NR}$ can be traded for the matching coefficients 
Im $f_{\rm EM}({}^3S_1)$  and Im $g_{\rm EM}({}^3S_1)$ respectively.

\subsubsection{Poincar\'e/reparametrization  invariance}
The QCD Lagrangian is invariant under Lorentz boosts. 
However, the NR expansion has destroyed the manifest invariance of the EFT under 
Lorentz boosts. Since the EFT is equivalent to QCD 
at any order of the strong-coupling and NR expansion, 
the invariance under Lorentz boosts is not lost, but must be somehow  
incorporated in the EFT. Indeed, it imposes specific constraints on the form of the EFT itself.

Constraints imposed by the relativistic invariance have been first worked out for 
HQET, which coincides with NRQCD in the 
bilinear sector of the heavy-quark fields \cite{Luke:1992cs,Manohar:1997qy}.
In HQET the realization of the relativistic 
invariance is called reparametrization invariance.
It imposes constraints on the Wilson coefficients of the EFT. For instance:
\be
c_k = c_4  = 1, 
\qquad
2 \, c_F  - c_S - 1 = 0, 
\qquad
\label{poinv2}
\ee
where we have dropped the explicit indication of the flavor index.

An alternative derivation consists of imposing the Poincar\'e algebra 
on the generators $H$, ${\bf P}$, ${\bf J}$ and ${\bf K}$ 
of time translations, space translations, 
rotations, and Lorentz-boosts transformations of NRQCD \cite{Brambilla:2003nt}.
The idea originates from \cite{Dirac:1949cp}, and has been used to constrain the form of 
the relativistic corrections to phenomenological potentials in 
\cite{Foldy:1960nb,Krajcik:1974aa,Sebastian:1979rr}. It was applied to
NR EFTs in \cite{Brambilla:2003nt,Vairo:2003gx}.
In a field theory, the Poincar\'e  algebra has to be understood among fields 
quantized in accordance with the canonical equal-time commutation relations.\footnote{
More precisely, the algebra imposes relations among the bare fields and coupling constants.
These relations are preserved in the renormalized theory if Poincar\'e invariance 
is not broken by quantum effects.}
The translation and rotation generators ${\bf P}$ and ${\bf J}$ may be derived from the NRQCD Lagrangian 
or by matching to the QCD generators. They are exact, because translational and 
rotational invariance have not been explicitly broken in going to the EFT. 
The Lorentz-boost generators may be obtained by matching order by order in $1/m$ to 
the Lorentz-boost generators of QCD. They depend on some specific matching coefficients 
independent of those in the Lagrangian. 
The  NRQCD  Poincar\'e generators satisfy the Poincar\'e algebra 
if Eq.~(\ref{poinv2}) is satisfied for each flavor 
up to ${\cal O}(1/m)$ and ${\cal O}(\bfnabla^2 \,\bfnabla /m^2)$ 
(plus some other constraints 
on the matching coefficients appearing in ${\bf K}$)\cite{Brambilla:2003nt}.
Therefore, at the considered order, we get the same result as from reparametrization 
invariance. The calculation of constraints specific to NRQCD, i.e. involving 
4-fermion operators, has not been done in either approach yet.  This would correspond to going to higher 
orders in $1/m$.

In general, we will constrain the matching coefficients of the kinetic 
energy in accordance with Eq.~(\ref{poinv2}). Occasionally, however, we will keep 
them explicit for tracking purposes.

\subsection{Matching}
\label{NRmatching}
The calculation of the Wilson coefficients of NRQCD is done
through a  procedure called {\it matching}. In a matching calculation suitable
renormalized QCD and renormalized NRQCD Green's functions (or matrix elements) 
are imposed to be equal for scales below $\nu_{NR}$ at the desired
order of $\als$ and  $1/m$. In particular, the  expansion of Green's
functions in external energies $E$ and three-momenta $p$ must be equal. This fixes
the matching coefficients, which will depend on the renormalization schemes
used in QCD and in NRQCD.  It extraordinarily simplifies calculations if
these expansions are done {\it before} the loop integrals are performed. However, doing
so may introduce IR divergences and for the  equality between QCD and NRQCD Green
functions to remain valid  the same IR regulator must be used in
both theories.  It is very convenient to use DR as an IR regulator as
well as an UV one. This is so because all loop integrals in the NRQCD
calculations will be scaleless and can be set to zero, as we will argue
below. Let us advance what will happen. Schematically \cite{Manohar:1997qy},
one has
\be
A_{eff}\left( {1 \over \epsilon_{UV}} - {1 \over \epsilon_{IR}} \right)
\ee
in the EFT, which is zero if $\epsilon_{UV}=\epsilon_{IR}$
in DR.
Therefore, we only have to calculate loop integrals in QCD that depend on a
single scale ($m$). Typically we get
\be
A {1 \over \epsilon_{UV}} + B {1 \over \epsilon_{IR}}  + \left( A+B
\right) \ln {\nu \over m} + D
\,.
\ee
Since the full and the effective theory share the same IR behavior
$B=-A_{eff}$. Moreover the UV divergences are absorbed in the
coefficients of the full and effective theory. In this way the difference
between the full and the effective theory reads
\be
\left( A+B \right) \ln{\nu\over m}+D
\,,
\ee
which provides the one-loop contribution to the matching coefficients for the effective theory.
It is implicitly in this procedure that the same renormalization scheme is used
for  both UV and IR divergences in NRQCD. In the QCD calculation both the UV and IR divergences can also be
renormalized in the same way, for instance using the  $\MS$ scheme, which
is the standard one for QCD calculations. This fixes the UV
renormalization scheme in NRQCD in which the Wilson coefficients have
been calculated. This means that for these Wilson coefficients to be
consistently used in a NRQCD  calculation, this calculation must be
carried out in the same scheme, for instance in DR and in the $\MS$ scheme.

The matching calculation can be carried out in any gauge, since both
the QCD and NRQCD Lagrangians are manifestly gauge
invariant. However, since most of the times one is matching
gauge-dependent Green functions, the same gauge must be chosen in  QCD
and NRQCD. Using different gauges or, in general, different ways to
carry out the matching procedure, may lead to apparently  different
results for the matching coefficients (within the same regularization
and renormalization scheme). These results must be  related by local
field redefinitions, or, in other words, if both matching calculations
had agreed to use the same {\it minimal} basis  of operators
beforehand, the results would have coincided. If the matching is
carried out as described above, it is more convenient  to choose a
covariant gauge (i.e. Feynman gauge), since only QCD calculations,
which are manifestly covariant, are to be carried out.

In the procedure described above, one may be worried about the fact
that the NR  propagator $1/(p^0 -{\bf p}^2/m)$ contains
the scale $m$, which spoils the usual argument (used in HQET for instance)
that loop integrals in the EFT contain no scales once
one has expanded in the external energies and three-momenta. 
Let us argue in the following paragraphs that the procedure is indeed correct.

Consider first the single quark (antiquark) sector. In any diagram in
NRQCD, one can always choose the momenta flowing along the heavy quark
(antiquark) line in the same direction. Then all heavy quark
propagators will have poles in either the lower or the upper half of the complex
plane  only. Then, if all integrals over the energies flowing through
the heavy quark propagators are carried out by closing the contour
around  the opposite half-complex plane, these energies will be
substituted by linear dependencies in the three-momenta in the
NR  quark (antiquark) propagators. These linear
dependencies dominate over quadratic dependencies of the kinetic terms
both in the  IR and in the UV. The latter is so because
$\nu_{NR}$ is always smaller than $m$. Hence the kinetic term can be
expanded and  the integrals become dimensionless. In fact, in DR the
kinetic term not only can but {\it must} be expanded, since this is
the only way to implement that three-momenta must remain smaller
than $\nu_{NR}$.

Consider next the quark-antiquark sector. Any fixed-order loop
calculation may contain  heavy quark-antiquark irreducible diagrams
(meaning diagrams  which cannot be disconnected by cutting an internal
quark line and an internal antiquark line) and heavy quark-antiquark
reducible ones.

Consider first a quark-antiquark irreducible diagram. The fact that at
any  point of an internal quark propagator there is always  at least
one gluon propagator (or two light quark propagators) in addition to an
antiquark propagator allows to choose all momenta  flowing both along
the quark and along the antiquark propagator in the same
direction. Hence the poles of both the quark and  antiquark
propagators are in the same complex half-plane, and therefore the
argument put forward for the single-quark sector also holds here.

Consider finally a quark-antiquark reducible diagram. It can always be
written as a series of  2PI diagrams linked by a quark and by an
antiquark  propagator. Let us choose the center of mass momentum to be zero
and focus on one such 2PI block.  If $p$ ($p'$) is the momentum
flowing along the incoming (outgoing) quark line, then $-p$ ($-p'$) is
necessarily the momentum flowing along the  incoming (outgoing)
antiquark line. $p^0$ (${p^0}'$) has  two relevant scalings, namely
$p^0\sim |{\bf p}|$ and $p^0\sim {\bf p}^2/2m$. If the scaling $p^0\sim 
|{\bf p}|$ occurs, then  kinetic terms $\sim {\bf p}^2/2m$ can be neglected 
in the 2PI diagram and no further scale $m$ will be introduced. If the scaling 
$p^0\sim {\bf p}^2/2m$ occurs, then  $k^0$ can be neglected in the gluon propagators
and the only dependence in $p^0$ can be reduced to either the quark or
antiquark propagator.  Furthermore, the internal energies in the 2PI
diagram eventually take the value of the three-momenta $|{\bf p}|$ and hence
$p^0$ and  ${\bf p}^2/2m$ can be expanded. Hence in either case, no extra
scales $m$ are introduced in the 2PI diagrams and they can be set to zero. 
Consider now the link between two 2PI diagrams. If $p^0\sim |{\bf p}|$, 
the kinetic term can be expanded and no further scale $m$ is introduced. 
If   $p^0\sim {\bf p}^2/2m$, no further dependence on $p^0$ in the 2PI diagrams exists,
and hence the integral over $p^0$ can be trivially  done inducing a
$m/{\bf p}^2$ propagator so that the $m$ dependence factorizes trivially. In
summary, 2PR diagrams also become  scaleless and can be set to zero.

One might be worried about the appearance of pinch singularities when
the kinetic terms are expanded in the links. Let us argue that  they
are of no concern. Recall first that pinch singularities blow up only
after the limit $\eta\rightarrow 0$ is taken, where $i\eta$ defines
the causal propagator. We prescribe to take this limit at the end of
the calculation. If no other dependence on $p^0$ existed, we could
carry out all integrals except the one over $p^0$. Since
they have no scale, as argued before, and they contain  no pinch
singularities, they can safely be set to zero, and hence the net
result is zero. If there are further dependencies on $p^0$,  by
fraction decomposition one can always isolate the pinch singularity in
a term $1/(p^0 +i\eta)(-p^0+i\eta)$ with no further  dependencies on
$p^0$ (plus other terms with no pinch singularity) and proceed as
above.

Let us finally note that this matching procedure corresponds to taking the purely hard 
contribution in the threshold expansion for the NRQCD matching coefficients.

In order to address the matching calculation, 
we also need the relation between the QCD and NRQCD quark (antiquark) fields:
\be
Q_1(x) \rightarrow Z_1^{1\over 2}{1+\gamma_0 \over 2} e^{-im_1t}\psi (x) \quad ,
\quad Q_2(x) \rightarrow Z_2^{1\over 2} {1- \gamma_0 \over 2} e^{im_2t}\chi (x)
\ee
At one loop, one obtains for the wavefunction renormalization constants
\be
Z_i = 1 + C_F {\al \over \pi} \left( {3 \over 4}\ln{m_i^2\over \nu^2} -1
\right) + {\cal O}\left(\left({\al \over \pi}\right)^2 \right)\quad ,\quad i=1,2
\,.
\ee
Notice also that the states are differently normalized in relativistic 
($\left< {\bf p}\vert {\bf p}'\right>=(2\pi)^3 2 \sqrt{{\bf p}^2 + m^2} 
\delta^3({\bf p}-{\bf p}')$) or
NR ($\left< {\bf p}\vert {\bf p}'\right>=(2\pi)^3 \delta^3({\bf p}-{\bf p}')$) theories. 
Hence, in order to compare the S-matrix elements between the
two theories, a factor $(2\sqrt{{\bf p}^2 + m^2})^{1/2}$ has to
be introduced for each external fermion.

For the single quark (antiquark) sector as well as for the purely gluonic sector, 
the matching coefficients have been obtained at one loop up to 
${\cal O}(1/m^2)$ in the background Feynman gauge by  \cite{Manohar:1997qy}. They read (similarly for $1 \rightarrow 2$)
\begin{eqnarray}
c_F^{(1)} &=& 1+{\als\over 2 \pi}
\left[ C_F + \left(1-\ln{m_1 \over \nu}\right)C_A \right]
\nn \,,\\
c_D^{(1)} &=& 1+{\als\over \pi} \left[ \left(
{8\over 3} \ln {m_1 \over \nu} \right)C_F +
\left({1 \over 2} + {2 \over 3} \ln {{m_1} \over \nu} \right)C_A
       \right]-{4 \over 15}{\als \over \pi}T_F\left({m_1^2+m_2^2 \over m_2^2}\right)
\nn \,,\\
c_S^{(1)} &=& 1+{\als\over \pi} \left[ C_F +
\left(1 -  \ln {m_1 \over \nu} \right)C_A \right] \,, \nn
\\
c_1^{g\,(1)} &=& { \als\over 360 \pi} T_F  \,,
\end{eqnarray}
($\nu=\nu_{NR}$). 
The complete ${\cal O}(\als^2)$ correction to $c_F$ is also known \cite{Czarnecki:1997dz}. 

\begin{figure}
\hspace{-.1in}
\epsfxsize=3.6in
\centerline{\epsffile{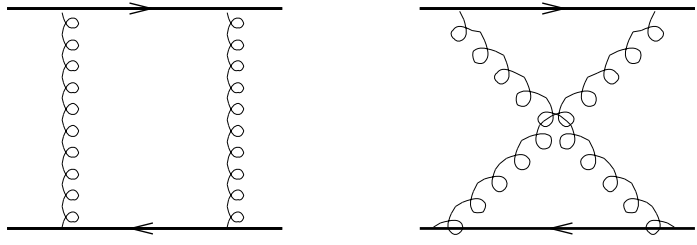}}
\caption {\it Relevant one-loop diagrams for the matching of the 4-fermion
operators at order ${\cal O}(1/m^2)$ 
for the case of unequal masses. The incoming and outgoing particles are on-shell and exactly at
rest.}
\label{figunmass}
\end{figure}

For the quark-antiquark sector, they have been obtained at one loop up to 
${\cal O}(1/m^2)$ in \cite{Pineda:1998kj}.
For the non-annihilation diagrams, which are displayed in Fig.~\ref{figunmass}, 
it is convenient to use the following basis
\bea
{\cal L}_{\psi\chi} &=&
  {d_{ss} \over m_1 m_2} \psi_1^{\dag} \psi_1 \chi_2^{\dag} \chi_2
+
  {d_{sv} \over m_1 m_2} \psi_1^{\dag} {\bfsigma} \psi_1
                         \chi_2^{\dag} {\bfsigma} \chi_2
\nn
\\
&&
+
  {d_{vs} \over m_1 m_2} \psi_1^{\dag} {\rm T}^a \psi_1
                         \chi_2^{\dag} {\rm T}^a \chi_2
+
  {d_{vv} \over m_1 m_2} \psi_1^{\dag} {\rm T}^a {\bfsigma} \psi_1
                         \chi_2^{\dag} {\rm T}^a {\bfsigma} \chi_2
\,,
\label{lag1}
\eea
which is equivalent to the one in Eq.~(\ref{Lhh}). The relation between them can be found 
(in four dimensions) in \cite{Pineda:1998kj}. 
In this basis, for the case of the quark and the antiquark having 
arbitrary flavor, the matching coefficients at one loop read in  
Feynman gauge
\bea
d_{ss}&=&
  - C_F \left({C_A \over 2} -C_F \right)
    {\als^2 \over m_1^2-m^2_2}
\left\{m_1^2\left(  \ln{m^2_2 \over  \nu^2}
                   + {1 \over 3} \right)
       -
       m^2_2\left(  \ln{m^2_1 \over  \nu^2}
                   + {1 \over 3} \right)
\right\}
\,,\label{dss}
\\
d_{sv}&=& C_F \left({C_A \over 2} -C_F \right)
   {\als^2 \over m_1^2-m^2_2}
m_1 m_2\ln{m^2_1 \over m^2_2}
\,,
\label{dsv}
\\
\label{dvs}
d_{vs}&=&
- {2 C_F \als^2 \over m_1^2-m^2_2}
  \left\{m_1^2\left( \ln{m^2_2 \over \nu^2}
                   + {1 \over 3} \right)
       -
       m^2_2\left(  \ln{m^2_1 \over  \nu^2}
                   + {1 \over 3} \right)
  \right\}
\\
\nonumber
&&+ { C_A \als^2 \over 4 (m_1^2-m^2_2)}
 \Biggl[
  3\left\{m_1^2\left( \ln{m^2_2 \over \nu^2}
                   + {1 \over 3} \right)
       -
       m^2_2\left(  \ln{m^2_1 \over \nu^2}
                   + {1 \over 3} \right)
  \right\}
\\
&&
\nonumber
\quad\quad\quad\quad
  +
   { 1 \over m_1m_2}
   \left\{m_1^4\left( \ln{m^2_2 \over \nu^2}
                   + {10 \over 3} \right)
       -
       m^4_2\left(  \ln{m^2_1 \over  \nu^2}
                   + {10 \over 3} \right)
  \right\}
 \Biggr]
\,,
\\
\label{dvv}
d_{vv}&=&
 {2 C_F \als^2 \over m_1^2-m^2_2}
        m_1 m_2\ln{m^2_1 \over m^2_2}
\\
\nonumber
&&+
{ C_A \als^2 \over 4 (m_1^2-m^2_2)}
    \Biggl[
  \left\{m_1^2\left( \ln{m^2_2 \over  \nu^2}
                   + 3 \right)
       -
     m^2_2\left(  \ln{m^2_1 \over \nu^2}
                   + 3 \right)
  \right\}
    -
  3 m_1 m_2\ln{m^2_1 \over m^2_2}
    \Biggr]
\,.
\eea
($\nu=\nu_{NR}$). The $\nu$-independent pieces of $d_{vv}$ depend on the prescription 
for reducing the $D$-dimensional Dirac matrices to Pauli matrices. 
Note that we have used the prescription for the dimensionally regulated 
spin matrices of \cite{Pineda:1998kj}, which 
differs from the more standard 't Hooft-Veltman scheme.

\begin{figure}
\hspace{-.1in}
\epsfxsize=4.2in
\centerline{\epsffile{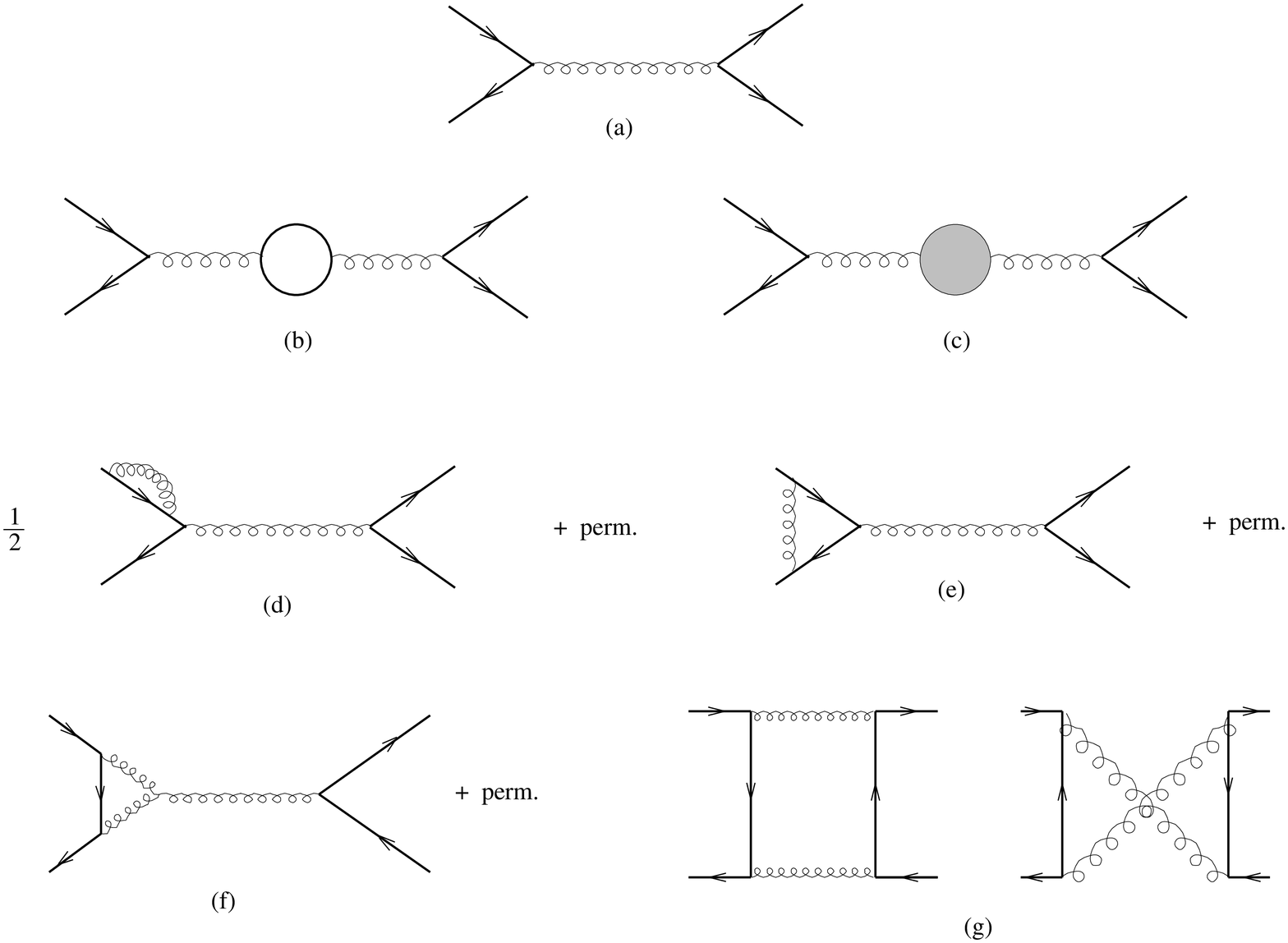}}
\caption {\it Relevant diagrams to the matching for the 4-fermion
operators at order ${\cal O}(1/m^2)$ and one loop that only appear for the
equal mass case. The incoming and outgoing
particles are on-shell and exactly at rest.}
\label{figeqtree}
\end{figure}

The contribution of the diagrams in Fig.~\ref{figunmass} to the case of equal flavor is obtained by taking the limit $m_1\rightarrow m_2=m$. 
Explicit formulas for this case can be found in  \cite{Pineda:1998kj}.
In this case, however, annihilation processes are allowed and they should
be taken into account. The relevant annihilation diagrams up to one loop are displayed in Fig.~\ref{figeqtree}.
One obtains:
\bea
f_1({}^1S_0)&=& \als^2 C_F\left({C_A\over 2}-C_F\right)
               \left(2-2\ln2 + i \pi\right) \,,\\
f_1({}^3S_1)&=& 0 \,,\\
f_8({}^1S_0)&=& {\als^2 \over 2}
               \left(-{3 \over 2}C_A+4\,C_F \right)
                \left(2-2\ln2 + i \pi\right) \,,\\
f_8({}^3S_1)&=& -\pi\als(m)\Biggl[1+{\als\over\pi}\Biggl(
T_{F} \left[{n_f\over 3}\left( 2\ln2-{5\over 3}-i\pi\right)
-{8\over 9}\right] +{109\over 36}C_A  -4\,C_F\Biggr)\Biggr]
\,.
\eea
Recall that we have to add to the annihilation contributions above the contributions
(\ref{dss})-(\ref{dvv}) in the $m_1\rightarrow m_2=m$ limit.
Note that imaginary contributions appear, which are relevant for the calculation of inclusive decay widths. 
The calculation of corrections of higher order in $\als$ to the imaginary parts 
of the 4-fermion matching coefficients has a long history 
\cite{Bodwin:1994jh,Petrelli:1997ge,Barbieri:1980yp,Barbieri:1981xz,Mackenzie:1981sf,
Maltoni:1999aa,Czarnecki:1998vz,Beneke:1997jm,Barbieri:1979be,Hagiwara:1980nv}.
An updated list of them and a summary of the state of the art can be found in \cite{Vairo:2003gh}.  
No further matching calculations beyond the order reported here 
have been carried out for the real part of 4-quark operators.

The $\nu$-dependence of the matching coefficients is eventually traded for a lower 
scale ($|{\bf p}|$, $E$, $\lQ$). This may introduce large logarithms, which ought to be summed 
up. This is discussed in sec.~\ref{sec:NRRG}. 
When higher order terms in $\als$ are calculated, it may occur that large numerical 
factors lead to poor convergence of the perturbative series. This is often related to 
 so-called renormalon singularities, which are discussed in ch.~\ref{secrenormalons}.

\subsection{Renormalization group}
\label{sec:NRRG}
Once the EFT has been built, one may try to perform its RG improvement. 
This has proven to be a non-trivial task for NRQCD, which is related to 
the fact that different kinds of degrees of freedom are encoded in the same fields.
In other words, the soft and US physics have not been disentangled 
at the NRQCD Lagrangian level. This means that obtaining the 
RG improvement at the NRQCD level becomes a not very well-defined problem. 
We will see later on that the introduction of pNRQCD, which does factorize
soft and US physics, indicates how this problem must be posed. 
Indeed, it is possible to obtain some results at this level 
(in fact it is even convenient), which will be used afterwards in order 
to obtain the RG equations in potential NRQCD (in the weak-coupling regime). 
The NRQCD matching coefficients are functions of
$\nu_{NR}=\{\nu_p,\nu_s\}$. It is convenient to restrict ourselves to derive RG 
equations with respect to the scale $\nu_s$, since the RG equations with respect 
to the scale $\nu_p$ are obtained in a much simpler way using pNRQCD.  

The matching coefficients of the terms bilinear in the heavy quark fields and of the pure
gluonic terms are just functions of $\nu_s$, i.e. 
$c=c(\nu_s,m)\equiv c(\nu_s)$. 
This is due to the fact that UV behavior of the Green functions in this sector 
is only sensitive to the energy 
and not to the three-momentum of the heavy quarks, as it
can also be seen by explicit computations.
Therefore, the anomalous dimensions can be computed 
using the static propagator for the heavy quark, and coincide 
with those obtained for HQET. The complete LL running of
these matching coefficients in the basis of operators (\ref{Lg}-\ref{Lhl}) 
has been calculated by \cite{Bauer:1998gs} in the (background) Feynman gauge (some partial previous 
results already existed in the literature \cite{Eichten:1990vp,Falk:1990pz,Blok:1996iz}). 
For the case of the only non-trivial matching coefficient at ${\cal O}(1/m)$, $c_F$, also a 
NLL evaluation is available \cite{Amoros:1997rx,Czarnecki:1997dz}, 
which we explicitly display to illustrate the typical structure of the RG improved 
matching coefficients: 
\bea
c_F(m_i)&=& 
z^{-{\gamma_0\over 2}}
\left[ 1 + \frac{\als(\nu_h)}{4\pi}
  \left(c_1+\frac{\gamma_0}{2}\ln\frac{\nu_h^2}{m_i^2}\right) 
  + \frac{\als(\nu_h) - \als(\nu_s)}{4\pi}\left(
    \frac{\gamma_1}{2\beta_0} - \frac{\gamma_0\beta_1}{2\beta_0^2}
  \right) + \dots \right] 
\,,
\nonumber\\
\eea
where  $z=\left(\als(\nu_s)/\als(\nu_h)\right)^{1/\beta_0}$, 
$\nu_h\sim m_i$ is the hard matching scale,  $c_1 = 2(C_A+C_F)$ 
and the one- and two-loop anomalous dimensions read
\be
   \gamma_0 = 2 C_A \,, \qquad
   \gamma_1 = \frac{68}{9}\,C_A^2 - \frac{52}{9}\,C_A T_F\,n_f
   \,.
\ee

Complications appear when the 4-heavy-quark operators, $\{f\}$, are considered. 
As we have mentioned, they depend on both cutoffs: $\nu_p$ and $\nu_s$. 
Nevertheless, at one loop, all the
dependence of the matching coefficients is only due to $\nu_s$, 
i.e. $f(\nu_p,\nu_s,m)\equiv
f(\nu_p,\nu_s) \simeq f(\nu_s)$. The dependence on $\nu_p$ appears 
at two loops or higher and will be discussed in ch.~\ref{sec:pNRQCDII}. 
In any case, if one restricts oneself 
to the purely soft running (i.e $\nu_s$-dependence only), it still makes sense 
to consider the (soft) RG running of the NRQCD matching coefficients 
including the 4-heavy fermion operators. In this approximation, 
one can always perform the computation with static propagators for the 
heavy quarks and order by order in $1/m$.

Formally, we can write the NRQCD Lagrangian as an expansion in $1/m$:
\be
{\cal L}_{\rm NRQCD} =\sum_{n=0}^{\infty}{1\over m^n}\lambda_nO_n
, 
\ee
where the RG equations of the matching coefficients read
\be
\nu_s {d \over d \nu_s}\lambda=B_{\lambda}(\lambda)
.
\ee
The RG equations have a triangular structure (the standard
structure one can see, for instance, in HQET RG equations):
\bea
\nu_s {d \over d \nu_s}\lambda_0&=&B_0(\lambda_0)
\,,
\nn
\\
\nu_s {d \over d \nu_s}\lambda_1&=&B_1(\lambda_0)\lambda_1
\,,
\nn
\\
\nu_s {d \over d
\nu_s}\lambda_2&=&B_{2(2,1)}(\lambda_0)\lambda_2+B_{2(1,2)}(\lambda_0)\lambda_1^2
\,,
\eea
$$
\cdots\,,
$$
where the different $B$'s can be expanded into a power series in $\lambda_0$
($\lambda_0$ corresponds to the marginal operators (renormalizable
interactions)). For NRQCD we have $\lambda_0=\als$ and
$\lambda_{1}=\{c_k, c_F\}$, $\lambda_{2}=\{c_1^g, c_D, c_S,
\{c^{ll}\},\{c^{hl}\}, \{f\} \}$.
 
As we have already mentioned, the LL running
for the $\{c\}$ in Feynman gauge can be read off from the 
results of \cite{Bauer:1998gs}. The LL running of the $\{f\}$ in Feynman gauge
can be found in \cite{Pineda:2001ra}.

At this stage, we would like to stress that we are working in a
non-minimal basis of operators for the NRQCD Lagrangian. Consequently,
the values of (some of) the matching coefficients are ambiguous (only some
combinations with physical meaning are unambiguous) and 
could depend upon the gauge in which the
calculation has been performed. At the practical level, this means that 
they will depend on
the specific basis of operators we have taken for the NRQCD Lagrangian
and on the procedure used (in particular on the gauge). Therefore, if
working in a non-minimal basis, one should be careful to do the
matching using the same gauge for all the operators (or at least for
those that are potentially ambiguous).
This affects the running of $c_D$, $f_8({}^1S_0)$ and $c_1^{hl}$. 
Indeed, it has been shown in \cite{Bauer:1998gs} that $c_D$ can be 
absorbed into $f_8({}^1S_0)$ and $c_1^{hl}$ by using the equations of motion (${\bf D}\cdot {\bf E}^a  =
g (\psi^\dagger T^a \psi + \chi^\dagger T^a \chi + \sum_{j=1}^{n_f} 
\bar{q}_j \gamma^0 T^a q_j)
$).

Let us illustrate the point by considering the running of 
$c_D$ and $f_8({}^1S_0)$ in the equal mass case and without light quarks. 
In Feynamn gauge we obtain:
\bea
\nu_s {d\over d\nu_s}c_D&=&{\als\over 4 \pi}\left[{4 C_A \over
    3}c_D-\left( {2 C_A \over 3}+{32 C_F \over 3}\right)c_k^2-{10 C_A
    \over 3}c_F^2\right],
\label{cDRG}
\\
\nu_s {d\over d\nu_s}f_8({}^1S_0)&=&4\left(C_F-C_A\right)\als^2c_k^2 
+ { 3 \over 2}\als^2C_Ac_D
\,,
\eea
while in Coulomb gauge we have: 
\bea
\nu_s {d\over d\nu_s}c_D({\rm Coulomb})&=&{\als\over 4 \pi}\left[{22 C_A 
\over
   3}c_D-\left( {32 C_A \over 3}+{32 C_F \over 3}\right)c_k^2-{10 C_A
        \over 3}c_F^2\right]
\,,
\nn\\
\nu_s {d\over d\nu_s}f_8({}^1S_0)({\rm Coulomb})&=&\left(4C_F-{3 C_A \over
2}\right)\als^2c_k^2 
\label{RGeqhcoulomb}
\,.
\eea
Clearly, the running of $c_D$ and $f_8({}^1S_0)$ is gauge 
dependent, but the running of the combination $\als c_D+(1/\pi) 
f_8({}^1S_0)$ is not, reflecting the fact that 
$c_D$ can be absorbed into $f_8({}^1S_0)$ by means of a suitable field 
redefinition.

\subsection{Applications: spectrum and inclusive decay widths}
\label{sec:NRapplications}
NRQCD has been applied over the last twelve years to a large number
of observables related to heavy quarkonium physics. 
Here we will only briefly discuss two kinds of observables: 
spectra and inclusive decay widths. 
Concerning the spectra, we will just mention the state of the art for what concerns the lattice
determination of the bottomonium levels.  We will keep a continuum EFT point
of view, since a discussion of lattice NRQCD is beyond the scope of the present
work (see \cite{Kronfeld:2003sd,Lepage:2005eg} for some recent
reviews). We will, however, give a more detailed
discussion of the inclusive decay width. We have chosen these observables because 
they are amenable to rather clean theoretical derivations. They will also be 
addressed in the following sections dedicated to pNRQCD.

Before proceeding, we have to establish a power counting for NRQCD.
As was mentioned in sec.~\ref{sec:powerNRQCD}, since the scales $(E,|{\bf p}|,\lQ)$ remain
dynamical, it is not possible to give a homogeneous counting for each operator.
In other words, in contrast to pNRQCD, we will not be able to
disentangle the contributions coming from the different 
scales. In order to be on the safe side, we have to 
assume the most conservative counting where each operator counts like $(mv)^d$, 
$d$ being its dimension, with the exception of $iD_0$ that counts like  $mv^2$ 
($v \sim  |{\bf p}|/m \sim  E/|{\bf p}|$).\footnote{
In principle, at least another scale is relevant for quarkonium: 
$\sqrt{m\, \lQ}$. Since this scale is larger than $E$, $|{\bf p}|$ and $\lQ$, it may, 
in principle, change  our counting. We will discuss this 
in ch.~\ref{sec:SCR}.
} 
To count matrix elements of color singlet operators 
between quarkonium states is rather simple. Since the quarkonium states are
normalized, it is sufficient to count the dimension of the gluon field
operators and covariant derivatives. For color octet operators\footnote{This 
  applies to the pure octet content of the octet operators ($O_8$, ${\cal  P}_8$, ...) 
  defined in Eq.~(\ref{ImLpsichi}), which, starting from ${\cal 
  O}(1/m^4)$, may also contain singlet parts.}, one has to take into account 
that they give a non-vanishing contribution between quarkonium states
if at least two extra $1/m$ operators are inserted. Hence, using the above rules, 
one has to add two to the dimension of the gluon field operators and covariant derivatives.
This counting, which we will call the ``conservative counting'', differs from 
the  ``original counting'' of NRQCD introduced in \cite{Lepage:1992tx}.
We refer to sec.~\ref{sec:powerNRQCD} for further details.

\begin{figure}
\makebox[-16truecm]{\phantom b}
\put(100,0){\epsfxsize=8truecm \epsfbox{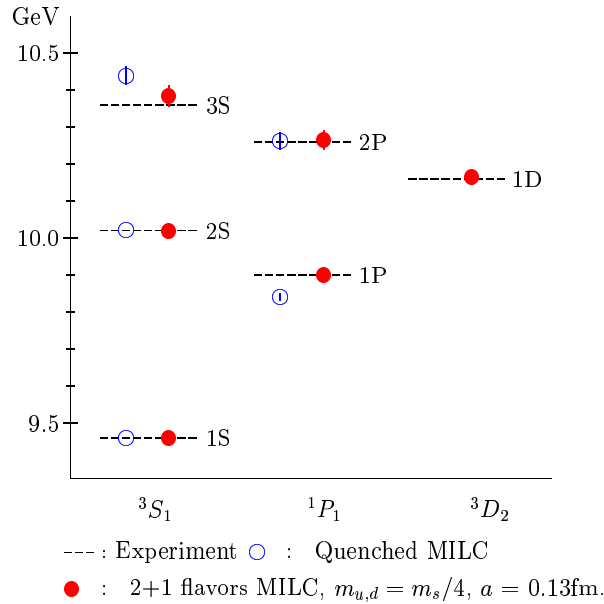}}
\vspace{2mm}
\caption{\it Radial and orbital splittings in the $\Upsilon$ system 
from lattice QCD in the quenched approximation (open circles) and including 
 dynamical $u, d$ and $s$ quarks (filled circles). The lattice spacing 
has been fixed on the radial splitting between $\Upsilon(2S)$ and $\Upsilon(1S)$.
The $b$ quark mass has been tuned to get the $\Upsilon$ mass correct. 
Figure taken from \protect\cite{Lepage:2004mq}.}
\label{figupslat}
\end{figure}

\subsubsection{Spectra}
The idea to put NRQCD on the lattice has been a very early one
\cite{Lepage:1992tx}. The advantages with respect to full QCD are obvious. 
The lattice spacing $a$ and the dimension $L$ of the lattice 
have to fulfill the requirement: $1/a \gg Q \gg q \gg 1/L$, where $Q$ is the
largest and $q$ the smallest scale of the system under study. In full QCD 
we have $Q \sim m$ while in NRQCD $Q \sim |{\bf p}| \ll m$. NRQCD, therefore, 
does not require such a fine lattice as full QCD, which means that much more
economical simulations are sufficient. The drawback is that the continuum scaling
window will not be reached and much more care has to be taken in order to
extrapolate from the discrete simulations to the continuum physics.

Some recent results obtained in the lattice version of NRQCD can be found in 
\cite{Lepage:2004mq}. For what concerns the heavy quarkonium spectra, as a
consequence of the rather precise data, all levels below the open flavor 
threshold have been obtained from multiexponential fits to suitable correlation
functions. In Fig.~\ref{figupslat}, we show some recent quenched and unquenched
results for the radial and orbital splittings in the bottomonium system \cite{Gray:2002vk}.

Let us comment on the theoretical limits on the precision of the lattice
results. We will neglect all (indeed rather serious) 
complications and uncertainties connected with the numerical simulations 
and the continuum extrapolations.
The version of the NRQCD Lagrangian used in all lattice simulations contains, apart
from the Yang--Mills term, only bilinear terms in the heavy quark fields.
The matching coefficients are taken at tree level. This is due to the fact 
that their calculation in a lattice regularization scheme turns out to be quite 
cumbersome so that, up to now, only  some preliminary numerical 
estimates are available for a few of them\cite{Trottier:1997bn}.
As a consequence, regardless of how many operators have been added to the 
bilinear sector of the Lagrangian, the theoretical limit on the precision of
the radial splittings is of relative order $\als \, v^2 \simeq 0.2 \times 0.1 \simeq
2\%$ in the original power counting of \cite{Lepage:1992tx}
($\als \, v \simeq 0.2 \times 0.3 \simeq 6\%$ in the conservative counting 
introduced above), while for the fine and hyperfine splittings it is of relative order 
$\als \simeq 0.2 \simeq 20\%$. We have assumed for the bottomonium case
$v^2\simeq 0.1$ and $\als(m_b) \simeq 0.2$.\footnote{It seems too optimistic to
replace $\als$ with $\als/\pi$ as suggested in \cite{Lepage:1992tx}, 
since several $\als$ corrections appear with
large coefficients (compare with the explicit expressions 
given in the previous section and with the discussion in 
\cite{Brambilla:1998vm}). Moreover large logarithms could also deteriorate the convergence.} 
In any case, the precision in the radial splittings is 
rather good, while it is worse by an order of magnitude in the fine and
hyperfine splittings. In the charmonium case, $v^2 \simeq 0.3$ and $\als(m_c)
\simeq 0.35$, which means that the theoretical limit on the precision of
the radial splittings is not smaller than $10\%$ in the original counting 
($20\%$ in the conservative counting). In order to improve the
present precision, it is, therefore, crucial to calculate the one loop
corrections to the Wilson coefficients in the NRQCD Lagrangian in a consistent lattice
regularization and renormalization scheme. In this sense, the recent work 
by \cite{Becher:2002if,Becher:2003fu} seems to be rather promising.
Note that at order $\als v^4$ ($\als v^3$  in the conservative counting),
$1/m^2$ corrections to the Yang--Mills sector of the NRQCD Lagrangian 
and 4-fermion operators also have to be taken into account.

\subsubsection{Inclusive decay widths}
\label{NRdecay}
Let us consider heavy quarkonia made out of a quark 
and an antiquark of the same flavor ($m_1$ $=$ $m_2$ $=$ $m$).
Annihilation processes happen in QCD at the scale of the mass $m$. 
Therefore, integrating out these scales in the matching from 
QCD to NRQCD produces imaginary terms in the matching coefficients 
of the 4-four fermion operators of the 
NRQCD Lagrangian as we have seen in sec.~\ref{NRmatching}. 
Therefore, the annihilation width of a heavy quarkonium H into light particles 
is given by \cite{Bodwin:1994jh}:
\be
\Gamma({\rm H} \to \hbox{light particles})= 
2\, {\rm Im} \, \langle {\rm H}|  {\cal L}_{\psi\chi}  |{\rm H} \rangle,
\label{imagnrqcd}
\ee
where $| {\rm H} \rangle$ is a normalized eigenstate of the NRQCD Hamiltonian with the 
quantum numbers of the considered quarkonium in its centre-of-mass frame
\footnote{This expression only holds at LO in the imaginary terms. 
The exact expression, which has not been necessary for applications so far, reads 
$ \Gamma({\rm H} \to \hbox{light particles})=
-2\, {\rm Im} \left( \langle {\rm \tilde H}|  
H |{\rm H} \rangle /\langle {\rm \tilde H}|{\rm H} \rangle \right)$, 
where $H$ is the NRQCD Hamiltonian and $|{\rm \tilde H} \rangle $ ($\not=|{\rm  H} \rangle $ in general) 
is the corresponding eigenstate of $H^\dagger$.}.
In Eq.~(\ref{Lhh}), we have given ${\cal L}_{\psi\chi}$ up to order $1/m^2$, here we
will need it up to order $1/m^4$:
\bea
\label{ImLpsichi}
{\cal L}_{\psi\chi} 
&=& {f_1({}^1S_0) \over m^2} \, O_1({}^1S_0) 
+ {f_1({}^3S_1) \over m^2} \, O_1({}^3S_1) 
+ {f_8({}^1S_0) \over m^2} \, O_8({}^1S_0) 
+ {f_8({}^3S_1) \over m^2} \, O_8({}^3S_1) 
\nn \\
&& 
+ {f_1({}^1P_1)   \over m^4}  O_1({}^1P_1)
+ {f_1({}^3P_{0}) \over m^4}  O_1({}^3P_{0})
+ {f_1({}^3P_{1}) \over m^4}  O_1({}^3P_{1}) 
+ {f_1({}^3P_{2}) \over m^4}  O_1({}^3P_{2})
\nn \\
&& 
+ {g_1({}^1S_0)   \over m^4}  {\cal P}_1({}^1S_0)
+ {g_1({}^3S_1)   \over m^4}  {\cal P}_1({}^3S_1) 
+ {g_1({}^3S_1,{}^3D_{1}) \over m^4}  {\cal P}_1({}^3S_1,{}^3D_{1})
\nn \\
&& 
+ [ O_1 \rightarrow O_8, {\cal P}_1 \rightarrow {\cal P}_8, f_1 \rightarrow f_8, g_1 
\rightarrow g_8],
\eea
where the explicit expressions for the operators in the first 
line can be found in (\ref{def4fops}) and for the remaining operators in \cite{Bodwin:1994jh}.

The NRQCD factorization formula for the inclusive heavy quarkonium 
annihilation width into light particles reads ($d_n$ denotes the dimension 
of the generic 4-fermion operator $O^{(n)}$): 
\bea
& &\Gamma({\rm H}\to LH) = 
\sum_n {2 \, {\rm Im} \, f^{(n)} \over m^{d_n - 4}} 
\langle {\rm H} | O^{(n)} |{\rm H} \rangle, 
\label{factannlh}
\\
& &\Gamma({\rm H}\to EM) = \sum_n {2 \, {\rm Im} \, 
f_{\rm EM}^{(n)} \over m^{d_n - 4}}
\langle {\rm H} |O^{(n)}_{\rm EM} |{\rm H} \rangle,
\label{factannem}
\eea
where we have distinguished between electromagnetic decay widths and 
decay widths into light hadrons ($LH$). 
Let us comment on the electromagnetic decay widths.
The information needed in order to describe
decays into hard electromagnetic particles is encoded in the
electromagnetic contributions to the matching coefficients that we denote by 
$f_{\rm EM}$, $g_{\rm EM}$, ... We do not use a special symbol to denote 
the purely hadronic component of the matching coefficients, which is the dominant one.
The purely electromagnetic component of the inclusive decay width 
may be singled out by projecting the 4-fermion operators onto the QCD vacuum
state $ |\hbox{vac}\rangle$ 
according to $\psi^\dagger \cdots \chi \, \chi^\dagger \cdots \psi \to 
\psi^\dagger \cdots \chi \, |\hbox{vac}\rangle \, 
\langle \hbox{vac}|\, \chi^\dagger \cdots \psi$.  
The projected operators are denoted by $O_{\rm EM}$, ${\cal P}_{\rm EM}$, $\cdots$.
For instance $O_{\rm EM}(^1S_0)=\psi^\dagger \chi \, 
|\hbox{vac}\rangle \, \langle \hbox{vac}|\, \chi^\dagger \psi$.
The inclusive annihilation width into light hadrons may be obtained 
from the full annihilation width by switching off the electromagnetic 
interaction. The factorization formulas (\ref{factannlh}) and
(\ref{factannem}) have been rigorously proven, also diagrammatically, 
in \cite{Bodwin:1994jh}.

Working out Eqs.~(\ref{factannlh}) and (\ref{factannem}),  
the explicit expressions for the decay widths of $S$- and $P$-wave
quarkonium up to ${\cal O}({\rm Im} \, f \times mv^5)$ are  
\bea
&&\hspace{-8mm}
\Gamma(V_Q (nS) \rightarrow LH) = {2\over m^2}\Bigg( 
{\rm Im\,}f_1(^3 S_1)   \langle V_Q(nS)|O_1(^3S_1)|V_Q(nS)\rangle
\nn
\\
&&
+ {\rm Im\,}f_8(^3 S_1) \langle V_Q(nS)|O_8(^3S_1)|V_Q(nS)\rangle
+ {\rm Im\,}f_8(^1 S_0) \langle V_Q(nS)|O_8(^1S_0)|V_Q(nS)\rangle 
\nn
\\
&&
+ {\rm Im\,}g_1(^3 S_1)
{\langle V_Q(nS)|{\cal P}_1(^3S_1)|V_Q(nS)\rangle \over m^2}
+ {\rm Im\,}f_8(^3 P_0)
{\langle V_Q(nS)|O_8(^3P_0)|V_Q(nS)\rangle \over m^2}
\nn
\\
&&
+ {\rm Im\,}f_8(^3 P_1)
{\langle V_Q(nS)|O_8(^3P_1)|V_Q(nS)\rangle \over m^2}
+ {\rm Im\,}f_8(^3 P_2)
{\langle V_Q(nS)|O_8(^3P_2)|V_Q(nS)\rangle \over m^2}\Bigg),
\label{gammaV}
\\
&&\hspace{-8mm}
\Gamma(P_Q (nS) \rightarrow LH) = {2\over m^2}\Bigg( 
{\rm Im\,}f_1(^1 S_0)   \langle P_Q(nS)|O_1(^1S_0)|P_Q(nS)\rangle
\nn
\\
&&
+ {\rm Im\,}f_8(^1 S_0) \langle P_Q(nS)|O_8(^1S_0)|P_Q(nS)\rangle
+ {\rm Im\,}f_8(^3 S_1) \langle P_Q(nS)|O_8(^3S_1)|P_Q(nS)\rangle 
\nn
\\
&&
+ {\rm Im\,}g_1(^1 S_0)
{\langle P_Q(nS)|{\cal P}_1(^1S_0)|P_Q(nS)\rangle \over m^2}
+ {\rm Im\,}f_8(^1 P_1)
{\langle P_Q(nS)|O_8(^1P_1)|P_Q(nS)\rangle \over m^2} \Bigg),
\label{gammaP}
\\
&&\hspace{-8mm}
\Gamma(\chi_Q(nJS)  \rightarrow LH)= 
{2\over m^2}\Bigg( {\rm Im \,}  f_1(^{2S+1}P_J) 
{\langle \chi_Q(nJS) | O_1(^{2S+1}P_J ) | \chi_Q(nJS) \rangle \over m^2}
\nn
\\
&&
+ {\rm Im \,} f_8(^{2S+1}S_S) \langle \chi_Q(nJS) | O_8(^1S_0 ) | \chi_Q(nJS) \rangle\Bigg),
\label{gammachi}
\\
&&\hspace{-8mm}
\Gamma(V_Q (nS) \rightarrow e^+e^-)= {2\over m^2}\Bigg( 
{\rm Im\,}f_{ee}(^3 S_1)   \langle V_Q(nS)|O_{\rm EM}(^3S_1)|V_Q(nS)\rangle
\nn
\\
&&
+ {\rm Im\,}g_{ee}(^3 S_1)
{\langle V_Q(nS)|{\cal P}_{\rm EM}(^3S_1)|V_Q(nS)\rangle \over m^2}\Bigg),
\label{gammaVem}
\\
&&\hspace{-8mm}
\Gamma(P_Q (nS) \rightarrow \gamma\gamma)= {2\over m^2}\Bigg( 
{\rm Im\,}f_{\gamma\gamma}(^1 S_0)   
\langle P_Q(nS)|O_{\rm EM}(^1S_0)|P_Q(nS)\rangle
\nn
\\
&&
+ {\rm Im\,}g_{\gamma\gamma}(^1 S_0)
{\langle P_Q(nS)|{\cal P}_{\rm EM}(^1S_0)|P_Q(nS)\rangle \over m^2} \Bigg),
\label{gammaPem}
\\
&&\hspace{-8mm}
\Gamma(\chi_Q(nJ1)  \rightarrow \gamma\gamma)= 
2 {\rm Im \,}  f_{\gamma\gamma}(^3P_J) 
{\langle \chi_Q(nJ1) | O_{\rm EM}(^3P_J ) | \chi_Q(nJ1) \rangle \over m^4}
\qquad {\rm for} \; J=0,2\,,
\label{gchiem}
\eea
where the symbols $V$ and $P$ stand for the vector 
and pseudoscalar $S$-wave heavy quarkonium and the symbol $\chi$ for  
the generic $P$-wave quarkonium (the states $\chi(n10)$ and $\chi(nJ1)$
are usually called $h((n-1)P)$ and $\chi_J((n-1)P)$, respectively). 

Let us comment on Eqs.~(\ref{gammaV})-(\ref{gchiem}). The first obvious
observation is that in the hadronic decay widths, besides singlet 
also octet matrix elements occur. In the case of the hadronic $P$-wave decay widths they are
of the same order as the singlet matrix
elements. This means that a description of heavy quarkonium in terms of a
color-singlet bound state of a heavy quark and antiquark necessarily fails at
some point: for $P$-wave decay this point is the leading order! There is
another way to understand the role of the octet matrix elements. The singlet matching
coefficients are plagued by IR divergences. The coefficients 
${\rm Im}\,f(^3P_0)$  and ${\rm Im}\,f(^3P_2)$ are IR divergent at NLO \cite{Barbieri:1976fp}.
These divergences are precisely canceled by the octet contributions \cite{Bodwin:1992ye}. 
Therefore, the inclusion of the octet matrix elements is crucial to make Eq.~(\ref{gammachi})
physical, i.e. independent of the cut-off.
For what concerns $S$-wave decays, let us note that in the original 
NRQCD power counting of \cite{Lepage:1992tx} the octet matrix elements are ${\cal O}(v^4)$ suppressed 
compared with the leading order. This is not so within the conservative power counting
adopted here, where they are ${\cal O}(v^2)$. This may 
be of phenomenological relevance for $\Gamma(V \rightarrow LH)$, since ${\rm Im}\,
f_1( ^3S_1)$ is ${\cal O}(\als)$-suppressed with respect to ${\rm Im} \,f_8( S)$. 

Despite the fact that the NRQCD factorization formulas for inclusive decay
widths are theoretically solid and have provided a solution 
to the long-standing problem of the cancellation of
the IR divergences, their practical relevance in calculating inclusive or
electromagnetic decay widths of quarkonia has been rather limited.
This is mainly due to the following reasons:\\
{\bf (1)} NRQCD matrix elements may be fitted on the experimental decay data \cite{Maltoni:2000km} or
calculated on the lattice \cite{Bodwin:2001mk,Bodwin:1996tg}. The matrix elements of singlet
operators can be linked at leading order to the Schr\"odinger wavefunctions at the
origin \cite{Bodwin:1994jh} and, therefore, may be evaluated by means of potential models \cite{Eichten:1995ch}.
In general, however, NRQCD matrix elements, in particular of higher dimensionality, 
are poorly known or completely unknown. 
\\
{\bf (2)} The formulas depend on a large number of matrix elements.
In the bottomonium system, 14 $S$- and $P$-wave states lie below the open 
flavor threshold 
($\Upsilon(nS)$ and $\eta_b(nS)$ with $n=1,2,3$; $h_b(nP)$ and $\chi_{bJ}(nP)$ 
with $n=1,2$ and $J=0,1,2$) and in the charmonium system 8 
($\psi(nS)$ and $\eta_c(nS)$ with $n=1,2$; $h_c(1P)$ and
$\chi_{cJ}(1P)$ with $J=0,1,2$). For these states, 
Eqs.~(\ref{gammaV})-(\ref{gchiem}) describe the decay widths 
into light hadrons and into photons or $e^+e^-$ in terms of 46 
NRQCD matrix elements (40 for the $S$-wave decays and $6$ for the $P$-wave
decays). More matrix elements are needed if higher-order operators have to be included. 
Indeed, it has been discussed in \cite{Ma:2002ev} and \cite{Bodwin:2002hg} 
that higher-order operators, not included in
Eqs.~(\ref{gammaV})-(\ref{gchiem}), even if parametrically suppressed,  
may turn out to give sizable contributions to the decay widths. 
This may be the case, in  particular, for charmonium, where $v^2 \sim 0.3$, so that relativistic corrections 
are large, and for $P$-wave decays where the above formulas provide 
only the leading-order contribution in the velocity expansion.
In fact it was pointed out in \cite{Ma:2002ev,Vairo:2002iw} that if no special cancellations  
among the matrix elements occur, then the order $v^2$ relativistic corrections 
to the electromagnetic decays $\chi_{c0} \to \gamma\gamma$ and $\chi_{c2}\to \gamma\gamma$ 
may be as large as the leading terms.
Finally, it was noted in \cite{Maltoni:2000km} that the relevance of higher-order 
matrix elements may be enhanced (or suppressed) by the multiplying matching coefficients.
\\
{\bf (3)} The convergence of the perturbative series of the 4-fermion 
matching coefficients is often poor (see, for instance, the examples 
in \cite{Vairo:2002iw}). This limits, in general, the reliability and 
stability of the results. Some classes of large perturbative contributions 
have been resummed for $S$-wave annihilation decays in
\cite{Bodwin:2001pt,Bodwin:1998mn,Braaten:1998au}, improving the convergence of
the series.

\section{Potential NRQCD. The physical picture}
\label{sec:pNRQCDI}
Of the full hierarchy of scales in heavy quarkonium, NRQCD only takes advantage
of the fact that $m$ is much larger than the remaining ones 
($|{\bf p}|$, $E$, $\lQ$, ...). This means that if we are interested in 
physics at the scale of the binding energy $E$,  NRQCD
contains degrees of freedom that can never appear as physical states at that
scale.   These are, in particular, light degrees of freedom of energy $\sim |{\bf p}|
\gg E$ and heavy quarks with energy fluctuations of the same order.
Therefore, within the philosophy  of EFTs,  these degrees
of freedom  should better be integrated out. The implementation  of this idea
gives rise to a new effective theory called pNRQCD \cite{Pineda:1998bj}.  The
appropriate description of the remaining degrees of freedom and how this
integration can actually be carried out will clearly depend on the relative
size of $\lQ$ compared to the scales $|{\bf p}|$ and $E$. We consider the different
possibilities in the next two sections.  In pNRQCD, it is the large scale $|{\bf p}|$
that limits the UV cut-off of the energy fluctuations.  Even if its typical
value in a bound state can be associated with $mv$, its fluctuations  may reach
up to the scale $m$, which is the UV cut-off for the three-momentum
fluctuations of the heavy quarks, $|{\bf p}|$.

\subsection{Weak-coupling regime}
If $|{\bf p}| \gg \lQ$, the integration of degrees of freedom of energy scale $|{\bf p}|$ 
can be done in perturbation theory. Hence, we do not expect a qualitative change 
in the degrees of freedom, but only a lowering of their energy cut-off.  Let us call
the resulting EFT pNRQCD'.  pNRQCD' is thus defined by the same particle
content as NRQCD and the cut-offs $\nu_{pNR}=\{\nu_p,\nu_{us}\}$, where
$\nu_p$ is the cut-off of the relative three-momentum of the heavy quarks and
$\nu_{us}$ is the cut-off of energy fluctuations of the heavy quarks and of
the four-momenta of the gluons and light quarks. They satisfy the following
inequalities: $|{\bf p}| \ll \nu_p \ll m$ and ${\bf p}^2/m \ll \nu_{us} \ll
|{\bf p}|$.  The Wilson coefficients of pNRQCD' will then depend on ${\bf p}$
and ${\bf p}'$, the three-momenta of the heavy quark  and antiquark
respectively,  usually through the combination ${\bf k}= {\bf p}-{\bf p}'$.
Hence, non-local terms (potentials) in real space are produced. Indeed, these
potentials encode the non-analytic behavior in the momentum  transfer ${\bf k}$ of the
heavy quark, which is  of the order of the
relative three-momentum of the heavy quarks. This is again a peculiar feature
of pNRQCD which had not been observed in any EFT before.  It provides an
appealing interpretation of the usual potentials in quantum mechanics within
an EFT framework.

In order to take advantage of the fact that the 
three-momentum of the heavy quarks is always larger than the 
four-momentum of the light degrees of freedom, it is very convenient 
to use fields in which the relative coordinate 
(conjugate to the relative momentum) appears explicitly.
We   define the centre-of-mass coordinate of the $Q$-$\bar Q$ system 
${\bf R} \equiv ({\bf x}_1+{\bf x}_2)/2$ and the relative coordinate ${\bf r} \equiv {\bf x}_1-{\bf x}_2$. 
A $Q$-$\bar Q$ state can be decomposed into a singlet state $S({\bf r},{\bf R},t)$ and an octet 
state $O({\bf R},{\bf r},t)$, in relation to color gauge transformation 
with respect to the centre-of-mass coordinate. (We notice that in QED only 
the state analogous to the singlet appears). 
The gauge fields are evaluated in ${\bf R}$ and $t$, i.e.  $A_\mu = A_\mu({\bf R},t)$: 
they do not depend on ${\bf r}$. This is due to the fact that, since the typical size of ${\bf r}$ 
is the inverse of the soft scale, gluon fields are multipole expanded with respect to this variable. 

If the binding energy $E$ is larger than or of the same order as $\lQ$, 
we will have accomplished our goal and the EFT we are looking for, 
namely pNRQCD in the weak-coupling regime, coincides with pNRQCD'. If, on 
the contrary, $\lQ \gg E$, we still have to integrate out the energy scale $\lQ$, 
and its associated three-momentum scale $\sqrt{\lQ m}$ in order to obtain pNRQCD. 
This cannot be done perturbatively in $\als$ anymore, but one can definitely continue 
exploiting the hierarchy of scales, as will be discussed in the following section.

\subsection{Strong-coupling regime}
For illustration purposes, let us first consider the  particular case $|{\bf p}| \gg
\lQ \gg E$,  which directly links to the discussion in the previous
section. We have to figure out  what happens to the pNRQCD' degrees of freedom
after integrating out those of energy $\sim \lQ$.
Below the scale $\lQ$, it is better to think  in terms of hadronic degrees of
freedom,  which are color singlet states. Hence the octet field is not acceptable 
in the final EFT and must be
integrated out.  Since it couples to gluons of energy $\sim \lQ$ it is also
expected that it develops a  mass gap of the same order. Therefore in pure
gluodynamics the only degree of freedom  left is the singlet field interacting
with a potential, which  also has non-perturbative contributions from the
integration of degrees of freedom of order $\lQ$.  In real QCD,
pseudo-Goldstone bosons, which have masses smaller than $\lQ$,  should also be
included. These are the expected degrees of freedom of pNRQCD in the 
strong-coupling regime \cite{Brambilla:1999xf}.

In the general case $|{\bf p}|\gtrsim \lQ$, we cannot integrate out energy 
degrees of freedom at the scale $|{\bf p}|$ in perturbation theory in $\als$. 
Still the relevant energy scales are at a lower scale $E\ll |{\bf p}| \sim \lQ$ 
and one can in principle build an EFT at that scale, as we have done above in 
a particular case. This is pNRQCD in the strong-coupling regime.  At scales 
$E\ll \lQ$, QCD becomes strongly coupled and it is again better to think in 
terms of hadronic degrees of freedom, which are color singlet states. Hence 
the most likely degrees of freedom in this regime
are a singlet field and pseudo-Goldstone bosons.  This is supported  by our
knowledge of the static limit of QCD as will be argued below.

\begin{figure}
\hspace{-0.1in} \epsfxsize=3.8in \centerline{\epsffile{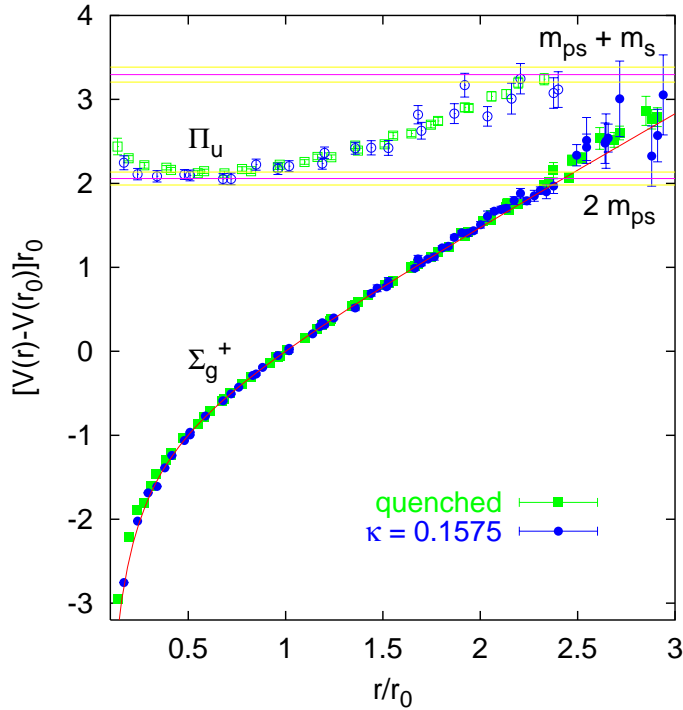}}
\caption {\it Mass gap between the singlet and hybrid fields. From \cite{Bali:2000vr}.}
\label{Masgap}
\end{figure}

In the static limit, there is an energy gap between the ground state and the first excited state. 
In the non-static case there will be a set of states 
$\{n_{\rm us}\}$ whose energies $E_{n_{\rm us}}$ lie much below the energy of the first excited state in the static case.
We denote these states as US. The aim of pNRQCD is to describe the behaviour of the
US states. Therefore, all the physical degrees of freedom in NRQCD with energies larger than 
$E_{n_{\rm us}}$ can be integrated out in order to obtain
pNRQCD. The available lattice calculations of the static spectrum (see Fig.~\ref{Masgap}) clearly show
that from small to moderately large values of ${\bf r}$ there is an energy  gap between 
the ground state and higher excitations. 
The ground state energy is known as the static QCD potential. 
If the binding energy of the heavy quarkonium state we are interested in is much 
lower than the first excitation of the static limit, we can integrate out all 
higher excitations of this limit and keep only the ground state, 
which will be represented by a singlet field 
whose static energy is given by the static QCD potential. 

Note finally, that for heavy quarkonium states whose binding energy is close to or 
above the region where higher excitations occur, the use of pNRQCD is not 
justified and one should stay at the NRQCD level. In the case of real QCD, 
the heavy-light meson pair threshold plays the role of a higher excitation.

\section{Potential NRQCD. The weak-coupling regime}
\label{sec:pNRQCDII}

\subsection{pNRQCD: the degrees of freedom}
The degrees of freedom of pNRQCD in the
weak-coupling regime ($|{\bf p}| \gg E \gtrsim \lQ $) are a quark-antiquark pair,
gluons and light quarks with the cut-offs
$\nu_{pNR}=\{\nu_p,\nu_{us}\}$. $\nu_p$ is the cut-off of the relative
three-momentum of the heavy quarks and $\nu_{us}$ is the cut-off of 
the energy of the heavy quark-antiquark pair and of the four momentum 
of the gluons and light quarks. They satisfy the following
inequalities: $|{\bf p}| \ll \nu_p \ll m$ and ${\bf p}^2/m \ll
\nu_{us} \ll |{\bf p}|$.

The degrees of freedom of pNRQCD can be represented by the same fields as in NRQCD.
The main difference with respect to the
NRQCD Lagrangian will be that now non-local terms in space (namely,
potentials) are allowed. This representation is suitable for explicit
perturbative matching calculations. However, in order to establish a power counting,
it is more convenient to represent the quark-antiquark pair by a wavefunction
field
\be
 \Psi ({\bf x}_1, {\bf x}_2, t)_{\alpha\beta}
 \sim
 \psi_{\alpha} ({\bf x}_1, t) \chi_{\beta}^\dagger ({\bf x}_2, t)
 \sim
 {1 \over N_c}\delta_{\alpha\beta}\psi_{\sigma} ({\bf x}_1, t)
 \chi_{\sigma}^\dagger ({\bf x}_2, t)
 +
 {1 \over T_F} T^a_{\alpha\beta}T^a_{\rho\sigma}\psi_{\sigma} ({\bf x}_1, t)
 \chi_{\rho}^\dagger ({\bf x}_2, t)
 \,.
 \label{Psipsichi}
 \ee
This can be rigorously achieved in a NR system: (i) time is
universal, and hence one can constrain oneself to calculating correlators in
which the time coordinate of the quark field coincides with the time
coordinate of the antiquark field, (ii) since particle and antiparticle numbers
are separately conserved, if we are interested in the one-heavy-quark 
one-heavy-antiquark sector, there is no loss of generality if we project our
 theory to that subspace of the Fock space, which is described by the wave
function field $\Psi ({\bf x}_1, {\bf x}_2, t)$. Furthermore, this wave
function field can be uniquely decomposed into singlet and octet field components with
homogeneous (US) gauge transformations with respect 
to the centre-of-mass coordinate:
\be
\Psi ({\bf x}_1 ,{\bf x}_2 , t)= 
P\bigl[e^{ig\int_{{\bf x}_2}^{{\bf x}_1} {\bf A} \cdot 
                   d{\bf x}} \bigr]\;{\rm S}({{\bf r}}, {{\bf R}}, t)
+P\bigl[e^{ig\int_{{\bf R}}^{{\bf x}_1} {\bf A} \cdot  d{\bf x}}
\bigr]\; {\rm O} ({\bf r} ,{\bf R} , t) \;
P\bigl[e^{ig\int^{{\bf R}}_{{\bf x}_2} {\bf A} \cdot d{\bf x}}
\bigr]
\,.
\label{SinOct}
\ee
$P$ stands for path ordered. Under (US) color gauge transformations ($g({\bf R} ,t)$), we have
\be
{\rm S} ({\bf r} ,{\bf R} , t)\rightarrow {\rm S} ({\bf r} ,{\bf R} , t)
\,, \qquad
O ({\bf r} ,{\bf R} , t)\rightarrow g({\bf R} ,t)O ({\bf r} ,{\bf R} ,
t)g^{-1}({\bf R} ,t)
\label{Ogaugetrans}
\,.
\ee
Using these fields has the advantage that the relative coordinate ${\bf r}$
is explicit, and hence the fact that ${\bf r}$ is much smaller than the
typical  length of the light degrees of freedom
 can be easily implemented via a multipole expansion. This implies that the gluon
fields will always appear evaluated at the centre-of-mass coordinate. Note
that this is nothing but translating to real space the constraint $\nu_p \gg
\nu_{us}$. In addition, if we restrict ourselves to the singlet field only, we
are left with a theory which is totally equivalent to a quantum-mechanical
Hamiltonian. The whole theory however will contain singlet-to-octet
transitions mediated by the emission of an US gluon, which cannot be
encoded in any quantum-mechanical Hamiltonian.

\subsection{Power counting}
\label{pNRweakcounting}
The power counting of the pNRQCD
Lagrangian is easier to establish when it is written in terms of singlet and
octet fields. Since quark and antiquark particle numbers are separately
conserved, the Lagrangian will be bilinear in these fields and, hence, we only
have to estimate the size of the terms multiplying those bilinears. $m$ and
$\als (m)$, inherited from the hard matching coefficients, have well-known
values. Derivatives with respect to the relative coordinate $i\bfnabla_{\bf
r}$ and $1/r \sim k$ (the transfer momentum) must be assigned the soft scale 
$\sim |{\bf p}|$. 
Time derivatives
$i\partial_0$, centre-of-mass derivatives $i\bfnabla_{\bf R}$, and the fields
of the light degrees of freedom must be assigned the US scale $E\sim
{\bf p}^2/m$. The $\als$ arising in the matching calculation from NRQCD, namely
those in the potentials, must be assigned the size $\als (1/r)$ and those
associated with the light degrees of freedom (gluons, at lower orders) the size $\als
(E)$. If $\lQ$ did not exist (like in QED) this would provide a homogeneous
counting in which each term has a well-defined size. If $E\sim \lQ$  (recall
that then $\als (E)\sim 1$) this is also true, but calculations at the
US scale cannot be done in perturbation theory in $\als (E)$
anymore. If $E\gg \lQ$, the counting becomes inhomogeneous (i.e. it is not
possible to assign a priori a unique size to each term) since the light
degrees of freedom may have contributions both at the scale $E$ and at the
scale $\lQ$ (see sec.~\ref{pNRweakObservables}). Nevertheless, 
the largest size a term may have can be estimated
identically as before.

\subsection{Lagrangian and symmetries}
\label{sec:pNRLagW}
The degrees of freedom of pNRQCD can be arranged in several ways and so accordingly 
can the pNRQCD Lagrangian. We first write it in terms of quarks and gluons, 
which allows a smooth connection with the NRQCD chapter. One of the 
most distinct features of the pNRQCD Lagrangian is the appearance of the terms
$V$, non-local in $r$, as matching coefficients of 4-fermion operators:
\bea
\label{lpot}
L_{\rm pot} &=& -\int d^3{\bf x}_1 d^3{\bf x}_2 \;
\psi^{\dagger} (t,{\bf  x}_1)  \chi(t,{\bf  x}_2)
\;
V({\bf r}, {\bf p}_1, {\bf p}_2, {\bf S}_1,{\bf S}_2)
\\
\nn
&&
\times (\hbox{US gluon fields}) 
\;
\chi^\dagger(t,{\bf  x}_2) \psi(t,{\bf  x}_1)
\,,
\eea
where ${\bf p}_j = -i \bfnabla_{{\bf x}_j}$, for $j = 1,2$, and ${\bf S}_1= {\bfsigma}_1/2$, 
${\bf S}_2={\bfsigma}_2/2$ act on the fermion and antifermion, respectively 
(the fermion and antifermion spin indices are contracted with the 
indices of $V$, which are not explicitly displayed). Typically, US 
gluon fields show up at higher order.
With this new term the pNRQCD Lagrangian can be written in the following way
\be
L_{\rm pNRQCD}=L_{\rm NRQCD}^{\rm US}+L_{\rm pot}
\,,
\label{lagpnrqcda}
\ee
where $L_{\rm NRQCD}^{\rm US}$ has the form of the NRQCD Lagrangian, 
but all the gluons must be understood as US.
This way of writing the pNRQCD Lagrangian is advantageous for calculating the
matching potentials straightforwardly by means of standard 
Feynman diagram techniques. 
On the other hand, for the study of heavy quarkonium, 
it is convenient, before calculating physical quantities,  
to project the above Lagrangian onto the quark-antiquark sector 
of the Fock space. This makes  the 
multipole expansion explicit at the Lagrangian level, and it may also be useful 
at the matching level, depending on how it is done. An example is the 
matching via Wilson loops discussed in sec.~\ref{sec:matchingII}.
The projection onto the quark-antiquark sector is easily  done at the
Hamiltonian level by projecting onto the subspace spanned by
\be
\int d^3{\bf x}_1 d^3{\bf x}_2 \, \Psi ({\bf x}_1,{\bf x}_2 ) \, \psi^{\dagger}({\bf x}_1)
\chi({\bf x}_2) \vert \hbox{US gluons} \rangle
\,,
\ee
where $\vert \hbox{US gluons}\rangle$ is a generic state belonging to 
the Fock subspace with no quarks and antiquarks but an arbitrary number of
US gluons. The pNRQCD Lagrangian then has the form:
\bea
L_{\rm pNRQCD} &=&
\int d^3{\bf x}_1 \, d^3{\bf x}_2 \; 
{\rm Tr}\, \left\{\Psi^{\dagger} (t,{\bf
  x}_1 ,{\bf x}_2 ) \left(
iD_0  +{{\bf D}_{{\bf x}_1 }^2\over 2\, m_1}+{{\bf D}_{{\bf x}_2 }^2\over 2\,
  m_2} + \cdots \right)\Psi (t,{\bf x}_1 ,{\bf x}_2 )\right\}
\nn
\\
&-&
\int d^3 x \; {1\over 4} G_{\mu \nu}^{a}(x) \,G^{\mu \nu \, a}(x) + 
\int d^3 x \;
\sum_{i=1}^{n_f} \bar q_i(x) \, i \dsl \,q_i(x)
+ \cdots
\nn
\\
&+& 
\int d^3{\bf x}_1 \, d^3{\bf x}_2 \; 
{\rm Tr} \left\{ \Psi^{\dagger} (t,  {\bf x}_1,{\bf x}_2)\,
V( {\bf r}, {\bf p}_1, {\bf p}_2, {\bf S}_1,{\bf S}_2)
\right.
\nn
\\
&&
\left.
\times 
(\hbox{US gluon fields}) \,
\Psi(t, {\bf x}_1, {\bf x}_2 )
\right\},
\label{lagpnrqcdb} 
\eea
where the first two lines stand for the NRQCD Lagrangian  
projected onto the quark-antiquark sector and 
\be
iD_0 \Psi (t,{\bf x}_1 ,{\bf x}_2)
= i\partial_0\Psi (t,{\bf x}_1 ,{\bf x}_2 ) -g A_0(t,{\bf x}_1)\,  
\Psi (t,{\bf x}_1 ,{\bf x}_2) + \Psi (t,{\bf x}_1 ,{\bf x}_2)\, g A_0(t,{\bf
  x}_2).
\ee 
The dots in Eq.~(\ref{lagpnrqcdb}) stand for higher terms in the $1/m$ expansion.
The last two lines contain the 4-fermion terms specific of pNRQCD. 
In general also US gluon fields may appear 
there, but the leading term (in $\als$, $1/m$ and in the multipole expansion) 
is simply given by the Coulomb law (one gluon exchange):
\be
{\als \over |{\bf x}_1 - {\bf x}_2|}\, {\rm Tr} \, \biggl( 
T^{a} \, \Psi^{\dagger}  (t,{\bf x}_1 ,{\bf x}_2 ) \, T^{a}\Psi(t,{\bf x}_1 ,
{\bf x}_2 )\biggr).
\label{coulcol}
\ee

We can enforce the gluons to be US by multipole expanding them in ${\bf
  r}$. In the case of the covariant derivatives in (\ref{lagpnrqcdb}) this 
  corresponds to:
\bea 
&&
iD_0 \Psi (t,{\bf x}_1 ,{\bf x}_2) = 
i\partial_0\Psi (t,{\bf x}_1 ,{\bf x}_2) -[g A_0(t,{\bf R}), \Psi (t,{\bf x}_1
  ,{\bf x}_2)]
\nn
\\
&& \qquad\qquad
- {1\over 2} {\bf r}^i \, (\partial_i g A_0(t,{\bf R})) \, \Psi (t, {\bf x}_1
  ,{\bf x}_2)
- {1\over 2} {\bf r}^i \Psi (t, {\bf x}_1 ,{\bf x}_2) \, 
(\partial_i g A_0(t,{\bf R})) 
+ {\cal O}(r^2),
\label{D0multipole}
\\
&&
i{\bf D}_{{\bf x}_{1(2)}}  \Psi (t,{\bf x}_1 ,{\bf x}_2) = 
\left(+(-) i\bfnabla_{\bf r} + {i\over 2} \bfnabla_{\bf R} 
+ g {\bf A}(t,{\bf R}) +(-) {{\bf r}^i\over 2} \, (\partial_i g {\bf A}(t,{\bf R}))
\right) \Psi (t,{\bf x}_1 ,{\bf x}_2) 
\nn
\\
&& \qquad\qquad
+ {\cal O}(r^2) .
\label{Dvecmultipole}
\eea 
From now on, all the gluon (and light-quark) fields will be understood as functions of $t$ and ${\bf R}$. 
We will not always explicitly display this dependence.
According to the power counting given in the previous section, the multipole
expansion makes explicit the size of each term in the Lagrangian. 
On the other hand, expansions like (\ref{D0multipole}) and (\ref{Dvecmultipole}) 
spoil the manifest gauge invariance of the Lagrangian. 
This may be restored by introducing singlet and octet fields as in 
Eq.~(\ref{SinOct}). We choose the following normalization with respect to color:
\be
{\rm S} = { S 1\!\!{\rm l}_c / \sqrt{N_c}} \quad , \quad\quad\quad 
{\rm O} = O^a { {\rm T}^a / \sqrt{T_F}}.
\label{SSOO}
\ee
We will not always explicitly display their dependence on ${\bf R}$, {\bf r} and $t$
in the following.
After multipole expansion, the pNRQCD Lagrangian may be organized as an
expansion in $1/m$ and $r$ (and $\als$). The most general pNRQCD Lagrangian density, 
compatible with the symmetries of QCD, that can be constructed
with a singlet field, an octet field and US gluon fields up to order $p^3/m^2$ 
(see sec.~\ref{pNRweakcounting}) has the form:
\bea
& & \!\!\!\!\!\!\!
{\cal L}_{\rm pNRQCD} = \int d^3{\bf r} \; {\rm Tr} \,  
\Biggl\{ {\rm S}^\dagger \left( i\partial_0 
- h_s({\bf r}, {\bf p}, {\bf P}_{\bf R}, {\bf S}_1,{\bf S}_2) \right) {\rm S} 
+ {\rm O}^\dagger \left( iD_0 
- h_o({\bf r}, {\bf p}, {\bf P}_{\bf R}, {\bf S}_1,{\bf S}_2) \right) {\rm O} \Biggr\}
\nn
\\
& &\qquad\qquad 
+ V_A ( r) {\rm Tr} \left\{  {\rm O}^\dagger {\bf r} \cdot g{\bf E} \,{\rm S}
+ {\rm S}^\dagger {\bf r} \cdot g{\bf E} \,{\rm O} \right\} 
+ {V_B (r) \over 2} {\rm Tr} \left\{  {\rm O}^\dagger {\bf r} \cdot g{\bf E} \, {\rm O} 
+ {\rm O}^\dagger {\rm O} {\bf r} \cdot g{\bf E}  \right\}  
\nn
\\
& &\qquad\qquad 
- {1\over 4} G_{\mu \nu}^{a} G^{\mu \nu \, a} 
+  \sum_{i=1}^{n_f} \bar q_i \, i \dsl \, q_i 
\,,
\label{Lpnrqcd}
\\
& &
\nn 
\\
& &
h_s({\bf r}, {\bf p}, {\bf P}_{\bf R}, {\bf S}_1,{\bf S}_2) = 
\{ c_S^{(1,-2)}(r), {{\bf p}^2 \over 2\, m_{\rm red}} \} 
+ c_S^{(1,0)}(r) 
{{\bf P}_{\bf R}^2 \over 2\, m_{\rm tot}} + 
V_s({\bf r}, {\bf p}, {\bf P}_{\bf R}, {\bf S}_1,{\bf S}_2), 
\\
& & 
h_o({\bf r}, {\bf p}, {\bf P}_{\bf R}, {\bf S}_1,{\bf S}_2) = 
\{ c_O^{(1,-2)}(r), {{\bf p}^2 \over 2\, m_{\rm red}} \} 
+  c_O^{(1,0)}(r) 
{{\bf P}_{\bf R}^2 \over 2\, m_{\rm tot}} + 
V_o({\bf r}, {\bf p}, {\bf P}_{\bf R}, {\bf S}_1,{\bf S}_2), 
\\
&&
\nn
\\
&& V_s = 
V^{(0)}_s + {V^{(1,0)}_s \over m_1}+{V^{(0,1)}_s\over m_2}
+ {V^{(2,0)}_s \over m_1^2}+ {V^{(0,2)}_s\over m_2^2}+{V^{(1,1)}_s \over m_1m_2},
\label{V1ovm2}
\\
&& V_o = 
V^{(0)}_o + {V^{(1,0)}_o \over m_1}+{V^{(0,1)}_o\over m_2}
+ {V^{(2,0)}_o \over m_1^2}+ {V^{(0,2)}_o\over m_2^2}+{V^{(1,1)}_o \over m_1m_2},
\eea
where $iD_0 {\rm O} \equiv i \partial_0 {\rm O} - g [A_0({\bf R},t),{\rm O}]$, 
${\bf P}_{\bf R} = -i{\bf D}_{\bf R}$, ${\bf p} = -i\bfnabla_{\bf r}$,
$m_{\rm red} = m_1 \, m_2/ m_{\rm tot}$ and $m_{\rm tot} = m_1 + m_2$. 
When acting between singlet fields, the color trace reduces 
${\bf P}_{\bf R}$ to $-i\bfnabla_{\bf R}$. According to the order 
at which we are working,  
the potentials have been displayed up to terms of order $1/m^2$.
The static and the $1/m$ potentials are real-valued functions of $r$ only.
The $1/m^2$ potentials have an imaginary part proportional to
$\delta^{(3)}({\bf r})$ and a real part that may be decomposed as (we drop the labels 
$s$ and $o$ for singlet and octet, which have to be understood):
\bea
&&
V^{(2,0)}=V^{(2,0)}_{SD}+V^{(2,0)}_{SI}, \qquad 
V^{(0,2)}=V^{(0,2)}_{SD}+V^{(0,2)}_{SI}, \qquad 
V^{(1,1)}=V^{(1,1)}_{SD}+V^{(1,1)}_{SI},
\label{decomSDSI}
\\
\nn
\\
&& 
V^{(2,0)}_{SI}={1 \over 2}\left\{{\bf p}_1^2,V_{{\bf p}^2}^{(2,0)}(r)\right\}
+{V_{{\bf L}^2}^{(2,0)}(r)\over r^2}{\bf L}_1^2 + V_r^{(2,0)}(r),
\label{v20sistrong}
\\
&&
V^{(0,2)}_{SI}={1 \over 2}\left\{{\bf p}_2^2,V_{{\bf p}^2}^{(0,2)}(r)\right\}
+{V_{{\bf L}^2}^{(0,2)}(r)\over r^2}{\bf L}_2^2 + V_r^{(0,2)}(r),
\\
&&
V^{(1,1)}_{SI}= -{1 \over 2}\left\{{\bf p}_1\cdot {\bf p}_2,V_{{\bf p}^2}^{(1,1)}(r)\right\}
-{V_{{\bf L}^2}^{(1,1)}(r)\over 2r^2}({\bf L}_1\cdot{\bf L}_2+ {\bf L}_2\cdot{\bf L}_1)+ V_r^{(1,1)}(r),
\\
\nn
\\
&&
V^{(2,0)}_{SD}=V^{(2,0)}_{LS}(r){\bf L}_1\cdot{\bf S}_1, 
\\
&&
V^{(0,2)}_{SD}=-V^{(0,2)}_{LS}(r){\bf L}_2\cdot{\bf S}_2,
\\
&&
V^{(1,1)}_{SD}=
V_{L_1S_2}^{(1,1)}(r){\bf L}_1\cdot{\bf S}_2 - V_{L_2S_1}^{(1,1)}(r){\bf L}_2\cdot{\bf S}_1
+ V_{S^2}^{(1,1)}(r){\bf S}_1\cdot{\bf S}_2 + V_{{\bf S}_{12}}^{(1,1)}(r){\bf
  S}_{12}({\hat {\bf r}}),
\label{v11sdstrong}
\eea
where, ${\bf S}_1=\bfsigma_1/2$, ${\bf S}_2=\bfsigma_2/2$, ${\bf L}_1 \equiv {\bf r} \times {\bf p}_1$, ${\bf L}_2 \equiv {\bf
  r} \times {\bf p}_2$ and ${\bf S}_{12}({\hat {\bf r}}) \equiv 
3 {\hat {\bf r}}\cdot \bfsigma_1 \,{\hat {\bf r}}\cdot \bfsigma_2 - \bfsigma_1\cdot \bfsigma_2$.
The pNRQCD Lagrangian density at order $r^2/m^0$, $r^0/m$,
$(r/m) \, {\bf P}_{\bf R}$ and $(r^0/m^2) \, {\bf P}_{\bf R}$ 
and the corresponding matching coefficients at tree level can be found 
in \cite{Brambilla:2003nt}.

For the case $m_1=m_2=m$, the potential has the following structure,
\bea
V(r)&=&V^{(0)}(r) +{V^{(1)}(r) \over m}+{V^{(2)} \over m^2}+\cdots,  
\label{ppot}
\\
V^{(2)}&=&V^{(2)}_{SD}+V^{(2)}_{SI},\nn\\
V^{(2)}_{SI}
&=&
{1\over 8}\{ {\bf P}_{\bf R}^2, V^{(2)}_{{\bf p}^2,{\rm CM}}(r)\}
+
{({\bf r}\times {\bf P}_{\bf R})^2\over 4 r^2} V^{(2)}_{{\bf L}^2,{\rm CM}}(r)
\nn
+{1 \over 2}\left\{{\bf p}^2,V_{{\bf p}^2}^{(2)}(r)\right\}
+{V_{{\bf L}^2}^{(2)}(r)\over r^2}{\bf L}^2 + V_r^{(2)}(r),
\\
V^{(2)}_{SD} &=&
{({\bf r}\times {\bf P}_{\bf R})\cdot ({\bf S}_1 - {\bf S}_2) \over 2}
V^{(2)}_{LS,{\rm CM}}(r)
+
V_{LS}^{(2)}(r){\bf L}\cdot{\bf S} + V_{S^2}^{(2)}(r){\bf S}^2
 + V_{{\bf S}_{12}}^{(2)}(r){\bf S}_{12}({\hat {\bf r}}), \nn
\eea
${\bf S}={\bf S}_1+{\bf S}_2$ and ${\bf L}={\bf r}\times {\bf p}$.
Other forms of the potential can be brought to the one above by using 
unitary transformations, or the relation
\be
- \left\{ {1 \over r},{\bf p}^2 \right\} + 
{1 \over r^3}{\bf L}^2 + 4\pi\delta^{(3)}({\bf r})
= - {1 \over r} \left( {\bf p}^2 + 
{ 1 \over r^2} {\bf r} \cdot ({\bf r} \cdot {\bf p}){\bf p} \right).
\ee

From Eq.~(\ref{Lpnrqcd}) we see that the relative coordinate ${\bf r}$ plays the role 
of a continuous parameter, which specifies different fields. Moreover, we note
that the Lagrangian is now in an explicitly gauge invariant form. This is a 
consequence of the transformation properties (\ref{Ogaugetrans}) of 
the singlet and octet fields and of the fact that the 
gluon fields depend on $t$ and ${\bf R}$ only. The functions 
$V_s$, $V_o$, $c_S^{(1,-2)}$, $c_O^{(1,-2)}$, $c_S^{(1,0)}$, $c_O^{(1,0)}$, 
$V_A$ and $V_B$ are the matching coefficients of the effective theory.
At leading order it follows  from (\ref{D0multipole})  that
$V_A = V_B = 1$, from (\ref{Dvecmultipole}) that 
$c_S^{(1,-2)} = c_O^{(1,-2)} = c_S^{(1,0)} = c_O^{(1,0)} =1$, 
and from (\ref{coulcol}) that 
$V_s^{(0)} = -C_F \,\als /r$ and $V_o^{(0)} = 1/(2\,N_c) \,\als /r$.

Equations (\ref{lagpnrqcda}), (\ref{lagpnrqcdb}) and (\ref{Lpnrqcd}) provide
three different ways to write the pNRQCD Lagrangian. We have also shown how to
derive one from the other. This works (and is useful) at tree level.  In
general, each form of the pNRQCD Lagrangian may be constructed independently of the others by
identifying the degrees of freedom, using symmetry arguments and matching directly to NRQCD.

The expressions for the currents in pNRQCD are equal to those of NRQCD
with the replacements: NR $\rightarrow$ pNR and $\nu \rightarrow
\nu_{pNR}$. In particular, this applies to Eqs.~(\ref{NRcurrent}) and (\ref{NRcurrenta}). As in
NRQCD, most of the physical information can be extracted from the
imaginary part of the potentials, which are proportional to the imaginary
part of the NRQCD 4-fermion mathing coefficients.
Therefore, the imaginary part of the (singlet or octet) potential
will have the following structure (with only local potentials,
delta functions or derivatives of delta functions):
\be
\label{imhpert}
{\rm Im} \, V =
{{\rm Im} \, V^{(2)} \over m^2} +{{\rm Im} \, V^{(4)} \over m^4}  +
\cdots \;,
\ee
where the explicit expressions for ${\rm Im} \, V^{(2)}$ and ${\rm Im}
\, V^{(4)}$
are:
\bea
{\rm Im} \, V^{(2)}&=&-{C_A \over 2}\delta^{(3)}({\bf r})
\Bigg(
4\, {\rm Im} \, f_1^{\rm pNR}(^1 S_0)
-2\,{\bf S}^2\left({\rm Im}\, f_1^{\rm pNR}(^1 S_0)-{\rm Im}\,
f_1^{\rm pNR}(^3 S_1)\right)
\nn
\\
&&
\qquad\qquad
+ 4\,{\rm Im}\, f_{\rm EM}^{\rm pNR}(^1 S_0)
-2\,{\bf S}^2\left({\rm Im} \, f_{\rm EM}^{\rm pNR}(^1 S_0)-{\rm Im}\,
f_{\rm EM}^{\rm pNR}(^3 S_1)\right)
\Bigg),
\label{imh2pert}
\eea
\bea
&&
{\rm Im} \, V^{(4)}
=
C_A \, {\cal T}^{ij}_{SJ} \bfnabla_{\bf r}^i\delta^{(3)}({\bf
r})\bfnabla_{\bf r}^j \,
\left({\rm Im}\,f_1^{\rm pNR}({}^{2S+1}P_J)+{\rm Im}\,f_{\rm
    EM}^{\rm pNR}({}^{2S+1}P_J)\right)
\nn
\\
&&\qquad\qquad
+{C_A\over 2}\,
\Omega^{ij}_{SJ}\bigg\{ \bfnabla_{\bf r}^i\bfnabla_{\bf r}^j
, \delta^{(3)}({\bf r}) \bigg\} \,
\left({\rm Im}\,g_1^{\rm pNR}({}^{2S+1}S_J)
+{\rm Im}\,g_{\rm EM}^{\rm pNR}({}^{2S+1}S_J)\right)
,
\label{imh4pert}
\eea
\bea
{\cal T}^{ij}_{01} &=& \delta_{ij}(2\one-{\bf S}^2), \label{defT1}\\
{\cal T}^{ij}_{10} &=& {1\over 3} {\bf S}^i \, {\bf S}^j, \\
{\cal T}^{ij}_{11} &=& {1\over 2}\epsilon_{ki\ell}\, \epsilon_{kj\ell'} \,
{\bf S}^\ell \, {\bf S}^{\ell'}, \\
{\cal T}^{ij}_{12} &=&
\left({\delta_{ik}{\bf S}^\ell + \delta_{i\ell}{\bf S}^k \over 2}
- {{\bf S}^i\delta_{k\ell}\over 3}\right)
\left({\delta_{jk}{\bf S}^\ell + \delta_{j\ell}{\bf S}^k \over 2}
- {{\bf S}^j\delta_{k\ell}\over 3}\right), \label{defT2} \\
\Omega^{ij}_{00} &=& \delta_{ij}(2\one-{\bf S}^2), \qquad
\Omega^{ij}_{11} = \delta_{ij} \, {\bf S}^2,
\label{defOmega}
\eea
and we have omitted the labels singlet/octet in the matching coefficients for simplicity.
Note that we use a notation for the matching coefficients similar to the one used 
in NRQCD, but this does not imply that the matching coefficients are equal.

\subsubsection{Discrete symmetries and Poincar\'e invariance}
The pNRQCD Lagrangian is invariant under charge conjugation plus $1
\leftrightarrow 2$ exchange (\ref{cca}), time reversal (\ref{trb})
and parity (\ref{ppc}). In particular singlet, octet and gluon 
fields transform under these as:
\bea
 \!\!\!\!\!\!\!\!\!\!\!\!\!\!\!\! 
&& 
{\rm S}({\bf r}, {\bf R},t)\! \rightarrow \!\sigma^2  {\rm S}(-{\bf r}, {\bf
  R},t)^T  \sigma^2, 
\,\,
{\rm O}({\bf r}, {\bf R},t)\! \rightarrow \sigma^2  {\rm O}(-{\bf r}, {\bf
  R},t)^T  \sigma^2, 
\,\,
A_\mu({\bf R},t) \rightarrow - A_\mu({\bf R},t)^T,
\label{cca}
\\
 \!\!\!\!\!\!\!\!\!\!\!\!\!\!\!\!
 &&  
{\rm S}({\bf r}, {\bf R},t) \rightarrow \!\sigma^2 \, {\rm S}({\bf r}, {\bf
  R},-t) \, \sigma^2, 
\>\>
{\rm O}({\bf r}, {\bf R},t) \rightarrow \sigma^2 \, {\rm O}({\bf r}, {\bf
  R},-t) \, \sigma^2, 
\!\!\!\quad
A_\mu({\bf R},t) \rightarrow A^\mu({\bf R},-t),
\label{trb}
\\
\!\!\!\!\!\!\!\!\!\!\!\!\!\!\!\!
 && 
{\rm S}({\bf r}, {\bf R},t) \! \rightarrow \!- {\rm S}(-{\bf r}, -{\bf R},t), 
\quad\>
{\rm O}({\bf r}, {\bf R},t) \rightarrow - {\rm O}(-{\bf r}, -{\bf R},t),
\!\!\!\!\qquad 
A_\mu({\bf R},t) \rightarrow A^\mu(-{\bf R},t).
\label{ppc}
\eea
Singlet and octet field transformations may be derived from Eq.~(\ref{Psipsichi}).

The discrete symmetries constrain the form of the Lagrangian. As 
an example we observe that the charge conjugate of
$\int d^3{\bf r}\; {\rm Tr} \left\{  {\rm O}^\dagger {\bf r} \cdot g{\bf E} \,
{\rm O} \right\}$ is 
$\int d^3{\bf r}\; {\rm Tr} \left\{  {\rm O}^\dagger {\rm O} {\bf r} \cdot g{\bf E}  \right\}$ 
and, therefore, only the sum of the two appears in the Lagrangian. For 
a similar reason the term 
$\int d^3{\bf r}\; {\rm Tr} \left\{  {\rm S}^\dagger ({\bf r} \times {\bf p} \cdot g{\bf
  B}) \,{\rm S} \right\}/m$ cannot appear, while the combination
$\int d^3{\bf r}\; {\rm Tr} \left\{  
{\rm O}^\dagger ({\bf r} \times {\bf p}  \cdot g{\bf B})\,{\rm O} 
- {\rm O}^\dagger {\rm O} \, ({\bf r} \times {\bf p}  \cdot g{\bf B}) 
\right\}/m$ is possible.

As in NRQCD, also the form of the pNRQCD Lagrangian may be constrained 
by imposing the Poincar\'e  algebra of the generators $H$, ${\bf P}$, ${\bf J}$ and ${\bf K}$ 
of time translations, space translations, 
rotations, and Lorentz boosts of the EFT \cite{Brambilla:2003nt}.
$H$ is the pNRQCD Hamiltonian. The translation and
rotation generators ${\bf P}$ and ${\bf J}$ may be derived from the pNRQCD
Lagrangian or by matching to the NRQCD generators. They are exact, because
translational and rotational invariance have not been broken in
going to the EFT.  The Lorentz-boost generators may be obtained by matching to
the Lorentz-boost generators of NRQCD.  As can be seen from the explicit
expressions given in \cite{Brambilla:2003nt}, they depend on some specific
matching coefficient independent of those in the Lagrangian. The tree-level
matching may be performed by multipole expanding the NRQCD Lorentz-boost
generators and projecting onto singlet and octet two-particles states.  Loop
corrections can, in principle, be calculated as has been done for the matching
coefficients of the pNRQCD Lagrangian.

Imposing the Poincar\'e algebra on the above generators 
constrains the form of the pNRQCD Lagrangian. For the constraints 
on the Lorentz-boost generators, see \cite{Brambilla:2003nt}. 
For what concerns the Lagrangian, the constraints 
\bea
&&
c_{S}^{(1,0)} = c_{O}^{(1,0)} = 1 
\label{kin}
\eea
fix the centre-of-mass kinetic energy to be equal to ${\bf P}_{\bf
  R}^2/4m$. The coefficient of the kinetic energy ${\bf
  p}^2/m$,  $c_{S}^{(1,-2)}$, is not fixed by 
Poincar\'e invariance.  However, one may argue that, because no other momentum-dependent operator 
than the kinetic energy of NRQCD, $- \psi^\dagger \, \bfnabla^2/(2m) \, \psi + 
\chi^\dagger \, \bfnabla^2/(2m) \, \chi$, may contribute to the kinetic energy of
pNRQCD, the coefficients $c_{S}^{(1,0)}$ and $c_{S}^{(1,-2)}$
have to be equal. It follows then that also $c_{S}^{(1,-2)}=1$\footnote{
One may also obtain $c_{S}^{(1,-2)}=1$ by a direct non-perturbative  matching computation,
as it has been done in \cite{Brambilla:2000gk}. The relevant steps of that
calculation are reproduced in Eqs. (\ref{vsnrqcd})-(\ref{g10}).
The kinetic energy operator may be read from the ratio of the $1/m$ Green
function (\ref{g10}) and the zeroth-order one (\ref{vsnrqcd}).} 
(analogously for $c_{O}^{(1,-2)}$).
In the singlet and octet potential sectors we obtain:
\be
{V_{LS,{\rm CM}} \over V^{(0)\prime}} =  -{1\over 2r},
\label{ex0}
\qquad
V_{{\bf L}^2,{\rm CM}} + {r \, V^{(0)\prime} \over 2} = 0,
\qquad
V_{{\bf p}^2,{\rm CM}} + V_{{\bf L}^2,{\rm CM}} + {V^{(0)} \over 2}= 0\,,
\ee
where $V^{\prime}=d\,V/dr$. 
We will come back to the relations between the singlet 
potentials in the strong-coupling regime in sec.~\ref{secrealpnrqcdpot}.
Finally, in the singlet-octet and octet-octet sectors of the Lagrangian,  
the chromoelectric fields are constrained to enter in the combination
\be 
{\bf r}\cdot \left(g{\bf E} + 
{1\over 2} \left\{ {{\bf P}_{\bf R}\over 2m} \times, g {\bf B}\right\} \right),
\ee 
i.e. like in the Lorentz force. Further constraints can be found in \cite{Brambilla:2003nt}.

\subsection{Feynman rules}
\label{sec:pNRFR}
\begin{figure}[htb]
\makebox[-5cm]{\phantom b}
\put(-150,1){\epsfxsize=13truecm \epsfbox{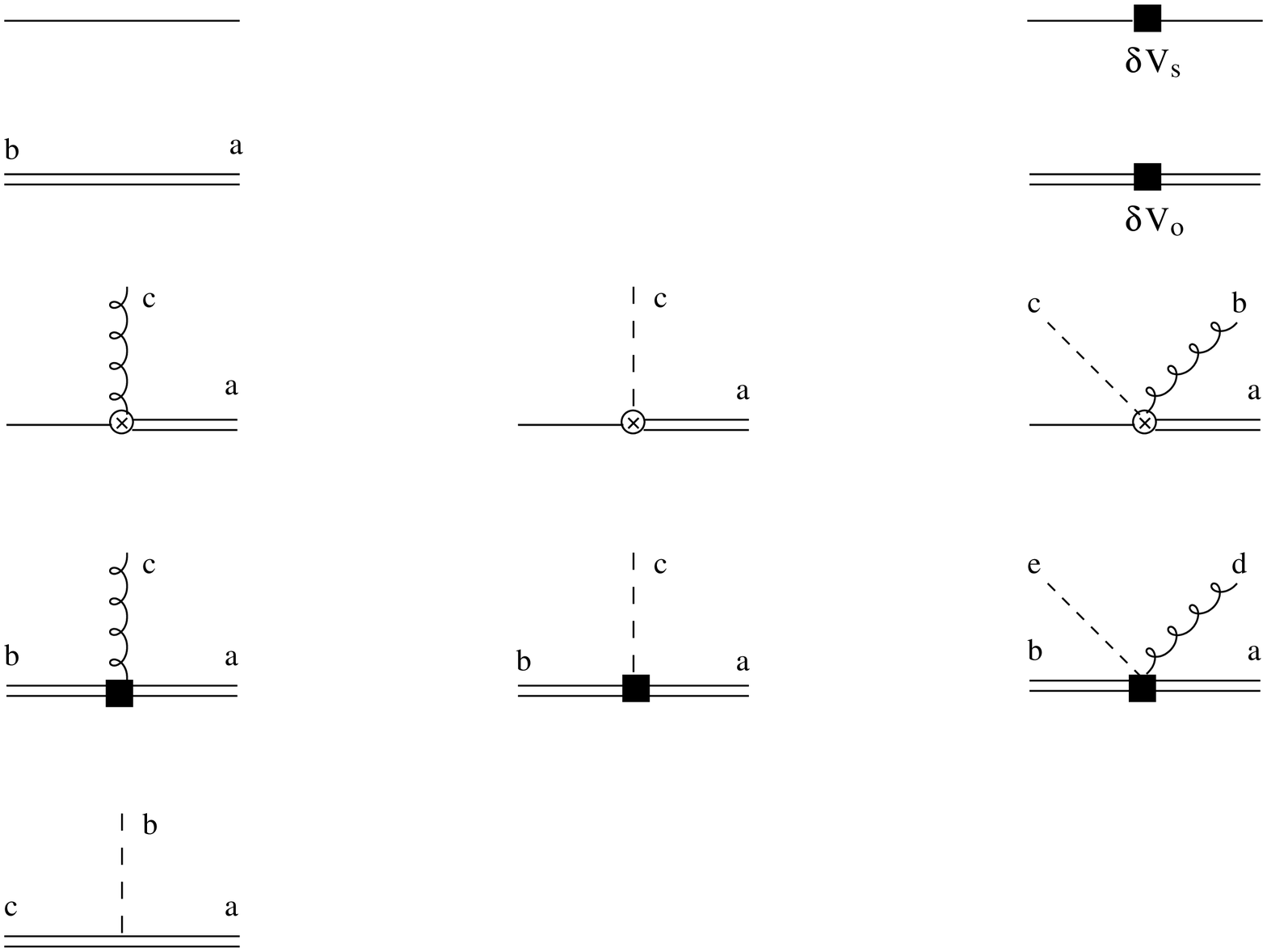}}
\put(-75,270){\tiny $=-iG_c(E)=\displaystyle{{i \over 
\displaystyle{E-h_s^{(0)}}}
=
{i \over \displaystyle{E-{\bf p}^2/m-C_F \als/ r}}}$}
\put(-75,225){\tiny $=-iG_c^{o}(E)\delta_{ab}=\displaystyle{{i 
\delta_{ab}\over 
\displaystyle{E-h_o^{(0)}}}
={i \delta_{ab} \over \displaystyle{E-{\bf p}^2/m-(1/(2N_c)) \als/ r}}}
$}
\put(225,270){\tiny $= - i \delta V_s $}
\put(225,225){\tiny $= - i \delta V_o $}
\put(75,153){\tiny $= g V_A \displaystyle\sqrt{T_F\over N_c} 
\delta_{ca}{\bf r} \cdot{\bf P}$}
\put(-75,153){\tiny $= -g V_A \displaystyle\sqrt{T_F\over N_c} 
\delta_{ca}{\bf r} P^0$}
\put(225,153){\tiny $=-ig^2V_A \displaystyle\sqrt{T_F\over N_c} {\bf 
r}f_{abc}$}
\put(75,75){\tiny $= g \displaystyle{V_B\over 2} d^{abc} {\bf r} \cdot 
{\bf P}$}
\put(-75,75){\tiny $=-g \displaystyle{V_B\over 2} d^{abc} {\bf r} P^0$}
\put(225,75){\tiny $=-ig^2\displaystyle{V_B\over 2} d^{abc} f_{cde}{\bf 
r}$}
\put(-75,0){\tiny $=  g f^{abc}$}
\caption{ \it Propagators and vertices of the pNRQCD Lagrangian 
(\ref{Lpnrqcd}). Dashed lines represent longitudinal gluons and 
curly lines transverse gluons. $P^\mu$ represents the gluon 
incoming momentum.}
\label{pnrqcdfig}
\end{figure}

The Feynman rules of pNRQCD for the 
static limit were given in \cite{Brambilla:1999xf} in terms 
of the time variable and background gluon fields. However, 
for computations in pNRQCD using Feynman diagrams, it is sometimes more useful to 
consider the Feynman rules in US momentum space (even if preserving
the relative distance ${\bf r}$ between the heavy quarks in position space). 
The propagator of the singlet is 
\be
{i \over \displaystyle{E-h_s}} \,.
\ee
This expression contains subleading terms in the velocity expansion. 
In order to have homogeneous power counting, it is 
convenient to expand it about the Coulomb Green function, $G_c$, 
defined in Fig.~\ref{pnrqcdfig}, which scales as $1/(mv^2)$, 
and similarly for the octet. The complete set of Feynman rules 
at the order displayed in (\ref{Lpnrqcd}) is shown in Fig.~\ref{pnrqcdfig}.

\subsection{Matching: diagrammatic approach}
\label{pNRmatchingI}
We discuss here how the matching between NRQCD and pNRQCD (in the formulation
of Eq.~(\ref{lagpnrqcda})) within a diagrammatic approach is made along the
lines of \cite{Pineda:1998bj,Pineda:1998ie,Pineda:1998kn}. This procedure is
specially convenient for obtaining the potentials order by order in $\als$, since
the whole technology of Feynman diagrams can be used.

A practical way to obtain the matching coefficients of pNRQCD is by enforcing
2- and 4-fermion Green functions with arbitrary US external gluons to
be equal to those of NRQCD at any desired order in $E/k$.  It is convenient to
expand the energy of the external quark and the energy and momenta
of the US gluons around zero {\it before} carrying out the loop integrals so that
the integrals become homogeneous in the soft scale and hence are easier to
evaluate.  This may produce IR divergences which are most conveniently (but
not necessarily) regulated in DR, in the same way as the UV divergences
are. Since the IR behavior of NRQCD and pNRQCD is the same, these IR
divergences will cancel out in the matching, provided the same IR regulator is
used in both theories. The UV divergences of NRQCD must be renormalized in the
$\MS$ scheme if we want to use the matching coefficients of the NRQCD Lagrangian
computed themselves in the $\MS$ scheme. We still have a choice in the
renormalization scheme of pNRQCD. However, it is most advantageous to use
again the $\MS$ scheme. Indeed, with this choice we can blindly subtract any
divergence regardless of whether it is UV or IR in the matching calculation. For the UV
divergences of NRQCD and pNRQCD, this just corresponds to our choice of scheme, and for the
IR divergences this is possible since, as long as we use the same treatment in both
theories, their IR behavior is the same. This allows to set integrals with
no scale equal to zero.

Notice that we demand off-shell Green functions in NRQCD and pNRQCD to be
equal and not on-shell Green functions (or on-shell matrix elements) as it is
usual in many matching calculations, in particular in matching
calculations from QCD to NRQCD. This is due to the fact that we are eventually
interested in bound states, and particles in a bound state are typically
off-shell. Being more precise, the equations of motion at lowest order are not
those of the free particles.  The equations of motion of pNRQCD (with
potential terms included), or local field redefinitions, may be consistently
used later on to remove time derivatives in higher order terms and to write the
pNRQCD Lagrangian in a standard form, in the philosophy advocated by
\cite{Scherer:1995wi} (see also \cite{Balzereit:1998jb}). It has actually been
checked by \cite{Pineda:1998kn}\footnote{However, there is still some freedom
in the choice of the wavefunction field, due to time independent unitary
transformations which commute with the leading terms in the pNRQCD
Lagrangian. Therefore, in general, it is not to be expected that the standard
forms of the pNRQCD Lagrangian calculated with different gauges coincide, but
that they are only related by one such unitary transformation. This explains the
different expressions for the potential that one may find in the literature
but which still lead to the same physics.} that this procedure produces gauge
independent results at ${\cal O}(m\als^4)$ in the computation of the
positronium spectrum.

The remaining important ingredient to carry out the matching efficiently is
the use of static (HQET) propagators for the fermions.  This can be justified
as follows. When $p^0\sim |{\bf p}|$ we are in the kinematical region we wish to
integrate out, and the cut-offs of both NRQCD and pNRQCD ensure that the
kinetic term ${\bf p}^2/2m$ will always be subleading with respect to the energy
irrespectively of the value of $|{\bf p}|$.  This fact is not automatically
implemented in DR.  When DR is used, the correct UV behavior of NRQCD is only
obtained when expanding about the static propagator.  When $p^0\sim {\bf p}^2/2m$,
we are in a kinematical region which still exists in pNRQCD, and hence it
should not be integrated out. The simplest way to avoid this kinematical
region is, again, by expanding the kinetic term.  After all these
simplifications the computations in the NRQCD side reduce to diagrams with
only one scale inside loops. In short, one would have (where $E$ generically
denotes the external momentum or the kinetic term ${\bf p}^2/m$)
\be
\int d^Dq \, f(q,k,E)= \int d^Dq \, f(q,k,0)+ {\cal O}\left({E \over k} \right)
\,.
\ee
Now we are in a position to prove that no pNRQCD diagram containing a loop
contributes to the matching calculation. Consider first the 2-fermion Green
function with an arbitrary number of US legs.  For potential terms to
contribute we need at least a 4-fermion Green function and hence we only have
to care about US gluons.  If we put a momentum $\sim |{\bf p}|$ in
the fermion line, this momentum cannot flow out through any external US
gluon line (by definition of US).  Then it must flow through the
fermion line, which is a series of static propagators insensitive to the
momentum flowing through them. Hence upon expanding about external fermion energies and
external energies and momenta of the US gluons there is no scale left in any
of the integrals and therefore any loop contribution vanishes. In fact, exactly the
same argument can be used for the NRQCD calculation. Then we conclude
that the terms bilinear in fermions are exactly the same in NRQCD and
pNRQCD. However, we have to keep in mind that the latter (by definition) must
be understood as containing US gluons only.

Consider next the 4-fermion Green function in pNRQCD containing several
potential terms but no US gluons. Since no energy can flow through the
potentials and the static propagators are insensitive to the momentum, upon
expanding about the US external energy, the integrals over internal energies
have no scale.  However, these integrals have IR (pinch) singularities, which
are not regulated by standard DR.  How to rigorously deal with them is
discussed in sec.~\ref{Pinch}.  Since the IR behavior of pNRQCD and NRQCD is
the same, the same kind of integrals appear in the NRQCD calculation. If we
{\it consistently} set them to zero we obtain the correct potential terms. It
is important to keep in mind that the Wilson coefficients compensate the
different UV behavior of the effective theory (pNRQCD) with respect to that of
the more 'fundamental' theory (NRQCD). Hence they are not sensitive to the details
of the IR behavior, which legitimates the prescription above.  Then any loop
diagram in pNRQCD with no US gluons can be set to zero.  This still holds if
an arbitrary number of US gluon lines is included in the diagram. Indeed, any
potential line in the diagram may now also contain US momenta from the gluon
lines. These, however, can be expanded about zero since they are (by
definition) much smaller than the momentum transfer in the potential. Hence
the integrals over US gluon energies and momenta contain no scale (again upon
expanding the US external energy in the fermion static propagators) and can
also be set to zero. In short, loops in pNRQCD will have the following
structure in general:
\be
\int d^Dq \, f(q,E)= \int d^Dq \, f(q,0)+{\cal O} \left({E \over k} \right)= 0\,.
\ee
In summary, we can directly identify the potential terms from a
 calculation in NRQCD. We would like to stress again the similarity in 
the procedure
 with the matching between QCD and NRQCD as carried out before.
 The potential terms in pNRQCD play the role of Wilson
coefficients in the matching procedure.
As a summary for the practitioner, the final set of rules are the following:

\begin{itemize}
\item
Compute (off-shell) NRQCD Feynman diagrams within an expansion in
$\als$, $1/m$ and $E$. In case loops appear, they have to be computed using
static propagators for the heavy quark and antiquark, which makes the
integrals depend on $k$ only.
\item
Match the resulting expression to the {\it tree level} expression in pNRQCD
(i.e. the potentials that appear in the pNRQCD Lagrangian) to the required
order in $\als$, $1/m$ and $E$.
\item
In case pinch singularities appear, one must isolate them in expressions which are identical 
to those which appear in the pNRQCD computation and set them to zero. Or, alternatively, one may just
subtract the pNRQCD diagrams with the same pinch singularity, as discussed in
sec.~\ref{Pinch} below.
\end{itemize}

Let us mention here, that when this procedure is used to match local NRQCD
4-fermion operators, these do not get any loop correction.  Indeed, due to the
use of HQET propagators, all NRQCD integrals become scaleless and hence
vanish. We often say that they are {\it inherited} in pNRQCD.

A word of caution is necessary concerning the procedure above. It heavily relies on
the fact that there are no further scales other than $m$, $k$ and $E$.  If,
for instance, an energy scale $m'$ such that $E\ll m'\sim k \ll m$ enters the
game, it would be convenient to take $\nu_{us} \ll m'$ rather than $\nu_{us}
\ll k$ and hence $\nu_{p} \ll \sqrt{m \,m'}$ rather than $\nu_{p} \ll
m$. Then, in the matching calculation we should also integrate out quarks
with energy $\sim m'$ and three-momentum $\sim \sqrt{m \,m'}$, which cannot be
done anymore in the static approximation.  A careful analysis of the
integration regions along the lines of the threshold expansion discussed below
should be carried out in this case. Incidentally, this situation is of physical relevance
for the $\Upsilon (1S)$ system, where the charm quark mass plays the role of $m'$.

It is also possible to perform the matching to pNRQCD using the threshold
expansion \cite{Beneke:1998zp}.  This possibility has been followed by several
groups \cite{Beneke:1999qg,Kniehl:2001ju,Kniehl:2002br}.  Typically (although
not always), the procedure consists in taking one specific diagram of NRQCD
and splitting it in the different existing regions of momenta. According to
this terminology the modes (and correspondingly the regions of momenta) that
appear in NRQCD are the following:

\medskip

\noindent
(i) {\bf soft modes}. Quarks and gluons with energy and three-momenta of
${\cal O}(mv)$ (the quarks are off-shell in this situation).\\
(ii) {\bf potential modes}. Quarks and gluons with energy of ${\cal O}(mv^2)$
and three-momenta of ${\cal O}(mv)$ (the gluons are off-shell in this 
situation).\\
(iii) {\bf US modes}. Quarks and gluons with energy 
and three-momenta of ${\cal O}(mv^2)$ (in practice, it does not seem there are 
quarks in this situation).

\medskip

Integrating out soft modes and potential gluons corresponds to matching NRQCD to pNRQCD.
In some cases, it is customary to perform the matching using (free) on-shell quarks.
This has the consequence that loops in pNRQCD do not vanish (since the energy is not
left as a free parameter in which one can expand) and have to be
subtracted accordingly. In addition, the on-shell condition may set 
to zero some terms in the (off-shell) potential. When these terms 
enter in a NRQCD subdiagram of a higher-loop matching calculation, 
they may give rise to new contributions to the potential due to 
quark potential loops. 
This never occurs if the procedure described above is used.
In any case, the potentials obtained by using different methods can  
be related to each other by unitary transformations.

\subsubsection{Pinch singularities}
\label{Pinch}
\begin{figure}
\hspace{-0.1in}
\epsfxsize=3.8in
\centerline{\epsffile{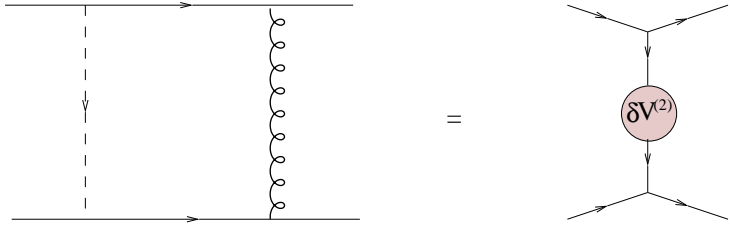}}
\caption {\it Matching between NRQCD and pNRQCD without considering pinch 
singularities. The dashed and 
curly lines represent the longitudinal and transverse 
gluon exchange respectively.}
\label{pinch1}
\end{figure}

\begin{figure}
\hspace{-0.1in}
\epsfxsize=5.5in
\centerline{\epsffile{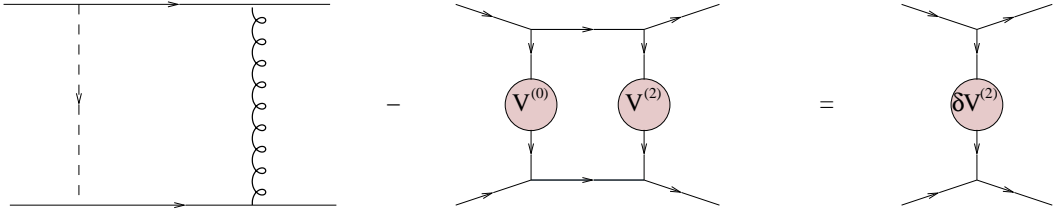}}
\caption {\it Matching between NRQCD and pNRQCD taking into account pinch 
singularities. In pNRQCD the loop regulates the pinch singularity.}
\label{pinch2}
\end{figure}

Let us now discuss the issue of the so-called pinch singularity.
We illustrate this discussion with the diagram (in the Coulomb gauge) in Fig.~\ref{pinch1}.
Actually, such a diagram appears in the computation of the positronium
spectrum at ${\cal O}(m\al^5)$ faced in \cite{Pineda:1998kn}. The one-loop
integral of this diagram reads 
\be
I \sim \int  {d^D\, q \over (2\pi)^D} {1 \over ({\bf q-k})^2}
{1 \over  q^0+i\epsilon }
\,
{1 \over -q^0 +i\epsilon }{1 \over q^2}\left(\delta_{ij}-{{\bf q}^i{\bf 
q}^j  \over {\bf q}^2} \right)\left( \cdots \right)
\,,
\ee
where $\left( \cdots \right)$ stands for a $q_0$ independent term.
We see that it has two singularities at $q_0=\pm i\epsilon$. 
This is usually referred to as the pinch singularity.
The rigorous procedure to set the prescription
to eliminate the pinch singularity comes from the matching computation. 
Previously we have mentioned that loops in pNRQCD 
could be set to zero, as far as the matching computation
was concerned, but that required that the same kind of 
pinch singularity diagrams were set to zero in NRQCD.
The implementation of this idea can be translated into 
a simple prescription: since for any NRQCD diagram with a pinch 
singularity, there must be a pNRQCD diagram with the same pinch 
singularity, just subtract it (see Fig.~\ref{pinch2}).   
Therefore, the actual integral to be computed reads
\be
I \sim \int  {d^D\, q \over (2\pi)^D} {1 \over ({\bf q-k})^2}
{1 \over  q^0+i\epsilon }
\,
{1 \over -q^0 +i\epsilon }
\left(
{1 \over q^2}+{1 \over {\bf q}^2}
\right)
\left(\delta_{ij}-{{\bf q}^i{\bf q}^j  \over {\bf q}^2} \right)
\left( \cdots \right)
\,.
\ee
We can see how the pinch singularity disappears, and the resulting 
integral provides new contributions to the potential only.

Pinch singularities also appear in computations using the threshold expansion.
We have seen here that understanding the pinch singularities within the EFT 
framework provides a consistent prescription to eliminate them in each case.

\subsubsection{Potentials}
\label{potentials}
The general structure of the potentials has been given in sec.~\ref{sec:pNRLagW}. 
We will focus on the equal mass case, Eq.~(\ref{ppot}). 
By dimensional analysis, $V^{(1)}$ scales like $1/r^ 2$, $ V^{(2)}_{{\bf p}^2}$ like $1/r$, 
$V^{(2)}_r$ like $1/r^3$ or $\delta^{(3)} ({\bf r})$, and so on. They read
\bea
V^{(0)}_s(r) &
=& - C_F {\alpha_{V_s}(r) \over r}, \\
{V^{(1)}_s } (r)&
=& -{C_FC_A D^{(1)}_s \over 2r^2},\\
V_{{\bf p}^2,s}^{(2)}(r)&=&- { C_F D^{(2)}_{1,s} },\\
V_{{\bf L}^2,s}^{(2)}(r)&=& { C_F D^{(2)}_{2,s} \over 2 }{1 \over r},\\
V_{r,s}^{(2)}(r)&=&{\pi C_F D^{(2)}_{d,s} }
\delta^{(3)}({\bf r}),\\
 V_{S^2,s}^{(2)}(r)&=&{4\pi C_F D^{(2)}_{S^2,s} \over 3}
 \delta^{(3)}({\bf r}),\\
V_{LS,s}^{(2)}(r)&=&{ 3 C_F D^{(2)}_{LS,s} \over 2 }{1 \over r^3},\\
V_{{\bf S}_{12},s}^{(2)}(r)&=& { C_F D^{(2)}_{S_{12},s} \over 4 }{1 \over r^3},
\label{V2}
\eea
where $\al_{V_s}$ and the various $D$s depend logarithmically on $r$ and the
renormalization scale $\nu_{pNR}$.
In order to obtain the spectrum at order $m\als^ 4$, $\al_{V_s}$ has to be 
calculated to order $\als^ 3$ (two loops), $V_s^{(1)}$ to order $\als^ 2$ (one loop) 
and the remaining potentials to order $\als$ (tree level). They read
\bea
&&{\alpha}_{V_s} =\als(r)
\left\{1+\left(a_1+ 2 {\gamma_E \beta_0}\right) {\als(r) \over 4\pi}\right.
\nonumber\\
&&\qquad\qquad 
+\left. 
\left[\gamma_E\left(4 a_1\beta_0+ 2{\beta_1}\right)+\left( {\pi^2 \over 3}+4 \gamma_E^2\right) 
{\beta_0^2}+a_2\right] {\als^2(r) \over 16\,\pi^2}
\right\},
\label{newpot0}\\ 
& &D^{(1)}_s=\als^2(r)\quad , \quad
D^{(2)}_{1,s}=
D^{(2)}_{2,s}=
D^{(2)}_{d,s}= 
D^{(2)}_{S^2,s}=
D^{(2)}_{LS,s}=
D^{(2)}_{S_{12},s}=\als(r)
\,.
\label{Dsten2}
\eea
$a_1$ was computed by \cite{Fischler:1977yf} and $a_2$ by
\cite{Peter:1997ig,Schroder:1998vy}.  If one wishes to have the spectrum to
one order higher, namely $m\als^ 5$, all these potentials must be calculated
to one more power in $\als$.  For $\al_{V_s}$, only the logarithmic
contributions are known \cite{Brambilla:1999qa,Kniehl:1999ud} (Pad\'e
approximant \cite{Chishtie:2001mf} and renormalon based \cite{Pineda:2001zq} 
estimates are also available).  $V_s^{(1)}$ was calculated by \cite{Kniehl:2001ju} (the
logarithmic corrections were computed by
\cite{Kniehl:1999ud,Brambilla:1999xj}) and the complete $V_s^{(2)}$ have been
computed over the years
\cite{Gupta:1981pd,Gupta:1982qc,Buchmuller:1981aj,Pantaleone:1985uf,
Titard:1993nn,Pineda:1998kn,Brambilla:1999xj,Manohar:2000hj,Kniehl:2002br} 
and can be found in \cite{Kniehl:2002br}. Several comments are in order 
concerning these calculations.

{\bf 1.} The potentials in the matching calculation appear naturally in momentum
space, and so they are given in many of the references above. The real space
potentials, which are better suited for bound-state calculations,
are obtained by Fourier transforming the momentum space potentials. At lower
orders, it is enough to take the Fourier transform in $3$-dimensions in the
sense of distributions \cite{Titard:1993nn}. At higher orders, it must be taken
in $d$ dimensions, as discussed below.

{\bf 2.} In different papers, the results displayed for each of the potentials may
vary, even if the same basis (\ref{ppot}) is used.  This does not mean {\it a
  priori} that there are inconsistencies. The basis (\ref{ppot}) is
overcomplete and hence apparently different results may be related to each
other by unitary transformations.  In particular $V_s^ {(1)}$ can be totally
reshuffled into $1/m^2$ potentials.

{\bf 3.} In earlier papers, the potentials were calculated directly from QCD
without expanding in the kinetic energy. In that case there are contributions
from the pNRQCD side to the matching calculation due to the fact that the
kinetic term in the pNRQCD Hamiltonian cannot be expanded anymore.  
In this framework, the integrals involved in the
calculation have more than one scale and hence are harder to evaluate.

{\bf 4.} In higher-order calculations, quantum-mechanical perturbation theory
requires regularization and renormalization.  The UV divergences are
renormalized by local potentials inherited from NRQCD and the scale dependence
is compensated by the one in the NRQCD matching coefficients. In order to use the
NRQCD matching coefficients obtained in sec.~\ref{NRmatching}, the potentials
must be kept in $d$ dimensions.  This is not important as far as the
soft/US factorization is concerned (it amounts to a change of
subtraction scheme), but it becomes so when the calculation is sensitive to
divergences due to the hard/potential factorization. This occurs at order
$m\als^6$ for the spectrum and in ${\cal O}(\als^2)$ corrections for the
current.  Note that any loop correction to a given (e.g. Coulomb) potential 
slightly changes its functional form (it gets multiplied by $(r\nu)^{(4-D)}$
for each loop). The expressions for the potentials in $3$ dimensions
calculated at higher orders display {\it small} logarithms, which eventually cancel
out in the full calculation, in addition to the {\it large} logarithms, which
eventually become $\ln \als$, as discussed in \cite{Kniehl:2002br}
(note that in \cite{Brambilla:1999xj} only the large logarithms were displayed).

{\bf 5.} The octet potential is also known at two loop accuracy \cite{Kniehl:2004rk},
\be
\label{alVopert}
V^{(0)}_o(r) \equiv  \left({C_A\over 2} -C_F\right) { \al_{V_o}(r) \over r} ,
\quad \al_{V_o}(r)=\al_{V_s}(r)-\left( {3 \over 4}-{\pi^2 \over 16}\right)C_A^2\als^3
+{\cal O}(\als^4) .
\ee

{\bf 6.} At order $m\als^ 5$ for the spectrum and at ${\cal O}(\als^3)$ for the current 
US loops start to contribute. This implies that $V_A(r)$ is also needed. 
At tree level we have 
\be
V_A(r)=V_B(r)=1.
\ee

{\bf 7.} For the case $m_1\not=  m_2$, the $1/m^ 2$ potentials have only been calculated in 
the scheme described in sec.~\ref{pNRmatchingI} for QED \cite{Pineda:1998kn}. 
Earlier calculations both for QCD \cite{Gupta:1982qc} and QED 
\cite{Gupta:1989aa} exist, which have been carried out by matching directly 
the fundamental theory to a quantum-mechanical Hamiltonian. 

{\bf 8.} RG improved expressions for the potential can also be obtained. 
They are discussed in sec.~\ref{RGreview}.

Finally, we would like to briefly discuss the matching of currents
and the imaginary pNRQCD potential.
Integrating out the soft scale when matching local currents produces
scaleless integrals, which are zero in DR. This means that the matching
coefficient remains the same at the matching scale. If we take the
electromagnetic vector current as an example, the matching condition reads
$b^v_{\rm 1, pNR}(\nu_p,\nu_{us}=\nu_s)=b^v_{\rm 1, NR}(\nu_p,\nu_s)$.
In the case of $b^v_{1,\rm NR}$, only a dependence on $\nu_p$ appears
(at least at low orders).  An equivalent discussion applies to the
imaginary terms of the Lagrangian for which the general matching
condition Im $f^{\rm pNR}(\nu_p,\nu_{us}=\nu_s)=$Im$f(\nu_p,\nu_s)$ holds.
Nevertheless, one should keep in mind that the expressions for the
matching coefficients will change once their
running is considered (see sec.~\ref{RGreview}).

\subsection{Matching: Wilson loop approach}
\label{sec:matchingII}
We will discuss here an other way to perform the matching to pNRQCD.
We will sometimes denote it as Wilson-loop matching.
With respect to the previously discussed procedure, it is characterized by
the following points.
\begin{itemize}
\item[(a)]{It is done in coordinate space.}
\item[(b)]{It is done with the pNRQCD Lagrangian in the form
  of Eq.~(\ref{Lpnrqcd}). This means that the degrees of freedom that appear most
  naturally in the pNRQCD part of the matching are singlet and octet fields.}
\item[(c)]{As a consequence of (b), only one time appears in the
  computation.}
\item[(d)]{For what concerns the gluon fields, they appear in the NRQCD part 
of the matching procedure in terms of Wilson-loop amplitudes. Therefore, 
the formulation will be explicitly gauge invariant at each step.}
\item[(e)]{ Gauge-invariant expressions can be obtained for the 
potentials that encode all the corrections in $\als(1/r)$, for a given 
order in $1/m$ and the multipole expansion. }
\end{itemize}
The results obtained within this matching procedure 
will be equivalent (up to field redefinitions) to those obtained in the previous section. 

From point (d) and (e) above, it is clear that the
Wilson-loop matching is well suited to be generalized to non-perturbative
cases. Therefore, it provides us  with a bridge 
between the weak-coupling matching procedure of this section and the 
strong-coupling one of ch.~\ref{sec:SCR}. There, the language will be exactly 
the one introduced here in the safe framework of perturbative QCD.

In the following, we will define our interpolating fields, set the basis of the matching,
and illustrate the procedure by  discussing the static matching up to and
including order $r^2$ in the multipole expansion. We will closely follow \cite{Brambilla:1999xf}, to
which we refer for more details of the original derivation.

\subsubsection{Interpolating fields}
Our aim is to match, in coordinate space, amplitudes defined in terms of the 
fields of NRQCD with amplitudes defined in terms of the fields that appear in 
the pNRQCD Lagrangian (\ref{Lpnrqcd}), i.e.~$A_{\mu}$, $S$ and $O^a$ 
fields. Therefore, we need to identify some interpolating fields in
NRQCD that have the same quantum numbers and the same transformation
properties as $S$ and $O^a$. The correspondence is not one-to-one. 
Given an interpolating field in NRQCD
there are an infinite number of combinations of singlet and octet
operators with US fields that have the same quantum numbers 
and, therefore, a non-vanishing overlap with the NRQCD operator. Fortunately, the operators in pNRQCD 
can be organized according to the counting of the multipole expansion. 
For instance, for the singlet we have  
\be
\chi^\dagger({\bf x}_2,t) \phi({\bf x}_2,{\bf x}_1;t) \psi({\bf x}_1,t) 
\rightarrow  \sqrt{Z^{(0)}_s(r)} S({\bf r},{\bf R},t) 
+ \sqrt{Z_{E,s}(r)} \, r \, {\bf r}\cdot g{\bf E}^a({\bf R},t) O^a({\bf r},{\bf R},t) + \dots,  
\label{Sdef}
\ee
and  for the octet 
\bea
\chi^\dagger({\bf x}_2,t) \phi({\bf x}_2,{\bf R};t) T^a \phi({\bf R},{\bf x}_1;t) \psi({\bf x}_1,t) 
&\rightarrow& \sqrt{Z^{(0)}_o(r)} O^a({\bf r},{\bf R},t) 
\nn\\
&& \hspace{-5mm}
+ \sqrt{Z_{E,o}(r)} \, r \, {\bf r}\cdot g{\bf E}^a({\bf R},t) S({\bf r},{\bf R},t) + \dots, 
\label{Odef}
\eea
where 
\be
\phi({\bf y},{\bf x};t)\equiv P \, \exp \left\{ i \displaystyle 
\int_0^1 \!\! ds \, ({\bf y} - {\bf x}) \cdot g{\bf A}({\bf x} - s({\bf x} - {\bf y}),t) \right\}.
\label{schwinger}
\ee
The arrows are a reminder that the two operators act on different Hilbert spaces and
that the equalities hold only inside Green functions. The factors $Z$ are
normalization factors. From Eqs.~(\ref{Sdef}) and (\ref{Odef}), it follows that 
the operators on the left-hand side overlap at leading order in the multipole
expansion with the singlet and octet fields respectively. 

The matching for the octet in Eq.~(\ref{Odef}) does not make use of 
a gauge-invariant operator. In a perturbative matching this is not
problematic, since  $V_o$ is gauge invariant order by order in $\als$. 
However, if one aims at taking advantage of non-perturbative
techniques, it is preferable to work with a manifestly gauge-invariant quantity. 
The simplest solution consists in substituting the $T^a$ 
color matrix on the left-hand side of Eq.~(\ref{Odef}) by a local gluonic operator
$H^a({\bf R},t)\,T^a$ with the right transformation properties, an example
being $g{\bf B}^a({\bf R},t)\,T^a$.
All $H^a({\bf R},t)\,T^a$ with the right transformation properties will give
in the weak-coupling regime the same potential, corresponding to the perturbative
octet potential. In the strong-coupling regime, where octet quark-antiquark fields 
do not exist as independent degrees of freedom, they identify 
different degrees of freedom and, hence, different potentials, corresponding to the different symmetry
properties of $H^a$. We will come back to this in full detail in ch.~\ref{sec:static}.

\begin{figure}[htb]
\makebox[0cm]{\phantom b}
\epsfxsize=4truecm \epsfbox{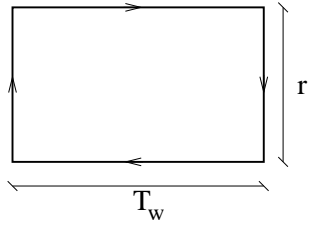}
\caption{ \it A graphical representation of the static Wilson loop. We adopt the convention 
that the time direction is from the left to the right. Therefore, the quark trajectories 
are represented by horizontal lines and the equal-time endpoint Wilson lines 
by shorter vertical lines.}
\label{wilsonfig}
\end{figure}

\subsubsection{Matching at ${\cal O}(r^0,1/m^0)$}
\label{weak00}
In order to get $V^{(0)}_s$ and $Z^{(0)}_s$, 
we choose the following Green function in NRQCD \cite{Susskind:1976pi,Brown:1979ya}:
\bea
G_{\rm NRQCD}  &=& 
\langle {\rm vac} \vert  \chi^\dagger(x_2) \phi(x_2,x_1) \psi(x_1) 
\psi^\dagger(y_1)\phi(y_1,y_2) \chi(y_2) \vert {\rm vac} \rangle 
\nn\\
&=& \delta^3({\bf x}_1 - {\bf y}_1) \delta^3({\bf x}_2 - {\bf y}_2) 
\langle W_\Box \rangle + \dots \;, 
\label{vsnrqcdbis}
\eea
where the dots stand for higher-order corrections in the $1/m$ expansion.
The quantity $W_\Box$ is the rectangular Wilson loop \cite{Wilson:1974sk}
with corners  $x_1 = (T_W/2,{\bf r}/2)$, $x_2 = (T_W/2,-{\bf r}/2)$, 
$y_1 = (-T_W/2,{\bf r}/2)$ and  $y_2 = (-T_W/2,-{\bf r}/2)$:
\be
W_\Box \equiv P \,\exp\left\{{\displaystyle - i g \oint_{r\times T_W} \!\!dz^\mu A_{\mu}(z)}\right\}.
\ee
A graphical representation that we will often use is given in Fig.~\ref{wilsonfig}.
We also define
\be 
\langle \cdots \rangle \equiv 
\langle{\rm vac}|{\rm Tr} \left\{ \cdots \right\}|{\rm vac}\rangle
=
\int {\cal D} A \, {\cal D} q \, {\cal D} \bar{q} \; e^{-iS^{(0)}} \; 
{\rm Tr} \left\{ \cdots \right\}, 
\label{pathaverage}
\ee
where $S^{(0)}$ is the pure Yang--Mills plus light-quark action of QCD and the path-integral 
is over all light fields. 

Equation (\ref{Sdef}) states that the leading overlap of the Green function (\ref{vsnrqcdbis}) 
is with the singlet propagator in pNRQCD. Indeed, in pNRQCD 
we get in the static limit and at the zeroth order in the multipole expansion:
\be
G_{\rm pNRQCD} = Z_s^{(0)}(r) \delta^3({\bf x}_1 - {\bf y}_1) \delta^3({\bf x}_2 -
{\bf y}_2) e^{-iT_W V_s^{(0)}(r)}.
\label{vspnrqcd}
\ee
In order to single out the soft scale, we consider the large $T_W$ limit of the 
Wilson loop (equivalent to setting $E \rightarrow 0$):
\be
{i\over T_W}\ln  \langle W_\Box \rangle = u_0(r) + i {u_1(r)\over T_W} + {\cal
  O}\left( {1\over T^2_W}\right) 
\quad \hbox{for} \quad T_W\to\infty \,   , 
\label{sexp}
\ee
then from the matching condition $G_{\rm NRQCD} = G_{\rm pNRQCD}$ we obtain:
\bea
V_s^{(0)}(r) &\equiv& -C_F {\alpha_{V_s}(r) \over r} = u_0(r), 
\label{vs1}\\
\ln Z^{(0)}_s(r) &=& u_1(r). \label{zs1}
\eea
The matching does not rely on any perturbative expansion in 
$\als$. However, since we are concerned with the weak-coupling 
situation, the quantities on the right-hand side of Eqs.~(\ref{vs1}) and
(\ref{zs1}) can be evaluated expanding order by order in $\als$. 
At LO in $\als$ we have
\bea
V^{(0)}_s(r) &=& -C_F {\als \over r} \quad \hbox{or} \quad 
\alpha_{V_s} =  \als, 
\label{vs1leading}\\
Z^{(0)}_s(r) &=& N_c. 
\label{zs1leading}
\eea

In order to get $V^{(0)}_o$ and $Z^{(0)}_o$ one proceeds in a similar way.
We choose the NRQCD Green function: 
\bea
G_{\rm NRQCD}^{ab} &=& 
\langle {\rm vac}  \vert  
\chi^\dagger(x_2) \phi\left({\bf x}_2,{{\bf x}_1+{\bf x}_2\over 2};{T_W\over 2}\right) 
T^a \phi\left({{\bf x}_1+{\bf x}_2\over 2},{\bf x}_1;{T_W\over 2}\right)
\psi(x_1) 
\nn\\
& & \qquad\qquad 
\times \psi^\dagger(y_1)\phi\left({\bf y}_1,{{\bf y}_1+{\bf y}_2\over 2};-{T_W\over 2}\right) 
T^b \phi\left({{\bf y}_1+{\bf y}_2\over 2},{\bf y}_2;-{T_W\over 2}\right)
\chi(y_2) \vert {\rm vac} \rangle
\nn\\
&=&
\delta^3({\bf x}_1 - {\bf y}_1) \delta^3({\bf x}_2 - {\bf y}_2) 
\langle T^a W_\Box T^b \rangle 
+ \dots \;,
\label{vonrqcd}
\eea
where in the last line the color matrices are understood as inserted 
in the static Wilson loop at the points (${\bf R},T_W/2$) and (${\bf R},-T_W/2$). 
The dots stand for higher-order corrections in the $1/m$ expansion.

Equation (\ref{Odef}) states that the leading overlap of the Green function (\ref{vonrqcd}) 
is  with the octet propagator in pNRQCD. Indeed, in pNRQCD we obtain 
in the static limit and at zeroth order in the multipole expansion:
\be
G_{\rm pNRQCD}^{ab} = Z^{(0)}_o(r) \delta^3({\bf x}_1 - {\bf y}_1) \delta^3({\bf x}_2 - {\bf y}_2) 
e^{-iT_WV^{(0)}_o(r)} \langle \phi_{ab}^{\rm adj}(T_W/2,-T_W/2) \rangle,  
\label{vopnrqcd}
\ee
where the Wilson line 
$$
\phi(T_W/2,-T_W/2) \equiv \phi(T_W/2,{\bf R},-T_W/2,{\bf R}) 
= P\, \exp \left\{ - ig \displaystyle \int_{-T_W/2}^{T_W/2} \!\! dt \, A_0({\bf R},t) \right\}
$$
is evaluated in the adjoint representation. 
As in the singlet case, we define
\be
{i\over T_W}\ln  {\langle T^a W_\Box T^b \rangle \over 
\langle \phi_{ab}^{\rm adj}(T_W/2,-T_W/2)\rangle}
= v_0(r) + i {v_1(r)\over T_W} + {\cal O}\left( {1\over T_W^2}\right) 
\quad \hbox{for} \quad T_W\to\infty \,   .
\label{soexp}
\ee
From the matching condition $G_{\rm NRQCD}^{ab} = G_{\rm pNRQCD}^{ab}$ we obtain:
\bea
V^{(0)}_o(r) &\equiv& \left({C_A\over 2} -C_F\right) {\alpha_{V_o}(r) \over r} = v_0(r) , 
\label{vo1}\\
\ln Z^{(0)}_o(r) &=& v_1(r). 
\label{zo1}
\eea
Again, the formulas above do not rely on any expansion in $\als$.
However, in the weak-coupling situation, the quantities on the right-hand side
of Eqs.~(\ref{vo1}) and (\ref{zo1}) can be expanded order-by-order in $\als$. 
At LO in $\als$, we obtain
\bea
V^{(0)}_o(r) &=& \left({C_A\over 2} - C_F\right) {\als \over r} \quad \hbox{or} \quad 
\alpha_{V_o} =  \als, 
\label{vo1leading}\\
Z^{(0)}_o(r) &=& T_F. 
\label{zo1leading}
\eea 

Note that, despite the octet-matching procedure being gauge dependent, 
the octet static potential obtained in this way is not at any finite order in perturbation theory 
(it corresponds to the pole of the octet static propagator). All the gauge
dependence goes into the normalization factor $Z^{(0)}_o$. In this respect, it is
worthwhile to observe that the string $\langle \phi_{ab}^{\rm adj}(T_W/2,-T_W/2)\rangle$ does 
not give contributions to the potential at any finite order in perturbation
theory, but it does to $Z^{(0)}_o$.

\subsubsection{Matching at ${\cal O}(r^1,1/m^0)$ and ${\cal O}(r^2,1/m^0)$}
At ${\cal O}(r)$, there are no additional contributions to the singlet and octet matching potentials 
and to the normalization factors. At this order in the multipole expansion one finds $V_A$ and $V_B$. 
In the weak-coupling regime at LO in $\als$, they are  
\be
V_A(r) = 1,  \qquad\qquad V_B(r) = 1.
\ee
At ${\cal O}(r^2)$, one finds the next-to-leading contributions to the singlet and
octet static potentials and to the singlet static normalization factor.

\begin{figure}[htb]
\makebox[0cm]{\phantom b}
\epsfxsize=12truecm \epsfbox{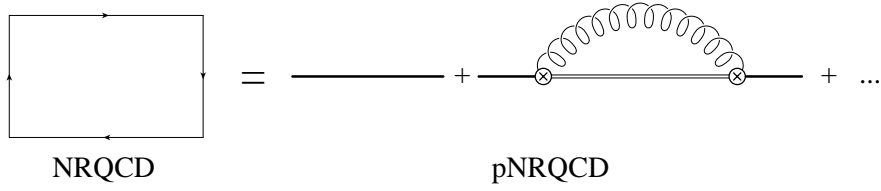}
\vspace{0.2cm}
\caption{ \it The matching of $V^{(0)}_s$ and $Z^{(0)}_s$ 
at NLO in the multipole expansion. On the left-hand side is the Wilson loop in NRQCD, on the right-hand  
side are the pNRQCD propagators. The 1st and 2nd term on the right-hand side 
symbolically represent the 1st and 2nd term in Eq.~(\ref{vspnrqcdus}). 
}
\label{figmats}
\end{figure}

The NLO correction in the multipole expansion to 
the singlet static propagator (\ref{vspnrqcd}) is given by (see Fig.~\ref{figmats}) 
\bea
& &G_{\rm pNRQCD}
= Z^{(0)}_s(r) \delta^3({\bf x}_1 - {\bf y}_1) \delta^3({\bf x}_2 - {\bf y}_2) 
e^{-iT_WV^{(0)}_s(r)} 
\label{vspnrqcdus}
\\
& &\times \left(1 
- {T_F\over N_c} V_A^2 (r)
\int_{-T_W/2}^{T_W/2} \! dt \int_{-T_W/2}^{t} \! dt^\prime \, 
e^{-i(t-t^\prime)(V^{(0)}_o-V^{(0)}_s)} 
\langle {\bf r}\cdot g{\bf E}^a(t) \phi^{\rm adj}_{ab}(t,t^\prime){\bf r}
\cdot g{\bf E}^b(t^\prime)\rangle
\right)\,,
\nn
\eea
where fields with only temporal arguments are evaluated in the centre-of-mass coordinate. 
From the matching condition $G_{\rm NRQCD}= G_{\rm pNRQCD}$, we obtain 
$Z^{(0)}_s$ and $V^{(0)}_s$  at NLO in the multipole
expansion:
\bea
V_s^{(0)}(r) &=& u_0(r)
+ {T_F \over N_c} V_A^2 (r) \, \lim_{T_W\to\infty}\, {i\over T_W}
\int_{-T_W/2}^{T_W/2} \!\!\! dt \int_{-T_W/2}^{t} \!\!\! dt^\prime \, 
e^{-i(t-t^\prime)(V^{(0)}_o-V^{(0)}_s)} 
\nn\\
&& \qquad\qquad\qquad\qquad\qquad\qquad\qquad  
\times 
\langle {\bf r}\cdot g{\bf E}^a(t) \phi^{\rm adj}_{ab}(t,t^\prime){\bf r}
\cdot g{\bf E}^b(t^\prime)\rangle, 
\label{vs1r2}\\
\nn \\
\ln Z^{(0)}_s(r) &=& u_1(r)
+ {T_F \over N_c} V_A^2 (r) 
\int_{-\infty}^{\infty} \! dt \int_{-\infty}^{t} \! dt^\prime \, 
e^{-i(t-t^\prime)(V^{(0)}_o-V^{(0)}_s)} 
\nn\\
&& \qquad\qquad\qquad\qquad\qquad\qquad\qquad  
\times
\langle {\bf r}\cdot g{\bf E}^a(t) \phi^{\rm adj}_{ab}(t,t^\prime){\bf r}
\cdot g{\bf E}^b(t^\prime)\rangle. 
\label{zs1r2}
\eea
Equations (\ref{vs1r2}) and (\ref{zs1r2}) do not rely on any perturbative 
expansion in $\als$. However, since we are considering the weak-coupling case, 
they can be evaluated order-by-order in $\als$
and one can obtain the leading logarithmic contribution to the static potential. 
This comes from the three-loop IR logarithmic 
divergence of the Wilson loop first noticed in \cite{Appelquist:1978es}
(see also \cite{Kummer:1996jz}). 
The calculation may be done in various ways, depending on how divergences
are regularized. Obviously the scheme adopted for calculating the Wilson loop 
must be the same as adopted for calculating the loop diagram in pNRQCD. This study 
has been performed by \cite{Brambilla:1999qa,Brambilla:1999xf} giving 
\bea
V^{(0)}_s(r,\nu_{us}) 
&=& -C_F {\alpha_{V_s}(r) \over r} = (u_0(r))_{\rm two-loops} 
-{C_FC_A^3\over 12} {\als\over r}{\als^3\over \pi}\ln ({r\nu_{us}}),
\label{vsus}\\
\ln Z^{(0)}_s(r,\nu_{us}) &=& 
(u_1(r))_{\rm two-loops} + {C_F\,C_A^2\over 2} {\als^3\over \pi}\ln ({r\nu_{us}}).
\label{zsus}
\eea
The two-loop expression for $u_0(r)$ is given by 
$-C_F \alpha_{V_s}(r)_{\rm two-loops}/  r$ 
and the two-loop expression for $\alpha_{V_s}$ can be found in Eq.~(\ref{newpot0}).
The contributions proportional to $\ln ({r\nu_{us}})$
in Eq.~(\ref{vsus}) and (\ref{zsus}) would be zero in QED. 
The fact that $\alpha_{V_s}$ depends on the IR behaviour of the theory
is, therefore, a distinct feature of QCD, more specifically, 
of the non-Abelian nature of QCD, which allows gluons to
interact with themselves at arbitrarily small energy scales.
We stress that, in order to match the normalization factor (\ref{zsus}), it is necessary 
to take into account contributions coming from the end-point Wilson lines, which   
can be considered irrelevant only at order $(1/T_W)^0$, i.e. for the potential
(note that this does not require any special assumption 
about the large-time behaviour of the gluon fields).

\begin{figure}[htb]
\makebox[0cm]{\phantom b}
\epsfxsize=12truecm \epsfbox{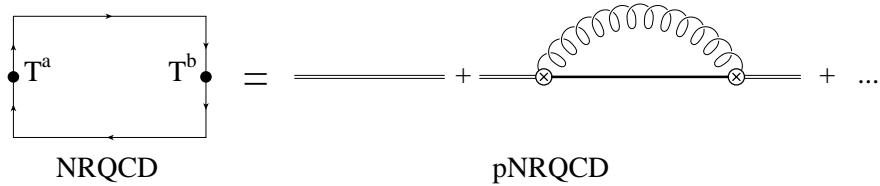}
\vspace{0.2cm}
\caption{ \it The matching of $V^{(0)}_o$ and $Z^{(0)}_o$ at NLO in the multipole 
expansion. On the left-hand side is the Wilson loop in NRQCD with color matrix insertions, 
on the right-hand side are the pNRQCD propagators.}
\label{figmato}
\end{figure}

\begin{figure}[htb]
\makebox[0cm]{\phantom b}
\epsfxsize=5truecm \epsfbox{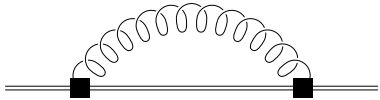}
\vspace{0.2cm}
\caption{ \it Octet self-energy graph proportional to $V_B^2$.}
\label{figzoo}
\end{figure}

The NLO correction to Eq.~(\ref{vopnrqcd}) in the multipole expansion 
comes from the graph shown in Fig.~\ref{figmato}.
We omit a term proportional to $V_B^2$ of the type shown in Fig.~\ref{figzoo} and terms which contain 
operators like ${\rm Tr}\{{\bf r}^i{\bf r}^j[{\bf D}^i,{\bf E}^j]{\rm O} {\rm O}^{\dagger}\}$, 
because in perturbation theory they neither contribute to the octet matching potential nor to the
normalization. The reason is that, differently from the non-perturbative 
regime where we may have dependencies on the scale $\lQ$, in perturbation
theory loops on octet lines are scaleless and vanish in DR. 
With an analogous calculation as in the singlet case we obtain at leading
logarithmic three-loop accuracy:
\be
V^{(0)}_o(r,\nu_{us}) = 
\left({C_A\over 2} -C_F\right) {\alpha_{V_o}(r) \over r} = (v_0(r))_{\rm two-loops} 
+\left({C_A\over 2} -C_F\right) {C_A^3\over 12} {\als\over r}{\als^3\over \pi}\ln{r\nu_{us}}. 
\label{vous}
\ee
The two-loop expression for $v_0(r)$ is given by $(C_A - C_F/2) \alpha_{V_o}(r)_{\rm two-loops}/  r$ 
and for the two-loop expression of $\alpha_{V_o}$, see the comments after Eq.~(\ref{alVopert}). 
Similarly also $Z^{(0)}_o$ may be calculated, but only in a specific gauge.

\subsubsection{Matching at order $r^0$, ($1/m$, $1/m^2$ and beyond)}
\label{1overmpert}
Following this method, one could consider $1/m$ corrections. If one 
works at LO in the multipole expansion, the singlet and 
octet fields decouple. If we further focus on the singlet sector, the 
computations would be similar to those that appear in sec.~\ref{secwilmatch} 
for the strong-coupling regime. This is so because we are actually 
performing the matching order by order in $1/m$ and to any order in $\als$. 
Therefore, the expressions obtained in the strong-coupling regime also hold here up to 
corrections due to US effects. This reasoning also applies 
to what in ch.~\ref{sec:SCR} is called the "quantum-mechanical matching"
(see sec.~\ref{seconestep}), where explicit expressions in terms of
Wilson-loop amplitudes for the real and imaginary parts of the pNRQCD potentials are 
derived. Those expressions are also valid here, in the perturbative regime, 
if they are understood to be at ${\cal O}(r^0)$ in the multipole expansion. Note that 
the Wilson loops multiplying delta functions of ${\bf r}$ or derivatives of them 
are zero in the perturbative regime, since they become dimensionless objects 
and vanish in DR. In particular, this applies to the 
gluonic correlators which appear in the imaginary part of the potential. 
Finally, we note that non-analytic terms due to the scale $\sqrt{m\,\lQ}$ 
do not appear here since for $ \lQ \siml E$, this three-momentum scale has not been 
integrated out.

\subsection{Observables: spectrum and inclusive decay widths}
\label{pNRweakObservables}
We have finally built the pNRQCD Lagrangian, and are 
in the position to calculate observables with it. We will mainly consider 
observables (being the theoretically cleanest ones) that in pNRQCD only  
involve the calculation of the NR propagator (Green function) of the system
projected onto the colorless sector of a quark-antiquark pair (let us call $P_s$
the corresponding projector) and the gluonic vacuum,
\be
\Pi(E,{\bf r},{\bf r}') \equiv i\int d t \, d^3{\bf R} \,e^{iEt}
\langle {\rm vac} |
T\{ S({\bf r}', {\bf 0}, 0)
S^{\dagger}({\bf r}, {\bf R}, t) \}
| {\rm vac}\rangle= \langle {\bf r}'|G_s(E)|{\bf r} \rangle\,,
\ee
\be
\label{Gsfull}
G_s(E)\equiv P_s \langle {\rm vac} |{1 \over H-E} | {\rm vac} \rangle P_s=G_c(E)+\delta G_s
\,,
\ee
where $H$ is the pNRQCD Hamiltonian, $G_c$ the Coulomb Green function, 
defined in Fig.~\ref{pnrqcdfig}, and $E$ the energy measured from the threshold $2m$. 

Besides the heavy quarkonium spectrum (i.e. the poles of the Green function),
we will consider inclusive (electromagnetic) decay widths, NR sum rules and
$t$-$\bar t$ production near threshold. For these only the normalization at
the origin will be important\footnote{Other observables that do not belong to
this category are semi-inclusive radiative decay widths, which have been
studied by \cite{GarciaiTormo:2004jw} and are considered in sec.~\ref{semi}, or
heavy quarkonium production, for which an analysis in the weak-coupling regime
is available \cite{Beneke:1999gq}.}, i.e. the object $\langle {\bf
r}=0|G_s(E)|{\bf r}=0\rangle$ has to be computed.

In pNRQCD, there are only potential and US loops. Within pNRQCD, talking 
about potential loops is nothing but talking about quantum-mechanical 
perturbation theory:
\begin{figure}[htb]
\makebox[0.0cm]{\phantom b}
\put(-200,1){$\delta G_s^{\rm pot.}=$}
\put(-150,-5){\epsfxsize=4.5truecm \epsfbox{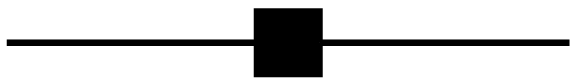}}
\put(-93,-20){$\delta V_s$}
\put(1,1){$
\displaystyle{~~+~~\cdots ~~\sim ~~ G_c\delta V_s G_c}~~+~~\cdots$,}
\label{figdeltaV}
\end{figure}

\noindent
where the black square represents a generic $\delta V_s$ correction to the 
singlet Coulomb Hamiltonian. 

\begin{figure}[htb]
\hspace{-0.1in}
\epsfxsize=4in
\centerline{\epsffile{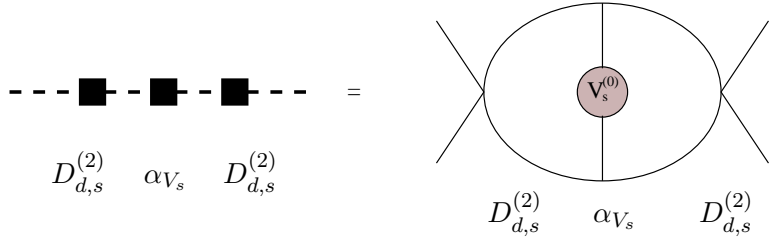}}
\bigskip
\put(-365,0){$D_{d,s}^{(2)}$}
\put(-330,0){$\alpha_{V_s}$}
\put(-300,0){$D_{d,s}^{(2)}$}
\put(-200,-15){$D_{d,s}^{(2)}$}
\put(-160,-15){$\alpha_{V_s}$}
\put(-120,-15){$D_{d,s}^{(2)}$}
\caption {{\it One possible contribution to the running of $D_{d,s}^{(2)}$ at
NLL. The first picture represents the calculation in 
terms of the free quark-antiquark propagator $G_c^{(0)}$ and the 
potentials (the small rectangles). The picture on the right 
is the representation within a more standard diagrammatic
interpretation in terms of quarks and antiquarks. The delta potentials
are displayed as local interactions and the Coulomb potential as an
extended object in space (but not in time).}}
\label{obs12}
\end{figure}

US loops can be computed using standard Feynman diagram techniques,
where it is sometimes convenient to work in momentum space for the US
momenta and in position space for the soft scale (this is certainly so if one
wants to do standard (finite) quantum-mechanical perturbation theory, although
it is clearly possible to do it in momentum space). We illustrate the
procedure with the first US contribution to $G_s$:

\begin{figure}[htb]
\makebox[0.0cm]{\phantom b}
\put(-200,1){$\delta G_s^{\rm us}=$}
\put(-150,-2){\epsfxsize=5truecm \epsfbox{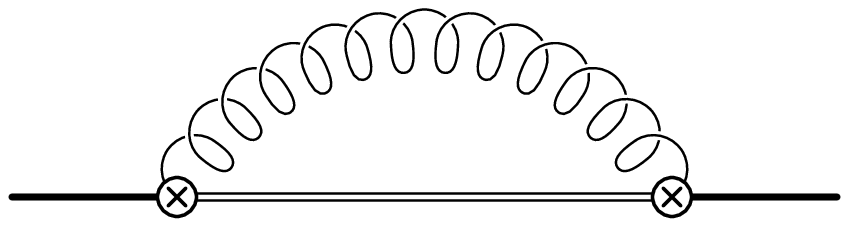}}
\put(1,1){$\displaystyle{\sim  G_c(E)\int { d^{d}{\bf k} 
\over (2\pi)^{d}}
{\bf r}\,{k \over k+h_o^{(0)}-E}
{\bf r}\,G_c(E)}$}
\end{figure}
\be
\label{figus}
\sim G_c(E)\,
{\bf r}\, (h_o^{(0)}-E)^3
\left\{
{1 \over \epsilon}+\gamma+\ln{(h_o^{(0)}-E)^2 \over \nu_{us}^2}+C
\right\}\,
{\bf r}\, G_c(E)
\,,
\vspace{3mm}
\ee
where $d=3+2\epsilon$. 
We can see that the result is UV divergent. This is not a problem 
in an EFT, where such divergences can (and should) 
be absorbed in the matching coefficients of the EFT, i.e. 
in the potentials. Moreover, there are other sources of logarithmic 
UV divergences, proportional to  $\ln\nu_p$, coming from potential loops. 
They show up by either 
going to high enough orders  in quantum-mechanical perturbation theory
(for instance if we are interested in  computing the spectrum at 
${\cal O}(m\als^6)$), 
\be
G_c(E)\delta V_s G_c(E) \cdots \delta V_s G_c (E)
\,,
\ee
or by inserting sufficiently singular operators in the computation (as it is 
the case for the renormalization of the matching coefficient of the 
electromagnetic current). These divergences can be absorbed in the
matching coefficients of the local potentials (those proportional 
to $\delta^{(3)}({\bf r}$) or its derivatives) or 
in the matching coefficients associated with the currents. 
Let us explain how this works in detail. Since the singular behavior of
the potential loops appears for $|{\bf p}| \gg \als/r$, a perturbative
expansion in $\als$ is allowed in $G_c(E)$, which can be approximated by the 
free propagator:

\begin{figure}[htb]
\makebox[0.5cm]{\phantom b}
\put(-150,1){\epsfxsize=3truecm \epsfbox{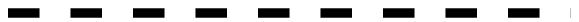}}
\put(-50,1){$
\label{Gc0}
\equiv G_c^{(0)}(E)=\displaystyle{
{1 \over \displaystyle{E-{\bf p}^2/m}}} \,.$}
\end{figure}
\noindent 
Therefore, a practical simplification follows from the fact that the Coulomb potential, $-C_F
{\als/r}$, can be considered a perturbation as far as the computation of the
$\ln \nu_p$ UV divergences is concerned.  Moreover, each
$G_c^{(0)}$ produces a potential loop and one extra power of $m$ in the
numerator, which kills the powers of $1/m$ in the different
potentials. This allows the mixing of potentials with different powers
of $1/m$. One typical example is the diagram in Fig.~\ref{obs12}, 
which corresponds to  
\be
G_c^{(0)}(E)
{\pi C_F D^{(2)}_{d,s} \over m^2}\delta^{(3)}({\bf r})
G_c^{(0)}(E)
C_F {\al_{V_s} \over r}
G_c^{(0)}(E)
{\pi C_F D^{(2)}_{d,s} \over m^2}\delta^{(3)}({\bf r})
G_c^{(0)}(E)
\,.
\ee
The relevant computation reads
\bea
&&
\langle{\bf r}=0|
G_c^{(0)}(E)
C_F {\al_{V_s} \over r}
G_c^{(0)}(E)
|{\bf r}=0\rangle
\\
\nn
&&
\qquad
\sim 
\int \frac{
{\rm d}^d p' }{ (2\pi)^d } \int \frac{ {\rm d}^d p }
{ (2\pi)^d } \frac{ m }{{\bf p}'^2 - mE } 
C_F
\frac{ 4\pi\alpha_{V_s} }{ {\bf q}^2 } \frac{ m }{{\bf p}^2-m E } 
\sim
- C_F\frac{m^2\alpha_{V_s}}{16\pi}  
\frac{ 1 }{\epsilon },
\eea
where ${\bf q}={\bf p}-{\bf p}'$. This divergence can be absorbed in 
$D_{d,s}^{(2)}$ contributing to its running as follows 
\be
\label{eqDd}
\nu_p {d \over d\nu_p}D_{d,s}^{(2)}(\nu_p) \sim 
\alpha_{V_s}(\nu_p)D_{d,s}^{(2)2}(\nu_p)+\cdots
\,.
\ee

It is particularly appealing how the EFT framework gives a
solution to the problem of the UV divergences one finds in standard
quantum-mechanical perturbation theory calculations. When potential
divergences are found it can be more convenient to work in a momentum
representation (see for instance \cite{Czarnecki:1999mw}).  Nevertheless, it
is also possible to handle the UV divergences in position space
\cite{Yelkhovsky:2001tx}.  Either way, the computation should be performed
in the same scheme used to compute the potentials (see
sec.~\ref{pNRmatchingI} for details).

\subsubsection{Heavy quarkonium mass}
After this discussion and taking into account the power counting rules 
given in \ref{pNRweakcounting} one can obtain the different observables
up to some order in $v \sim \als$. For instance, the level of precision of 
the perturbative computation for the heavy quarkonium mass
\be
\label{Mnlj}
M_{nlj}^{\rm pert.}=2m+\sum_{m=2}^{\infty}A_{nlj}^{(m)}\als^m
\,,
\ee
is as follows (some results were actually computed prior to the 
existence of pNRQCD). The ${\cal O}(m\als^2)$ result is nothing but the 
positronium-like result with the proper color factor. The ${\cal O}(m\als^3)$
contribution was computed by \cite{Billoire:1979ih}. The ${\cal O}(m\als^4)$ term was computed by
\cite{Pineda:1998hz,Melnikov:1998pr,Penin:1998kx,Pineda:1998ja}, 
the one at ${\cal O}(m\als^5\ln\als)$ by 
\cite{Brambilla:1999xj,Kniehl:1999mx,Hoang:2001rr}, the NNNLO large-$\beta_0$ result 
by \cite{Kiyo:2000fr,Hoang:2000fm}, and the computations that complete the NNNLO
result for the ground state (but without the static potential 
three-loop coefficient) by \cite{Kniehl:2002br,Penin:2002zv}\footnote{The application 
of pNRQED (the QED version of pNRQCD) and, in general, of factorization with DR, 
has also led to a plethora of results for the spectra of positronium 
\cite{Pineda:1998kn,Czarnecki:1998zv,Melnikov:1999uf,Kniehl:2000cx,Melnikov:2000zz}.}. 
Logarithms have also been resummed for the heavy quarkonium mass 
(we refer to sec.~\ref{RGreview} for details).

In principle, for the bottomonium ground state, finite charm mass effects
have to be taken into account, since the soft scale is of the order of the
charm mass. They can be found in \cite{Melles:2000dq,Eiras:2000rh,
Hoang:2000fm,Wang:2004bk}.

So far, non-perturbative effects have not been discussed. Therefore, it was 
implicitly assumed that $\lQ \ll mv^2$, which makes them 
relevant at ${\cal O}(m\als^5)$, where US modes appear for the 
first time. This assumption may be
reasonable for $t$--$\bar t$ systems, but for bottomonium and charmonium
it is more questionable. In the situation $\lQ \simeq mv^2$, one 
cannot compute using perturbation theory at the US scale. In this 
situation (which may be relevant for bottomonium), the energy of the 
heavy quarkonium reads as follows
\be
\label{Mnljbis}
M_{nlj}=2m+\sum_{m=2}^{\infty}A_{nlj}^{(m)}(\nu_{us})\als^m+
\delta M_{nlj}^{\rm US}(\nu_{us})
\,,
\ee
where the $\nu_{us}$ scale dependence of the different pieces
cancels in the overall sum (for the perturbative sum, this dependence first appears
in  $A_{nlj}^{(5)}$) and ($E_n \equiv A_{nlj}^{(2)}\als^2$)
\begin{equation}
\delta M_{nlj}^{\rm US}(\nu_{us}) \simeq \delta M_{nl}^{\rm US} (\nu_{us}) = 
{T_F \over 3 N_c}  \int_0^\infty \!\! dt 
\langle n,l |{\bf r} e^{-t(h_o^{(0)}-E_n)} {\bf r}| n,l \rangle \langle g{\bf E}^a(t) 
\phi(t,0)^{\rm adj}_{ab} g{\bf E}^b(0) \rangle(\nu_{us}), 
\label{energyUS}
\end{equation}
for which one can think of several possibilities depending on the relative size between
$mv^2$ and $\lQ$. In the limit $mv^2 \gg \lQ$, the result obtained by 
\cite{Penin:2002zv} is the combination 
\be
A_{nlj}^{(5)}(\nu_{us})\als^5+\delta M_{nlj}^{\rm
US}(\nu_{us})|_{\rm {\cal O}(\als^5)\; pert.}
\,.
\ee 
The expression for the non-perturbative object looks similar to 
Eq.~(\ref{energyUS}) but with an UV cutoff, $\Lambda$, such that 
$mv^2 \gg \Lambda \gg \lQ$. Therefore we have 
\be 
\delta M_{nlj}^{\rm US}(\nu_{us})=\delta M_{nlj}^{\rm pert.,US}(\nu_{us};\Lambda)
+\delta M_{nlj}^{\rm US}(\Lambda)
\,.
\ee
The study of the non-perturbative effects in this limit, often called Voloshin--Leutwyler limit,
has a long history starting from \cite{Voloshin:1979hc,Leutwyler:1981tn}. 
$\delta M_{nlj}^{\rm US}(\Lambda)$ reads
(this expression follows by Fourier transforming to energy space Eq.~(\ref{energyUS}) 
and setting $\nu_{us}=\Lambda$) 
\be
\label{deltan}
\nonumber
\delta M_{nlj}^{\rm US}(\Lambda)
= {g^2 \over 6N_c}
\langle {\rm vac} \vert E^{a}_j(0)
\langle n,l \vert
{\bf r}  
\left[
{1 \over E_n - h_o^{(0)} -iD_0^{\rm adj}}
\right]_{ab} 
{\bf r} \vert n,l \rangle
E^{b}_j (0) \vert  {\rm vac} \rangle \,.
\ee
A notation closer to the one used by \cite{Voloshin:1979hc} can be obtained by going 
to a Hamiltonian formulation (for instance, fixing the gauge $A_0=0$).
This corresponds to replace $iD_0^{\rm adj} \rightarrow H^{(0)}$, 
where $H^{(0)}$ is defined in Eq.~(\ref{H0}) and the physical states are
constrained to satisfy the Gauss law (projected to the octet sector)
\be
{\bf D}\cdot {\bfPi}^a \vert {\rm phys} \rangle = 
\left(
\int d^3R \;
{\rm Tr} \left\{  {\rm O}^\dagger  [g T^a, \,{\rm O}] \right\}
+\bar{q}\gamma^0T^aq
\right)
\vert {\rm phys} \rangle, 
\label{gausslawpNRQCD}
\ee
where $\bfPi^a$ is the canonical momentum conjugated to ${\bf A}^a$.
As far as we do not study the fine and hyperfine splittings (see \cite{Leutwyler:1981tn,
Curci:1982fc,Campostrini:1986hy,Kramer:1990ek,Titard:1994id,Pineda:1996nw} for such 
studies in the Voloshin--Leutwyler limit),  the corrections do not
depend on $j$ (total angular momentum) and $s$ (spin) so we will not
display these indices in the states.
The octet propagator mixes low $ {\cal O} (iD_0^{\rm adj} \sim \lQ$) 
and high energies $ {\cal O}(h_o^{(0)} \sim E_n \sim mv^2$). Therefore,  an 
operator product expansion can be performed whose expansion parameter is of order
\be
\left( {iD_0^{\rm adj} \over E_n - h_o}\right)^2 \sim
\left( {\lQ \over m \beta_n^2} \right)^2,
\ee
and one obtains
\be
\delta M_{nl}^{\rm US}(\Lambda) = \sum_{r=0}^{\infty} C_r O_r\equiv 
\sum_{r=0}^{\infty}\delta E^{(r)}_{nl}\,,
\ee
where
\be
C_r =
\langle n,l \vert
{\bf r} \left( {1 \over E_n - h_o} \right)^{2r+1}
{\bf r} \vert n,l \rangle\,,
\ee
\be
\label{Or}
O_r =
{g^2 \over 54}
\langle {\rm vac} \vert
{\rm Tr} \left(
[D_{0}(0),[...[D_{0}(0),{\bf E}(0)]...]
[D_{0}(0),[...[D_{0}(0),{\bf E}(0)]...]
\right) \vert  {\rm vac} \rangle\,,
\ee
and the trace is in the adjoint representation.
$\delta E_{nl}^{(0)}$ has been obtained by \cite{Leutwyler:1981tn,
Voloshin:1982uv,Pineda:1997uk} and 
$\delta E^{(1)}_{nl}$ by \cite{Pineda:1997uk}. For further details, we 
refer to these works.

What we have discussed applies for $t\bar{t}$ production near threshold. 
In the case of bottomonium or charmonium, it is more likely that the
kinematical situations $mv^2 \sim \lQ$ (where the whole functional
form of the chromoelectric correlator is needed) or $\lQ \gg mv^2$ apply. 
This last situation is discussed in ch.~\ref{sec:SCR}. 
A phenomenological analysis is presented in  sec.~\ref{Groundstatemass}.

\subsubsection{Inclusive decay widths}
It is rather easy, after the matching has been performed, 
to calculate in pNRQCD the inclusive decay width of a heavy quarkonium 
H into light particles. This is the imaginary part of the singlet propagator 
pole in the complex plane and may be calculated as (at LO in 
Im ${H}$)
\be
\Gamma({\rm H} \to \hbox{ light particles})= 
- 2\,  \langle n,l,s,j|{\rm Im} \, H  |n,l,s,j \rangle.
\label{imag}
\ee
The imaginary part of the pNRQCD Hamiltonian has been written in 
Eqs.~(\ref{imh2pert}) and  (\ref{imh4pert}). It depends on delta (or derivatives 
of delta) potentials and does not mix singlet and octet fields.  
The states $|n,l,s,j \rangle$ are the eigenstates of the pNRQCD Hamiltonian. 
For electromagnetic inclusive decays, Im$\,f_{\rm EM}^{\rm pNR}({}^3S_1)$ is 
needed (or equivalently the matching coefficient of the electromagnetic 
current, $b^v_{1,\rm pNR}$) for the decay into $e^+e^-$ and 
Im$f_{\rm EM}^{\rm pNR}({}^1S_0)$, for the decay into $\gamma\gamma$. 
The first matching coefficient is known at present with two-loop 
accuracy \cite{Kallen:1955,Beneke:1997jm,Czarnecki:1998vz} in a closed analytic  
form. For the second, besides the one-loop result by \cite{Harris:1956}, 
a semi-analytic two-loop result was obtained by \cite{Czarnecki:2001gi}.
Apart from the electromagnetic matching coefficients, the relevant 
calculation is that of the residue of the NR propagator at the origin:
\be
{\
\lower-2.2pt\vbox{\hbox{\rlap{
{Res}}\lower9pt\vbox
{\hbox{$\scriptscriptstyle{E=E_{\rm pole}}$}}}}\
}
\langle {\bf r} =0 |G_s(E)|  {\bf r} =0 \rangle
=|\phi_n^{(0)}|^2
\left(1+\delta \phi_n\right)^2
\,,
\ee
where 
\be
|\phi_n^{(0)}|^2=
{1 \over \pi} \left({m C_F \als \over 2n}\right)^3\equiv \rho_n\,,
\ee
and $E_{\rm pole}$ is the energy for which $G_s(E)$ has a pole. 
Explicit expressions for the purely perturbative computation 
at NNLO can be found in \cite{Melnikov:1998ug,Penin:1998kx}. Note 
that at this order the LO expressions for Im $g_{\rm EM}^{\rm pNR}({}^3S_1)$ and 
Im $g_{\rm EM}^{\rm pNR}({}^1S_0)$ are also 
needed. Therefore, with NNLO precision, the electromagnetic decays can 
be written in the following way
\bea
\Gamma(V_Q (nS) \rightarrow e^+e^-) &=&
{4C_A \over m^2}\rho_n
\left[
{\rm Im} f_{\rm EM}^{\rm pNR}({}^3S_1)
(1+\delta \phi_n)^2
+{\rm Im} g_{\rm EM}^{\rm pNR}({}^3S_1){E_n \over m}
\right]
\,,
\\
\Gamma(P_Q (nS) \rightarrow \gamma\gamma) &=&
{4C_A \over m^2}\rho_n
\left[
{\rm Im} f_{\rm EM}^{\rm pNR}({}^1S_0)
(1+\delta \phi_n)^2
+{\rm Im} g_{\rm EM}^{\rm pNR}({}^1S_0){E_n \over m}
\right]
\,,
\eea
where $V$ and $P$ stand for the vector and pseudoscalar heavy
quarkonium. Some higher order corrections are also known. The 
${\cal O}(\als^3\ln\als)$ term has been computed by \cite{Kniehl:2002yv,Hoang:2003ns}, 
the ${\cal O}(\als^3\ln^2\als)$ term by \cite{Kniehl:1999mx}\footnote{A major 
progress has also been obtained in QED for positronium decays using these 
techniques, see \cite{Melnikov:2000fi,Kniehl:2000dh}.}. 
For RG improved expressions, see sec.~\ref{RGreview}.

For the non-perturbative corrections, a discussion similar to the 
mass case applies concerning the relative size between $\lQ$ and $mv^2$. Near
the pole $E_n$, we have the expansion (we only consider non-perturbative 
corrections in what follows)
\bea
\nonumber
&&\langle {\bf r} =0 |G_s(E)|  {\bf r} = 0 \rangle =
{ \rho_n + \delta \rho^{\rm np}_n \over E_n +\delta E_{n0}^{\rm np} -E} 
+ {\cal O}((E_n +\delta E_{n0}^{\rm np} -E)^0) \\ && =
{ \rho_n \over E_n -E}- { \rho_n \delta E_{n0}^{\rm np} \over (E_n -E)^2}+
{ \delta \rho_n^{\rm np} \over E_n -E} + {\cal O}((E_n -E)^0)+{\cal O}(\delta^{\rm np} E_{n0}^2) \,.
\eea
On the other hand, one obtains
\be
\langle {\bf r}=0 \vert
G_s(E)\vert {\bf r}=0 \rangle \simeq \langle {\bf r}=0 \vert
G_c(E)\vert {\bf r}=0
\rangle + \langle {\bf r}=0 \vert \delta G_s^{\rm np}(E) \vert {\bf r}=0
\rangle \,,\ee
where
\bea
&&
\langle {\bf r}=0 \vert \delta G_s^{\rm np}(E) \vert {\bf r}=0 \rangle 
={g^2 \over 18}
\langle {\rm vac} \vert E^{a}_j(0)
\langle {\bf r}=0 \vert
{1 \over h_s^{(0)} -E }
{\bf r}  
\left[
{1 \over h_o^{(0)} +iD_0^{\rm adj} -E }
\right]_{ab}
{\bf r}
\nn\\
&&\qquad \times 
{1 \over h_s^{(0)} -E }
\vert {\bf r}=0 \rangle
 E^{b}_j(0) \vert {\rm vac} \rangle
=- { \rho_n \delta E_{n0}^{\rm np} \over (E_n -E)^2}+
{ \delta \rho^{\rm np}_n \over E_n -E} + {\cal O}((E_n -E)^0)\,.
\eea
Proceeding in the same way as before, we can factorize $mv^2$ from $\lQ$ effects:
\be
\langle {\bf r}=0 \vert \delta G_s^{\rm np}(E) \vert {\bf r}=0 \rangle =
\sum_{r=0}^{\infty} C_r^{G} O_r\,,
\ee
where
\bea
\nonumber
&C_{r}^G &=
\langle {\bf r}=0 \vert
{1 \over h_s^{(0)} -E }
{\bf r}  \left({1 \over h_o^{(0)} -E }\right)^{2r+1}
{\bf r}
{1 \over  h_s^{(0)}-E }
\vert {\bf r}=0 \rangle
\\
&&
= { A_{-2}^{(r)} \over (E_n -E)^2}+
 { A_{-1}^{(r)} \over (E_n -E)}+
{\cal O}((E_n -E)^0),
\eea
and $O_r$ is defined in Eq.~(\ref{Or}).
Now, from these expressions, we can read off the observables we are
interested in, namely
\be
\label{drho}
\delta \rho_n^{\rm np} \equiv \sum_{r=0}^{\infty} \delta \rho_n^{(r)} =
\sum_{r=0}^{\infty} A_{-1}^{(r)} O_r\,,
\qquad \qquad
\delta E_{n0}^{\rm np}={ -1 \over \rho_n} \sum_{r=0}^{\infty} A_{-2}^{(r)}O_r\,.
\ee
This also provides a new method to obtain the energy corrections for $l=0$
states, which can be used to check the results of the previous
subsection. $\delta \rho_n^{(0)}$ and $\delta E^{(0)}_{n0}$ were 
calculated by \cite{Voloshin:1982uv} and $\delta \rho_n^{(1)}$ by 
\cite{Pineda:1997uk}. We refer to these works for further details. 

NR sum rules and $t$-$\bar t$ production near threshold will be discussed 
in secs.~\ref{sec:sumrules} and \ref{tt} respectively. For those, the relevant
objects to be computed are again $\langle {\bf r}=0 \vert G_s (E) \vert {\bf
r}=0 \rangle$, but for arbitrary energy $E \sim mv^2$,  
and the electromagnetic matching coefficients considered before.
Finally, it is also possible to obtain RG-improved expressions,  
which we consider in the next section.

\subsection{Renormalization group}
\label{RGreview}
Schematically, we can write the pNRQCD Lagrangian as an expansion in 
$r$ and $1/m$ in the following way
\be
{\cal L}_{\rm pNRQCD} =\sum_{n=-1}^{\infty}r^n{\tilde V}_n O_n
+{1 \over m}\sum_{n=-2}^{\infty}r^n{\tilde V}_n^{(1)}O_n^{(1)}
+{\cal O}\left({1 \over m^2}\right),
\ee
where ${\tilde V}_{n}^{(\ell)}$ (${\tilde V}_n^{(0)} \equiv {\tilde V}_n$) are
dimensionless constants (in four dimensions).
Since they reabsorbe the divergences of the EFT in the way explained in
sec.~\ref{pNRweakObservables}, they will depend on $\nu_p$ and $\nu_{us}$.
One can obtain RG improved expressions for ${\tilde V}_{n}^{(\ell)}$ in
the following way.

One first performs the matching from QCD to NRQCD. The latter depends on
some matching coefficients: $c(\nu_s)$ and $f(\nu_p,\nu_s)$, which can
be obtained order by order in $\als$ (with $\nu_p=\nu_s$) following the
procedure described in sec.~\ref{NRmatching}. In sec.~\ref{sec:NRRG}, we discussed
the procedure to get the running of $c$ and the soft ($\nu_s$)
running of $f$ at any finite order (basically using HQET techniques).
Nevertheless, the running of $f(\nu_p,\nu_s)$ is more complicated beyond
one loop since a dependence on $\nu_p$ appears.
As we will see, it can be obtained within pNRQCD.

The second step is the matching from NRQCD to pNRQCD. The latter depends
on some matching coefficients (potentials), which typically have the
following structure: ${\tilde V}(c(\nu_s), f(\nu_p,\nu_s),\nu_s, \nu_{us},r)$. 
These potentials can be obtained order by order in
$\als$ following the procedure described in secs.~\ref{pNRmatchingI}
and \ref{sec:matchingII}. The integrals in the matching calculation
depend on a factorization scale $\nu$, which corresponds either to $\nu_s$ or to $\nu_{us}$.
In an explicit calculation, they may be distinguished looking at 
the UV and IR behavior of the diagrams: UV divergences are proportional
to $\ln\nu_s$, which are such to cancel the $\nu_s$ scale dependence inherited from
the NRQCD matching coefficients, and IR divergences are proportional to $\nu_{us}$. In
practice, however, as far as we only want to perform a matching
calculation at some given scale $\nu=\nu_s=\nu_{us}$
(or when working order by order in $\als$ without attempting any resummation
of logarithms), it is not necessary to distinguish between $\nu_s$ and $\nu_{us}$.

The third step is to obtain the RG equations of the potentials. 
$\nu_s$ provides us with the starting point of the
RG evolution with respect to $\nu_{us}$ (up to a constant of order
one).  The running with respect to $\nu_{us}$ can then be obtained
following the procedure described by \cite{Pineda:2000gz,Pineda:2001ra}. Formally, the
RG equations of the matching coefficients due to the $\nu_{us}$-dependence read
\be
\label{nusRG}
\nu_{us} {d \over d \nu_{us}}{\tilde V}=B_{{\tilde V}}({\tilde V}).
\ee
From a practical point of view, one can organize the RG equations
within an expansion in $1/m$ and $\als(\nu_{us})$.
At ${\cal O}(1/m^0)$, the analysis corresponds to the study of the static
limit of pNRQCD, which has been carried out by \cite{Pineda:2000gz}.
Since ${\tilde V}_{-1}\not= 0$, there are
relevant operators (super-renormalizable terms) in the Lagrangian
and the US RG equations lose the triangular structure that we enjoyed
for the RG equations of $\nu_s$. Still, if ${\tilde V}_{-1} \ll 1$, the RG
equations can be obtained as a double expansion in ${\tilde V}_{-1}$ and
${\tilde V}_0$, where the latter corresponds to the marginal operators (renormalizable interactions).
At short distances ($1/r \gg \lQ$), the static limit of pNRQCD is
in this situation. Specifically, we have ${\tilde V}_{-1}=\{\al_{V_s},\al_{V_o}\}$, 
that fulfills ${\tilde V}_{-1} \sim \als(r) \ll 1$;
${\tilde V}_0=\als(\nu_{us})$  and ${\tilde V}_{1}=\{V_A,V_B\} \sim 1$.
Therefore, we can calculate the anomalous dimensions order by
order in $\als(\nu_{us})$. In addition, we also have an expansion in
${\tilde V}_{-1}$. Moreover, the specific form of the pNRQCD
Lagrangian severely constrains the RG equations' general structure.
Therefore, for instance, the leading non-trivial RG equation for $\al_{V_s}$ reads
\be
\label{alVsrunning}
\nu_{us} {d\over d\nu_{us}}\al_{V_{s}}
=
{2 \over 3}{\als\over\pi}V_A^2\left( \left({C_A \over 2} -C_F\right)\al_{V_o}+C_F\al_{V_s}\right)^3
+{\cal O}({\tilde V}_{-1}^4{\tilde V}_0,{\tilde V}_0^2{\tilde V}_{-1}^3)
\,.
\ee
At higher orders in $1/m$ the analysis has been carried out by \cite{Pineda:2001ra}.
The same considerations as for the static limit apply
here as far as the non-triangularity of the RG equations is concerned.
In general, one has the structure
\be
\nu_{us} {d\over d\nu_{us}}{\tilde V}_{n}^{(\ell)}
\sim
\sum_{\{n_i\}\{\ell_i\}}{\tilde V}_{n_1}^{(\ell_1)}{\tilde V}_{n_2}^{(\ell_2)}\cdots{\tilde V}_{n_j}^{(\ell_j)}\,, 
\quad {\rm with}\quad \sum_{i=1}^j \ell_i=\ell\;,\, \sum_{i=1}^j n_i=n\,,
\ee
and one has to pick up the leading contributions from all possible
terms. Actually, as far as the NNLL heavy quarkonium mass is concerned,
the relevant US running can be obtained by computing the
diagram displayed in Eq.~(\ref{figus}) (one also has
to consider the running of $V_A$, which happens to be zero). Working in
DR, one should note that the potentials have to be understood in $D$ dimensions (see for instance
Eq. (3.1) of \cite{Schroder:1999sg}). Therefore, powers of $g_B^2$ (the bare
coupling) have dimensions and have to be compensated by powers of $k^{2\epsilon}$
in $\al_{V_{s}}$. This means that the US divergences ($1/\epsilon$ poles)
generated by the right-hand-side of Eq.~(\ref{alVsrunning}) are absorbed
by the terms in $\al_{V_{s}}$ proportional to $g_B^8$ or to a higher power. Finally, 
by solving Eq. (\ref{nusRG}) between $\nu_s$ and $\nu_{us}$, we
will have ${\tilde V}(c(\nu_s),f(\nu_p,\nu_s),\nu_s,\nu_{us},r)$, where the
running with respect to $\nu_{us}$ is known. Note that the running 
with respect $\nu_s$ is also known, since we demand the potential to 
be independent of it:
\be
\nu_{s}{d\over d\nu_{s}}{\tilde V}=0
\,,
\ee
which can be solved by setting $\nu_s=1/r$. Therefore, one can also deduce the 
dependence of ${\tilde V}$ on $r$. 

The final step is to obtain the RG equation for $\nu_p$.
In pNRQCD, integrals over the relative three-momentum of the heavy quarks occur.
When these integrals are finite no dependence on $\nu_p$ occurs and 
one has $|{\bf p}| \sim 1/r \sim m\als$ and
${\bf p}^2/m \sim m\als^2$. Therefore, one can lower $\nu_{us}$ down to
$\sim m\als^2$ reproducing the results obtained by \cite{Pineda:2001ra}. In general, 
the integrals over ${\bf p}$ 
are divergent, and the structure of the logarithms is dictated by the UV
behavior of ${\bf p}$ and $1/r$. This means that we cannot replace
$1/r$ and $\nu_{us}$ by their physical expectation values but rather
by their cutoffs within the integral over ${\bf p}$, i.e. $\nu_p$.
Therefore, besides the explicit dependence on $\nu_p$  of the
potential, which appears in $f$, the potential also implicitly depends on $\nu_p$
through the requirement $1/r \sim |{\bf p}| \ll \nu_p$, and also through
$\nu_{us}$, since $\nu_{us}$ has to fulfill ${\bf p}^2/m \ll \nu_{us}\ll |{\bf p}|$ 
in order to ensure that only soft degrees of freedom
have been integrated out for a given $|{\bf p}|$. This latter requirement holds 
if we fix the final point of the evolution of the ultrasoft RG equation to
be $\nu_{us}=\nu_p^2/m$. At this stage, a single cutoff,
$\nu_p$, exists and the correlation of cutoffs becomes manifest.
Therefore, for the RG equation for $\nu_p$, the anomalous dimensions
of ${\tilde V}(c(1/r),f(\nu_p,1/r),1/r,\nu_p^2/m,r)$ is at LO the same as
the one of ${\tilde V}(c(\nu_p),f(\nu_p,\nu_p),\nu_p,\nu_p^2/m,\nu_p)$.\footnote{
Roughly speaking, this result can be thought of as expanding $\ln r$ around $\ln \nu_p$ 
in the potential i.e. 
\be 
{\tilde V}(c(1/r),f(\nu_p,1/r),1/r,\nu_p^2/m,r) 
\simeq {\tilde V}(c(\nu_p),f(\nu_p,\nu_p),\nu_p,\nu_p^2/m,\nu_p)
+\ln(\nu_pr)r{d \over d r} {\tilde V}\bigg|_{1/r=\nu_p} + \cdots 
\,.
\ee 
The $\ln(\nu_pr)$ terms give subleading contributions to the
anomalous dimension when introduced in divergent integrals over ${\bf p}$. 
An explicit example of this type of corrections appears in the 
computation of the hyperfine splitting of the heavy quarkonium at NLL \cite{Kniehl:2003ap,Penin:2004xi}.} 
It appears through the divergences induced by the
iteration of the potentials in the way explained in \cite{Pineda:2001et}
and sec.~\ref{pNRweakObservables}. In particular, the computation of the 
anomalous dimension can be organized within an expansion in $\als$ and using the free propagators $G_c^{(0)}$. 
Finally the running will go from $\nu_p \sim m$ down to $\nu_p \sim m\als$.
A similar discussion applies to the running of the matching coefficients
of the currents (or, in other words, of the imaginary terms of the potential).
This completes the procedure to obtain the RG equations for the hard,
soft and US scales. An example is given below. 

This line of investigation has led to several new results on heavy 
quarkonium physics. They can be summarized as follows (we omit all numerical
coefficients that may be found in the quoted literature):
\begin{itemize}
\item
The NNLL correction to the heavy quarkonium energy 
\cite{Pineda:2001ra}, i.e. corrections of order 
\be
\delta E \sim m\als^4+m\als^5\ln\als+m\als^6\ln^2\als+\cdots\,.\\
\ee
\item
The LL \cite{Pineda:2001ra} (first obtained in \cite{Hoang:2001rr}) 
and NLL \cite{Kniehl:2003ap,Penin:2004xi} correction to the heavy quarkonium 
hyperfine splitting
\bea
\delta E_{HF} 
&\sim& m\als^4+m\als^5\ln\als+m\als^6\ln^2\als+\cdots\\
&+&m\als^5+m\als^6\ln\als+m\als^7\ln^2\als+m\als^8\ln^3\als+\cdots \,.
\nn
\eea
\item
The NLL \cite{Pineda:2001et} correction to the inclusive electromagnetic decays 
(this result can be applied to ${\bar t}$-$t$ 
production at threshold or NR sum rules since the running 
of the electromagnetic current matching coefficient is  
the only non-trivial object that appears in the NLL running)
\bea
\Gamma(V_Q (nS) \rightarrow e^+e^-) 
&\sim& 
m\als^3(1+\als^2\ln\als+\als^3\ln^2\als+\cdots)\,,
\\
\nn
\Gamma(P_Q (nS) \rightarrow \gamma\gamma) 
&\sim& 
m\als^3(1+\als^2\ln\als+\als^3\ln^2\als+\cdots)\,,
\eea
and for the ratio the NNLL correction \cite{Penin:2004ay}
\be
{\Gamma(V_Q (nS) \rightarrow e^+e^-) 
\over
\Gamma(P_Q (nS) \rightarrow \gamma\gamma) }
\sim
1+\als^2\ln\als+\als^3\ln^2\als+\cdots
+\als^3\ln\als+\als^4\ln^2\als+\cdots \,.
\ee
\end{itemize}

The resummation of logarithms using EFTs was first addressed within the 
vNRQCD framework \cite{Luke:1999kz} (see also \cite{Manohar:1999xd,
Manohar:2000kr,Hoang:2001rr,Hoang:2002yy,Hoang:2003ns}), where the 
relevance of the cutoff correlation for the RG was first realized. 
Nevertheless, the early formulations of this theory had some 
problems (in particular concerning the treatment of US modes), 
which led to incorrect results for the heavy quarkonium mass at NNLL 
\cite{Hoang:2001rr} and the electromagnetic current matching 
coefficient at NLL \cite{Manohar:2000kr}. 
They have been resolved by \cite{Hoang:2002yy} and their 
results now agree with those obtained 
in pNRQCD \cite{Pineda:2001ra,Pineda:2001et}.
The application of the RG to QED bound states 
has also been considered in both formalisms, see 
\cite{Manohar:2000rz,Pineda:2001et,Pineda:2002bv,Penin:2004ay}.

\subsubsection{An example}
Finally, we illustrate the method in the simplest possible situation
where all the scales appear. We consider the corrections to the heavy quarkonium 
spectrum for the non-equal mass case in the limit where one of the masses ($m_2$) 
goes to infinity, and in the Abelian limit with zero light flavors 
($C_F \rightarrow 1$, $C_A \rightarrow 0$, $T_F \rightarrow 1$, $n_f \rightarrow 0$). 
This is nothing but the hydrogen atom case. We will compute some NNNLL corrections 
to the Lamb shift of ${\cal O}(m\als^8\ln^3\als)$. They were first computed 
using the RG by \cite{Manohar:2000rz}. We will follow here the discussion in 
\cite{Pineda:2002bv,Pineda:2001et}. In this limit, $\als$ does not run and we can 
neglect the four-fermion matching coefficients, since they are suppressed by powers 
of $1/m_2$. Therefore, we only have to consider the running of the matching coefficients 
of the heavy-quark bilinear terms. At ${\cal O}(1/m^2)$, $c_D$ is the only matching 
coefficient with non-trivial running. By solving Eq.~(\ref{cDRG}) in this limit, one obtains
\be
c_D(\nu_s)=1-{8\over 3}
{\als \over \pi}\ln{\nu_s \over m} \,.
\label{cDRGQED}
\ee
At the pNRQCD level, we have to consider first the US RG running of $D_{d,s}^{(2)}$, 
which follows from Eq. (\ref{figus}).
It reads (we already use that $V_A=1$ and $c_S^{(1,-2)}=1$)
\be
\label{DdsRGUS}
\nu_{us} {d\over d\nu_{us}}D_{d,s}^{(2)}= -{4 \over 3}
{\als^2\over \pi} \,.
\ee
By using the initial matching condition:
\be
D^{(2)}_{d,s}(\nu_s)= \als{c_D(\nu_s) \over 2}
\,,
\ee
we can solve Eq.~(\ref{DdsRGUS}). The solution reads
\be
\label{DdQED}
D^{(2)}_{d}(\nu_{us})= {\als \over 2}
\left( 1-{8\over 3}
{\als \over \pi}\ln{\nu_{us} \over m}\right) \,,
\ee
which gives the full NNLL contribution to the spectrum, of ${\cal O}(m\als^5\ln\als)$ 
and nothing else. 
At NNNLL, we can obtain the ${\cal O}(m\als^8\ln^3\als)$ contribution
from Eq.~(\ref{eqDd}), which is due to the diagram in Fig.~\ref{obs12}.
This is because the ${\cal O}(m\als^8\ln^3\als)$ term has the highest possible
power of logarithm that could appear from a NNNLL evaluation of the energy and that, in
order to achieve such power, it is necessary to mix with its NNLL logarithms. As we have
seen, the latter only appear in the LL evaluation of $D_{d}^{(2)}$ (\ref{DdQED}), 
which, indeed, only produces a single logarithm. The other point is that the NLL
evaluation of the potentials only produces single logarithms unless mixed with LL
running. Therefore, the diagrams with the highest possible power of
$D_{d}^{(2)}$ will give the highest possible power of logarithm of the spectrum
at NNNLL. Thus, we only have to solve Eq. (\ref{eqDd}) (note the replacement 
$\nu_{us}=\nu_p^2/m$), which in the limit considered here reads
\be
\label{eqDdmh}
\nu_p {d \over d\nu_p}D_{d,s}^{(2)}(\nu_p) = 
\als D_{d,s}^{(2)2}(\nu_p)+\cdots \,.
\ee
The solution is
\be
\delta D_{d}^{(2)} =
{64 \over 27}
\als^3 \left({\als \over \pi}\right)^2\ln^3{\nu_p \over m} \,.
\ee

\section{Renormalons and the definition of the heavy quark mass}
\label{secrenormalons}

\subsection{The pole mass and static singlet potential renormalon}
\label{secmas}
The pole mass of a heavy quark can be related to the $\MS$ mass by the series
\be
\label{series}
m = m_{\MS} + \sum_{n=0}^\infty r_n \als^{n+1}\,, 
\ee
where $\als \equiv \als(\nu)$, $m_{\MS}$ is calculated at the normalization point $\nu=m_{\MS}$ 
(in this way logarithms that are not associated with the
renormalon are effectively resummed)
and the first three coefficients $r_0$, $r_1$ and $r_2$ are known 
\cite{Gray:1990yh,Melnikov:2000qh,Chetyrkin:1999qi}.
The pole mass is also known to be IR finite and scheme-independent 
at any finite order in $\als$ \cite{Kronfeld:1998di}. We then define the Borel transform 
\be
\label{borel}
m = m_{\MS} + \int\limits_0^\infty d t \,e^{-t/\als}
\,B[m](t)
\,,
\qquad B[m](t)\equiv \sum_{n=0}^\infty 
r_n \frac{t^n}{n!} . 
\ee
We will denote by renormalons the singularities on the real axis of the Borel
plane.\footnote{We will not consider singularities due to instantons \cite{LeGuillou:1990nq}.}
The behavior of the perturbative expansion of Eq.~(\ref{series}) at large orders is dictated by 
the closest renormalon to the origin of its Borel transform, which happens to be located at
$t=2\pi/\beta_0$ \cite{Beneke:1994sw,Bigi:1994em,Neubert:1995wq}.
Being more precise, the behavior of the Borel transform near the
closest renormalon at the origin reads (we define $u={\beta_0 t /(4 \pi)}$)
\be
B[m](t(u))=B[\delta m_{\RS}](t(u))+({\rm term\;analytic\;at}\;u=1/2),
\ee
where
\be
B[\delta m_{\RS}](t(u)) \equiv N_m\nu {1 \over
(1-2u)^{1+b}}\left(1+c_1(1-2u)+c_2(1-2u)^2+\cdots \right).
\ee
This dictates the behavior of the perturbative expansion at large orders to be 
\be
\label{generalm}
r_n \stackrel{n\rightarrow\infty}{=} N_m\,\nu\,\left({\beta_0 \over 2\pi}\right)^n
\,{\Gamma(n+1+b) \over
\Gamma(1+b)}
\left(
1+\frac{b}{(n+b)}c_1+\frac{b(b-1)}{(n+b)(n+b-1)}c_2+ \cdots
\right).
\ee
The different $b$, $c_1$, $c_2$, etc ... can be obtained from the procedure used by \cite{Beneke:1995rs}. 
The coefficients $b$ and $c_1$ were computed by \cite{Beneke:1995rs}, and $c_2$ by 
\cite{Beneke:1998ui,Pineda:2001zq}. They read 
\be
b={\beta_1 \over 2\beta_0^2}\,,
\qquad
\qquad
c_1={1 \over 4\,b\beta_0^3}\left({\beta_1^2 \over \beta_0}-\beta_2\right),
\ee
and 
\be
c_2={1 \over b(b-1)}
{\beta_1^4 + 4 \beta_0^3 \beta_1 \beta_2 - 2 \beta_0 \beta_1^2 \beta_2 + 
   \beta_0^2 (-2 \beta_1^3 + \beta_2^2) - 2 \beta_0^4 \beta_3 
\over 32 \beta_0^8}
\,.
\ee
 Approximate determinations for $N_m$ have been obtained  
 by \cite{Pineda:2001zq,Lee:2002sn,Cvetic:2003wk} (see also \cite{Pineda:2002se}).

One can think of performing the same analysis with the singlet 
static potential in the situation where $\lQ \ll 1/r$. Its perturbative
expansion reads 
\be
V_s^{(0)}(r;\nu_{us})=\sum_{n=0}^\infty V_{s,n}^{(0)} \als^{n+1}.
\ee
The potential, however, is not an IR safe object, since it depends on the IR cutoff $\nu_{us}$, which first appears 
at ${\cal O}(\als^4)$ (for more details see sec.~\ref{sec:matchingII}). 
Nevertheless, these US logarithms are not
associated with the first IR renormalon, since they also appear in momentum
space (see also the discussion below), so they will not be 
considered further in this section. 

We now use the observation that the first IR renormalon of the singlet
static potential cancels with (twice) the renormalon of 
the pole mass. This has been proven in the (one-chain) large $\beta_0$
approximation in \cite{Pineda:1998id,Hoang:1998nz} and at any loop (disregarding
possible effects due to $\nu_{us}$) in \cite{Beneke:1998rk}. It can also be 
argued to hold from an EFT approach where any renormalon ambiguity
should cancel between operators and matching coefficients. 
Let us consider, for instance, the situation  $1/r \gg \lQ$. 
If we understand the quantity $2m+V_s^{(0)}$ as an
observable up to ${\cal O}(r^2\lQ^3,\lQ^2/m)$ renormalon (and/or
non-perturbative) contributions, then this proves the (first
IR) renormalon cancellation at any loop (as well as the
independence of this IR renormalon of $\nu_{us}$). 

One can now read off the asymptotic behavior of the static potential from the one
of the pole mass and work analogously. We define the Borel transform 
\be\label{borelb}
V_s^{(0)} = \int\limits_0^\infty d t \,e^{-t/\als}\,B[V_s^{(0)}](t)
\,,
\qquad 
B[V_s^{(0)}](t)\equiv \sum_{n=0}^\infty 
V_{s,n}^{(0)} \frac{t^n}{n!} . 
\ee
The closest renormalon to the origin is located at 
$t=2\pi/\beta_0$. This dictates the behavior of the perturbative expansion at large orders to be 
\be\label{generalV}
V_{s,n}^{(0)} \stackrel{n\rightarrow\infty}{=} N_{V_s}\,\nu\,\left({\beta_0 \over 2\pi}\right)^n
 \,{\Gamma(n+1+b) \over
 \Gamma(1+b)}
\left(
1+\frac{b}{(n+b)}c_1+\frac{b(b-1)}{(n+b)(n+b-1)}c_2+ \cdots
\right)
,
\ee
and the Borel transform near the singularity reads
\be
B[V_{s}^{(0)}](t(u))=N_{V_s}\nu {1 \over
(1-2u)^{1+b}}\left(1+c_1(1-2u)+c_2(1-2u)^2+\cdots \right)+({\rm
analytic\; term}).
\ee
In this case, by {\it analytic term} we mean an analytic function up to 
the next IR renormalon at $u=3/2$ \cite{Aglietti:1995tg}. 

For $N_{V_s}$ some approximate determinations exist \cite{Pineda:2001zq,Lee:2002sn} 
(see also \cite{Pineda:2002se}). 
Actually, the best determinations come from $N_m$ using the cancellation 
of the pole mass and static singlet potential renormalon, i.e.
\be
2N_m+N_{V_s}=0\,.
\ee

\subsection{Renormalon-subtracted scheme and power counting}
\label{secdefRS}
In EFTs with heavy quarks, the inverse of the heavy
quark mass becomes one of the expansion parameters (and of the matching
coefficients). A natural choice in the past has been the pole mass because
it is the natural definition in processes where the particles
eventually measured in the detectors correspond to the fields in
the Lagrangian (as in QED). This is not the case in QCD.
One consequence of this is that the pole mass suffers from renormalon 
singularities. Moreover, since these renormalon singularities lie close 
to the origin of the Borel plane and perturbative calculations have gone
very far for systems with heavy quarks, they manifest themselves as a poor convergence of 
the perturbative series. It is then natural to try to define a new mass parameter, 
which replaces the pole mass, but is still adequate for threshold problems. 
Several choices have been proposed in the literature: the 
kinetic mass \cite{Bigi:1994em}, the PS mass \cite{Beneke:1998rk}, the 1S mass
\cite{Hoang:1998ng}, the ${\overline {\rm PS}}$ mass
\cite{Yakovlev:2000pv} and the $\RS$ mass \cite{Pineda:2001zq}. All of them 
achieve the renormalon cancellation and share the following structure:
\be
m_{\rm X}=m-\delta m_{\rm X}
\,,
\ee
where X$=\{$PS, 1S, ...$\}$ and $\delta m_{\rm X}$ is an object such that 
\be
B[\delta m_{\rm X}]=B[\delta m_{\RS}]+({\rm analytic\;term\;at}\;u=1/2)
\,.
\ee
The different definitions have different analytic terms. $\delta m_{\rm kin}$ is defined as 
the self energy of a static quark computed with a hard cutoff, $\delta m_{\PS}$ is 
defined as $1/2$ the self energy of the Coulomb potential computed with a hard cutoff 
much smaller than $1/r$, $\delta m_{\overline{\PS}}$ is 
defined as the soft part of the heavy quark self energy computed with a hard cutoff, 
$\delta m_{\rm 1S}$ is $1/2$ the perturbative 
binding energy of the ground state of heavy quarkonium (note that 
in this case the renormalon cancellation is achieved between different powers of $\als$). 
We will not discuss further all these threshold masses. 
Instead, we will focus on one, the $\RS$ mass, 
which better matches with the analyses of the previous
section. In any case, a large part of the discussion also holds when replacing $\RS \to $ X. 
It should be noticed that, 
since different masses implement the renormalon cancellation in
different ways, different systematic errors appear.
For instance, the major error in the $\RS$ mass comes from $N_m$ (see Eq.~(\ref{generalm})).
For the kinetic and $\PS$ masses, it seems difficult to compute higher-order terms.
The $\PS$ and 1S masses depend on the US scale at NNNLO, which may be
problematic once this precision is needed (for instance in $B$ physics).
Finally, the 1S mass assumes the ground state of heavy quarkonium to be mainly 
a perturbative system.  Therefore, having at disposal several masses  
may help to have a better handle on the errors, e.g. in the
extraction of the $\MS$ quark masses. 

The $\RS$ definition tries to cancel the poor perturbative behavior
associated with the renormalon, which is due to the non-analytic terms in
$1-2u$ in the Borel transform of the pole mass. 
These terms also exist in the effective theory.  Therefore,
the procedure followed by \cite{Pineda:2001zq} was to subtract the pure renormalon 
contribution in the new mass definition\footnote{
One could also choose not to include terms proportional to 
$c_n$ for $n \ge 2$, since these terms actually go to zero for $u \rightarrow 1/2$ 
for the physical values of $b \sim 0.4$.}, which was called RS mass, 
$m_{\RS}$, and reads
\be
\label{mrsvsmpole}
m_{\RS}(\nu_f)=m-\sum_{n=1}^\infty  N_m\,\nu_f\,\left({\beta_0 \over
2\pi}\right
)^n \als^{n+1}(\nu_f)\,\sum_{k=0}^\infty c_k{\Gamma(n+1+b-k) \over
\Gamma(1+b-k)}
\,,
\ee
where $c_0=1$. We expect that with this renormalon free definition, the 
coefficients multiplying the expansion parameters in the effective
theory calculation will have a natural size and that the same holds for the coefficients 
multiplying the powers of $\als$ in the perturbative expansion relating $m_{\RS}$ to $m_{\MS}$. 
Therefore, we do not lose accuracy if we first obtain $m_{\RS}$ and later on we
use the perturbative relation between $m_{\RS}$ and
$m_{\MS}$ in order to obtain the latter. Nevertheless, since we will work order by order in
$\als$, in the relation between $m_{\RS}$ and
$m_{\MS}$ it is important to expand everything in terms of
$\als$, specifically $\als(\nu_f)$,  
in order to achieve the renormalon cancellation order by order in
$\als$. Then, the
perturbative expansion in terms of the $\MS$ mass reads 
\be
m_{\RS}(\nu_f)=m_{\MS} + \sum_{n=0}^\infty r^{\RS}_n\als^{n+1}\,,
\ee
where $r^{\RS}_n=r^{\RS}_n(m_{\MS},\nu,\nu_f)$. These $r^{\RS}_n$ are
the ones expected to be of natural
size (or at least not to be artificially enlarged by the first IR renormalon).
 
These definitions significantly improve the convergence of the 
perturbative series in comparison with the pole mass. We refer 
to \cite{Pineda:2001zq} for numerical details.

The shift from the pole mass to the RS mass  affects the explicit 
expression of the effective Lagrangians. In particular, in HQET, at
LO, a residual mass term appears in the Lagrangian
\be
{\cal L}=\bar h \left(iD_0-\delta m_{\RS}\right)h+{\cal O}\left({1 \over
m_{\RS}}\right) \,,
\ee
where $\delta m_{\RS}=m-m_{\RS}$ and similarly for the NRQCD
Lagrangian.  

For pNRQCD in the situation where $\lQ \ll
m\als$, if we consider the LO in $1/m$, the residual mass 
term is absorbed in the
static potential (in going from NRQCD to pNRQCD, one runs down the scale
$\nu_f$ to $\nu_f \siml m\als$). We can then,
analogously to the RS mass, define a singlet static RS potential 
\be
V^{(0)}_{s,\RS}(\nu_f)=V^{(0)}_s+2\delta m_{\RS}
\,,
\ee
where the coefficients multiplying the perturbative series should be of ${\cal O}(1)$
(provided that we expand $V^{(0)}_s$ and $\delta m_{\RS}$ in the same parameter,
namely $\als$). Notice also the trivial fact that the scheme dependence
of $m_{\RS}$ cancels with the scheme dependence of $V_{\RS}$. 
This definition significantly improves of
the perturbative expansion in the potential. For a numerical analysis 
we refer to \cite{Pineda:2001zq,Pineda:2002se}.  

The pNRQCD Lagrangian in the weak-coupling regime 
in the RS scheme is formally equal to the one in 
the on-shell scheme (see Eq.~(\ref{Lpnrqcd})) with the modifications 
$m_{1(2)} \rightarrow m_{1,\RS(2,\RS)}$, $V \rightarrow V_{\RS}$ and so on. 
Note in particular that now the expansion is in terms of $1/m_{\RS}$\footnote{ 
Note that the definition of the RS scheme in the octet sector is more involved, 
since there are some renormalons left at $u=1/2$ in $V^{(0)}_{o}+2\delta m_{\RS}$. 
The reason is that, even at LO in $1/m$, $2m+V_o^{(0)}$ is not an
 observable. This is due to the fact that there is still interaction
 with low energy gluons. Therefore, one expects $2m+V_o^{(0)}$ to
 be ambiguous by an amount of ${\cal O}(\lQ)$. We will elaborate on this in the 
 next chapter.}. 
One can then compute observables along the lines of sec.~\ref{pNRweakObservables} 
(at the practical level, one could work in the on-shell scheme and
do the above replacement to go to the RS scheme). For instance, one would obtain the following expression for  
the heavy quarkonium spectrum (see Eq.~(\ref{Mnlj})):
\be
\label{MnljRS}
M_{nlj}=2m_{\RS}
+\sum_{m=2}^{\infty}A_{nlj}^{m,\RS}(\nu_{us})\als^m+\delta M_{nlj}^{\rm US}(\nu_{us})
\,,
\ee
where the $\nu_{us}$ scale dependence of the different pieces
cancels in the overall sum (for the perturbative sum, this dependence first appears
in  $A_{nlj}^{5,\RS}$). 

We expect that by working with the RS scheme the coefficients multiplying
the powers of $\als$ will now be of natural size and, therefore, the
convergence is improved compared with the on-shell scheme. Actually, 
this seems to be the case. See ch.~\ref{sec:phen} for details and a phenomenological discussion. 

Finally, we would like to discuss some theoretical issues
(see also the discussion in \cite{Beneke:1998ui}). First, once one
agrees to give up using the pole mass as an expansion parameter, one
may still wonder why not to use the $\MS$ mass instead. There are
several answers to this question. The first is that 
due to the fact that there is another scale, $m\als$, besides $m$, one would
not achieve the renormalon cancellation order by order in $\als$ but rather 
between different orders in $\als$, jeopardizing in this way  
the convergence of the perturbative expansion. This can 
be resolved by using the upsilon expansion \cite{Hoang:1998ng}. 
Nevertheless, some other problems may remain. On the one hand, working with
$m_{\MS}$ would mean introducing a large shift in the pNRQCD
Lagrangian of ${\cal O}(m\als)$, and
therefore jeopardizing the power counting rules.\footnote{This is
certainly so for $t$-$\bar t$ physics. Nevertheless, for bottom, 
the ${\cal O}(m\als)$ term does not seem to be that large numerically,
being much smaller than the typical values
of the soft scale in the $\Upsilon(1S)$. Therefore, it may happen that
working with the $\MS$ mass does not destroy the power counting rules
of pNRQCD (or HQET) at the practical level.}
Furthermore, by expanding everything in terms of $\als$,
we may introduce a potentially large logarithm, $\ln{m / \nu}$
(note that we cannot minimize this logarithm except at the
price of introducing another large logarithm, $\ln (m\als / \nu)$).

\section{(p)NRQCD: the static limit}
\label{sec:static}
Although NRQCD and pNRQCD were originally designed to study $Q$-$\bar Q$ 
systems of large
but finite mass, it is very interesting to consider their static limit 
(where
$m \rightarrow \infty$ while keeping all the other scales finite).
On the one hand the static energy spectra are the main ingredient for 
the potentials
both in the strong and in the weak-coupling regime.
On the other hand the study of the energy spectrum is interesting by
itself. For instance, a linear dependence on $r$ for the ground state 
energy at long distances is usually considered a proof of confinement.
The abundant lattice data (at least of quenched simulations) makes it
possible to study quantitatively
for which distances the potentials are in the perturbative or
non-perturbative regime, providing a controlled framework
to discern when to use the weak- or the strong-coupling version of pNRQCD.
To answer this question the proper handling of the renormalon singularities
will be crucial.

\subsection{NRQCD in the static limit}
\label{hybrids}
The Hamiltonian associated with the Lagrangian (\ref{LagNRQCD}) is
\begin{eqnarray}
H &=& H^{(0)}+{\cal O}(1/m),
\label{HH}\\
H^{(0)} &=&
\int d^3{\bf x} {1\over 2}\left( {\bfPi}^a{\bfPi}^a +{\bf B}^a{\bf B}^a 
\right)
-\sum_{j=1}^{n_f} \int d^3{\bf x}\, \bar{q}_j \, i {\bf D }\cdot 
{\bfgamma} \, q_j \, ,
\label{H0}
\end{eqnarray}
and the physical states are constrained to satisfy the Gauss law:
\be
{\bf D}\cdot {\bfPi}^a \vert {\rm phys} \rangle =
g (\psi^\dagger T^a \psi + \chi^\dagger T^a \chi + \sum_{j=1}^{n_f} 
\bar{q}_j \gamma^0 T^a q_j)
\vert {\rm phys} \rangle.
\label{gausslaw}
\ee

\begin{figure}
\epsfxsize=0.85\columnwidth
\includegraphics[width=0.6\columnwidth]{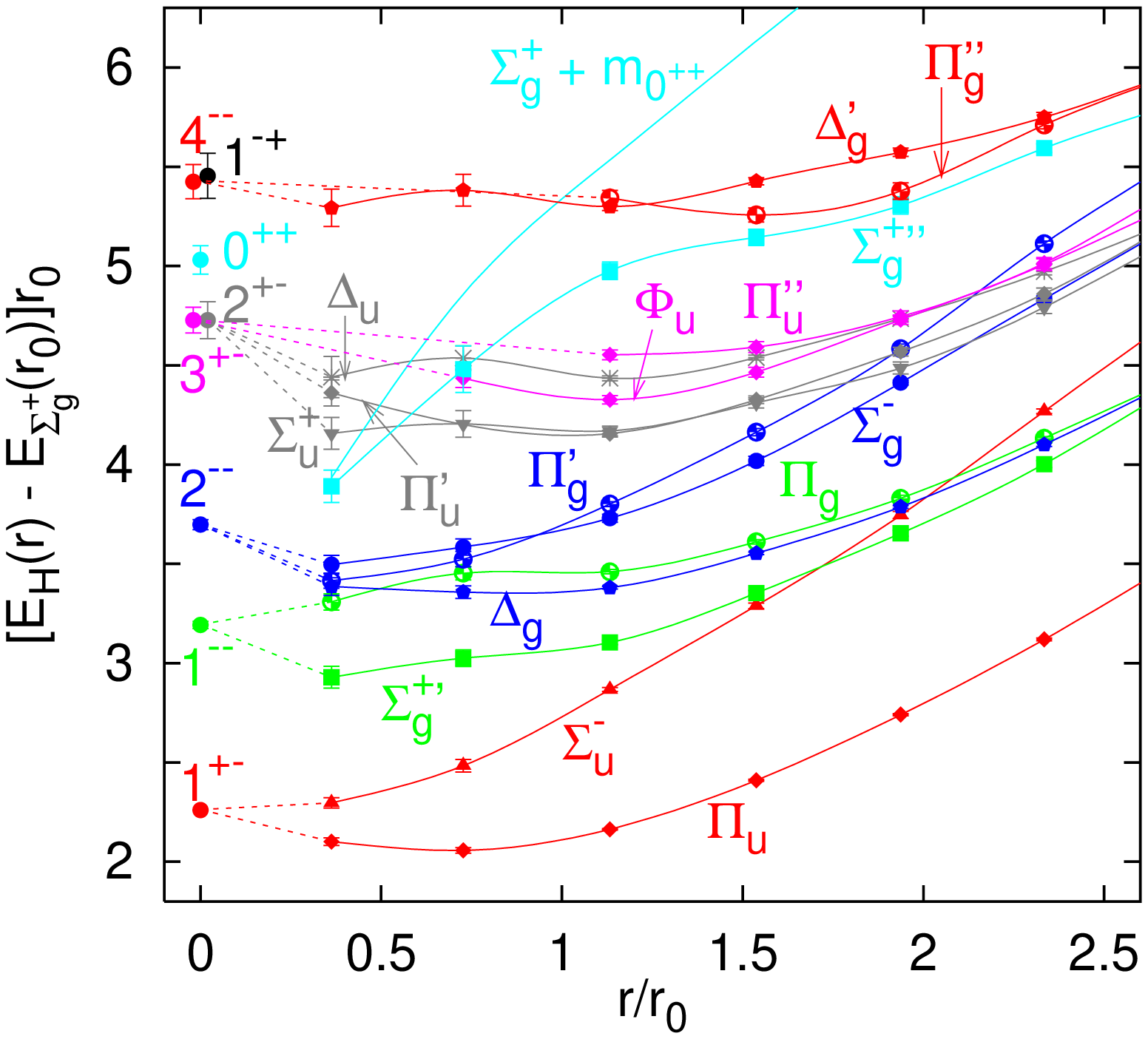}
\caption{\label{hybspect} \it
Different hybrid potentials \cite{Juge:2002br} at a
lattice spacing $a_{\sigma}\approx 0.2$ fm $\approx 0.4\,r_0$,
where $r_0\approx 0.5$ fm is the scale for which
  $-r_0^2dV/dr|_{r=r_0}=1.65$ \cite{Sommer:1993ce}, in
comparison with the gluelump spectrum \cite{Foster:1998wu} (circles, 
left-most
data points). The gluelump spectrum
has been shifted by a constant to adjust the $1^{+-}$ state
with the $\Pi_u$ and $\Sigma_u^-$ potentials at short distance.  In
addition, we include the sum of the ground state ($\Sigma_g^+$)
potential and the scalar glueball mass $m_{0^{++}}$ \cite{Bali:1993fb,
Morningstar:1999rf,Lucini:2001ej}.
The lines are drawn to guide the eye. From \cite{Bali:2003jq}.}
\end{figure}

We are interested in the one-quark--one-antiquark sector of the Fock space.
In the static limit it is spanned by
\begin{equation}
\vert \underbar{n}; {\bf x}_1 ,{\bf x}_2  \rangle^{(0)}
\equiv  \psi^{\dagger}({\bf x}_1) \chi_c^{\dagger} ({\bf x}_2)
|n;{\bf x}_1 ,{\bf x}_2\rangle^{(0)},\qquad \forall {\bf x}_1,{\bf x}_2\,,
\label{basis0}
\end{equation}
where $|\underbar{n}; {\bf x}_1 ,{\bf x}_2\rangle^{(0)}$ is a 
gauge-invariant
(since it satisfies the Gauss law) eigenstate (up to a phase) of $ H^{(0)}$
with energy $E_{n}^{(0)}({\bf x}_1 ,{\bf x}_2)$. For convenience, we use 
here the field
$\chi_c ({\bf x})=i\sigma^2 \chi^{*} ({\bf x})$, instead of
$\chi ({\bf x})$, because it is the one to which a particle interpretation
can easily be given: it corresponds to a Pauli spinor that annihilates a 
fermion in
the $3^*$ representation of color $SU(3)$ with the standard, particle-like,
spin structure.
$|n;{\bf x}_1 ,{\bf x}_2\rangle^{(0)}$ encodes the gluonic content of the
state, namely it is annihilated by $\chi_c({\bf x})$ and $\psi ({\bf x})$
for all ${\bf x}$.
It transforms as a $3_{{\bf x}_1}\otimes 3_{{\bf x}_2}^{\ast}$ under 
color $SU(3)$.
The normalizations are taken as follows
\bea
&& ^{(0)}\langle m;{\bf x}_1 ,{\bf x}_2|n;{\bf x}_1 ,{\bf 
x}_2\rangle^{(0)}  =\delta_{nm},
\\
&& ^{(0)}\langle \underbar{m}; {\bf x}_1 ,
{\bf x}_2|\underbar{n}; {\bf y}_1 ,{\bf y}_2\rangle^{(0)} =\delta_{nm}
\delta^{(3)} ({\bf x}_1 -{\bf y}_1)\delta^{(3)} ({\bf x}_2 -{\bf y}_2)\,.
\eea
We have made explicit that the positions ${\bf x}_1$ and ${\bf x}_2$ of 
the quark
and antiquark respectively are good quantum numbers for the static solution
$|\underbar{n};{\bf x}_1 ,{\bf x}_2 \rangle^{(0)}$
(since there are no spatial derivatives in the Lagrangian), whereas $n$ 
generically denotes the remaining
quantum numbers. We also choose the basis such that $T|\underbar{n};{\bf 
x}_1
,{\bf x}_2 \rangle^{(0)}= |\underbar{n};{\bf x}_1
,{\bf x}_2 \rangle^{(0)}$ where $T$ is the time-reversal operator.
The ground-state energy $E_0^{(0)}({\bf x}_1,{\bf x}_2)$ can be
associated with the static potential of the heavy quarkonium in some
circumstances (see sec.~\ref{seconestep}). The remaining energies
$E_n^{(0)}({\bf x}_1,{\bf x}_2)$,
$n\not=0$, are usually associated with the potential used in order to 
describe hybrids (they may also correspond to heavy quarkonium or
heavy hybrids plus glueballs). They can be computed on the lattice (see \cite{Juge:2002br} and
also Fig.~\ref{hybspect}). Translational invariance implies that 
$E_n^{(0)}({\bf x}_1,{\bf x}_2) = E_n^{(0)}(r)$.
This means that they are functions of $r$ and the only other scale in 
the system, $\lQ$.

In static NRQCD, the gluonic excitations between static quarks have the 
same symmetries
as in a diatomic molecule (see \cite{Messiah:1979eg}).
In the centre-of-mass system, these correspond to the symmetry group 
$D_{\infty
  h}$ (substituting the parity generator by CP).
According to that symmetry, the mass eigenstates are classified in terms 
of the
angular momentum
along the quark-antiquark axis ($|L_z| = 0,1,2, \dots$ to which one 
gives the traditional
names $\Sigma, \Pi, \Delta, \dots$), CP (even, $g$, or odd, $u$), and
the reflection properties with respect to
a plane that passes through the quark-antiquark axis
(even, $+$, or odd, $-$).
Only the $\Sigma$ states are not degenerate with respect to the 
reflection symmetry.

\subsection{Static pNRQCD in the weak-coupling regime}
\label{sec:pNRQCDstatic}
In the static limit pNRQCD has the same symmetries as NRQCD.
In this section we will discuss some general properties of the 
short-distance behaviour
of the static energies that can be straightforwardly derived within this 
EFT.
We will follow \cite{Brambilla:1999xf}.

In the limit $\lQ \ll 1/r$ and at LO in $1/m$, the spectrum 
of the theory
can be read off from 
the Lagrangian (\ref{Lpnrqcd}). In particular, the LO 
solution corresponds
to the zeroth order
of the multipole and $1/m$ expansions. At this order,
while the singlet decouples from the octet and the gluons,
the octet is still coupled to gluons.
We call gluelumps the states made of an adjoint source in the presence of a
gluonic field,
\be
H({\bf R},{\bf r},t) \equiv H^a({\bf R},t)O^a({\bf R},{\bf r},t).
\ee
These, in turn, correspond to the gluonic excitations between static 
quarks in the short-distance limit,
for which there is abundant
non-perturbative data available from lattice simulations (see Fig. 
\ref{hybspect}).
Depending on the glue operator $H^a$ and
its symmetries, the gluelump operator $O^a H^a$ describes a specific 
gluonic
excitation between static quarks and its static energy, $V_H$.

In static pNRQCD at lowest order in the multipole expansion, besides the 
symmetries
of static NRQCD, extra symmetries for the gluonic excitations between 
static quarks appear.
The glue dynamics no longer involves the relative coordinate ${\bf r}$. 
Therefore,
the glue associated with a gluonic excitation between static quarks 
acquires a spherical symmetry.
In the centre-of-mass system, gluonic excitations between static quarks 
are, therefore, classified
according to representations of $O(3) \otimes{\rm C}$, which we 
summarize by $L$, the angular momentum, CP
and reflection with respect to a plane passing through the 
quark-antiquark axis.
Since this symmetry group is larger than that of NRQCD, several gluonic 
excitations
between static quarks are expected to be approximately degenerate in 
pNRQCD,
i.e. in the short-distance limit $r \ll 1/\lQ$. We illustrate this point in
Tab. \ref{tab3}
where all operators, $H$, up to dimension 3 are built and classified 
according
to their quantum numbers in NRQCD and pNRQCD. In Tab. \ref{tab3} all the 
operators
are evaluated at the centre-of-mass coordinates.
$\Sigma_g^+$ is not displayed since it corresponds to the singlet state.
The prime indicates excited states of the same quantum numbers.
The operators chosen for the $\Pi$ and $\Delta$ states are not 
eigenstates of the
reflection operator.
This is not important since these states are degenerate with respect to 
this symmetry.
 From the results of Tab. \ref{tab3} the following degeneracies are 
expected in
the short-distance limit:
\begin{eqnarray}
&&\Sigma_g^{+\, \prime} \sim \Pi_g\;; \qquad
\Sigma_g^{-} \sim \Pi_g^{\prime} \sim \Delta_g\,;
\nonumber
\\
&&
\Sigma_u^{-} \sim \Pi_u\;; \qquad
\Sigma_u^{+} \sim \Pi_u^{\prime} \sim \Delta_u \,.
\label{dege}
\end{eqnarray} 
Similar observations have also been made by \cite{Foster:1998wu}. In 
pNRQCD they
emerge in a quite clear
and straightforward way and one can explicitly write down the relevant 
operators.
For higher excitations the expected degeneracies have been obtained by 
\cite{Bali:2003jq}.
We will discuss them further when comparing with lattice data in sec. 
\ref{gluelumplatt}.

\begin{table}[htb]
\makebox[6cm]{\phantom b}
\begin{center}
\begin{tabular}{|c|c|c|}
\hline
Gluelumps       & $~$ & $~$ \\
$ O^a H^a$      & $L=1$ & $L=2$ \\
\hline
$\Sigma_g^{+\, \prime}$ & ${\bf r}\cdot{\bf E} \;,
             {\bf r}\cdot({\bf D}\times {\bf B})$  & $~$  \\\hline
$\Sigma_g^-$ & $~$ & $({\bf r}\cdot {\bf D})({\bf r}\cdot {\bf B}) $ 
\\\hline
$\Pi_g$ & ${\bf r}\times{\bf E}\;,
             {\bf r}\times({\bf D}\times {\bf B}) $ & $~$ \\\hline
$\Pi_g^{\prime}$ & $~$ & ${\bf r}\times(({\bf r}\cdot{\bf D}) {\bf B}
              +{\bf D}({\bf r}\cdot{\bf B}))$  \\\hline
$\Delta_g$ & $~$ & $({\bf r}\times {\bf D})^i({\bf r}\times{\bf B})^j
             +({\bf r}\times{\bf D})^j({\bf r}\times{\bf B})^i$ \\\hline
\hline
$\Sigma_u^{+}$ & $~$ & $({\bf r}\cdot {\bf D})({\bf r}\cdot {\bf 
E})$\\\hline
$\Sigma_u^-$ & ${\bf r}\cdot{\bf B} \;,
             {\bf r}\cdot({\bf D}\times {\bf E})$ & $~$ \\\hline
$\Pi_u$ & ${\bf r}\times{\bf B}\;,
             {\bf r}\times({\bf D}\times {\bf E})$ & $~$ \\\hline
$\Pi_u^{\prime}$ & $~$ & ${\bf r}\times(({\bf r}\cdot{\bf D}) {\bf E}
              +{\bf D}({\bf r}\cdot{\bf E})) $ \\\hline
$\Delta_u$ & $~$ & $({\bf r}\times {\bf D})^i({\bf r}\times{\bf E})^j
             +({\bf r}\times{\bf D})^j({\bf r}\times{\bf E})^i$ \\\hline
\end{tabular}
\end{center}
\caption{ \it Operators $H$ for the $\Sigma$, $\Pi$ and $\Delta$ gluonic 
excitations
between static quarks in
pNRQCD up to dimensions 3. The covariant derivative is understood in the
adjoint representation. ${\bf D}\cdot{\bf B}$ and ${\bf D}\cdot{\bf E}$ 
do not appear, the first because it
is identically zero after using the Jacobi identity, while the second gives
vanishing contributions after using the equations of motion. From 
\cite{Brambilla:1999xf} }
\label{tab3}
\end{table}
 
So far only the symmetries of pNRQCD at lowest order in the
multipole expansion have been used. In fact one can go beyond that and 
predict the shape
of the static energies by calculating the singlet and gluelump (static 
hybrid)
 correlators
\begin{equation}
\langle {\rm vac}|H({\bf R},{\bf r}, T/2) H^{\dagger}
({\bf R}^{\prime},{\bf r}^{\prime}, -T/2)|{\rm vac} \rangle
\sim \delta^3({\bf R}-{\bf R}^{\prime})\delta^3({\bf r}-{\bf 
r}^{\prime})\,e^{-iTV_H(r)}
\,,
\end{equation}
\begin{equation}
\langle {\rm vac}|S({\bf r},{\bf R}, T/2) S^{\dagger}
({\bf R}^{\prime},{\bf r}^{\prime}, -T/2)|{\rm vac} \rangle
\sim \delta^3({\bf R}-{\bf R}^{\prime})\delta^3({\bf r}-{\bf 
r}^{\prime})\,e^{-iTV_s^{(0)}(r)}
\,,
\end{equation}
for large $T$. At lowest order in the multipole expansion
the spectrum of the singlet state reads\footnote{By taking 
the arbitrary subtraction constant as twice the pole mass of a
heavy quark in Eqs (\ref{Es}), (\ref{EH}), these equations become
renormalon free.}
\be
\label{Es}
E_s(r)=2m +V_s^{(0)}(r)+{\cal O}(r^2)
\,.
\ee
For the static hybrids, the spectrum reads ($V_H=V_o^{(0)}(r)+\Lambda_H$)
\be
\label{EH}
E_H(r)=2m+V_o^{(0)}(r)+\Lambda_H+{\cal O}(r^2)
\,,
\ee
where
\be
\Lambda_H\equiv \lim_{T\to\infty} \frac{i}{T}\ln
\langle H^a(T/2)\phi_{ab}^{\rm adj}(T/2,-T/2)H^b(-T/2) \rangle
\label{LH}
\,.
\ee
Note that Eq.~(\ref{LH}) allows us to relate the correlation length of some
gluonic correlators to the behavior of the spectrum of the static 
hybrids at
short distances. Note also that $\Lambda_H$ is the same for operators 
corresponding to 
states that are degenerate.

The potentials $V_s^{(0)}$ and $V_o^{(0)}$ can be computed within 
perturbation theory. One could then
perform a detailed comparison with lattice data. We will see that in 
order to do so we will have to
deal first with the renormalon ambiguities in the way explained in ch.~\ref{secrenormalons}.

\subsection{The singlet static potential at short distances versus lattice}
\label{singletstatic}
\begin{figure}[ht]
\hspace{-0.1in}
\epsfxsize=3in
\centerline{
\put(30,120){$r_0(V_{\RS}(r)-V_{\RS}(r')+E_{latt.}(r'))$}
\epsffile{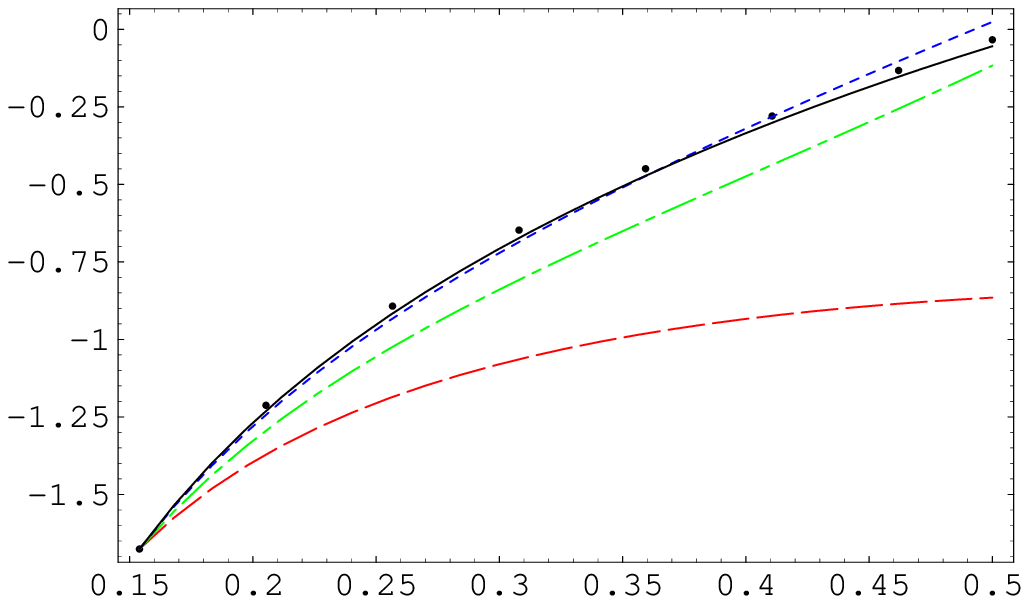}
\put(15,1){$r/r_0$}
}
\caption {{\it \label{potRSlattr} Plot of
$r_0(V_{\RS}(r)-V_{\RS}(r')+E_{latt.}(r'))$ versus r at tree (dashed line),
one-loop (dash-dotted line), two-loop (dotted line) and three-loop level
(estimate) plus the RG expression for the US logarithms (solid line) compared
with the lattice simulations $E_{latt.}(r)$ \cite{Necco:2001xg}.  For the
scale of $\als(\nu)$, we set $\nu=1/r$. Further, 
$\nu_f=\nu_{us}=2.5\,r_0^{-1}$,
$\Lambda_{\MS}=0.602\,r_0^{-1}$ \cite{Capitani:1998mq}, and
$r'=0.15399\,r_0$. From \cite{Pineda:2002se}.}}
\end{figure}

In the last years, lattice simulations \cite{Bali:1997am,Necco:2001xg} have
improved their predictions at short distances allowing very accurate
comparisons between perturbation theory and lattice simulations.  In 
order to
perform this comparison, we cannot work in the on-shell scheme due to the
presence of the renormalon, which destroys the convergence of the 
perturbative
series.  Therefore, schemes were introduced to make the renormalon
cancellation explicit.  In \cite{Sumino:2001eh,Recksiegel:2001xq} the 
renormalon
cancellation is achieved order by order in $\als$ by expanding both $m$ and
$V_s^{(0)}$ in terms of the same $\als(\nu)$.  A potential problem of this
method is the appearance of large logarithms in the mass expansion.  In
\cite{Necco:2001gh} lattice data were shown to agree with perturbation 
theory
at short distances if the force was used instead of the potential.  It was
shown in \cite{Pineda:2002se} that this is equivalent to working in a
renormalon-free scheme, and a first quantitative comparison of the 
(quenched)
lattice data with the renormalon-subtracted potential 
$V_{s,\RS}^{(0)}(r)$ was
done (see also \cite{Lee:2002sn}).  This analysis allowed to put 
quantitative
bounds on non-perturbative effects at short distances and, in particular, it
ruled out a linear potential with slope $\sigma=0.21 \, {\rm GeV}^2$ at 
short
distances (see also \cite{Pineda:2003jv}). It also showed that today 
lattice data
are precise enough to be sensitive to three-loop perturbation theory
(see Fig.~\ref{potRSlattr}).  Overall, up to distances of around
0.4-0.5$\,r_0$, perturbation theory is convergent with small errors
and agrees with lattice data in all of the previous analyses.  For 
larger distances the
analysis of \cite{Pineda:2002se} shows agreement with the lattice data 
(within
errors) up to distances of $\sim 0.8 \, r_0$ if large logarithms are 
resummed.
In \cite{Recksiegel:2002um} it was argued that, by fine-tuning the
renormalization scale, agreement with lattice data can be reached 
(within errors)
up to $3\,r_0$. Nevertheless, for such large distances, the use of
perturbation theory is quite doubtful. Therefore, further studies are needed
to see whether this agreement is purely accidental or a theoretical
explanation can be given.

\subsection{Gluelumps versus lattice}
\label{gluelumplatt}
We compare the predictions of pNRQCD for the static hybrids
in the weak-coupling regime with lattice data. We first explore 
at which distances the expected degeneracies start to be fulfilled and
whether the gluelump mass and hybrid potential splittings agree
with each other. See Fig.~\ref{hybspect}.
On a qualitative level the short-distance data are consistent
with the expected degeneracies. In any case, at best, one can possibly
imagine perturbation theory to be valid for the left-most two data
points. With the exception of the $\Pi_u$, $\Pi_u'$ and $\Phi_u$
potentials there are also no clear
signs for the onset of the short-distance $1/r$ behaviour with a
positive coefficient as expected from perturbation
theory. Furthermore, most of the gaps within multiplets of hybrid
potentials,
which at LO depend on the
size of the non-perturbative $r^2$ term, are still quite significant, 
even at
$r=0.4\,r_0$; for instance, the difference between the $\Sigma_u^-$ and
$\Pi_u$ potentials at this distance is about 0.28 $r_0^{-1}\approx 110$ MeV.

\begin{figure}
\includegraphics[width=0.5\columnwidth]{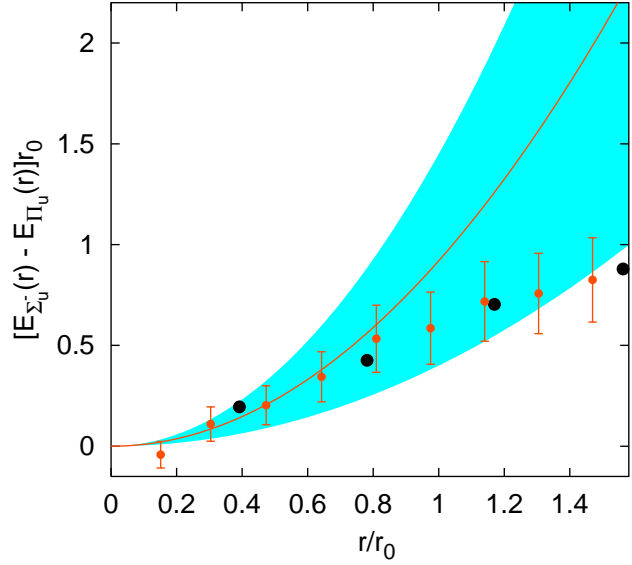}
\caption{\label{r2} \it
Splitting between the $\Sigma_u^-$ and the $\Pi_u$
potentials, extrapolated to the continuum limit,
and comparison with a quadratic fit to the
$r\siml 0.5\,r_0$ data points ($r_0^{-1}\approx 0.4$ GeV). The big
circles correspond to the data of \cite{Juge:2002br}.
The errors in this case are smaller than the symbols. The smaller circles 
correspond to the data of \cite{Bali:2003jq}. From 
\cite{Bali:2003jq}.}
\end{figure}

 From the above considerations it is clear that
for a more quantitative study one needs lattice data at
shorter distances. These have been provided by \cite{Bali:2003jq} for the
lowest two gluonic excitations, $\Pi_u$ and $\Sigma_u^-$. We display their
differences in the continuum limit in Fig.~\ref{r2}.
We see how these approach zero at small $r$, as expected
from the short-distance expansion. pNRQCD predicts that the next effects 
should
be of ${\cal O}(r^2)$ (and renormalon-free).
The lattice data are fitted rather well by
a $\Delta E_{\Pi_u-\Sigma_g^+} =A_{\Pi_u-\Sigma_u^-}r^2$ ansatz for 
short distances, with slope
(see Fig.~\ref{r2}),
\be
\label{eqa}
A_{\Pi_u-\Sigma_u^-}=0.92^{+0.53}_{-0.52}\,r_0^{-3}
\,,
\ee
where the error is purely statistical (lattice), the systematic
error being negligible. We remark that within the
framework of static pNRQCD and to second order in the
multipole expansion, one can relate
the slope $A_{\Pi_u-\Sigma_u^-}$ to gluonic correlators of QCD.  

One can go beyond these analyses and use lattice data plus the
knowledge of the (perturbative) octet potential to obtain numerical
values for gluelump masses in a particular scheme.  However,
analogously to the situation with the static singlet potential, the
convergence of the perturbative series of the octet potential is bad.
The solution to this problem comes again from
working in a RS scheme properly generalized to the hybrid case. The 
hybrid energy reads
\be
\label{EHRS}
E_H(r)=2m_{\RS}(\nu_f)+V_{o,\RS}(r;\nu_f)+\Lambda_H^{\RS}(\nu_f)+{\cal 
O}(r^2)
\,.
\ee
In the RS scheme the octet potential reads
\be
V_{o,\RS}(\nu_f)=V_o-\delta V_{o,\RS}
=\sum_{n=0}^\infty V^{\RS}_{o,n}\als^{n+1} \,,
\ee
where
\be
\delta
V_{o,\RS}=
\sum_{n=1}^\infty N_{V_o}\,\nu_f\,\left({\beta_0 \over
2\pi}\right )^n \als^{n+1}(\nu_f)
\sum_{k=0}^\infty
c_k{\Gamma(n+1+b-k) \over \Gamma(1+b-k)} \,. 
\ee
This specifies the gluelump mass which reads
\be
\label{lamrs}
\Lambda_H^{\RS}(\nu_f)=\Lambda_H -\delta\Lambda_{\RS}(\nu_f),
\ee
where
\be
\delta\Lambda_{\RS}(\nu_f)
=
\sum_{n=1}^\infty
N_{\Lambda}\,\nu_f\,\left({\beta_0 \over 2\pi}\right )^n
\als^{n+1}(\nu_f)
\sum_{k=0}^\infty c_k{\Gamma(n+1+b-k) \over
\Gamma(1+b-k)} \,. 
\ee
Note that factorization requires
\be
2N_m+N_{V_o}+N_{\Lambda}=0.
\ee
$N_m$ is already known and $N_{V_o}$ can also be obtained approximately
from low orders in perturbation theory following the same procedure as
in sec.~\ref{secmas}. 
One now has a convergent series in perturbation theory
and can obtain absolute values for the masses of the gluelumps, in 
particular
for the lowest gluelump using the splitting of the $\Sigma_g^+$ and the 
$\Pi_u$
potential. Then using the lattice data of \cite{Foster:1998wu},
it is possible to obtain the absolute values for the masses of all 
gluelump excitations
in a given scheme (in this case, the RS scheme). 
The results are summarized in Table \ref{tablegluelumps}.
For a comparison with other determinations, see \cite{Bali:2003jq}.

\begin{table}
\caption{\label{tablegluelumps} \it
Absolute values for the gluelump masses in the continuum limit
in the RS scheme at $\nu_f=2.5\,r_0^{-1}\approx 1$ GeV,
in $r_0$ units and in GeV. Note that an additional
uncertainty of about 10\% should be added to the last column to account for
the quenched approximation. We also display examples of creation operators
$H$ for these states. The curly braces denote complete
symmetrization of the indices. From \cite{Bali:2003jq}}

\begin{ruledtabular}
\begin{tabular}{c|ccc}
$J^{PC}$&$H$&$\Lambda_H^{\RS}r_0$&$\Lambda_H^{\RS}$/GeV\\\hline
$1^{+-}$&$B_i$                           &2.25(39)&0.87(15)\\
$1^{--}$&$E_i$                           &3.18(41)&1.25(16)\\
$2^{--}$&$D_{\{i}B_{j\}}$                &3.69(42)&1.45(17)\\
$2^{+-}$&$D_{\{i}E_{j\}}$                &4.72(48)&1.86(19)\\
$3^{+-}$&$D_{\{i}D_{j}B_{k\}}$           &4.72(45)&1.86(18)\\
$0^{++}$&${\mathbf B}^2$                 &5.02(46)&1.98(18)\\
$4^{--}$&$D_{\{i}D_jD_kB_{l\}}$          &5.41(46)&2.13(18)\\
$1^{-+}$&$({\mathbf B}\wedge{\mathbf E})_i$&5.45(51)&2.15(20)
\end{tabular}
\end{ruledtabular}
\end{table}

\section{Potential NRQCD. The strong-coupling regime}
\label{sec:SCR}
In this chapter, we discuss pNRQCD under the condition that $\lQ \gg E$.
We have called this situation the strong-coupling regime of pNRQCD
in ch.~\ref{sec:pNRQCDI}, where some general features of the
physical picture have already been discussed.
Since the EFT does not tell us anything about the non-perturbative
dynamics of QCD, we have to rely on some assumptions in order to identify
the relevant degrees of freedom. The assumptions will be minimal, 
supported by general
considerations and lattice data, but clearly we are on a less
solid ground here than in the weak-coupling regime.

\subsection{Degrees of freedom}
If we consider the case without light quarks, the physical states
made by a heavy quark and antiquark are heavy quarkonium states or hybrids
or both of them in the presence of glueballs. Quenched lattice data show
that the static energy of the lowest
state is separated by a gap of order $\lQ$ from the higher ones.
This feature is preserved in going to unquenched simulations (see Fig. 
\ref{Masgap}).
We assume that this feature is also preserved in the dynamical case of 
heavy quarks
with finite masses.
This leads to identify the heavy quarkonium with the solution of the 
Schr\"odinger
equation on which the static potential corresponds to the ground state
static energy.

Once light fermions have been incorporated, however, new gauge-invariant 
states
appear besides the heavy quarkonium, hybrids and glueballs. First, we 
have states
with no heavy quark content. Due to chiral symmetry, there is a mass gap,
of ${\cal O}(\Lambda_{\rm QCD})$, between
the Goldstone bosons, which are massless in the chiral limit, and the
rest of the spectrum. Therefore, the Goldstone bosons are
US degrees of freedom, while the rest of the spectrum is integrated out
at the scale $\lQ$.
Besides these, we also have bound states made of one heavy quark and 
light quarks. In
practice, we are considering the $Q \bar q$--$\bar Q q $ system. The
energy of this system is, according to the HQET counting rules
\cite{Neubert:1994mb}, $m_{Q \bar q}+m_{\bar Q q}=2m+2\,{\bar \Lambda}$.
Therefore, since ${\bar \Lambda} \sim \lQ$, we assume that also these states
are integrated out at the scale $\lQ$. This cannot be done for
heavy quarkonium states near threshold, since in this case there is no
mass gap between the heavy quarkonium and the creation of a $Q
\bar q$--$\bar Q q $ pair. Thus, if we want to study the heavy quarkonium
near threshold, we should include these degrees of freedom in the
spectrum (for a model-dependent approach to this situation see, for
instance, \cite{Eichten:1978tg}). We will assume here that the considered
heavy quarkonium states are safely far from threshold\footnote{
One may think of relaxing this condition in the large $N_c$ limit, where 
the mixing between the heavy quarkonium 
and the $Q \bar q$--$\bar Q q $ is suppressed by powers of $1/N_c$.}.

Summarizing, the degrees of freedom of pNRQCD in the regime $\lQ \gg E$ for
quarkonium states far from threshold are a singlet field $S$, describing 
the
heavy quarkonium state, and Goldstone boson fields. In the following, we
will not consider the Goldstone boson fields. If one switches off
the light fermions, only the singlet survives and pNRQCD reduces to
a pure two-particle NR quantum-mechanical system, usually referred to as 
a pure
potential model. 

\subsection{Power counting}
\label{secpowcoustr}
The structure of the pNRQCD Lagrangian under the above conditions is very
simple: it is just a bilinear in the singlet field. Therefore, 
establishing the
power counting means to estimate the size of the terms multiplying the 
bilinear.

The soft scale $|{\bf p}|$ must be assigned to $- i\bfnabla_{\bf r}$ and $1/r$,
the US scale $E\sim {\bf p}^2/m$ to the time derivatives $i\partial_0$ and 
$V_s^{(0)}$.
This last condition follows from the consistency of the theory that 
requires
the virial theorem to be fulfilled. In other words, all the terms in the 
Schr\"odinger
equation, 
\be
i\partial_0 \phi = E \, \phi = ({\bf p}^2/m +  V_s^{(0)})\phi
,
\ee
must count the same. Note that the normalization condition of the 
wavefunction
($\int d^3{\bf r} \, |\phi({\bf r})|^2 =1$) sets $|\phi|^2 \sim |{\bf p}|^3$.
In general, the $1/m$ corrections to the potential (real and
imaginary) will be a combination of $\als$ calculated at different scales,
derivatives with respect to the relative coordinate $- i\bfnabla_{\bf
r}$, $1/r$ and expectation values of the fields of the light degrees of 
freedom.
The quantities $m$ and $\als(m)$ are inherited from the
hard matching and have well-known values, in particular $\als(m) \ll 1$.
The strong-coupling constant also appears evaluated at the scales 
$\sqrt{m\,\lQ}$,
$1/r$, $\lQ$ and $E$. At the scale $\sqrt{m\,\lQ}$, which appears in loop
calculations (see below), $\als(m) \ll \als(\sqrt{m\,\lQ}) \ll 1$ since
$\sqrt{m\,\lQ} \gg \lQ$. At the scales $\lQ$ and $E$, $\als(\lQ) \sim 1$ 
and
$\als (E)\sim 1$ by definition of the strong-coupling regime.
If $|{\bf p}| \sim \lQ$, then also $\als (1/r)\sim 1$. If $|{\bf p}| \gg \lQ \gg E$, then
$\als(\sqrt{m\,\lQ}) \ll \als (1/r) \ll 1$.
In the situation $|{\bf p}| \sim \lQ$ the expectation values of the fields of the
light degrees of freedom depend on ${\bf r}$ and $\lQ$, while in the
situation $|{\bf p}| \gg \lQ \gg E$, the $1/r \sim |{\bf p}|$ dependence factorizes and
the expectation values of the fields of the light degrees of freedom depend
only on $\lQ$\footnote{This is certainly so for states with low 
principal quantum
number $n$. For higher excitations one should keep in mind that
${\bf p}$ and $1/r$ could scale differently with $n$.}.
In both cases their natural counting is $\lQ$ to the power of their 
dimension.

\subsection{Lagrangian and symmetries}
\label{sec:pNRLagS}
The pNRQCD Lagrangian (without Goldstone bosons) is given by:
\be
L_{\rm pNRQCD} = \int d^3 {\bf R}  \int d^3 {\bf r}  \;
S^\dagger \big( i\partial_0 - h_s({\bf x}_1,{\bf x}_2, {\bf p}_1, {\bf p}_2,
{\bf S}_1,  {\bf S}_2) \big) S,
\label{pnrqcdstrong}
\ee
where
\be
h_s({\bf x}_1,{\bf x}_2, {\bf p}_1, {\bf p}_2, {\bf S}_1,  {\bf S}_2) =
{{\bf p}_1^2\over 2m_1} + {{\bf p}_2^2\over 2m_2}
+ V_s({\bf x}_1,{\bf x}_2, {\bf p}_1, {\bf p}_2, {\bf S}_1,  {\bf S}_2),
\ee
${\bf p}_j = -i\bfnabla_{{\bf x}_j}$, ${\bf r} = {\bf x}_1-{\bf x}_2$,
${\bf R} = ({\bf x}_1+{\bf x}_2)/2$ and ${\bf S}_j$ is the spin operator of
particle $j$. In the following, as long as not
stated differently, we will assume $m_1\neq m_2$. However, we will not 
exploit
a possible hierarchy between the two masses, which, for our purpouses, are
of the same order $\sim m \gg \lQ$.
The potential $V_s$ contains a real and an imaginary part.
The real part is responsible for the binding, the imaginary part for the
decay width of the heavy quarkonium state. The imaginary part of $V_s$ 
comes from the imaginary parts
of the matching coefficients of the 4-fermion operators of NRQCD.
The potential $V_s$ is, in general, a non-perturbative quantity, even if, to
some degree, it may contain pieces calculable in perturbation theory, like,
for instance, the matching coefficients of NRQCD, or in general any
contribution coming from scales larger than $\lQ$.
It is the aim of the matching procedure, which we will discuss in the 
following sections,
to provide the factorization formulas and the exact expressions for the 
non-perturbative pieces.
These may be eventually calculated on the lattice or in QCD vacuum 
models, which will be
the subject of sec.~\ref{sec:latmod}.

The symmetries of the singlet field are those already discussed for the 
pNRQCD Lagrangian
in the weak-coupling regime. In particular, the potential and the 
kinetic energies
satisfy the Poincar\'e invariance constraints (\ref{kin}) and (\ref{ex0})
(for the singlet potential). Note that Poincar\'e invariance may also 
constrain the natural
power counting discussed in sec.~\ref{secpowcoustr}.

\subsection{Matching: analytic and non-analytic mass terms}
\label{Sec:anandnoan}
Despite the fact that the strong-coupling Lagrangian (\ref{pnrqcdstrong})
looks quite simple, the matching procedure that leads to it may be
complicated. This is due to the fact that 
we have to integrate out, and, therefore, to make explicit, all the
degrees of freedom (or momentum regions) that appear in the range
from the hard to the US scale within a non-perturbative environment.

\begin{figure}[htb]
\makebox[-16cm]{\phantom b}
\put(120,0){\epsfxsize=7truecm \epsfbox{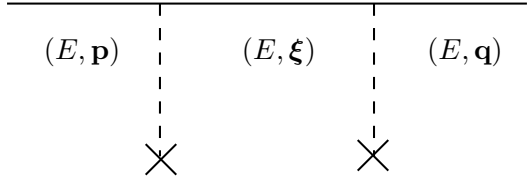}}
\put(135,45){\small $(E,{\bf p})$}
\put(210,45){\small $(E,{\bfxi})$}
\put(280,45){\small $(E,{\bf q})$}
\vskip 0.2truecm
\caption{ \it The incoming energy $E$ is of order $\lQ$, $p$ and $q$ of 
order
  $mv$. The vertex describes the interaction with an external potential 
$V$.}
\label{figsqrt}
\end{figure}
Since we are also integrating out $\lQ$, new momentum regions (apart 
from $\lQ$
itself) that do not appear in the weak-coupling matching show up.
Let us consider, for instance, the diagram of Fig.~\ref{figsqrt}.
Suppose that the incoming (outcoming) particle is an off-shell particle of 
energy $\sim
\lQ$ and three-momentum $p (q) \sim mv$ (for instance, an on-shell 
particle that
just emitted (absorbed) a soft gluon of energy $\lQ$). The diagram 
corresponds to
the integral:
\be
\int {d^3 \bfxi\over (2\pi)^3} V({\bf p}-\bfxi) {1\over E - {\bfxi^2 / m}
  +i\epsilon} V(\bfxi - {\bf q}).
\ee
This integral also receives a contribution from the three-momentum 
region $\xi \sim \sqrt{m\,E}
\sim \sqrt{m\,\lQ}$. Since $\sqrt{m\,\lQ} \gg \lQ$, the potential
is perturbative, and since  $\sqrt{m\,\lQ} \gg p,q$, we may expand in 
$p$ and $q$
and the integral effectively reduces to:
\be
\als^2 \int {d^3 \bfxi\over (2\pi)^3} {1\over \xi^4} {1\over E - 
{\bfxi^2 / m}
  +i\epsilon} \sim \als^2 {1\over \lQ} {1\over \sqrt{m\,\lQ}},
\ee
where $\als$ is calculated at the (perturbative) scale  $\sqrt{m\,\lQ}$.
 From the above example we may draw the following conclusions.
First, in the strong-coupling regime new degrees of freedom show up
in loops, namely quark-antiquark pairs with relative three-momentum of 
order $\sqrt{m\,\lQ}$
and on-shell energy of order $\lQ$. Since the scale $\sqrt{m\,\lQ} \gg |{\bf p}|$
for $\lQ \gg E$, this is the largest scale below $m$ and, thus, the first
to be integrated out. The only reason this otherwise dominant
contribution to the potential is suppressed is that it appears only in 
loops.
Second, since we expand in the external momenta, which are small compared
to  $\sqrt{m\,\lQ}$, the effective interaction that arises is local.
Third, this kind of contribution is non-analytic in $m$.

It is convenient to split the potential (imaginary and real part) into a 
part
that gets contributions only from scales that are analytic in the mass,
$V^{1/m}$, and another, $V^{1/\sqrt{m}}$, that contains any contribution
coming from the scale $\sqrt{m\,\lQ}$:
\be
V_s = V^{1/m} + V^{1/\sqrt{m}}.
\label{V1m1sm}
\ee
We will often refer to $V^{1/\sqrt{m}}$ as the part of the potential that
is non-analytic in $1/m$. This is only true at LO, which
is, however, the order at which we will work here. The matching for the
$V^{1/m}$ part may be performed in a strict $1/m$ expansion. The matching
for the $V^{1/\sqrt{m}}$ part maybe done by integrating out quark-antiquark
pairs with relative three-momenta  of order $\sqrt{m\,\lQ}$.

We will next discuss the matching procedures for $V^{1/m}$ and 
$V^{1/\sqrt{m}}$
in the situations $|{\bf p}| \sim \lQ$  and $|{\bf p}| \gg \lQ \gg E$.
We will first consider the case $|{\bf p}| \sim \lQ$ in sec.~\ref{seconestep}.
The potential will be a function of ${\bf r}$ and $\lQ$.
This is the most general case. The particular case
$|{\bf p}| \gg \lQ \gg E$ may be derived from it by factorizing the potential in a
high-energy part that depends on $1/r \sim |{\bf p}|$ and a low-energy part
that depends on $\lQ$. In this case, however, it is more practical and 
consistent
with the general philosophy of the EFT to achieve  factorization 
directly by integrating
out the scales $|{\bf p}|$ and $\lQ$ in two different steps of the matching 
procedure. We will consider this
situation in sec.~\ref{sectwostep}. We note here that terms that come 
out local in the
situation $|{\bf p}| \sim \lQ$ are already factorized and, therefore, will be
reproduced (up to field redefinitions) in the situation $|{\bf p}| \gg \lQ \gg E$.
This is the case of the imaginary part of the potential, which comes 
from the
4-fermion contact terms of the NRQCD Lagrangian,
and the part of the potential that is non-analytic in $1/m$.

Finally, we would like to mention that soft light fermions will
not be explicitly considered in the matching computation. If we
want to incorporate them, the procedure would be analogous. One would have
to consider the matrix elements and Wilson loops with dynamical light 
fermions
incorporated and new terms appearing in the energies at ${\cal O}(1/m^2)$
due to operators involving light fermions that appear in the NRQCD 
Lagrangian
at ${\cal O}(1/m^2)$ and the Gauss law.

\subsection{Matching for  $|{\bf p}| \sim \lQ$}
\label{seconestep}
In sec.~\ref{hybrids}, we have discussed the static limit of NRQCD. The 
spectrum
consists of the static energies $E_0^{(0)} \ll E_1^{(0)} \ll \dots$.
We assume a gap of order $\lQ$ between  $E_0^{(0)}$ and the higher 
excitations.
pNRQCD is, by definition, the EFT that describes the lowest excitation 
of the NRQCD spectrum.
 From Eq.~(\ref{pnrqcdstrong}), it follows that pNRQCD in the static limit
consists of a singlet field $S$ with static energy $V^{(0)}$.
Since the static energy is an observable, the matching condition in the 
static limit
is:
\be
E_0^{(0)}(r) = V^{(0)}(r).
\label{matchstat}
\ee
Note that the left-hand side is a quantity defined in
NRQCD, while the right-hand side is a matching coefficient of pNRQCD.

We may think of generalizing the matching condition (\ref{matchstat}) to 
the non-static
case. Similarly to what we have done in sec.~\ref{hybrids},
we introduce the normalized eigenstates, $|\underbar{n}; {\bf x}_1 ,{\bf 
x}_2\rangle$,
and eigenvalues, $E_n({\bf x}_1 ,{\bf x}_2; {\bf p}_1, {\bf p}_2)$,
of the full NRQCD Hamiltonian $H$. They satisfy the equations
\bea
& & H |\underbar{n}; {\bf x}_1 ,{\bf x}_2\rangle = \int d^3x_1^\prime 
d^3x_2^\prime
|\underbar{n}; {\bf x}_1^\prime ,{\bf x}_2^\prime \rangle
E_n({\bf x}_1^\prime,{\bf x}_2^\prime, {\bf p}_1^\prime, {\bf 
p}_2^\prime, {\bf S}_1,{\bf S}_2)
\delta^{(3)}({\bf x}_1^\prime-{\bf x}_1)\delta^{(3)}({\bf 
x}_2^\prime-{\bf x}_2),
\nn\\
\label{schroe}
\\
& &
\langle \underbar{m}; {\bf x}_1 ,{\bf x}_2|\underbar{n}; {\bf y}_1 ,{\bf 
y}_2\rangle =
\delta_{nm} \delta^{(3)} ({\bf x}_1 -{\bf y}_1)\delta^{(3)} ({\bf x}_2 
-{\bf y}_2),
\label{norm}
\eea
where the states are labeled with the positions ${\bf x}_1$ and ${\bf x}_2$
of the static solution even if the position operator does not commute 
with $H$ beyond the static limit. The eigenvalues $E_n$ are, in general, 
functions of the momentum and spin operators and, therefore, should be understood as 
operators as well. We assume a gap of order $\lQ$ between  (the levels of) $E_0$
and (the levels of) $E_n$ for $n > 0$. Under this circumstance, and
arguing as in the static case above, it follows that the matching 
condition reads:
\be
E_0({\bf x}_1,{\bf x}_2, {\bf p}_1, {\bf p}_2, {\bf S}_1,  {\bf S}_2) =
h_s({\bf x}_1,{\bf x}_2, {\bf p}_1, {\bf p}_2, {\bf S}_1,  {\bf S}_2).
\label{matchnostat}
\ee
Again, this equation expresses the (real and imaginary parts of the)
pNRQCD Hamiltonian in terms of a quantity,
\be
E_0({\bf x}_1,{\bf x}_2, {\bf p}_1, {\bf p}_2, {\bf S}_1,  {\bf S}_2)
\delta^{(3)} ({\bf x}_1 -{\bf y}_1)\delta^{(3)} ({\bf x}_2 -{\bf y}_2)
=
\langle \underbar{0}; {\bf x}_1 ,{\bf x}_2| H |\underbar{0}; {\bf y}_1 
,{\bf y}_2\rangle, 
\label{E0average}
\ee
defined in NRQCD. The aim of the matching is to calculate this quantity. 
As discussed above,
it will contain a part that is analytic in $1/m$ and another that is not.

\subsubsection{Matching of the analytic terms: quantum-mechanical matching}
\label{analyticQM}
The analytic part of $E_0$ can be calculated, by definition, in a strict
$1/m$ expansion. The idea is to split the NRQCD Hamiltonian as
\be
H=H^{(0)}+H_I,
\ee
where $H^{(0)}$ is the static Hamiltonian, whose eigenstates and 
eigenvalues have
been discussed in sec.~\ref{hybrids}, and 
\be
H_I= {H^{(1,0)}\over m_1} + {H^{(0,1)}\over m_2}
+ {H^{(2,0)}\over m_1^2}  + {H^{(0,2)}\over m_2^2}  + {H^{(1,1)}\over 
m_1 m_2}
+ \cdots \; .
\ee
is the sum of all higher-order terms in the $1/m$ expansion of the NRQCD 
Hamiltonian.
Then solve Eq.~(\ref{schroe}) by doing quantum-mechanic
perturbation theory around the static solution.
Calculated in this way, the eigenstates (and eigenvalues) of 
Eq.~(\ref{schroe})
come out as expansions in powers of $1/m$:
\bea
&& |\underbar{n};{\bf x}_1, {\bf x}_2\rangle=|\underbar{n};{\bf x}_1,{\bf
  x}_2\rangle^{(0)}
+{1 \over m_1}| \underbar{n};{\bf x}_1, {\bf x}_2\rangle^{(1,0)}
+{1 \over m_2}| \underbar{n};{\bf x}_1, {\bf x}_2\rangle^{(0,1)}
\nn\\
&&~~~~~~~~~~~~
+{1 \over m_1^2}|\underbar{n};{\bf x}_1,{\bf x}_2\rangle^{(2,0)}
+{1 \over m_2^2}|\underbar{n};{\bf x}_1,{\bf x}_2\rangle^{(0,2)}
+{1 \over m_1 m_2}|\underbar{n};{\bf x}_1,{\bf x}_2\rangle^{(1,1)}
+\cdots \; .
\label{nexpansion}
\eea
A complete derivation can be found in the original literature
\cite{Brambilla:2000gk,Pineda:2000sz,Brambilla:2002nu}.\footnote{
A similar approach has been used in \cite{Szczepaniak:1997tk} in order 
to derive, from
the QCD Hamiltonian in the Coulomb gauge, the spin-dependent part of the
potential up to ${\cal O}(1/m^2)$.}
Here we only make a few remarks. First, the expressions
for $|\underbar{n};{\bf x}_1, {\bf x}_2\rangle^{(1,0)}$ and
$|\underbar{n};{\bf x}_1, {\bf x}_2\rangle^{(2,0)}$ look symilar to
the well-known formulas of time-independent perturbation theory in quantum
mechanics, the only difference being the fact that the energies
$E_n^{(0)}$ depend on spatial coordinates and that the matrix elements 
of $H^{(1,0)}$ and  $H^{(2,0)}$ are
operators in the quantum-mechanical sense.
Second, as usually done in quantum mechanics, we have set the relative 
phase between
$|\underbar{n};{\bf x}_1, {\bf x}_2\rangle$ and $|\underbar{n};{\bf 
x}_1, {\bf x}_2\rangle^{(0)}$
 to $1$ in Eq.~(\ref{nexpansion}). This choice is arbitrary. The freedom 
of choice
reflects the fact that the eigenvalues and eigenstates solution of 
Eq.~(\ref{schroe}) are defined
up to a unitary transformation $e^{iO_n}$ (with $O_n^{\dagger}=O_n$):
\bea
&&
|\underbar{n},{\bf x}_1,{\bf x}_2\rangle
\rightarrow
 \int d^3{\bf x}^{\prime}_1 \, d^3 {\bf x}^{\prime}_2\,
|\underbar{n},{\bf x}^{\prime}_1,{\bf x}^{\prime}_2\rangle e^{iO_n({\bf 
x}^{\prime}_1,
{\bf x}^{\prime}_2, {\bf p}^{\prime}_1,{\bf p}^{\prime}_2, {\bf 
S}_1,{\bf S}_2)}
\delta^{(3)} ({\bf x}^{\prime}_1 -{\bf x}_1)\delta^{(3)} ({\bf 
x}^{\prime}_2 -{\bf x}_2),
\label{unrqcd1}
\\
&& E_n({\bf x}_1,{\bf x}_2, {\bf p}_1, {\bf p}_2, {\bf S}_1,{\bf S}_2)
\rightarrow
\nn\\
&& \qquad
\int d^3{\bf x}^{\prime}_1 \, d^3 {\bf x}^{\prime}_2\,
 e^{iO_n({\bf x}^{\prime}_1,{\bf x}^{\prime}_2, {\bf p}^{\prime}_1,{\bf 
p}^{\prime}_2, {\bf S}_1,{\bf S}_2)}
E_n({\bf x}_1^\prime,{\bf x}_2^\prime, {\bf p}_1^\prime, {\bf 
p}_2^\prime, {\bf S}_1,{\bf S}_2)
 e^{-iO_n({\bf x}^{\prime}_1,{\bf x}^{\prime}_2, {\bf p}^{\prime}_1,{\bf 
p}^{\prime}_2, {\bf S}_1,{\bf S}_2)}
\nn\\
&& ~~~~~~~~~~~~~~~~~~~~~~~~~~~~~~~~~~~~~~~~~~~~ \times
\delta^{(3)} ({\bf x}^{\prime}_1 -{\bf x}_1)\delta^{(3)} ({\bf 
x}^{\prime}_2 -{\bf x}_2).
\label{unrqcd2}
\eea
Our choice preserves the power counting and allows us to obtain rather 
compact
expressions for the potentials. Third, from the expression for the
state $|\underbar{0};{\bf x}_1, {\bf x}_2\rangle$, the expression for 
the energy
$E_0$ may be derived straightforwardly, order by order in $1/m$, from 
Eq.~(\ref{E0average}).
Finally, the matching condition (\ref{matchnostat}) gives the pNRQCD 
Hamiltonian.

In order to transform the quantum-mechanical expressions into 
expressions that only
contain expectation values of gluon fields, the following steps are 
necessary.
\\\\
{\bf (1)}
{The first step is to integrate out the fermion fields. They appear in 
the matrix elements
of  $H^{(1)}$ and  $H^{(2)}$ either in the states (see 
Eq.~(\ref{basis0})) or in the Hamiltonian
itself as 2- or 4-fermion interaction terms.
In the first case, we have, for instance,
\bea
&&
{}^{(0)}\langle \underbar{n}; {\bf x}_1 ,{\bf x}_2| \;
\int d^3 \bfxi \; \psi^\dagger(\bfxi) \, O(\bfxi) \, \psi(\bfxi) \; 
|\underbar{m}; {\bf y}_1 ,{\bf y}_2\rangle^{(0)} =
\nn\\
&& ~~~~~~~~~~
{}^{(0)}\langle n; {\bf x}_1 ,{\bf x}_2| O({\bf x}_1) | m; {\bf x}_1 
,{\bf x}_2\rangle^{(0)} \;
\delta^{(3)} ({\bf x}_1 -{\bf y}_1)\delta^{(3)} ({\bf x}_2 -{\bf y}_2),
\eea
in the second case
\bea
&&
{}^{(0)}\langle \underbar{n}; {\bf x}_1 ,{\bf x}_2| \;
\int d^3 \bfxi \;
\psi^\dagger(\bfxi) \, O_A(\bfxi) \, \psi(\bfxi) \, 
\chi^\dagger_c(\bfxi) \, O_B(\bfxi) \, \chi_c(\bfxi)
\;  |\underbar{m}; {\bf y}_1 ,{\bf y}_2\rangle^{(0)} =
\\
&& ~~~
\delta^{(3)} ({\bf x}_1 -{\bf x}_2)\;
{}^{(0)}\langle n; {\bf x}_1 ,{\bf x}_2| O_A({\bf x}_1) \,
O_B({\bf x}_2) | m; {\bf x}_1 ,{\bf x}_2\rangle^{(0)} \;
\delta^{(3)} ({\bf x}_1 -{\bf y}_1)\delta^{(3)} ({\bf x}_2 -{\bf y}_2),
\nn
\eea
where $O$, $O_A$ and $O_B$ are combinations of gluon fields.
In the last case, the interaction is local ($\sim \delta^{(3)}({\bf r})$).
At this stage, the expressions only contain matrix elements of gluon 
fields on
the pure gluonic states $| n; {\bf x}_1 ,{\bf x}_2\rangle^{(0)} \equiv | n
\rangle^{(0)}$. At this point, it is also possible to use the
Gauss law (\ref{gausslaw}). It allows us to write all the terms of the type
$[{\bf D},g{\bf E}]$ in terms of $\delta^{(3)}({\bf r})$
times some color matrices, up to terms proportional to 
$\delta^{(3)}({\bf 0})$
that vanish in DR. We will assume to be working
in this regularization scheme from now on.
}
\\\\
{\bf (2)}
{Further simplifications may be achieved using the identities ($F_{1,2}
  \equiv F({\bf x}_{1,2})$):
\bea
&&  \hspace{-8mm}
{}^{\,\,(0)} \langle n | {\bf D}_{1} | n \rangle^{(0)} = \bfnabla_{1}, 
\qquad\qquad\qquad 
{}^{\,\,(0)} \langle n | {\bf D}_{c \,2} | n \rangle^{(0)} = 
\bfnabla_{2}, \\
&&  \hspace{-8mm}
{}^{\,\,(0)} \langle n | {\bf D}_{1} | j \rangle^{(0)} =
{ {}^{\,\,(0)} \langle n | g{\bf E}_1 | j \rangle^{(0)} \over
E_n^{(0)} - E_j^{(0)}}, \quad
{}^{\,\,(0)} \langle n | {\bf D}_{c\,2} | j \rangle^{(0)} =
-{ {}^{\,\,(0)} \langle n | g{\bf E}^{T}_2 | j \rangle^{(0)} \over
E_n^{(0)} - E_j^{(0)}} ~\forall \, n\neq j,\\
&&  \hspace{-8mm}
{}^{\,\,(0)} \langle n | g{\bf E}_1 | n \rangle^{(0)} =
-(\bfnabla_{1} E_n^{(0)}), \qquad~
{}^{\,\,(0)} \langle n | g{\bf E}^T_2 | n \rangle^{(0)} =
(\bfnabla_{2} E_n^{(0)}),
\eea
where ${\bf D}_{c}$ is the charge conjugate of ${\bf D}$.
The first equality follows from symmetry considerations, the second and 
the third may be derived
from ${}^{\,\,(0)} \langle n | [  H^{(0)}, {\bf D} ] | j \rangle^{(0)} = 
E_n^{(0)} {}^{\,\,(0)} \langle n | {\bf D} | j \rangle^{(0)}
- {}^{\,\,(0)} \langle n | {\bf D} | j \rangle^{(0)} E_j^{(0)}$
and the canonical commutation relations.
}
\\\\
{\bf (3)}
{The last step consists in rewriting the quantum-mechanical
expressions in terms of Wilson-loop amplitudes. We proceed in the following
way. We consider an interpolating state (in the Heisenberg representation)
that has a non-vanishing overlap with the ground state:
\be
\psi^\dagger({\bf x}_1)\phi({\bf x}_1,{\bf x}_2) \chi_c^\dagger({\bf x}_2)
\vert \hbox{vac} \rangle,
\label{interfield}
\ee
where $\phi$ may be everything that makes the above state overlap
with the ground state $| \underbar{0}; {\bf x}_1,{\bf x}_2 \rangle^{(0)}$.
We will use here the popular choice (\ref{schwinger}),
which assumes that the ground state has the $\Sigma_g^+$ quantum numbers.
We also define $\phi({\bf y},{\bf x};t=0) \equiv \phi({\bf y},{\bf x})$.
Then we have
\be
\psi^\dagger({\bf x}_1)\phi({\bf x}_1,{\bf x}_2) \chi_c^\dagger({\bf x}_2)
\vert \hbox{vac} \rangle =
\sum_n a_n({\bf x}_1,{\bf x}_2)\vert \underbar{n};{\bf x}_1,{\bf x}_2 
\rangle^{(0)},
\ee
or, without fermion fields,
\be
\phi({\bf x}_1,{\bf x}_2) \vert \hbox{vac} \rangle =
\sum_n a_n({\bf x}_1,{\bf x}_2)\vert n;{\bf x}_1,{\bf x}_2 \rangle^{(0)},
\label{andef}
\ee
with $a_0 \neq 0$. At this point, we define the Wilson-loop average
$\langle \cdots \rangle_\Box \equiv \langle \cdots W_\Box\rangle$. 
The gauge fields are, in general, localized on the static quark lines of the
Wilson loop. Therefore, $\langle \cdots \rangle_\Box$ is gauge invariant.
Inserting the identity operator  $\sum |n\rangle^{(0)} 
{}^{\,\,(0)}\langle n|$
into the Wilson-loop averages, from Eq.~(\ref{andef}) it follows that:
\bea
&& \langle W_\Box \rangle = \sum_n e^{-i E^{(0)}_n T_W} |a_n|^2,
\\
&& \langle F^{(1)}(t_1) \cdots F^{(n)}(t_n) \rangle_\Box  =
\sum_{n,m,s_1, \dots, s_{n-1}} a_n^* a_m  {}^{\,\,(0)}
\langle n | F^{(1)} |s_1 \rangle^{(0)}
\cdots {}^{\,\,(0)}\langle s_{n-1} | F^{(n)} |m \rangle^{(0)}
\nn\\
&& ~~~~~~~~~~~~~~~~~~~~~~~~~~~~
\times
e^{-i(E^{(0)}_n+E^{(0)}_m){T_W\over 2} }
e^{i (E^{(0)}_n-E^{(0)}_{s_1})t_1}
\cdots
e^{i (E^{(0)}_{s_{n-1}}-E^{(0)}_m)t_n},
\\
&&
\langle \! \langle F^{(1)}(t_1) \cdots F^{(n)}(t_n) \rangle\!\rangle \equiv
\lim_{T_W\rightarrow \infty} {\langle F^{(1)}(t_1) \dots F^{(n)}(t_n) 
\rangle_\Box
  \over \langle W_\Box \rangle}
\rangle^{(0)}
\nn
\\
&&
=
\sum_{s_1, \dots, s_{n-1}} {}^{\,\,(0)}\langle 0 | F^{(1)} |s_1 
\rangle^{(0)}
\cdots {}^{\,\,(0)}\langle s_{n-1} | F^{(n)} |0 \rangle^{(0)}
e^{i (E^{(0)}_0-E^{(0)}_{s_1})t_1}
\cdots
e^{i (E^{(0)}_{s_{n-1}}-E^{(0)}_0)t_n},
\label{FWoverW}
\eea
where $T_W/2 \ge t_1 \ge t_2 \ge \dots t_n \ge -T_W/2$ and 
$F^{(n)}$ are gluon fields localized on the static Wilson loop.
All the quantum-mechanical expressions obtained at the end of step (2) 
may be
expressed as combinations of
\be
\int_0^\infty dt_1 \cdots  \int_0^{t_{n-1}} dt_n \;
t_1^{j_1} \dots t_n^{j_n} \;
\langle \! \langle F^{(1)}(t_1) \dots F^{(n)}(t_n) \rangle\!\rangle_c,
\label{integralconn}
\ee
where $\langle \! \langle \cdots \rangle\!\rangle_c$ stands for  the 
connected
part of $\langle \! \langle \cdots \rangle\!\rangle$.
}

\subsubsection{Matching of the analytic terms: the real pNRQCD potential}
\label{secrealpnrqcdpot}
We give here and in the following section, the explicit formulas for the 
part
of the pNRQCD potential that is analytic in $1/m$.
For the real part, we will give formulas up to (and including) order 
$1/m^2$,
for the imaginary part up to (and including) order $1/m^4$.
The formulas are given in four dimensions. Divergences have been 
regularized, if necessary,
in DR. We have explicitly used the Gauss-law constraint (\ref{gausslaw}).
Note that we would need to generalize these formulas to $d$ dimensions, 
if we would like 
to work in an $\MS$-like scheme and consistently use the same scheme 
used for renormalizing 
the NRQCD matching coefficients.

Up to (and including) order $1/m^2$, the real part of the potential 
$V^{1/m}$ may be written as
in Eq.~(\ref{V1ovm2}) and the $1/m^2$ potentials may be decomposed in 
terms of their momentum
and spin content as in Eqs.~(\ref{decomSDSI})-(\ref{v11sdstrong}).
The different pieces are given by \cite{Brambilla:2000gk,Pineda:2000sz}:
\bea
&&
V^{(0)}(r) = \lim_{T_W\to\infty}{i\over T_W} \ln \langle W_\Box \rangle,
\label{v0strong}
\\
&&
V^{(1,0)}(r)=
-{1 \over 2} \int_0^{\infty}dt \, t \, \lla g{\bf E}_1(t)\cdot g{\bf 
E}_1(0) \rra_c,
\label{Em12strong}
\\
&&
V^{(0,1)}(r) = V^{(1,0)}(r),
\label{cc1}
\\
&&
V_{{\bf p}^2}^{(2,0)}(r)={i \over 2}{\hat {\bf r}}^i{\hat {\bf r}}^j
\int_0^{\infty}dt \,t^2 \lla g{\bf E}_1^i(t) g{\bf E}_1^j(0) \rra_c,
\label{vp20strong}
\\
&&
V_{{\bf L}^2}^{(2,0)}(r)={i \over 4}
\left(\delta^{ij}-3{\hat {\bf r}}^i{\hat {\bf r}}^j \right)
\int_0^{\infty}dt \, t^2 \lla g{\bf E}_1^i(t) g{\bf E}_1^j(0) \rra_c,
\label{vl20strong}
\\
&&
V_r^{(2,0)}(r)=  {\pi C_F \als c_D^{(1)} \over 2}  \delta^{(3)}({\bf r})
- {i c_F^{(1)\,2} \over 4} 
\int_0^{\infty}dt \,
\lla g{\bf B}_1(t)\cdot g{\bf B}_1(0) \rra_c
+ {1 \over 2}(\bfnabla_{\bf r}^2 V_{{\bf p}^2}^{(2,0)})
\nn
\\
\nn
&&
~~~~~~~~~~~
-{i \over 2}
\int_0^{\infty}dt_1\int_0^{t_1} dt_2 \int_0^{t_2}
dt_3\, (t_2-t_3)^2 \lla g{\bf
  E}_1(t_1)\cdot g{\bf E}_1(t_2) g{\bf E}_1(t_3)\cdot g{\bf E}_1(0) \rra_c
\\
\nn
&&
~~~~~~~~~~~
+ {1 \over 2}
\left(\bfnabla_{\bf r}^i
\int_0^{\infty}dt_1\int_0^{t_1} dt_2 \, (t_1-t_2)^2 \lla
g{\bf E}_1^i(t_1) g{\bf E}_1(t_2)\cdot g{\bf E}_1(0) \rra_c
\right)
\\
\nn
&&
~~~~~~~~~~~
- {i \over 2}
\left(\bfnabla_{\bf r}^i V^{(0)}\right)
\int_0^{\infty}dt_1\int_0^{t_1} dt_2 \, (t_1-t_2)^3 \lla
g{\bf E}_1^i(t_1) g{\bf E}_1(t_2)\cdot g{\bf E}_1(0) \rra_c
\\
&&
\nn
~~~~~~~~~~~
+ {1 \over 4}
\left(\bfnabla_{\bf r}^i
\int_0^{\infty}dt\, t^3
\lla g{\bf E}_1^i(t) g{\bf E}_1^j (0) \rra_c (\bfnabla_{\bf r}^j V^{(0)})
\right)
\\
&&
\nn
~~~~~~~~~~~
- {i \over 12}
\int_0^{\infty}dt \,t^4
\lla g{\bf E}_1^i(t) g{\bf E}_1^j (0) \rra_c
(\bfnabla_{\bf r}^i V^{(0)}) (\bfnabla_{\bf r}^j V^{(0)})
\\
& &
~~~~~~~~~~~
- {c_1^{g(1)} \over 4} f_{abc} \int d^3{\bf x} \,
\lla gG^a_{\mu\nu}({x}) G^b_{\mu\al}({x}) G^c_{\nu\al}({x}) \rra,
\label{vr20strong}
\\
&&
V_{{\bf p}^2}^{(0,2)}(r) =V_{{\bf p}^2}^{(2,0)}(r), \qquad
V_{{\bf L}^2}^{(0,2)}(r) =V_{{\bf L}^2}^{(2,0)}(r), \qquad
V_r^{(0,2)}(r)=V_r^{(2,0)}(r;m_2 \leftrightarrow m_1),
\label{cc2}
\\
&&
V_{{\bf p}^2}^{(1,1)}(r)=i{\hat {\bf r}}^i{\hat {\bf r}}^j
\int_0^{\infty}dt \, t^2
\lla g{\bf E}_1^i(t) g{\bf E}_2^j(0) \rra_c,
\label{vp11strong}
\\
&&
V_{{\bf L}^2}^{(1,1)}(r)=i
{\delta^{ij}-3{\hat {\bf r}}^i{\hat {\bf r}}^j \over 2}
\int_0^{\infty}dt \,t^2
\lla g{\bf E}_1^i(t) g{\bf E}_2^j(0) \rra_c,
\label{vl11strong}
\\
&&
\nn
V_r^{(1,1)}(r)=
-{1 \over 2}(\bfnabla_{\bf r}^2 V_{{\bf p}^2}^{(1,1)})
\,\delta^{(3)}({\bf r})
\\
\nn
&&
~~~~~~~~~~~
-i \int_0^{\infty}dt_1\int_0^{t_1} dt_2 \int_0^{t_2}
dt_3\, (t_2-t_3)^2 \lla g{\bf
  E}_1(t_1)\cdot g{\bf E}_1(t_2) g{\bf E}_2(t_3)\cdot g{\bf E}_2(0) \rra_c
\\
\nn
&&
~~~~~~~~~~~
+{1 \over 2}
\left(\bfnabla_{\bf r}^i
\int_0^{\infty}dt_1\int_0^{t_1} dt_2 (t_1-t_2)^2
\lla g{\bf E}_1^i(t_1) g{\bf E}_2(t_2)\cdot g{\bf E}_2(0) \rra_c \right)
\\
\nn
&&
~~~~~~~~~~~
+ {1 \over 2}
\left(\bfnabla_{\bf r}^i
\int_0^{\infty}dt_1\int_0^{t_1} dt_2 (t_1-t_2)^2
\lla g{\bf E}_2^i(t_1) g{\bf E}_1(t_2)\cdot g{\bf E}_1(0) \rra_c
\right)
\\
\nn
&&
~~~~~~~~~~~
- {i \over 2}
\left(\bfnabla_{\bf r}^i V^{(0)}\right)
\int_0^{\infty}dt_1\int_0^{t_1} dt_2  (t_1-t_2)^3
\lla g{\bf E}_1^i(t_1) g{\bf E}_2(t_2)\cdot g{\bf E}_2(0) \rra_c
\\
\nn
&&
~~~~~~~~~~~
- {i \over 2}
\left(\bfnabla_{\bf r}^i V^{(0)}\right)
\int_0^{\infty}dt_1\int_0^{t_1} dt_2  (t_1-t_2)^3
\lla g{\bf E}_2^i(t_1) g{\bf E}_1(t_2)\cdot g{\bf E}_1(0) \rra_c
\\
&&
\nn
~~~~~~~~~~~
+ {1 \over 4}
\left(\bfnabla_{\bf r}^i
\int_0^{\infty}dt \, t^3
\left\{
\lla g{\bf E}_1^i(t) g{\bf E}_2^j (0) \rra_c
+ \lla g{\bf E}_2^i(t) g{\bf E}_1^j (0) \rra_c
\right\}
(\bfnabla_{\bf r}^j V^{(0)})
\right)
\\
&&
~~~~~~~~~~~
- {i \over 6}
\int_0^{\infty}dt \, t^4
\lla g{\bf E}_1^i(t) g{\bf E}_2^j (0) \rra_c
(\bfnabla_{\bf r}^i V^{(0)}) (\bfnabla_{\bf r}^j V^{(0)}),
\nn
\\
&&
~~~~~~~~~~~
- {C_A\over 2} ({\rm Re}\, f_1(^1S_0) + 3\, {\rm Re}\, f_1(^3S_1)) 
\,\delta^{(3)}({\bf r}),
\label{vr11strong}
\\
&&
V_{LS}^{(2,0)}(r) = -{c_F^{(1)} \over r^2}i {\bf r}\cdot 
\int_0^{\infty}dt \, t \, 
\lla g{\bf B}_1(t) \times g{\bf E}_1 (0) \rra
+ {c_S^{(1)}\over 2 r^2}{\bf r}\cdot (\bfnabla_{\bf r} V^{(0)}),
\label{vls20strong}
\\
&&
V^{(0,2)}_{LS}(r)=V^{(2,0)}_{LS}(r; m_2 \leftrightarrow m_1),
\label{cc3}
\\
&&
V_{L_2S_1}^{(1,1)}(r)= - {c_F^{(1)} \over r^2}i {\bf r}\cdot
\int_0^{\infty}dt \, t \,
\lla g{\bf B}_1(t) \times g{\bf E}_2 (0) \rra, 
\label{vls11strong}
\\
&&
V_{L_1S_2}^{(1,1)}(r)=V_{L_2S_1}^{(1,1)}(r; m_1 \leftrightarrow m_2),
\label{cc4}
\\
&&
V_{S^2}^{(1,1)}(r)= {2 c_F^{(1)} c_F^{(2)} \over 3}i \int_0^{\infty} dt \, 
\lla g{\bf B}_1(t) \cdot g{\bf B}_2 (0) \rra
+2\, C_A \, ({\rm Re}\, f_1(^1S_0) - {\rm Re}\, f_1(^3S_1))\,
\delta^{(3)}({\bf r}),
\nn\\
\label{vs11strong}
\\
&&
V_{{\bf S}_{12}}^{(1,1)}(r)=
{c_F^{(1)} c_F^{(2)} \over 4}i {\hat {\bf r}}^i{\hat {\bf r}}^j
\int_0^{\infty} dt \,
\left[
\lla g {\bf B}^i_1(t) g {\bf B}^j_2 (0) \rra  - {\delta^{ij}\over 3}\lla 
g{\bf B}_1(t)
\cdot g{\bf B}_2 (0) \rra
\right].
\label{vst11strong}
\eea
Equations (\ref{cc1}), (\ref{cc2}),  (\ref{cc3}) and (\ref{cc4}) follow 
from
invariance under simultaneous charge conjugation and $m_1 \leftrightarrow
m_2$ exchange.

Equation (\ref{v0strong}) is the well-known formula that gives the static
potential in terms of the static Wilson loop
\cite{Susskind:1976pi,Brown:1979ya}. In the weak-coupling case, we have seen
that this formula gets corrections from US degrees of freedom (in that
case, US gluons). Here, by assumption, we do not have other US
degrees of freedom besides the heavy-quarkonium singlet field and, hence,
there are no corrections.  Once Goldstone bosons are taken into account, 
their
contribution will eventually correct Eq.~(\ref{v0strong}). Concerning the
power counting, for dimensional reasons, $V^{(0)}$ would count like $|{\bf p}|$. In
sec.~\ref{secpowcoustr}, we have argued, however, that the NR dynamics
constrains $V^{(0)}$ to count like $E$.  The extra suppression of order $E/|{\bf p}|
\sim v$ has to arise on dynamical grounds.  In the perturbative case, it
originates from the factor $\als \sim v$ in the potential. In the
non-perturbative case little can be said and some other mechanism must be
responsible.

Equation (\ref{Em12strong}) gives the $1/m$ corrections to the static
potential.  They have first been calculated in \cite{Brambilla:2000gk}. In
accordance with the power counting of sec.~\ref{secpowcoustr}, these
corrections are of the order $\lQ^2/m \sim E$ in the situation
$|{\bf p}|\sim \lQ$. Therefore,
they may, in principle, be as large as the static potential. In the
weak-coupling regime, the first non-vanishing contribution to $V^{(1,0)}$ is
of order $\als^2$ and gives $V^{(1,0)}(r) = - C_FC_A \als^2 /(4 r^2)$, which
is suppressed by $\als^2$ with respect to the static potential.

Equations (\ref{vp20strong}), (\ref{vl20strong}), (\ref{vp11strong}) and
(\ref{vl11strong}) are momentum-dependent $1/m^2$ potentials. They were 
first
derived in a quantum-mechanical path integral approach in
\cite{Barchielli:1988zs}. Equations (\ref{vr20strong}) and 
(\ref{vr11strong})
are momentum- and spin-independent $1/m^2$ potentials. Their calculation was
first done in \cite{Pineda:2000sz}. Note that they are necessary to 
solve the
ordering ambiguity that plagues the calculation of the momentum-dependent
potentials.  The momentum- and spin-independent $1/m^2$ potentials also
depend on some of the matching coefficients of NRQCD. The last term of
Eq.~(\ref{vr20strong}) comes from the $1/m^2$ corrections to the Yang--Mills
Lagrangian of NRQCD. It is somehow different from the other terms since the
fields are not localized on the Wilson-loop lines.  Moreover, it exhibits a
fictitious dependence on the time at which the operator insertion is 
made, which
disappears in the limit $T_W \rightarrow \infty$.  However, the term is 
not so
peculiar as it may appear if we notice that also $V^{(0)}$ could be written
in a similar way: $ V^{(0)}={1\over 2}\int d^3{\bf x} \, \lla \left(
{\bfPi}^a{\bfPi}^a +{\bf B}^a{\bf B}^a \right)({x}) \rra$.

Equation (\ref{vls11strong}) gives the spin-orbit, 
Eq.~(\ref{vs11strong}) the
spin-spin and Eq.~(\ref{vst11strong}) the spin-tensor $1/m^2$ potential. 
These
potentials were first derived in the approach that we will discuss in
sec.~\ref{secwilmatch} by \cite{Eichten:1981mw} and re-derived later by
several authors in similar or different approaches, for instance, by
\cite{Peskin:1983up,Gromes:1984ma,Barchielli:1988zs}.  All the early
derivations did not include the NRQCD matching coefficients, which were 
first
included by \cite{Chen:1995dg}, see also \cite{Brambilla:1998vm}.
\cite{Pineda:2000sz} corrected an error in the formula of the spin-orbit 
potential
$V_{L_2S_1}^{(1,1)}$ that may be found in the original papers
\cite{Eichten:1981mw,Gromes:1984ma,Barchielli:1988zs,Chen:1995dg}.
For a detailed analysis and comments on this, see
\cite{Pineda:2000sz,Brambilla:2001xk}.

In the $|{\bf p}|\sim \lQ$ regime, the leading terms contributing to the $1/m^2$
potentials are of the order $\lQ^3/m$. Not all the terms contribute, 
however,
to the same order.  Terms involving $\bfnabla_{\bf r} V^{(0)}$ have an extra
${\cal O}(v)$ suppression, coming from the specific counting of
$V^{(0)}$. Terms involving matching coefficients of NRQCD also have an
expansion in $\als$. Since the matching coefficients of the 4-fermion and of
the pure Yang--Mills operators of NRQCD start at order $\als$, terms 
involving
them are suppressed by a factor $\als$.  In particular, if we consider the
potentials with more terms, $V_r^{(2,0)}$ and $V_r^{(1,1)}$, only the 
terms in
the first three and four lines listed in Eqs.~(\ref{vr20strong}) and 
(\ref{vr11strong})
respectively are expected to contribute at LO.  In the 
weak-coupling regime,
there is an extra $\als$ suppression coming from the $g^2$ in the 
Wilson-loop
amplitudes and the $1/m^2$ potentials give the familiar $m\als^4$
relativistic, fine and hyperfine corrections to the perturbative spectrum.

The Poincar\'e invariance constraints (\ref{ex0})
become in the present case with different masses:
\bea
&&
V_{LS}^{(2,0)}(r)- V_{L_2S_1}^{(1,1)}(r)+ {V^{(0)\prime}(r)\over 2r} = 0,
\label{gromes}
\\
&&
V_{{\bf L}^2}^{(2,0)}(r) + V_{{\bf L}^2}^{(0,2)}(r) - V_{{\bf 
L}^2}^{(1,1)}(r)
+ {r\over 2} V^{(0)\prime}(r) =0,
\label{BBP1}
\\
&&
-2(V_{{\bf p}^2}^{(2,0)}(r) + V_{{\bf p}^2}^{(0,2)}(r))+ 2 V_{{\bf 
p}^2}^{(1,1)}(r)
- V^{(0)}(r)+ r  V^{(0)\prime}(r)  =0\,.
\label{BBP2}
\eea
These are general symmetry relations, independent of the dynamics.
However, thanks to the expressions for the potentials given above, they 
now impose
specific relations among the Wilson-loop amplitudes and the matching 
coefficients of NRQCD,
which can be tested independently. Taking at tree level the NRQCD 
matching coefficients,
Eq. (\ref{gromes}) has been proved by \cite{Gromes:1984ma}, and 
Eqs.~(\ref{BBP1})
and  (\ref{BBP2}) by \cite{Barchielli:1990zp}. A way to proceed is the 
following  \cite{Brambilla:2001xk}.
Consider a chromoelectric or a chromomagnetic field insertion in a 
static Wilson loop
and then apply an infinitesimal Lorentz boost  with velocity ${\bf v}$. 
The following identities hold:
\bea
&& \lla g  {\bf B} ({\bf x}_1,t) \rra^{\rm boosted}
+ \lla [{\bf v} \times g {\bf E}({\bf x}_1,t)]\rra^{\rm boosted}
- \lla g {\bf B} ({\bf x}_1,t) \rra =0,
\\
&& \lla  ig \hat{\bf v}\cdot {\bf E}({\bf x}_1,t)\rra
- \lla ig \hat{\bf v}\cdot {\bf E}({\bf x}_1,t)\rra^{\rm boosted} =0.
\eea
Expanding both equations at order $v$ and $v^2$ respectively,
and considering that the difference between the boosted and
the static Wilson loop corresponds to insertions of chromoelectric fields,
we obtain from the first equation
\be
- i \int_{0}^{\infty} dt \, t
\bigg[ \lla g{\bf B}({\bf x}_1,t)\times g {\bf E}({\bf x}_1,0)\rra
- \lla g{\bf B}({\bf x}_1,t)\times g{\bf E}({\bf x}_2,0)\rra \bigg]
+  \hat{\bf r} \, V^{(0)\prime}(r) =0,
\label{EB0dif3}
\ee
and from the second the Eqs.~(\ref{BBP1}) and (\ref{BBP2}).
These relations have also been tested on the lattice, as we will discuss 
in sec.~\ref{sec:latmod}.

Finally, we would like to emphasize that the freedom we noticed at the level
of NRQCD to perform a unitary transformation of the states and energies,
Eqs.~(\ref{unrqcd1}) and (\ref{unrqcd2}), is obviously preserved at the 
level
of pNRQCD. The effect of a unitary field redefinition $U$ of the singlet 
field
is to transform $h_s \to U^\dagger \, h_s \, U$, where $h_s$ is the pNRQCD
Hamiltonian.  This means that no special physical meaning is associated 
with a
single potential term, which may be reshuffled into another by means of a
suitable unitary transformation. In other words, differently from physical
observables, which are unambiguous, potentials depend on the specific 
scheme
adopted. The potentials listed in 
Eqs.~(\ref{v0strong})-(\ref{vst11strong}) are
given in the scheme defined by Eq.~(\ref{nexpansion}), which fixes to
$1$ the relative phase between $|\underbar{n};{\bf x}_1, {\bf 
x}_2\rangle$ and
$|\underbar{n};{\bf x}_1, {\bf x}_2\rangle^{(0)}$.  We refer to
\cite{Brambilla:2000gk} (see also \cite{Brambilla:2001xk}) for more details.

\subsubsection{Matching of the analytic terms: the imaginary  pNRQCD 
potential}
\label{secimpnrqcdpot}
Let us consider heavy quarkonia made of a quark and an antiquark of the same
flavor ($m_1$ $=$ $m_2$ $=$ $m$).  Annihilation processes happen in QCD at
the scale of the mass $m$.  Integrating them out in the matching from QCD to
NRQCD gives rise to imaginary contributions to the 4-fermion matching
coefficients.  Under the assumptions that led to 
Eq.~(\ref{pnrqcdstrong}), they are the only source of contribution to the imaginary
pNRQCD Hamiltonian, which can be calculated in the same
way as the real part.  In practice, the calculation reduces to picking up
from the right-hand side of Eq.~(\ref{E0average}) only the contributions 
that
involve 4-fermion operators.

 From the above general considerations, the imaginary part of the 
potential $V^{1/m}$ reads
\be
\label{imh}
{\rm Im} \, V^{1/m} =
{{\rm Im} \, V^{(2)} \over m^2} +{{\rm Im} \, V^{(4)} \over m^4}  + 
\cdots \;.
\ee
The functions ${\rm Im} \, V^{(2)}$ and ${\rm Im} \, V^{(4)}$ encode the 
information from the dimension six
and the dimension eight 4-fermion operators of NRQCD, respectively.
They will have the following structure
\be
\Big( \hbox{spin}\Big)
\times \Big( \hbox{delta}\Big)
\times \Big( \hbox{Im} f \Big)
\times \Big( \hbox{non-perturbative matrix element}\Big).
\ee
The first factor, which is one of the projectors  
(\ref{defT1})-(\ref{defOmega}),
accounts for the spin structure. The second is a delta function or
(for ${\rm Im} \, V^{(n>2)}$) consist of
derivatives of delta functions. This is due to the fact that the 4-fermion
operators are local. The third is the imaginary part
of a 4-fermion matching coefficient of NRQCD. Note that, in general, the 
potential 
may also depend on some real matching coefficients of NRQCD. Finally, 
the last
term is a matrix element that contains all soft gluons
integrated out from NRQCD. These matrix elements are Wilson amplitudes, 
like those
that appear in the real part of the pNRQCD potentials, taken in the 
$r \to 0$ limit, due to the delta function. In other words, they are 
non-local
(in time) correlators of gluonic fields $F$:
$\langle F^{(1)}(t_1, {\bf 0}) \phi(t_1,t_2)
\cdots F^{(n)}(t_n, {\bf 0}) \phi(t_n,t_1) \rangle$.
In the following we will omit the Wilson lines $\phi$ connecting the 
fields and the
spatial location of the fields, which is irrelevant.
The correlators that show up at order $1/m^2$ and $1/m^4$ are
encoded in the non-perturbative parameters ${\cal E}_1$,
${\cal E}_3$,  ${\cal B}_1$, ${\cal E}^{(2,t)}_3$ and ${\cal E}^{(2,{\rm
    EM})}_3$, where
\be
{\cal E}_n \equiv {1\over N_c}
\int_0^\infty dt \, t^n \,\langle g{\bf E}(t)\cdot g{\bf E}(0)\rangle,
\qquad\quad
{\cal B}_n \equiv {1\over N_c}
\int_0^\infty dt \, t^n \,\langle g{\bf B}(t)\cdot g{\bf B}(0)\rangle,
\label{EEBBcorr}
\ee
and the definitions of ${\cal E}^{(2,t)}_3$ and
${\cal E}^{(2,{\rm EM})}_3$, which involve four chromoelectric fields, 
can be found in \cite{Brambilla:2002nu}.

The explicit expression for ${\rm Im} \, V^{(2)}$ is equal to
Eq.~(\ref{imh2pert}), while ${\rm Im} \, V^{(4)}$ is given by
\cite{Brambilla:2001xy,Brambilla:2002nu}:
\bea
&&
{\rm Im} \, V^{(4)}
=
C_A \, {\cal T}^{ij}_{SJ} \bfnabla_{\bf r}^i\delta^{(3)}({\bf 
r})\bfnabla_{\bf r}^j \,
\left({\rm Im}\,f_1({}^{2S+1}P_J)+{\rm Im}\,f_{\rm
    EM}({}^{2S+1}P_J)\right)
\nn
\\
\nn
&&\qquad\qquad
+{C_A\over 2}\,
\Omega^{ij}_{SJ}\bigg\{ \bfnabla_{\bf r}^i\bfnabla_{\bf r}^j
+ {\delta_{ij}  \over 3}\, {\cal E}_1, \delta^{(3)}({\bf r}) \bigg\} \,
\left({\rm Im}\,g_1({}^{2S+1}S_J)
+{\rm Im}\,g_{\rm EM}({}^{2S+1}S_J)\right)
\\
\nn
&&\qquad\qquad
+{T_F\over 3}\, {\cal T}^{ii}_{SJ}\delta^{(3)}({\bf r})\,
{\rm Im}\,f_8({}^{2S+1}P_J) \, {\cal E}_1
\\
\nn
&&\qquad\qquad
+{T_F \over 9}
{\bfnabla_{\bf r}}\delta^{(3)}({\bf r}){\bfnabla_{\bf r}}
\Bigg(
4\,{\rm Im} \,f_8(^1 S_0)
-2\,{\bf S}^2 \left({\rm Im} \,f_8(^1 S_0)-{\rm Im}\,f_8(^3 S_1)\right)
\Bigg) \, {\cal E}_3
\\
\nn
&&\qquad\qquad
+2\,T_F\,c_F^2\,\delta^{(3)}({\bf r})
\Bigg(
{\rm Im} \,f_8(^3 S_1)+
{1 \over 6}{\bf S}^2
({\rm Im} \,f_8(^1 S_0)-3\,{\rm Im} \,f_8(^3 S_1))
\Bigg) \, {\cal B}_1
\nn
\\
\nn
&&\qquad\qquad
+{T_F \over 3}\delta^{(3)}({\bf r})
\Bigg(
4\,{\rm Im} \, f_8(^1 S_0)-2\,{\bf S}^2
\left({\rm Im} \,f_8(^1 S_0)-{\rm Im} \,f_8(^3 S_1)\right)
\Bigg) \, {\cal E}_3^{(2)}
\\
\nn
&&\qquad\qquad
-{C_A \over 3}\delta^{(3)}({\bf r})
\Bigg(
4\,{\rm Im} f_1(^1 S_0)-2\,{\bf S}^2\left({\rm Im} f_1(^1 S_0)-{\rm Im} 
f_1(^3 S_1)\right)
\Bigg) \, {\cal E}_3^{(2,t)}
\\
\nn
&&\qquad\qquad
- C_A{2 \over 9}
\left\{ \bfnabla_{\bf r}^2, \delta^{(3)}({\bf r})\right\}
\Bigg(
{\rm Im} \,f_1(^1 S_0)+{\rm Im} \,f_{\rm EM}(^1 S_0)
\\
\nn
&&\qquad\qquad\qquad\qquad
+{{\bf S}^2 \over 2}
\left( {\rm Im} \,f_1(^3 S_1)-{\rm Im} \,f_1(^1 S_0)
+{\rm Im} \,f_{\rm EM}(^3 S_1)-{\rm Im} \,f_{\rm EM}(^1 S_0)\right)
\Bigg) \, {\cal E}_3
\nn
\\
\nn
&&\qquad\qquad
-2\,C_A\,c_F^2\,\delta^{(3)}({\bf r})
\Bigg(
{\rm Im} \,f_1(^1 S_0)+{\rm Im} \,f_{\rm EM}(^1 S_0)
\\
\nn
&&\qquad\qquad\qquad\qquad
+{{\bf S}^2 \over 6}\left({\rm Im} \,f_1(^3 S_1)-3\,{\rm Im} \,f_1(^1 S_0)
+{\rm Im} \,f_{\rm EM}(^3 S_1)-3{\rm Im} \,f_{\rm EM}(^1 S_0)\right)
\Bigg) \,{\cal B}_1
\\
&&\qquad\qquad
-{C_A \over 3}\delta^{(3)}({\bf r})
\Bigg(
4\,{\rm Im}\, f_{\rm EM}(^1 S_0)-2\,{\bf S}^2\left({\rm Im}\, f_{\rm 
EM}(^1 S_0)-
{\rm Im} \,f_{\rm EM}(^3 S_1)\right)
\Bigg) \, {\cal E}_3^{(2,{\rm EM})}.
\label{imh4}
\eea
Note that there are more terms in (\ref{imh4}) than in (\ref{imh4pert}) 
due to the non-perturbative counting.
Similarly to the real case, the quantities ${\rm Im} \, V^{(2)}$, ${\rm Im} \, V^{(4)}$, ... 
are defined up to unitary transformations. A discussion and an explicit
example may be found in \cite{Brambilla:2002nu}.

\subsubsection{Matching of the analytic terms: direct matching of
  Wilson-loop amplitudes}
\label{secwilmatch}
In the previous sections we have performed the matching to pNRQCD, first, by
deriving quantum-mechanical expressions, then by translating them into
Wilson-loop amplitudes.
Hence, one may wonder whether it would be possible to
directly perform the matching to Wilson-loop amplitudes.
This is possible and simply consists in applying to the strong-coupling regime the
Wilson-loop matching used in ch.~\ref{sec:matchingII} for the weak-coupling
regime. The only difference will be that no US corrections will have to be
subtracted from the Wilson-loop amplitudes in this case.  It should be noted
that historically the first derivation of some of the heavy-quarkonium
potentials was done by direct computation of Wilson-loop amplitudes, namely
the static potential \cite{Susskind:1976pi,Brown:1979ya}, the $1/m^2$
spin-dependent potentials \cite{Eichten:1981mw,Gromes:1984ma} and the ${\bf
p}^i {\bf p}^j /m^2$ spin-independent potentials 
\cite{Barchielli:1988zs}.  In
the following, we will (re-)derive the heavy-quarkonium potential up to (and
including) order $1/m$ by directly matching Wilson-loop amplitudes to pNRQCD
Green functions \cite{Brambilla:2000gk}.

Let us consider the following Green function of NRQCD:
\bea
G_{\rm NRQCD}
&=& \langle \vac \vert  \chi_c({\bf x}_2,T_W/2)
\phi({\bf x}_2,{\bf x}_1;T_W/2)\psi({\bf x}_1,T_W/2)
\nn\\
& & \qquad\qquad\qquad
\times
\psi^\dagger({\bf y}_1,-T_W/2)\phi({\bf y}_1,{\bf y}_2;-T_W/2)
\chi_c^\dagger({\bf y}_2,-T_W/2) \vert \vac \rangle.
\label{nrqcdG}
\eea
Expanding $G_{\rm NRQCD}$ order by order in $1/m$,
$
\displaystyle
G_{\rm NRQCD} = G_{\rm NRQCD}^{(0)} + {1 \over m_1}G_{\rm NRQCD}^{(1,0)}
+{1 \over m_2}G_{\rm NRQCD}^{(0,1)} + \dots \;,
$
and integrating out the fermion fields we obtain
\bea
G_{\rm NRQCD}^{(0)} &=&
 \langle W_\Box \rangle \, \delta^{(3)}({\bf x}_1-{\bf 
y}_1)\delta^{(3)}({\bf x}_2-{\bf y}_2),
\label{vsnrqcd}\\
G^{(1,0)}_{\rm NRQCD} &=& {i\over 2} \int_{-T_W/2}^{T_W/2}\!\!dt\,
\langle {\bf D}_1^2(t) \rangle_\Box
\delta^{(3)}({\bf x}_1-{\bf y}_1) \delta^{(3)}({\bf x}_2-{\bf y}_2).
\label{DD}
\eea
For simplicity we will not display here and in the following the analogous
formulas for $G_{\rm NRQCD}^{(0,1)}$. From time reversal it follows
that $\langle {\bf B}_1(t) \rangle_\Box=-\langle {\bf B}_1(-t) 
\rangle_\Box$, which
eliminates the spin-dependent term in Eq.~(\ref{DD}).
After some algebra it follows that
\bea
&& \hspace{-5mm}
G^{(1,0)}_{\rm NRQCD}
= {i\over 2} \bigg\{{T_W \over 2} \bfnabla^2_{{\bf x}_1} \langle 
W_\Box\rangle
+ {T_W \over 2} \langle W_\Box\rangle \bfnabla^2_{{\bf x}_1}
+ T_W\langle {\bf O}_f(T_W/2)\cdot{\bf O}_i(-T_W/2)\rangle_\Box
\nonumber \\
& &
+ i g \!\!\int_{-T_W/2}^{T_W/2} \!\!\!\!\!dt \left( {T_W\over 2} -t\right)
\langle {\bf O}_f(T_W/2)\cdot{\bf E}(t)\rangle_\Box
- i g \!\!\int_{-T_W/2}^{T_W/2} \!\!\!\!\!dt \left( {T_W\over 2} +t\right)
\langle {\bf E}(t)\cdot{\bf O}_i(-T_W/2)\rangle_\Box
\nonumber \\
& &
+ {g^2\over 2}\int_{-T_W/2}^{T_W/2} \!\!\!\!\!dt 
\int_{-T_W/2}^{T_W/2}\!\!\!\!\! dt^\prime
\vert t -t^\prime \vert \langle {\bf E}(t) \cdot {\bf E}(t^\prime) 
\rangle_\Box \bigg\}
\delta^{(3)}({\bf x}_1-{\bf y}_1)\delta^{(3)}({\bf x}_2-{\bf y}_2),
\label{g10}
\eea
where the explicit form of the operators ${\bf O}_i$ and ${\bf O}_f$ 
does not
matter here and may be found in \cite{Brambilla:2000gk}.

As discussed in the previous section, the state $\psi^\dagger({\bf
  x}_1)\phi({\bf x}_1,{\bf x}_2) \chi_c^\dagger({\bf x}_2) \vert 
\hbox{vac} \rangle$ has
a non-vanishing overlap  with the NRQCD ground state $|\underbar{0}; 
{\bf x}_1,{\bf x}_2 \rangle$:
\be
Z^{1/2}({\bf x}_1,{\bf x}_2, -i{\bfnabla}_{{\bf x}_1},-i{\bfnabla}_{{\bf 
x}_2})
\delta^{(3)} ({\bf x}_1 -{\bf y}_1)\delta^{(3)} ({\bf x}_2 -{\bf y}_2)
= \langle \vac| \chi_c({\bf x}_2) \phi({\bf x}_2,{\bf x}_1)
\psi({\bf x}_1)| \underbar{0}; {\bf y}_1,{\bf y}_2 \rangle . 
\label{Z}
\ee
Since we are only interested in the analytic terms in $1/m$ here, also the
normalization factor $Z$ may be expanded in $1/m$:
\bea
Z({\bf x}_1,{\bf x}_2,-i{\bfnabla}_{{\bf x}_1}, ,-i{\bfnabla}_{{\bf x}_2})
&=& Z^{(0)}(r) + \left({1\over m_1}+ {1\over m_2}\right) Z^{(1)}(r)
\nn\\
& &
+ i Z^{(1,p)}(r) \, {\bf r}\cdot \displaystyle \bigg( 
{-i{\bfnabla}_{{\bf x}_1} \over m_1}
-{-i{\bfnabla}_{{\bf x}_2} \over m_2}
\bigg) + \dots.
\eea
The NRQCD ground state is the degree of freedom that we identify with the
singlet field of pNRQCD. Therefore, the Green function in pNRQCD
that matches $G_{\rm NRQCD}$ is:
\bea
G_{\rm pNRQCD} &=&
\langle \vac \vert 
Z^{1/2}({\bf x}_1,{\bf x}_2,-i{\bfnabla}_{{\bf x}_1},-i{\bfnabla}_{{\bf 
x}_2}) 
S({\bf x}_1,{\bf x}_2,T_W/2)
\nn
\\
&& \qquad \times
S^\dagger({\bf y}_1,{\bf y}_2,-T_W/2)
Z^{\dagger 1/2}({\bf y}_1,{\bf y}_2,-i{\bfnabla}_{{\bf 
y}_1},-i{\bfnabla}_{{\bf y}_2})
\vert \vac \rangle
\nn\\
&=&
Z^{1/2} \,\, e^{-i\,T_W\,h_s} \,\, Z^{\dagger 1/2}
\delta^{(3)}({\bf x}_1-{\bf y}_1)\delta^{(3)}({\bf x}_2-{\bf y}_2).
\label{pnrqcdG}
\eea
Matching Eq.~(\ref{pnrqcdG}) with Eq.~(\ref{nrqcdG}) we obtain at ${\cal 
O}(1/m^0)$:
\bea
V^{(0)} &=& \lim_{T_W\to\infty}{i\over T_W} \ln \langle W_\Box \rangle,
\label{v0strong2}
\\
\ln Z^{(0)} &=& \lim_{T_W\to\infty} \big( \ln \langle W_\Box \rangle + i
V^{(0)}T_W \big).
\eea
Equation (\ref{v0strong2}) coincides with Eq.~(\ref{v0strong}) and
is equivalent (up to US corrections) to the weak-coupling result of sec. 
\ref{weak00}.
Matching at ${\cal O}(1/m)$ we obtain
\bea
& &V^{(1,0)} +{1\over 2} (\bfnabla_{\bf r} V^{(0)})\cdot{\bf 
r}{Z^{(1,p)}\over
  Z^{(0)}} =
\nn\\
& &\qquad
 \lim_{T_W\rightarrow \infty} \Bigg(
- {1 \over 8} \left( { (\bfnabla_{\bf r} Z^{(0)}) \over Z^{(0)}}\right)^2
+ i {T_W \over 4}{(\bfnabla_{\bf r} Z^{(0)})\over Z^{(0)}} 
\cdot(\bfnabla_{\bf r} V^{(0)}) 
+ {T_W^2 \over 12}(\bfnabla_{\bf r} V^{(0)})^2
\nn\\
& &\qquad
- {g \over 4}\int_{-T_W/2}^{T_W/2} \!\!dt \left\{ \left( 1 -{2t\over 
T_W} \right)
{\langle {\bf O}_f(T_W/2)\cdot{\bf E}(t)\rangle_\Box
\over\langle W_\Box \rangle} 
-\left( 1 +{2t\over T_W} \right)
{\langle {\bf E}(t)\cdot{\bf O}_i(-T_W/2)\rangle_\Box
\over\langle W_\Box \rangle}  \right\}
\nn\\
& &\qquad
- {1 \over 2} {\langle{\bf O}_f(T_W/2){\bf
    O}_i(-T_W/2)\rangle_\Box \over\langle W_\Box \rangle} 
- {g^2\over 4 T_W}\int_{-T_W/2}^{T_W/2} \!\! dt \int_{-T_W/2}^{T_W/2}
\!\!dt^\prime \vert t -t^\prime \vert
{\langle {\bf E}(t) \cdot {\bf E}(t^\prime)\rangle_\Box
\over\langle W_\Box \rangle}   \Bigg).
\label{v1strong2}
\eea
 From Eq.~(\ref{v1strong2}) we cannot disentangle $V^{(1,0)}$ from 
$Z^{(1,p)}$.
This reflects, in the framework of the Wilson-loop matching, the freedom
to perform unitary field redefinitions on the pNRQCD Lagrangian.
Indeed, the Green function (\ref{pnrqcdG}) does not uniquely define 
$h_s$, but
only up to a unitary transformation of $Z$ and $h_s$. Note that 
Eq.~(\ref{Z})
allows one to calculate $Z^{(1,p)}$ only after a prescription that
fixes $|\underbar{0}; {\bf x}_1,{\bf x}_2 \rangle$, which
is defined up to a transformation (\ref{unrqcd1}), has been given. 
A possible choice of $Z^{(1,p)}$ is the one that fixes $V^{(1,0)}$ to the
value found in Eq.~(\ref{Em12strong}).  Here this choice appears
arbitrary and no obvious criteria to prefer it with respect to others 
seem to
be at hand.  Naturally, the same result would follow by calculating
$Z^{(1,p)}$ from Eq.~(\ref{Z}) with the ``quantum-mechanical'' prescription
(\ref{nexpansion}).

In the same way we could perform the matching at order $1/m^2$. In that 
case,
the Wilson amplitude to match would be the sum of all amplitudes made by an
insertion of a $1/m^2$ or of two $1/m$ NRQCD operators.  In order to fix the
ambiguity between $Z$ and $h_s$ at order $1/m^2$, we would have to give some
prescription for the $1/m^2$ terms in $Z$.  Again we have no obvious 
criteria
to guide us in the choice.  However, with a suitable prescription we would
reproduce the potentials (\ref{vp20strong})-(\ref{vst11strong}).

In concluding this section, we remark that there appear to be some 
advantages in using the quantum-mechanical matching rather than the direct matching
of Wilson-loop amplitudes.  The first one is that it provides a natural and
physical prescription for calculating the potentials and the normalization
factors. This prescription works for all orders in the $1/m$ expansion.  
It is physical because the potentials come out independent of the initial and
final interpolating fields, while all that dependence is encoded in the
normalization factor. Moreover, the power counting is preserved. The second
one is that the quantum-mechanical expressions come out manifestly finite in
the large-time limit. This is not obvious for an expression like
Eq.~(\ref{v1strong2}), which contains several divergent pieces that 
eventually cancel each other.  
Finally, we mention that the calculation of $V^{1/\sqrt{m}}$ 
using the direct matching of Wilson-loop amplitudes has not been addressed yet.

\subsubsection{Matching of the non-analytic terms}
\label{sec:strongnonan}
In this section, we will calculate $V^{1/\sqrt{m}}$, which is the part 
of the
potential that is non-analytic in $1/m$.  We will consider real and 
imaginary
contributions at the same time and, therefore, restrict ourself to the case
$m_1=m_2=m$.  In sec.~\ref{Sec:anandnoan}, we have shown that 
$V^{1/\sqrt{m}}$
arises from quark-antiquark pairs of relative three-momentum of order
$\sqrt{m\,\lQ}$. This momentum region shows up in loops where gluons of 
energy
$\lQ$ are involved (see Fig.~\ref{figsqrt}).  In the situation $p \sim 
\lQ$,
the scale $\sqrt{m\,\lQ}$ is the largest after $m$ and, therefore, the first
to be integrated out from NRQCD.

Following the procedure of \cite{Brambilla:2003mu}, it is convenient to go
through the following three steps:\\

\smallskip

\noindent
{\bf (1)} {The first step is to make explicit at the level of NRQCD
the existence of different degrees of freedom by
splitting the quark (antiquark) field into two: a semi-hard field for the
(three-momentum) fluctuations of ${\cal O}(\sqrt{m\,\lQ})$, $\psi_{sh}$
($\chi_{sh}$), and a potential field for the (three-momentum)
fluctuations of ${\cal O}(p)$, $\psi_p$ ($\chi_{p}$):
\be
\psi = \psi_p + \psi_{sh}, \qquad \qquad \chi = \chi_p + \chi_{sh}.
\ee
The NRQCD Lagrangian then reads
\be
L_{\rm NRQCD}=L^{sh}_{\rm NRQCD}+L^p_{\rm NRQCD}+L_{\rm mixing}+ L_g + L_l,
\ee
where the Lagrangians $L^{sh}_{\rm NRQCD}$ and $L^p_{\rm NRQCD}$ are 
identical to the NRQCD
Lagrangian expressed in terms of semi-hard and potential fields 
respectively,
the quantities $L_g$ and $L_l$ are the NRQCD Lagrangians for gluons and 
light
quarks respectively, and $L_{\rm mixing}$ contains the mixing terms.}\\

\smallskip

\noindent
{\bf (2)} {The second step is to integrate out gluons and quarks of 
energy or
three momentum of ${\cal O}(\sqrt{m \, \lQ})$. This leads to the EFT
NRQCD$^\prime$:
\bea
L_{\rm NRQCD} \to L_{\rm NRQCD^\prime} &=&L^{sh}_{\rm pNRQCD'}+L^p_{\rm 
NRQCD}+
{\rm Re} \, L_{\rm mixing}^{(0)} + {\rm Im} \, L_{\rm mixing}^{(0)}
\nn\\
& & \qquad\qquad\qquad
+ {\rm Re} \, L_{\rm mixing}^{(1)}
+ \cdots + L_g + L_l.
\eea
Let us discuss the different terms.\\\\
{\bf (2.1)} {$L^{sh}_{\rm pNRQCD'}$ comes from integrating out gluons and
quarks of energy or three momentum of ${\cal O}(\sqrt{m \, \lQ})$ from
$L^{sh}_{\rm NRQCD}$.  The scale $\sqrt{m \, \lQ} \gg \lQ$ is perturbative
and, therefore, we can use weak-coupling techniques.  If we further 
project onto
the quark-antiquark sector, the Lagrangian $L^{sh}_{\rm pNRQCD'}$ will
formally coincide with Eq.~(\ref{Lpnrqcd}).  The multipole-expanded 
gluons in
$L^{sh}_{\rm pNRQCD'}$ have (four) momentum much smaller than
$\sqrt{m\,\lQ}$.}\\\\
{\bf (2.2)}
{In order to simplify the calculation of $ L_{\rm mixing}$, we will
assume
\be
\sqrt{m\, \lQ} \gg m\,\als(\sqrt{m \, \lQ}) \, ,
\label{count}
\ee
which implies that, whenever a momentum of order  $\sqrt{m \, \lQ}$ flows
into a Coulomb potential (note that at the scale $\sqrt{m \, \lQ}$ the
potential is perturbative), the potential can be
expanded about the kinetic energy. If this is not the case, then a
Coulomb resummation is needed. Here we will avoid the technical 
complications
connected with this case. However, there may be situations where
this cannot be avoided. For instance, this may be
the case for the $\Upsilon$ system, if the following attribution of 
scales holds
for the $\Upsilon(1S)$: $p_{\Upsilon(1S)} \sim 
m_b\,\als(p_{\Upsilon(1S)})$ and
$\lQ \sim m_b\, \als^2(p_{\Upsilon(1S)})$, where $p_{\Upsilon(1S)}$ is 
the typical
momentum transfer of the $\Upsilon(1S)$ and $m_b$ the bottom quark mass.
In this case one would have
$\sqrt{m_b\, \lQ} \sim p_{\Upsilon(1S)} \sim m_b\,\als(\sqrt{m_b \, \lQ})$
instead of Eq. (\ref{count}).

\begin{figure}[htb]
\vskip 0.8truecm \makebox[-16truecm]{\phantom b}
\put(150,0){\epsfxsize=6truecm\epsffile{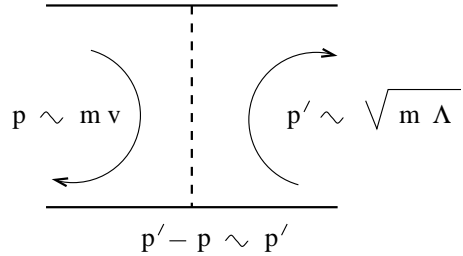}}
\caption{\it The Coulomb-exchange graph contributing to the leading mixing
interaction between semi-hard and potential fields.}
\label{figmix}
\end{figure}

The leading-order contribution to the real part of $L_{\rm mixing}$
comes from the one-Coulomb-exchange graph of Fig.~\ref{figmix}:
\bea
{\rm Re} \, L_{\rm mixing}^{(0)} &=& - \int d^3{\bf R} \, \int d^3{\bf 
r} \,
{\rm Tr}\left\{
J^\dagger({\bf R})
\, V_s^{(0)}({\bf r}) \, {\rm S}_{sh}({\bf R},{\bf r}) \right\} + {\rm H.c.}
\nn\\
&&
- \int d^3{\bf R} \, \int d^3{\bf r} \,
{\rm Tr}\left\{
J^\dagger({\bf R})
\, V_o^{(0)}({\bf r}) \, {\rm O}_{sh}({\bf R},{\bf r}) \right\} + {\rm 
H.c.}\;,
\label{mix0}
\\
J^\dagger({\bf R}) &\equiv& \chi_p({\bf R}) \psi^\dagger_p({\bf R}),
\eea
where ${\rm S}_{sh}$ and ${\rm O}_{sh}$ are semi-hard singlet and octet
quark-antiquark fields respectively.
The potentials $V_s^{(0)}$ and $V_o^{(0)}$ are perturbative:
$V^{(0)}_s = -C_F\, \als/r$ and  $V^{(0)}_o = 1/(2N_c)\,\als/r$.
The coupling constant is calculated at the semi-hard scale $\sqrt{m\,\lQ}$.

The leading contribution to the imaginary part of $L_{\rm mixing}$ may 
be read
off from the imaginary part of the pNRQCD Lagrangian at order $1/m^2$
in the weak-coupling regime:
\bea
{\rm Im} \, L_{\rm mixing}^{(0)} &=&
- \int d^3{\bf R}\,\int d^3{\bf r}\, 
{\rm Tr}\left\{ {\rm S}^\dagger_{sh}({\bf R},{\bf 0}) \,
{K_s\over m^2} \delta^{(3)}({\bf r})  \,
  J({\bf R}) \right\} + {\rm H.c.} 
\nn\\
&& - \int d^3{\bf R}\,\int d^3{\bf r}\, {\rm Tr}\left\{ {\rm 
O}^\dagger_{sh}({\bf R},{\bf 0}) \,
{K_o\over m^2} \delta^{(3)}({\bf r}) \,
 J({\bf R})\right\}  + {\rm H.c.}  \;,
\label{impotnp}
\eea 
where
\bea
K_s&=&-{C_A \over 2}
\Bigg(
4\, {\rm Im} \, f_1(^1 S_0)
-2\,{\bf S}^2\left({\rm Im}\, f_1(^1 S_0)-{\rm Im}\, f_1(^3 S_1)\right)
\nn \\
&&
\qquad\qquad
+ 4\,{\rm Im}\, f_{\rm EM}(^1 S_0)
-2\,{\bf S}^2\left({\rm Im} \, f_{\rm EM}(^1 S_0)-{\rm Im}\, f_{\rm 
EM}(^3 S_1)\right)
\Bigg),
\label{defKs}
\\
K_o&=&-{T_F\over 2}
\Bigg(
4\, {\rm Im} \, f_8(^1 S_0)
-2\,{\bf S}^2\left({\rm Im}\, f_8(^1 S_0)-{\rm Im}\, f_8(^3 S_1)\right) 
\Bigg).
\label{defKo}
\eea

The NLO term of the real part of $L_{\rm mixing}$
in the $p/\sqrt{m\,\lQ}$ expansion is given by
\bea
{\rm Re} \, L_{\rm mixing}^{(1)} &=& 
-
\int d^3{\bf R} \, \int d^3{\bf r} \,
{\rm Tr}\left\{
{\bf J}^\dagger({\bf R})\cdot {\bf r}
\, V_s^{(0)}({\bf r}) \, {\rm S}_{sh}({\bf R},{\bf r}) \right\} + {\rm H.c.}
\nn\\
&&
- \int d^3{\bf R} \, \int d^3{\bf r} \,
{\rm Tr}\left\{
{\bf J}^\dagger({\bf R})\cdot {\bf r}
\, V_o^{(0)}({\bf r}) \, {\rm O}_{sh}({\bf R},{\bf r}) \right\} + {\rm H.c.}
\label{mix1}
\;,
\\
{\bf J}^\dagger({\bf R}) &\equiv& \chi_p({\bf R}) {\vbfD \over 2}
\psi^\dagger_p({\bf R}),
\eea
which can be obtained by expanding the Coulomb potential of Fig. 
\ref{figmix}
in $p/p'$. In a similar way higher-order terms may be obtained.
Note that, as expected, the potential fields always appear as local currents
in $L_{\rm mixing}$.
}

\smallskip

\noindent
{\bf (3)} {The final step consists of integrating out degrees
of freedom of ${\cal O}(\lQ)$. This leads to the pNRQCD Lagrangian
(\ref{pnrqcdstrong}). How to calculate the analytic
part of the potential, $V^{1/m}$, has been discussed in secs.
\ref{analyticQM}, \ref{secrealpnrqcdpot} and \ref{secimpnrqcdpot}.
For the explicit computation of $V^{1/\sqrt{m}}$, we refer to 
\cite{Brambilla:2003mu}.
The results for ${\rm Re}\, V^{1/\sqrt{m}}(r)$ and $ {\rm Im}\, 
V^{1/\sqrt{m}}(r)$
turn out to be:
\bea
&& {\rm Re}\, V^{1/\sqrt{m}}(r) = (2\,C_F+C_A)^2 {4 \over
3\Gamma(9/2)} \pi \, \als^2 \, {\cal E}_{7/2}^E\; {\delta^{(3)}({\bf r})
\over m^{3/2}},
\label{Renonanal}
\\
&&
{\rm Im} \, V^{1/\sqrt{m}}(r) =
(2\,C_F+C_A) {4 \over 3\Gamma(7/2)} \, K_s\, \als\,
 {\cal E}^E_{5/2}\; {\delta^{(3)}({\bf r})\over m^{5/2}},
\label{Imnonanal}
\eea
where, in order to avoid the phase ambiguity in the definition of
the fractional power of a complex number, we have written the 
chromoelectric
correlator of Eq.~(\ref{EEBBcorr}) in Euclidean space
\be
{\cal E}^E_n \equiv
{1\over N_c} \int_0^\infty d\tau \, \tau^n \,\langle g{\bf E}(t)\cdot 
g{\bf E}(0)\rangle_E.
\ee
In accordance with the power counting of sec.~\ref{secpowcoustr}, 
Eq.~(\ref{Renonanal}) gives a contribution of order $p^3/m^2 \times 
m\als/\sqrt{m\,\lQ} \times \als$
and Eq.~(\ref{Imnonanal}) gives one of order $p^3/m^2 \times 
m\als/\sqrt{m\,\lQ} \times \lQ/m$.
Therefore, the correction (\ref{Renonanal}) is suppressed with respect 
to the largest
$1/m^2$ potentials calculated in sec.~\ref{secrealpnrqcdpot}.
The correction (\ref{Imnonanal}) is suppressed with respect to the 
imaginary part of the
$1/m^2$ potential, given in Eq.~(\ref{imh2pert}).
However, the relative size of it with respect to the imaginary part of the
$1/m^4$ potential, given in Eq.~(\ref{imh4}), depends on the size
of  $\als(\sqrt{m\,\lQ})$ about which no definite statement can be made 
at this point.
}

\subsection{Matching for $|{\bf p}|\gg \lQ \gg E$}
\label{sectwostep}
Although it is not clear whether quarkonia states fulfilling $|{\bf p}| \gg \lQ 
\gg E$
exist in nature, this situation is worth investigating. The reason is that
the calculation in the $|{\bf p}|\gg \lQ \gg E$ case
can be divided into two steps, the first of which
can be carried out by a perturbative calculation in $\als$.
The second step, even if it is non-perturbative in $\als$, admits a 
diagrammatic representation,
which makes the calculation somewhat more intuitive.

\subsubsection{pNRQCD$^\prime$}
We shall call pNRQCD$^\prime$ the EFT for energies below $|{\bf p}|$.
Since $|{\bf p}| \gg \lQ$, integrating out the energy scale $|{\bf p}|$, namely the
matching between NRQCD and pNRQCD$^\prime$, can be carried out 
perturbatively
in $\als$. The resulting EFT Lagrangian entirely coincides with the 
pNRQCD one in the
weak-coupling coupling regime, which at lower orders has
been displayed in Eqs.~(\ref{Lpnrqcd}) and (\ref{imhpert}).
Here we will need some higher-order terms in the multipole expansion (at 
tree level):
\bea
\delta{\mathcal L}_{\rm pNRQCD'} &=&
 {1 \over 8} {\rm Tr} \left\{\! {\rm O}^\dagger {\bf r}^i {\bf r}^j \, 
g {\bf D}^i {\bf E}^j \, {\rm O} - {\rm O}^\dagger {\rm O} {\bf r}^i 
{\bf r}^j
\, g {\bf D}^i {\bf E}^j \!\right\}
+  {1 \over 24} {\rm Tr} \left\{\! {\rm O}^\dagger {\bf r}^i {\bf r}^j
  {\bf r}^k \, g {\bf
    D}^i{\bf D}^j {\bf E}^k \,{\rm S}
+ \hbox{H.c.} \!\right\}
\nn \\
& &+ {c_F \over 2m} {\rm Tr} \left \{ {\rm O}^\dagger (\bfsigma_1 - 
\bfsigma_2) \cdot g{\bf B}\,{\rm S}
+ \hbox{H.c.} \right \} \, ,
\label{2steppnrqcd0}
\eea
where the traces are in color space only. S and O are chosen to
transform as a $1/2\otimes 1/2$ representation in spin space
(hence ${\bfsigma}_1 - {\bfsigma}_2 = {\bfsigma}_1\otimes \one_2 - \one_1
\otimes{\bfsigma}_2$).

\subsubsection{Matching pNRQCD to pNRQCD$^\prime$}
The matching of pNRQCD$^\prime$ to pNRQCD can no longer be done 
perturbatively in $\als$,
but it can, indeed, be done perturbatively in the following ratios of 
scales:
$\lQ/|{\bf p}|$ (multipole expansion), $\lQ/m$ and $E/ \lQ$.
Therefore, the basic skeleton of the calculation consists of an 
expansion in
$x= (\lQ/|{\bf p}|)^2$ and $y= (\lQ/m)^2$. This suggests writing the pNRQCD 
Hamiltonian as:
\be
h = h_s+  h_x + h_{x^2} + h_{y}+... \,.
\ee
The interpolating fields of pNRQCD$^\prime$ and pNRQCD will be related by:
\be
S\vert_{\rm pNRQCD'}=Z^{1\over 2} S\vert_{\rm pNRQCD}=\left( 1+ Z_x +Z_{x^2}
  + Z_y + ... \, \right)^{{1 \over 2}} S\vert_{\rm pNRQCD}
\, .
\ee
The matching calculation reads:
\bea
 & &
\int^{\infty}_{-\infty} dt \, e^{-iEt}\int d^3 {\bf R} \, \langle {\rm 
vac}\vert
T\{S({\bf x}, {\bf R}, t) S({\bf x}^{\prime}, {\bf 0}, 0)\}\vert {\rm vac}
\rangle\vert_{\rm pNRQCD'}
\cr & &
= \int^{\infty}_{-\infty} dt \,  e^{-iEt} \int d^3 {\bf R} \,
Z^{1\over 2}\langle {\rm vac}\vert T\{S({\bf x}, {\bf R}, t) S({\bf x}^{\prime}, {\bf 0},0)\}
\vert {\rm vac} \rangle
\vert_{\rm pNRQCD}{Z^{1\over 2}}^{\dagger}
\, .
\label{mtcal}
\eea
The right-hand side of the matching calculation has the following 
structure:
\bea
& &
{1\over E-h_s}+{1\over E-h_s}( h_x + h_{x^2} +h_{y} ) {1\over E-h_s}
+ {1 \over 2} \left( Z_x +Z_{x^2} +Z_y - { Z_x^2 \over 4} \right)
{1\over E-h_s}
\cr
&+& {1\over E-h_s}{1 \over 2} \left( Z_x +Z_{x^2} +Z_y
- {Z_x^2 \over 4} \right)^{\dagger}+
\left( {Z_x \over 2} \right) {1\over E-h_s}\left( {Z_x \over 2}
\right)^{\dagger}
\cr
&+& {1\over E-h_s}h_x{1\over E-h_s}h_x{1\over E-h_s}+
\left( {Z_x \over 2} \right){1\over E-h_s}h_x{1\over E-h_s}+{1\over
  E-h_s}h_x{1\over E-h_s}\left( {Z_x \over 2} \right)^{\dagger}  .\quad\quad
\label{wfpt}
\eea
Hence, once we have made sure that, up to contact terms, the left-hand 
side of
Eq.~(\ref{mtcal}) has exactly the same structure, we can easily identify 
the
contributions to the pNRQCD Hamiltonian from the second term of the 
expression
(\ref{wfpt}).

Let us illustrate how the calculation of the left-hand side of
Eq.~(\ref{mtcal}) proceeds by concentrating on the following contribution:
\bea
{1 \over E-h_s} \, {i \over N_c}  \int_0^{\infty} dt \, \langle i {\bf r}
\cdot g{\bf E} (t) \, e^{-i(h_o-E)t} \, i {\bf r} \cdot g{\bf E} (0) 
\rangle \, {1\over
  E-h_s} \, .
\label{2a}
\eea
One might naively think that the fact that $E/\lQ$ is small can be
implemented by expanding the exponential ($t$ takes the typical value of
$1/\lQ$) \cite{Brambilla:2001xy,Brambilla:2002nu}. However this is not 
entirely
correct. Whereas it is true that $h_o$, between the heavy quarkonium 
states we
are considering, has the size $E$, it may experience fluctuations of a 
larger
size, for instance $\sim \lQ$, since the cut-off of the relative three 
momentum
is only constrained to be smaller than $m$, and hence it may well reach 
values
$\sim \sqrt{m\,\lQ}$. Nevertheless, the energy $E$ can indeed always be
expanded, which guarantees that we will eventually get usual,
energy-independent, potentials. If $h_o$ could not be expanded, we would
obtain potentials which are non-trivial functions of $m$, $\lQ$ and  ${\bf
r}$. Fortunately, we can do much better by exploiting the fact that the
momenta, which prevent us from expanding, fulfill $|{\bf p}|\sim \sqrt{m\,\lQ} 
\gg \lQ$.
We shall proceed as follows. We split the relative momentum in two regions.
The first region fulfills $|{\bf p}| \ll  \sqrt{m\,\lQ}$ and hence $h_o$ can be 
expanded
and the second region contains the momentum fluctuations $\sim 
\sqrt{m\,\lQ}$.
\\\\
{\bf (1)} {The matching in the region $|{\bf p}| \ll  \sqrt{m\,\lQ}$.
\\\\
{\bf (1.1)}
{The real part of the potential. \\
At LO in the expansion, the exponential in Eq.~(\ref{2a}) 
reduces to $1$
and we obtain the leading non-perturbative correction to the Coulomb 
potential:
\be
\delta V_s=
-i{g^2 \over N_c}T_F
{r^2 \over 3} \int_0^\infty \!\! dt
\langle {\bf E}^a(t) \phi(t,0)^{\rm adj}_{ab}{\bf E}^b(0)\rangle.
\label{vsprime2}
\ee
This expression was first derived in \cite{Balitsky:1985iw}.
Higher orders in the $E/\lQ$ expansion can be easily
calculated. They induce contributions to potentials which are higher order
in $1/m$ as well as further contributions to the static potential.
Some of these have been calculated in \cite{Brambilla:1999xf}.}
\\\\
{\bf (1.2)}
{The imaginary part of the potential. \\
Since the imaginary parts, which are inherited from NRQCD, are contained in
local ($\delta^{(3)} ({\bf r})$, $\bfnabla \delta^{(3)} ({\bf r}) 
\bfnabla$,
 etc.) terms in the pNRQCD$^\prime$ Lagrangian, they tend
to vanish when being multiplied by the ${\bf r}$'{\small s} arising from 
the multipole expansion.
 Hence, for an imaginary part to contribute, it is necessary to have a 
sufficient number of derivatives
(usually arising from the $E/ \lQ$ expansion) in order to cancel all the 
${\bf
  r}$'s. Since derivatives are always accompanied by powers of $1/m$,
it implies that at a given order in $1/m$, only a finite number of terms in
the multipole expansion contributes.
We are only interested in collecting the imaginary parts that contribute
up to order $1/m^4$, in order to provide an independent
calculation to support the results of sec.~\ref{seconestep}.
Consider again the contribution of Eq.~(\ref{2a}).
The first imaginary terms arise at
${\cal O}\left(E/\lQ\right)$ from the ${\cal O}(1/m^4)$ parts of
the singlet and octet potentials displayed in Eq.~(\ref{imh4pert}):
\bea
& &
{i \over E-h_s} \,
\left(
{T_F {\cal T}_{S J}^{ii}\,{\rm Im}\,f_8 (^{2S+1} P_J) \over 3 N_c m^4 }
+{ {\cal T}_{S}\,\left({\rm Im}\,g_1 (^{2S+1} S_S)
+ {\rm Im}\,g_{\rm EM} (^{2S+1} S_S) \right)
\over
  m^4 } 
\right)
\nn\\
&& \qquad\qquad
\times \int_0^{\infty} dt \, t \, \langle g{\bf E}(t) \cdot g{\bf E} (0)
\rangle \, {\delta^{(3)} ({\bf r}) \over E-h_s} \, ,\quad\quad
\eea
where ${\cal T}_{SJ}^{ij}$ are defined in 
Eqs.~(\ref{defT1})--(\ref{defT2}) and
${\cal T}_S = \Omega^{ii}_{SS}/3$.
The calculation may be systematically extended to higher orders.
Details are given in \cite{Brambilla:2002nu}.
Here we just point out two subtleties. First,
ill-defined expressions arise in the calculation, from products of
distributions (both products of two delta functions and products of
delta functions with non-local potentials, which diverge as ${\bf 
r}\rightarrow
0$). It is most convenient to use DR in this case,
which sets all these terms to zero. This is shown in Appendix D of 
\cite{Brambilla:2002nu},
where the relation to other regularization schemes is also discussed. Second
there is a freedom in organizing the calculation, which may lead
to different forms of the potentials.
Let us consider, as an example, the term
\be
{1\over E-h_s}{\bf r} (E-h_s)^2 {\bf r}{1\over E-h_s}
\label{m2e2} \, .
\ee
If we decide to take one power $(E-h_s)$ to the right and one to the 
left we have
\be
{\bf r}^2 +{\bf r}[{\bf r}, h_s]{1\over E-h_s}+{1\over E-h_s}[h_s,{\bf r}]
{\bf r}+{1\over E-h_s}[h_s,{\bf r}][{\bf r}, h_s]{1\over E-h_s} \, ,
\label{m2e2sym}
\ee
which does not produce any imaginary part. However, an equally 
acceptable expression is
\be
{\bf r}^2+{1\over 2}[{\bf r},[{\bf r}, h_s]]{1\over E-h_s}
+{1\over E-h_s}{1\over 2}[[h_s,{\bf r}], {\bf r}]
+{1\over E-h_s}{1\over 2}\{[[{\bf r}, h_s],h_s],{\bf r}\}{1\over E-h_s} \, ,
\label{m2e2asym}
\ee
which does produce an imaginary part. The apparent paradox only reflects
the fact that expression (\ref{m2e2}) by itself (as well as others
that one may find in the calculation) does not determine uniquely its 
contribution
to the potential. It leads to contact terms,
wave-function normalization and potential, as is apparent in 
(\ref{m2e2sym})
and (\ref{m2e2asym}), but depending on how we decide to organize the
calculation, the terms associated with each of these pieces
change. For instance, when matched to (\ref{wfpt}), (\ref{m2e2sym}) gives
$
h_x= [h_s,{\bf r}][{\bf r}, h_s]
$,
$
Z_x={\bf r}[{\bf r}, h_s]
$,
whereas (\ref{m2e2asym}) gives
$
h_x= {1\over 2}\{[[{\bf r}, h_s],h_s],{\bf r}\}
$,
$
Z_x = {1\over 2}[{\bf r},[{\bf r}, h_s]] \, .
$
This should not be a surprise. It corresponds
to the freedom of making unitary
transformations in a quantum-mechanical Hamiltonian
already discussed in the previous sections, and does not affect
any physical observables.
In order to fix the contribution to the potential of any term once and 
forever, 
we use the  prescription described in detail in sec. V of 
\cite{Brambilla:2002nu}.
With this prescription, (\ref{m2e2}) gives rise to the potential
obtained in (\ref{m2e2sym}) and hence to no imaginary part.
Eventually, combining all the contributions, we obtain for the imaginary
part of the pNRQCD potential in the situation $p \gg \lQ \gg E$ the same
result, up to a unitary transformation, as obtained in sec. 
\ref{seconestep} for the situation
$p \sim \lQ$ and explicitly listed in Eqs.~(\ref{imh2pert}) and 
(\ref{imh4}).
The explicit form of the unitary transformation
can be found in \cite{Brambilla:2002nu}.}
\\\\
{\bf (2)}
{The matching in the region $|{\bf p}| \sim  \sqrt{m\,\lQ}$. \\
The contributions due to heavy quarks of three momentum of order
$\sqrt{m\,\lQ}$ may be calculated in a way very similar to sec. 
\ref{sec:strongnonan},
the main difference is that now potential and semi-hard degrees of freedom
need not be separated at the level of NRQCD, but of pNRQCD$^\prime$.
\\\\
{\bf (2.1)}
{The first step consists in rewriting the pNRQCD$^\prime$
  Lagrangian in terms of semi-hard fields $S_{sh}$ and $O_{sh}^a$ 
associated with
three-momentum fluctuations of ${\cal O}\left(\sqrt{m\,\lQ}\right)$ and
potential fields $S_p$ and $O_p^a$ associated with three-momentum 
fluctuations
of ${\cal O}(p)$:
\be
S = S_p + S_{sh},  \qquad\qquad O^a = O_p^a + O_{sh}^a.
\ee
The pNRQCD$^\prime$ Lagrangian then reads 
\be
L_{pNRQCD^\prime}= L^{sh}_{pNRQCD^\prime}+L^p_{pNRQCD^\prime}+L_{\rm
  mixing} + L_g + L_l,
\ee
where $L^{sh}_{pNRQCD^\prime}$ and $L^p_{pNRQCD^\prime}$ are identical
to the pNRQCD$^\prime$ Lagrangian in the heavy-quarkonium bilinear sector
except for the changes
$S$, $O^a$, $V_s$, $V_o$  $\rightarrow$ $S_{sh}$, $O_{sh}^a$, 
$V_s^{sh,sh}$, $V_o^{sh,sh}$
and $S$, $O^a$,
$V_s$, $V_o$  $\rightarrow$ $S_{p}$, $O_{p}^a$, $V_s^{p,p}$, 
$V_o^{p,p}$, respectively.
$L_g$ and $L_l$ are the parts of the pNRQCD$^\prime$
Lagrangian that contain only gluons and light quarks respectively,  
and $L_{\rm mixing}$ contains the mixing terms.
We recall that the gluons left dynamical have energies of ${\cal 
O}(\lQ)$ and that
analytic terms in ${\bf r}$ do not mix semi-hard and potential fields.
Therefore, the multipole expansion in (\ref{2steppnrqcd0})
is an expansion in either the scale ${\bf r}\sim
1/\sqrt{m\,\lQ}$ in $L^{sh}_{pNRQCD^\prime}$ or the scale ${\bf r}\sim 1/p$
in $L^p_{pNRQCD^\prime}$.}
\\\\
{\bf (2.2)}
{The second step consists in integrating out gluons and quarks
of energy and three momentum of  ${\cal O}\left(\sqrt{m\,\lQ}\right)$.
We will assume, as in  Eq.~(\ref{count}) and for the same reasons as 
discussed
there, that  $\sqrt{m\,\lQ} \gg m\,\als(\sqrt{m\,\lQ})$.
As an example, we consider the real part of the singlet-mixing term
due to the static Coulomb potential.
The matching works exactly as in paragraph (2.2) of sec. 
\ref{sec:strongnonan} and leads to
\bea
\label{twoexp}
&&{\rm Re} \, L_{\rm mixing}\Bigg|_{\rm Singlet} =
-\int d^3{\bf R}\,\int d^3{\bf r} \,
S^\dagger_p({\bf R},{\bf r})\, V^{(0)}_s({\bf r}) \, S_{sh}({\bf R},{\bf 
r})
+ {\rm H.c.}
\\
&&
\to 
-\int d^3{\bf R}\,\int d^3{\bf r}\,
\left(
S^\dagger_{p}({\bf R},{\bf 0})
+{\bf r}\cdot \bfnabla_{\bf r}S^\dagger_{p}({\bf R},{\bf 0})
+\cdots
\right)
\, V^{(0)}_s({\bf r}) \,
S_{sh}({\bf R},{\bf r})
+ {\rm H.c.}\,.
\nn
\eea
At the order of interest, we have $V^{(0)}_s = -C_F\, \als/r$ and
$\als=\als\left(\sqrt{m\,\lQ}\right)$.  Analogous results hold for the
real part of the octet-mixing term due to the static Coulomb
potential.

The leading contribution to the imaginary part of $L_{\rm mixing}$ is 
given by
\bea
&& {\rm Im} \, L_{mixing} = - \int d^3{\bf R}\,\int d^3{\bf r}\,  {\rm 
Tr} \, \left\{
{\rm S}^\dagger_{sh}({\bf R},{\bf 0}) \,
{K_s\over m^2} \delta^{(3)}({\bf r})  \, {\rm S}_p({\bf R},{\bf 0})  + 
{\rm H.c.} \right\} 
\nn\\
&& \quad
- \int d^3{\bf R}\,\int d^3{\bf r}\,  {\rm Tr} \, \left\{
{\rm O}^\dagger_{sh}({\bf R},{\bf 0}) \,
{K_o\over m^2} \delta^{(3)}({\bf r}) \, {\rm O}_p({\bf R},{\bf 0})  + 
{\rm H.c.}  \right\} ,
\label{impot}
\eea
where $K_s$ and $K_o$ have been defined in Eq.~(\ref{defKs})  and 
(\ref{defKo}),
respectively.}
\\\\
{\bf (2.3)}
{The final step  consists in integrating out from pNRQCD$^\prime$ all
fluctuations that appear at the energy scale $\lQ$.
These are light quarks and gluons of energy or three momentum  of
order $\lQ$, and singlet and octet fields of energy of order $\lQ$ or
three momentum of order $\sqrt{m\,\lQ}$. We will then be left with pNRQCD.
The part $V^{1/m}$ of the potential (see Eq.~(\ref{V1m1sm})) has been 
calculated in
paragraph (1) of this section. The part $V^{1/\sqrt{m}}$ of the 
potential develops a real and
an imaginary part. They turn out to be equal to Eq.~(\ref{Renonanal}) 
and (\ref{Imnonanal})
respectively, i.e. to the results obtained in the kinematical situation 
$p\sim \lQ$.
We refer to \cite{Brambilla:2003mu} for a detailed diagrammatical 
calculation.
}
}
\\\\
In summary, we have presented in this section a derivation of the pNRQCD
potential (real and imaginary) in a kinematical situation and with a 
technical
procedure that are quite different from the ones of sec. 
\ref{seconestep}. The
agreement of the results (up to unitary transformations) in the case when
the potentials are local (non-analytic and imaginary terms) is reassuring
and confirms in an explicit calculation what is expected in
sec.~\ref{Sec:anandnoan} on general grounds. Despite this, it should be 
noted
that the matching coefficients of the terms in the multipole expansion in
pNRQCD$^\prime$ (\ref{2steppnrqcd0}) were only calculated at tree level 
here,
whereas the expressions in sec.~\ref{seconestep} correspond to an all-order
result. This indicates that there must be a symmetry protecting these terms
against higher-loop corrections\footnote{For the LO term, the
non-renormalization was verified at one loop in \cite{Pineda:2000gz}.}.  
This
symmetry does not appear to be Poincar\'e invariance 
\cite{Brambilla:2003nt}.

\subsection{Potentials and spectra: lattice and models}
\label{sec:latmod}
The heavy-quarkonium spectrum is obtained by solving the Schr\"odinger
equation for the pNRQCD Hamiltonian $h_s$:
\be
h_s \,\phi_{njls}({\bf r})=E_{njls}\,\phi_{njls}({\bf r}).
\label{coordSchr}
\ee 
Since $h_s$ is known from Eqs.~(\ref{v0strong})-(\ref{vst11strong}), 
(\ref{imh2pert}), (\ref{imh4}), (\ref{Renonanal}) and (\ref{Imnonanal}), 
the Schr\"odinger equation (\ref{coordSchr}) is completely defined in terms of 
QCD quantities.

At LO, Eq.~(\ref{coordSchr}) becomes:
\be
h_s^{(0)} \phi^{(0)}_{njls}({\bf r}) =
\left({{\bf p}_1^2 \over 2 m_1} + {{\bf p}_2^2 \over 2 m_2} 
+V_{\rm LO}\right)\phi^{(0)}_{njls}({\bf r})
= E^{(0)}_{njls}\,\phi^{(0)}_{njls}({\bf r}).
\label{schn}
\ee
What $V_{\rm LO}$ is depends on the power counting.
We have argued in sec.\ref{secrealpnrqcdpot}
that in the situation $p \sim \lQ$ and in the most conservative power
counting, we could have 
$
V_{\rm LO}=V^{(0)}+{V^{(1)}/m}.
$
On the other hand, if $p \gg \lQ$, we have
$
V_{\rm LO}=V^{(0)}.
$
In both cases, at this order the potential is spin independent ($E_{njls}^{(0)} \equiv E_{nl}^{(0)}$)
and, therefore, the leading-order $S$- and $P$-wavefunctions read 
\be
\phi^{(0)}_{ns0s}({\bf r})=R^{(0)}_{n0}(r){1 \over \sqrt{4\pi}}|s\rangle_{\rm spin}
\qquad {\rm and} \qquad  
\phi^{(0)}_{nj1s}({\bf r})=R^{(0)}_{n1}(r) \;\langle {\bf \hat{r}}|js\rangle,
\ee
where 
$|s\rangle_{\rm spin}$ denotes the normalized spin component, 
$|{\bf \hat r}\rangle$ the normalized eigenstate of the position 
and $|js\rangle$ the $J$ (total angular momentum) and $S$ eigenstate such that
$
\langle {\bf \hat{r}}|j0\rangle=Y_j^m({\bf \hat{r}})|0\rangle_{\rm spin} \quad
(j=l=1)
$ and 
$
\langle {\bf \hat{r}}|j1\rangle={\cal Y}^1_{jm}({\bf \hat{r}})
$. 
The label $m$ denotes the third component of the angular momentum. 

At NLO, the $1/m^2$ potentials calculated in
sec.~\ref{secrealpnrqcdpot} have to be considered, except for those that may have some extra suppression.
 Also the contribution to the spectrum that comes from 
the $V^{1/\sqrt{m}}$ potential given in Eq.~(\ref{Renonanal}) 
turns out to be suppressed. Indeed, we have ($m^{\rm red}$ is the reduced mass) 
\be
\delta E_{njls}^{1/\sqrt{m}} = 
(2\,C_F+C_A)^2 {1 \over
3\Gamma(9/2)} \, \als^2 \, {\cal E}_{7/2}^E\;
{|R_{nl}(0)|^2 \over (2 \,m^{\rm red})^{3/2}}\delta_{l0},
\ee
which is of order $|{\bf p}|^3/m^2 \times  m\als/\sqrt{m\, \lQ} \times \als$, i.e. 
suppressed with respect to the contribution coming from the 
$1/m^2$ potentials of Eqs.~(\ref{v0strong})-(\ref{vst11strong}), which, in the
conservative counting is of order $p^3/m^2$.

We would like to emphasize that, in order to be consistent with the power counting, 
sub-leading terms in the expansion 
of the kinetic energy and the potential should be treated 
as perturbations when solving Eq.~(\ref{schn}). This differs from the common practice in potential models. 
In an EFT framework, the calculation of the spectrum is 
not plagued by the inconsistencies emerging in higher-order 
calculations in  potential models. 
It is, for instance, known that at second order in quantum-mechanical
perturbation theory the spin-dependent terms result in a contribution that is 
ill-defined. Regulating it requires the introduction of a cut-off (or DR).
A large cut-off gives rise to a linear and to a logarithmic divergence. 
These divergences can be renormalized by redefining the coupling constant 
of a delta potential \cite{Lepage:1997cs}.
On the other hand, when one matches QCD to NRQCD,
one expands in the energy and the three momentum. In general, this induces IR
divergences in the matching coefficients and, in particular, in the calculation of a matching coefficient
of a four fermion operator at two loops, which leads to the delta potential mentioned above. 
If one uses a consistent
regularization scheme both for the QCD-NRQCD matching calculation 
and the quantum-mechanical calculation in pNRQCD, the divergences exactly
cancel and eventually a totally consistent  scale independent  result is  obtained
(for a QED example see \cite{Czarnecki:1999mw,Czarnecki:1998zv}).
Notice that an EFT framework is crucial for understanding this 
second-order calculation and for making the result meaningful.

For a determination of the spectrum at order $p^3/m^2$ in the conservative
counting, one needs to consider, besides the static and the $1/m$ potential,
the ${\cal O} (1/m^2)$ potentials given in Eq.~(\ref{V1ovm2}), of which for
$V_r^{(2,0)}$ and $V_r^{(1,1)}$ only the terms in the first four lines of
Eqs.~(\ref{vr20strong}) and (\ref{vr11strong}) need to be considered. How can one 
get the explicit form of these potentials?  The EFT provides the expressions for 
such potentials in terms of Wilson-loop amplitudes typically involving
chromoelectric and chromomagnetic field insertions.  In the case of the
imaginary parts, they reduce to chromoelectric and chromomagnetic correlators.
These are low-energy objects that do not depend on the quarkonium state,
involve only integrations over gluon fields and light quarks, are gauge
invariant and perfectly suited for lattice calculations.  We emphasize that
the EFT approach greatly reduces the lattice effort necessary to produce heavy
quarkonium spectra and decay widths. This is for two reasons.  The first
reason is that the objects to be calculated on the lattice involve only
integrations over low energy gluons and light quarks. The second is that one
does not need to repeat a lattice evaluation for each quarkonium state (with
the problems related to the mass extraction of the excited states) but only
to extract the form of all the potentials with one simulation. These,
once inserted in the Schr\"odinger equation (\ref{schn}), will produce the
spectrum. One should check {\it a posteriori} which states in the obtained
spectrum fulfill the hypothesis of the strong-coupling regime. The ones that
do will be the ones for which the calculation is reliable.

\subsubsection{Potentials and spectrum from the lattice}
If DR is used in the continuum, the Wilson-loop amplitudes involved in the
static and $1/m$ potentials can be renormalized by the counterterms of light
degrees of freedom only, and hence they do not display a factorization scale
dependence. For the $1/m^2$ and higher potentials, counterterms involving
local potentials are also necessary, and the Wilson-loop amplitudes depend on
the factorization scale. In a physical observable, this scale dependence,
together with the one induced by the quantum-mechanical perturbation theory, will
cancel against the scale dependence of the NRQCD matching coefficients.  In
the strong-coupling regime, there are no US divergences, at least when
the US degrees of freedom (pseudo-Goldstone bosons) are neglected.

In a lattice regularization scheme, the situation is more complicated for
several reasons.  The Wilson-loop amplitudes contain additive $1/a$ dependent
self-energy contributions ($a$ being the lattice spacing), even in the static
case.  This dependence on $1/a$ is canceled by the quark-mass shift
and is removed by a suitable renormalization condition (see
the discussion on the static potential in ch.~\ref{sec:static} and below).
Moreover, large terms are generated having their origin in self-interactions
within the plaquette  as well as between plaquette and static propagator
(to higher orders). These affect all the Wilson-loop amplitudes. They would be
canceled by NRQCD matching coefficients calculated in a lattice
regularization.  Without those the scale dependence can be dramatic and
several {\it ad hoc} lattice recipes have been applied to get rid of it,
without actually calculating the matching coefficients, which would be the
definite cure.  In addition, the Wilson-loop amplitudes will generate $a$ and
$r$ dependent terms, which are specific to the lattice.  
On  top of this, Lorentz invariance is broken on the lattice. Thus,
order $a$ corrections to coefficients otherwise protected by Lorentz invariance may appear.

All these issues are related to the lattice regularization and
renormalization.  A proper treatment would require the calculation of both the NRQCD
matching coefficients and the Wilson-loop amplitudes in a proper lattice
regularization and renormalization scheme.  Also the Schr\"odinger equation
would need to be solved in the same scheme, due to the quantum-mechanical
divergences.  The NRQCD matching coefficients are known at different accuracy
in the continuum and in DR, see sec.~\ref{NRmatching}, but up to now no
calculation of the coefficients here relevant  exists within a lattice scheme,
apart from the one in  \cite{Trottier:1997bn}.  Another strategy would be
to use a non-perturbative renormalization \cite{Martinelli:1997zc} on both
parts in lattice regularization.  Alternatively, if the available $\MS$ NRQCD
matching coefficients are to be used, one should change the Wilson-loop
amplitudes from the lattice renormalization scheme to $\MS$.  This can be done
in lattice perturbation theory since the cut-off of these divergences is close
to $m$ \cite{Bodwin:2001mk}.  Then, the divergences arising in the
quantum-mechanical perturbation theory should also be $\MS$ renormalized.

A proper lattice treatment of pNRQCD has not been implemented so far.  NRQCD
matching coefficients were never considered in the lattice calculation of the
potentials with the exception of the work of \cite{Bali:1997am,Bali:2000gf}
where an estimate of the NRQCD matching coefficients was used.  Therefore,
this work may be considered the closest, up to now, to a lattice treatment of
pNRQCD.  We will mainly refer to it in the following.

The static potential is given only in terms of the static Wilson loop
(\ref{v0strong}) and it has been one of the first objects to be evaluated on
the lattice in relation to quark confinement \cite{Wilson:1974sk}. Today the
static potential is known with great accuracy
\cite{Bali:1997am,Bali:2000gf,Necco:2001xg,Luscher:2002qv}, even in the
unquenched case \cite{Bali:2000vr,Bolder:2000un}.  In Fig.~\ref{Masgap}, the
curve labeled $\Sigma_g^+$ displays the static potential data obtained in
\cite{Bali:2000vr} in units of $r_0\approx 0.5$ fm.  The squares refer to a
quenched simulation at $\beta = 6.2$ and the diamonds to unquenched
simulations at $\beta = 5.6$ with two  mass-degenerate quark flavors. The value of the mass
parameter is $\kappa = 0.1575$.  The physical units follow from a choice
of the lattice spacing $a$.  This is often fixed on the bottomonium spectrum
\cite{Bali:1997am,Bali:2000gf}. This procedure may potentially introduce large
uncertainties if the set of potentials at our disposal is not complete, if the
power counting not consistent or if, as is usually done, lower and higher
bottomonium states are fitted with the same confining potentials.  However,
such a determination seems to be numerically in agreement with others obtained
from the $m_\pi/m_\rho$ ratio.  The continuous curve in Fig.~\ref{Masgap}
represents the Cornell parametrisation $V^{(0)}(r)= -{e/ r} + \sigma r$ with
$e \approx 0.368$ and $\sigma \approx (445 ~\hbox{MeV})^2$.  An additive
self-energy contribution, associated with the static sources and  diverging in the
continuum limit, has been removed by normalising the data to $V(r_0)=0$.
This corresponds to the elimination of the static potential renormalon
described in ch.~\ref{secrenormalons}. As shown in ch.~\ref{sec:static}, QCD
perturbation theory perfectly agrees with the lattice data up to about 0.25 fm
(actually the analysis of \cite{Pineda:2002se} shows agreement up to 0.4 fm),
while from about 0.5 fm on the data is described very well by an effective 
string theory at NLO \cite{Luscher:2002qv}. However, this seems to be 
specific of the ground-state energy: the energy spectrum is still far from being
string-like at such distances \cite{Luscher:2004ib}. This is more
apparent for the excited-state energies \cite{Baker:2001xm}.
\begin{figure}[thb]
\makebox[-16truecm]{\phantom b}
\put(0,0){\epsfxsize=6.5truecm\epsffile{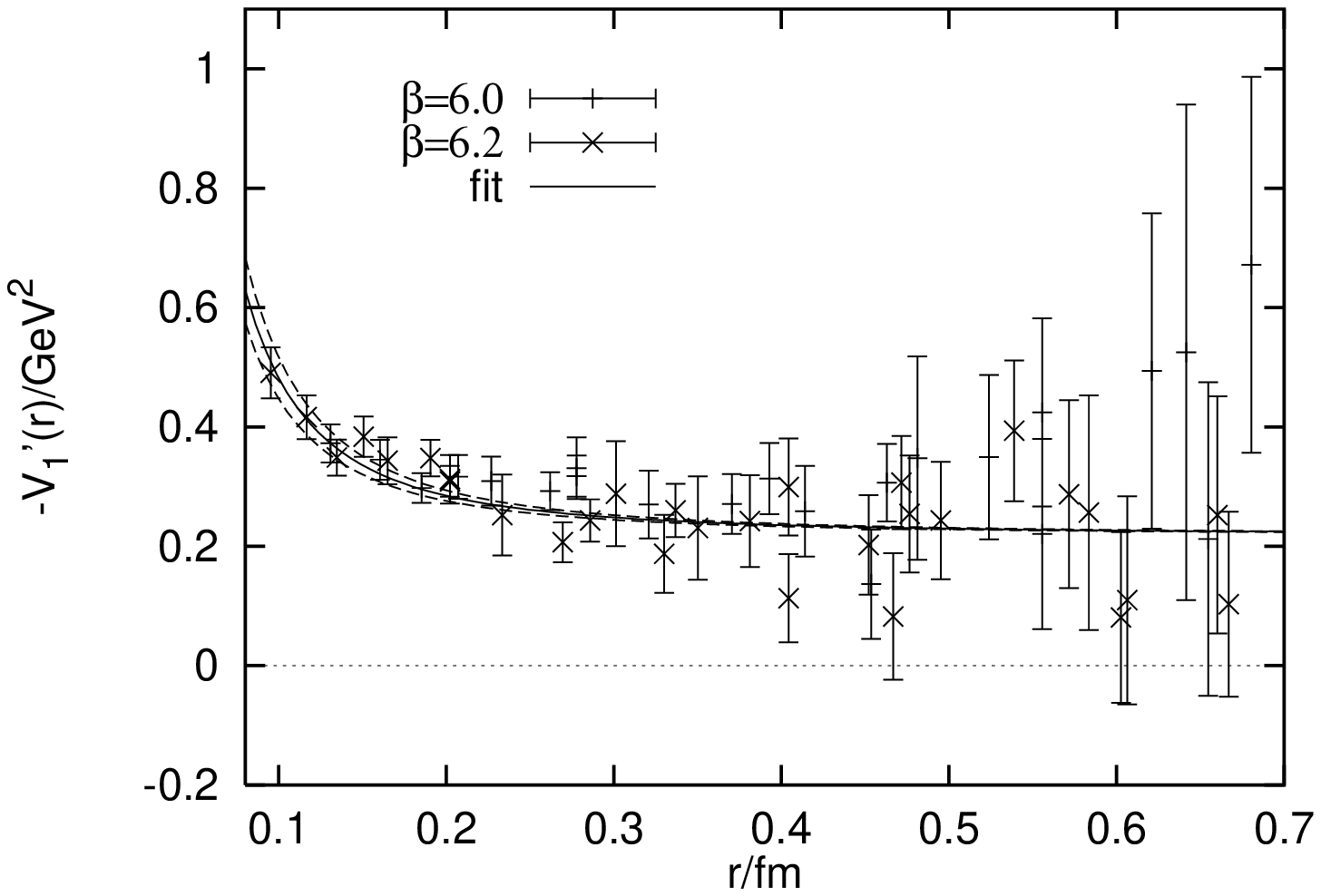}}
\put(0,120){\it a)}
\put(250,0){\epsfxsize=6.5truecm\epsffile{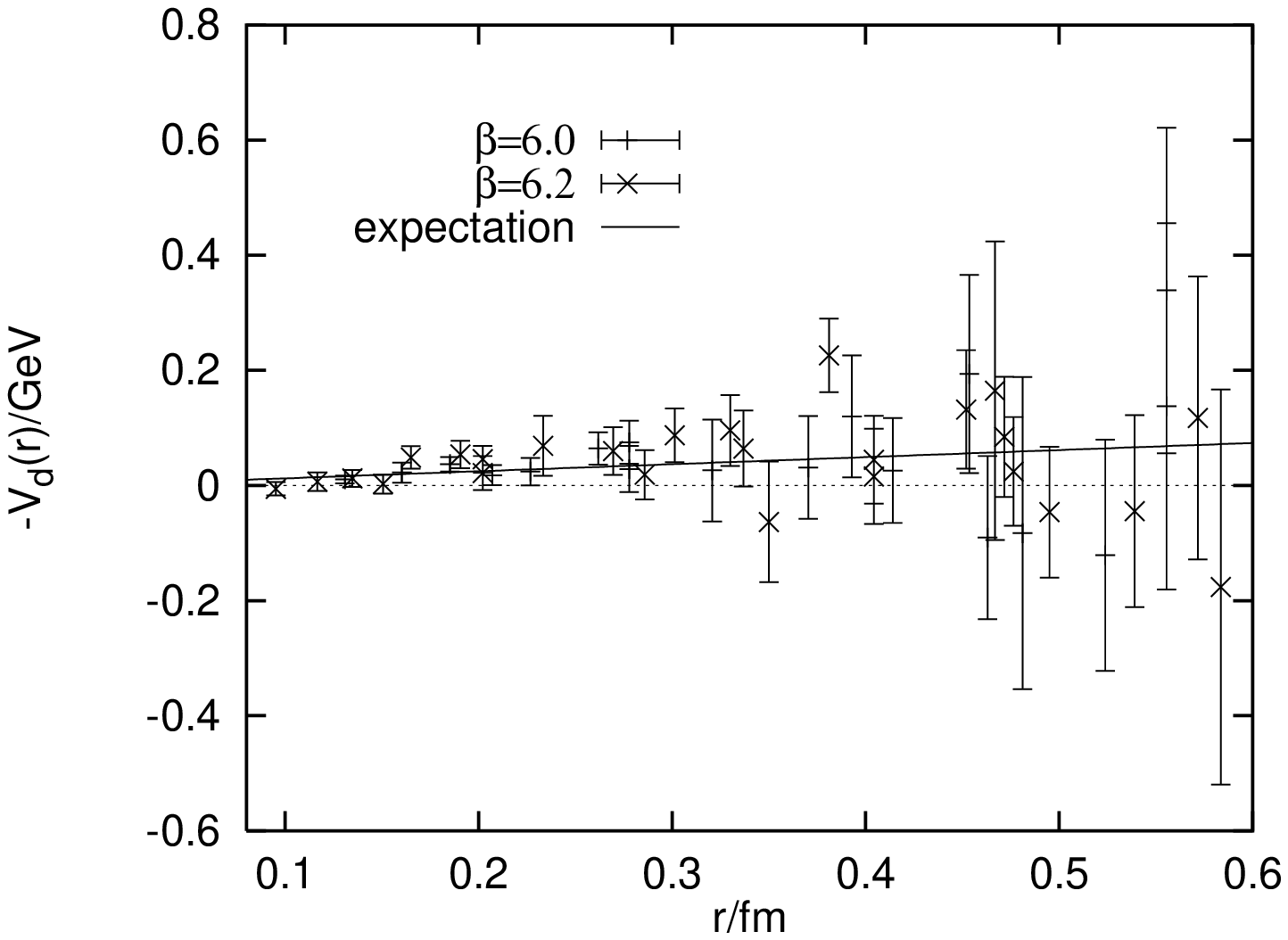}}
\put(250,120){\it b)}
\caption{\it a) The spin-orbit potential $-V_1'$ with the fit 
  $\sigma+h/r^2$ and b) the potential $V_d$ together with the curve
  $-\sigma/9\,r$.  The lattice simulations are quenched. The fitting
  parameters are $\sigma \approx (468 ~\hbox{MeV})^2$ and $h \approx
  0.067$. From \protect\cite{Bali:1997am}.
}
\label{figv1d}
\end{figure}

For what concerns the potential at order $1/m$, given in Eqs.~(\ref{Em12strong}) 
and (\ref{cc1}), no lattice evaluation is available yet. The spin-dependent 
$1/m^2$ potentials instead  have  a quite long record of calculations
\cite{Campostrini:1985uj,Michael:1986rh,Huntley:1987de,Born:1994cp,Bali:1997am}.
In the absence of a proper implementation of the NRQCD matching coefficients, 
the traditional way used to obtain lattice spacing and scale independent results 
for the spin-dependent Wilson loop potentials was proposed by
\cite{Huntley:1987de} and is based on the substitutions $\lla FF \rra
\rightarrow {\lla FF\rra / \langle FF \rangle}$, being $F$ the gluon field strength. 
The notations used by
\cite{Bali:2000gf} for the spin-dependent and momentum-dependent Wilson loop
potentials differ from what we presented in sec.~\ref{secrealpnrqcdpot}.  The
objects that were evaluated on the lattice are $V_1^\prime(r)$ (equal to $-r$
times the first term on the right-hand side of Eq.~(\ref{vls20strong}) with $c_F=1$), and
$V_2^\prime(r)$ (equal to $-r$ times the right-hand side of
Eq.~(\ref{vls11strong}) with $c_F=1$), for the spin-orbit, $V_3(r)$ (equal to the first
term on the right-hand side of Eq.~(\ref{vs11strong}) with $c_F=1$) for the spin-spin and
$V_4(r)$ (equal to the right-hand side of Eq.~(\ref{vst11strong}) with $c_F=1$) for the
tensor potential. All the lattice determinations of the spin-dependent
potentials use the correct expression for the spin-orbit potential (see
comments in sec.~\ref{secrealpnrqcdpot}).  An example is shown in
Fig.~\ref{figv1d}a.  For the momentum-dependent part, the objects evaluated on
the lattice are  $V_b = -2/3\; V_{{\bf L}^2}^{(1,1)} - V_{{\bf p}^2}^{(1,1)}$,
$V_c = -V_{{\bf L}^2}^{(1,1)}$, $V_d = V_{{\bf p}^2}^{(2,0)} + 2/3\; V_{{\bf
L}^2}^{(2,0)}$ and $V_e=V_{{\bf L}^2}^{(2,0)}$.  An example is shown in
Fig.~\ref{figv1d}b.  The spin-independent and momentum-independent potentials
at order $1/m^2$ have not yet been calculated.

The Poincar\'e invariance constraints (\ref{gromes})-(\ref{BBP2}) (which in
the above notation read $V_2'-V_1' = V_0'$, $V_b+2V_d = r V^{(0)\,\prime}/6 -
V^{(0)}/2$ and $V_c + 2V_e = -rV^{(0)\,\prime}/2$) have been used to test the
quality and the continuum limit of the lattice simulation in
\cite{Bali:1997am}.  The lattice data satisfy well the relations especially in
the short and medium range. For the long range, the data become noisy. We refer to
the original literature for more details.

More lattice plots may be found in \cite{Bali:1997am,Bali:2000gf}. In general  the
lattice curves appear to be  quite noisy for large interquark
separations. This calls for new determinations in a fully consistent lattice
renormalization context.  The lattice data have been compared with fits
motivated in the short range by the perturbative behaviour and in the long
range by QCD vacuum model calculations. We will briefly mention some of them
in the next subsection.

\subsubsection{QCD vacuum models}
The EFT has allowed us to systematically encode the low energy
contributions to the potentials into Wilson-loop amplitudes. These are very
convenient objects also for evaluation in a QCD vacuum model.  A QCD
vacuum model may be defined by the behaviour that it attributes to (not
necessarily static) Wilson-loop expectation values in the large-distance
region.  Once this is known, it is possible to obtain all Wilson-loop
amplitudes with field-strength insertions by means of functional derivatives
of the Wilson loop \cite{Migdal:1984gj,Brambilla:1993zw}.  In this way the
non-perturbative form of all the potentials is derived from only one
assumption on the Wilson loop behaviour. The lattice data on the potentials
can be compared with the expectations  from different QCD vacuum models.
We note  that a lot more knowledge may be gained here about the
mechanism of confinement. Indeed, while all the models predict confinement and
thus a linear increase of the static potential, the predictions for the
relativistic corrections to the static potential vary and may give
non-trivial information. We refer to \cite{Brambilla:1996aq} for calculations
within the Stochastic Vacuum Model \cite{Dosch:1988ha}, to \cite{Baker:1996mk, Baker:1996mu} 
for calculations inside Dual QCD (dual superconductor
mechanism of confinement) \cite{Baker:1991bc}, to \cite{Baker:1998jw} for a
comparison between the two, and to \cite{Brambilla:1998bt} for a comparison
also with the flux-tube model \cite{Isgur:1984bm} and the Bethe--Salpeter NR
reduction of a scalar confining kernel.  In \cite{Brambilla:1999ja} and
\cite{Brambilla:2000am}, one may find reviews of several QCD vacuum models and
results relevant for the non-perturbative behaviour of the potentials.

\subsection{Inclusive decay widths into light particles}
\label{secinclusivepnrqcd}
The inclusive decay width of a heavy quarkonium H into light particles reads (at leading
order in ${\rm Im} \, h_s$)
\be
\Gamma({\rm H} \to \hbox{ light particles})= - 2\,  \langle n,l,s,j|{\rm Im} \, h_s  |n,l,s,j \rangle.
\label{imagstrong}
\ee
The imaginary part of the pNRQCD Hamiltonian has been written in 
Eqs.~(\ref{imh2pert}),  (\ref{imh4}) and (\ref{Imnonanal}). 
The wavefunctions  $\phi_{njls}({\bf r}) = \langle {\bf
  r} | n,l,s,j \rangle$ have been discussed in sec.~\ref{sec:latmod}.
For the present purposes, a LO calculation is sufficient 
for $P$-wave wavefunctions, while a NLO analysis, which 
involves the $1/m^2$ potentials, is necessary for $S$-wave ones.

With the above specifications and from Eq.~(\ref{imagstrong}), we can now 
list the pNRQCD expressions for $S$- and $P$-wave decays.
We will proceed as follows. First, we will give the expressions 
for the matrix elements of NRQCD that appear in Eqs.~(\ref{gammaV})-(\ref{gchiem}).
We will distinguish between terms that are analytic in $1/m$ and terms that are not
($\langle {\rm H}| O | {\rm H} \rangle =  
\langle {\rm H}| O | {\rm H} \rangle^{1/m} + \langle {\rm H}| O | {\rm H} \rangle^{1/\sqrt{m}}$), 
since, as we have seen in the previous sections, they have been calculated in pNRQCD to 
different precision. Finally, we will explicitly give the decay widths in pNRQCD at
 the precision at which they are presently known.

The analytic contributions in $1/m$ to the NRQCD matrix elements have 
been calculated up to (once normalized to $m^0$) ${\cal O}(p^3/m^3 \times (\lQ^2/m^2, E/m))$ for 
$S$-wave \cite{Brambilla:2002nu} and up to ${\cal O}(p^5/m^5)$ 
for $P$-wave matrix elements \cite{Brambilla:2001xy}:
\bea
\label{O13S1}
&&\hspace{-8mm}
\langle V_Q(nS)|O_1(^3S_1)|V_Q(nS)\rangle^{1/m}=
C_A {|R^V_{n0}({0})|^2 \over 2\pi}
\left(1-{E_{n0}^{(0)} \over m}{2{\cal E}_3 \over 9}
+{2{\cal E}^{(2,t)}_3 \over 3 m^2 }+{c_F^2{\cal B}_1 \over 3 m^2 }\right),
\\
&&\hspace{-8mm}
\langle P_Q(nS)|O_1(^1S_0)|P_Q(nS)\rangle^{1/m}=
C_A {|R^P_{n0}({0})|^2 \over 2\pi}
\left(1-{E_{n0}^{(0)} \over m}{2{\cal E}_3 \over 9}
+{2{\cal E}^{(2,t)}_3 \over 3 m^2}+{c_F^2{\cal B}_1 \over m^2}\right),
\\
&&\hspace{-8mm}
\langle V_Q(nS)|O_{\rm EM}(^3S_1)|V_Q(nS)\rangle^{1/m}=
C_A {|R^V_{n0}({0})|^2 \over 2\pi}
\left(1-{E_{n0}^{(0)} \over m}{2{\cal E}_3 \over 9}
+{2{\cal E}^{(2,{\rm EM})}_3 \over 3 m^2}+{c_F^2{\cal B}_1 \over 3 m^2}\right),
\\
\label{OEM1S0}
&&\hspace{-8mm}
\langle P_Q(nS)|O_{\rm EM}(^1S_0)|P_Q(nS)\rangle^{1/m}=
C_A {|R^P_{n0}({0})|^2 \over 2\pi}
\left(1-{E_{n0}^{(0)} \over m}{2{\cal E}_3 \over 9}
+{2{\cal E}^{(2,{\rm EM})}_3 \over 3 m^2}+{c_F^2{\cal B}_1 \over m^2}\right),
\\
&&\hspace{-8mm}
\langle \chi_Q(nJS) | O_1(^{2S+1}P_J ) | \chi_Q(nJS) \rangle^{1/m} = 
\langle \chi_Q(nJS) | O_{\rm EM}(^{2S+1}P_J ) | \chi_Q(nJS) \rangle^{1/m}  
\nn\\
&&\qquad\qquad\qquad\qquad\qquad\qquad\quad
={3 \over 2}{C_A \over \pi} |R^{(0)\,\prime}_{n1}({0})|^2,
\label{chio1}
\\
&&\hspace{-8mm}
\langle V_Q(nS)|{\cal P}_1(^3S_1)|V_Q(nS)\rangle^{1/m}=
\langle P_Q(nS)|{\cal P}_1(^1S_0)|P_Q(nS)\rangle^{1/m}=
\nn\\
&&\hspace{-8mm}
\langle V_Q(nS)|{\cal P}_{\rm EM}(^3S_1)|V_Q(nS)\rangle^{1/m}=
\langle P_Q(nS)|{\cal P}_{\rm EM}(^1S_0)|P_Q(nS)\rangle^{1/m}
\nn\\
&&\qquad\qquad\qquad\qquad\qquad\qquad\quad
=C_A {|R^{(0)}_{n0}({0})|^2 \over 2\pi}
\left(m E_{n0}^{(0)} -{\cal E}_1 \right),
\label{P13S1}
\\
&&\hspace{-8mm}
\langle V_Q(nS)|O_8(^3S_1)|V_Q(nS)\rangle^{1/m}=
\langle P_Q(nS)|O_8(^1S_0)|P_Q(nS)\rangle^{1/m}
\nn\\
&&\qquad\qquad\qquad\qquad\qquad\qquad\quad
=C_A {|R^{(0)}_{n0}({0})|^2 \over 2\pi}
\left(- {2 (C_A/2-C_F) {\cal E}^{(2)}_3 \over 3 m^2 }\right),
\\
&&\hspace{-8mm}
\langle V_Q(nS)|O_8(^1S_0)|V_Q(nS)\rangle^{1/m}=
{\langle P_Q(nS)|O_8(^3S_1)|P_Q(nS)\rangle^{1/m} \over 3}
\nn\\
&&\qquad\qquad\qquad\qquad\qquad\qquad\quad
=C_A {|R^{(0)}_{n0}({0})|^2 \over 2\pi}
\left(-{(C_A/2-C_F) c_F^2{\cal B}_1 \over 3 m^2 }\right),
\\
&&\hspace{-8mm}
{\langle V_Q(nS)|O_8(^3P_J)|V_Q(nS)\rangle^{1/m} \over 2J+1}=
{\langle P_Q(nS)|O_8(^1P_1)|P_Q(nS)\rangle^{1/m} \over 9}
\nn\\
&&\qquad\qquad\qquad\qquad\qquad\qquad\quad
=
\,C_A {|R^{(0)}_{n0}({0})|^2 \over 2\pi}
\left(-{(C_A/2-C_F) {\cal E}_1 \over 9 }\right),
\\
&&\hspace{-8mm}
\langle \chi_Q(nJS)\vert O_8(^1S_0)\vert \chi_Q(nJS) \rangle^{1/m}
= {T_F\over 3}
{\vert R^{(0)\,\prime}_{n1}({0})\vert^2 \over \pi m^2} {\cal E}_3,
\label{matoct}
\eea
where the radial part of the vector $S$-wavefunction is 
$R_{n101}\equiv R^V_{n0}$ and the radial part of the pseudoscalar $S$-wavefunction is 
$R_{n000} \equiv R^P_{n0}$. The quantity $R^{(0)\,\prime}_{n1}$ is 
the derivative of the radial part of the LO $P$-wave wavefunction. 
Any other dimension-6 and dimension-8 $S$-wave matrix elements are  0 at the order considered here.

The non-analytic contributions in $1/m$ to the NRQCD matrix elements have 
been calculated up to (once normalized to $m^0$) 
${\cal O}(p^3/m^3 \times \lQ/m \times m\als/\sqrt{m\,\lQ})$ for 
$S$-wave matrix elements \cite{Brambilla:2003mu}:
\bea
\langle V_Q(nS)|O_1(^3S_1)|V_Q(nS)\rangle^{1/\sqrt{m}}&=&
\langle V_Q(nS)|O_{\rm EM}(^3S_1)|V_Q(nS)\rangle^{1/\sqrt{m}}
\nn\\
&=& C_A {|R^V_{n0}({0})|^2 \over 2\pi}
\left(1+  {4 (2\,C_F+C_A)\over 3\Gamma(7/2)} \, \; {\als\,
 {\cal E}^E_{5/2}\over m^{1/2}}\right),
\label{O3S1nonan}
\eea
\bea
\langle P_Q(nS)|O_1(^1S_0)|P_Q(nS)\rangle^{1/\sqrt{m}}&=&
\langle P_Q(nS)|O_{\rm EM}(^1S_0)|P_Q(nS)\rangle^{1/\sqrt{m}}
\nn\\
&=& C_A {|R^P_{n0}({0})|^2 \over 2\pi}
\left(1+  {4 (2\,C_F+C_A)\over 3\Gamma(7/2)} \, \; {\als\,
 {\cal E}^E_{5/2}\over m^{1/2}}\right).
\label{O1S0nonan}
\eea
All other matrix elements receive contributions which are 
${\cal O}(m\als/\sqrt{m \, \lQ})$ suppressed, under the condition 
(\ref{count}), with respect to those listed in Eqs.~(\ref{O13S1})-(\ref{matoct}). 

Some comments are in order. All matrix elements are factorized into a part that
is the wavefunction at the origin and a combination of gluon-field
correlators.  The wavefunction carries all the dependence on the state and
flavor content of the decaying heavy quarkonium (apart from the residual
dependence on $m$ and $n$ in the binding energy and on $m$ in the logarithms in
$c_F$) while the correlators only depend on the low-energy properties of QCD
and are in this sense universal. They may be calculated once and forever,
either by means of lattice simulations
\cite{Bali:2003jq,Foster:1998wu,Bali:1998aj,D'Elia:1997ne} or specific models
of the QCD vacuum \cite{DiGiacomo:2000va,Brambilla:2000ch,Baker:1998jw} or
extracted from experimental data \cite{Brambilla:2001xy} (see also
sec.~\ref{secphendecstr}). We emphasize that the factorization holds only if
$\lQ \gg E$, otherwise it would not be possible to disentangle the heavy
quarkonium, whose energy is $E$, from the non-perturbative gluons.

The factorization is also the reason for
 the reduction in the number of
non-perturbative parameters in going from NRQCD to pNRQCD. In pNRQCD these are
the wavefunctions and the gluon-field correlators. Among these only the wavefunctions 
depend on the specific heavy quarkonium state that we are
considering. As discussed at the beginning of the section, the wavefunction
may be calculated, in principle, in terms of QCD quantities by solving the
Schr\"odinger equation (\ref{coordSchr}). At the order at which they are
given, Eqs.~(\ref{O13S1})-(\ref{OEM1S0}) are sensitive to the difference
between the pseudoscalar and the vector $S$-wave wavefunction. For the other
$S$-wave operators, the difference is not important at the present level of
accuracy. The reduction in the number of parameters is more evident if we
consider ratios of matrix elements of hadronic operators and electromagnetic
ones. The wavefunction dependence drops out and we are left with a
combination of a few universal gluon-field correlators.  In
sec.~\ref{secphendecstr}, we will discuss the phenomenological relevance of
this for the calculation of bottomonium and charmonium inclusive decay widths.

Finally, we would like to recall that, apart from the matrix elements $\langle
V_Q(nS)|O_1(^3S_1)|V_Q(nS)\rangle$ and $\langle
P_Q(nS)|O_1(^1S_0)|P_Q(nS)\rangle$ that are affected at relative order $m/\lQ
\times m \als/\sqrt{m \, \lQ}$, all other matrix elements listed above receive
non-analytic contributions from the three-momentum scale $\sqrt{m\,\lQ}$ at
relative order $m\als/\sqrt{m \, \lQ}$ with respect to the leading piece.  It
may turn out that these contributions are numerically important, since the
suppression factor $m\als/\sqrt{m \, \lQ}$ may not be that small.  In this
case it would be important to have the leading non-analytic contributions
for all matrix elements. As long as this is not the case, non-analytic
contributions give the dominant source of uncertainty for the factorization
formulas (\ref{chio1})-(\ref{matoct}).

We conclude this section by giving the explicit formulas in pNRQCD for the electromagnetic 
and inclusive decay widths of heavy quarkonium into light particles at the present level 
of knwoledge. This means that $S$-wave decay widths are given up to and including  
${\cal O}({\rm Im}\,f \times p^3/m^2 \times \lQ/m \times m\als/\sqrt{m \, \lQ})$  and 
$P$-wave decay widths up to and including ${\cal O}({\rm Im}\,f \times p^5/m^4)$:
\bea
\Gamma(V_Q (nS) \rightarrow LH) &=& {C_A \over \pi}{|R^V_{n0}({0})|^2 
\over m^2}
\; {\rm Im\,}f_1(^3 S_1) \; 
\left(1+{4 \,(2\,C_F+C_A)\over 3\,\Gamma(7/2)} \, \; {\als\,
 {\cal E}^E_{5/2}\over m^{1/2}}\right), 
\label{hadrV}
\\
\Gamma(P_Q (nS) \rightarrow LH) &=& {C_A \over \pi}{|R^P_{n0}({0})|^2 
\over m^2}
\; {\rm Im\,}f_1(^1 S_0) \;
\left(1+{4 \, (2 \,C_F+C_A)\over 3 \,\Gamma(7/2)} \, \; {\als\,
 {\cal E}^E_{5/2}\over m^{1/2}}\right),
\\
\Gamma(\chi_Q(nJS)  \rightarrow LH) &=& 
{C_A\over \pi}{| R^{(0)\,\prime}_{n1} ({0}) |^2 \over m^4}
\Bigg[ 3 \, {\rm Im\,}\,   f_1(^{2S+1}P_J) 
+ {2  \,T_F\over 3  \,C_A } {\rm Im\,} \,  f_8(^{2S+1}{\rm{S}}_S) \, {\cal 
E}_3 \Bigg], 
\label{hadrchi}
\\
\Gamma(V_Q (nS) \rightarrow e^+e^-) &=& {C_A \over \pi}{|R^V_{n0}({0})|^2 
\over m^2}
\; {\rm Im\,}f_{ee}(^3 S_1) \;
\left(1+{4  \,(2 \,C_F+C_A)\over 3 \,\Gamma(7/2)} \, \; {\als\,
 {\cal E}^E_{5/2}\over m^{1/2}}\right),
\\
\Gamma(P_Q (nS) \rightarrow \gamma\gamma) &=&
{C_A \over \pi}{|R^P_{n0}({0})|^2 \over m^2}
\; {\rm Im\,}f_{\gamma\gamma}(^1 S_0) \;
\left(1+{4  \,(2 \,C_F+C_A)\over 3 \,\Gamma(7/2)} \, \; {\als\,
 {\cal E}^E_{5/2}\over m^{1/2}}\right),
\\
\label{electrchi}
\Gamma(\chi_Q(nJ1)  \rightarrow \gamma\gamma) &=& 
3 {C_A \over \pi}{|R^{(0)\,\prime}_{n1}({0})|^2 \over m^4} \; 
{\rm Im\,} f_{\gamma\gamma}(^3P_J) 
\qquad {\rm for} \; J=0,2\,. 
\eea

\section{Phenomenological applications}
\label{sec:phen}

\subsection{Determinations of $m_b$ and $m_c$ from the 1S resonances}
\label{Groundstatemass}
In this section we would like to give the state-of-the-art
determinations of the bottom and charm masses from 
the ground-state bottomonium and charmonium masses.

For precise determinations of those parameters, we need to be in a situation 
where the dynamics can be described by a weak-coupling analysis (at least in a first
approximation) and where  non-perturbative effects are
small. Therefore, the first question we should answer is 
if we are in such a dynamical situation.
For the bottomonium and charmonium systems, we believe that the masses
$m_b$ and $m_c$ are much larger than $\lQ$. 
This is not enough, however, since  we also need $mv \gg \lQ$. 
If this is the case then we are dealing (in a first approximation)
with a Coulomb-type bound state. In this situation we can apply the results 
of \ref{pNRweakObservables} once the renormalon cancellation
along the lines of ch.~\ref{secrenormalons} has been used. 
In other words, our starting point will be Eq.~(\ref{MnljRS}). 
Let us see whether the assumption $mv \gg \lQ$ 
is reasonable for bottomonium and charmonium ground states. The momentum transfer in
the first case is around $\siml 2$ GeV whereas in the second case it  is around $\siml 1$ GeV. 
The momentum transfer between the heavy quark and antiquark lies in the 
deep Euclidean domain. Therefore, 
the computation does not rely on local duality (at
least to low orders in perturbation theory). 
The assumption  $mv \gg \lQ$ then becomes equivalent to believing in 
perturbative calculations in the Euclidean domain in 
the above range of energies. We will report on work in which this
attitude is taken for the bottomonium ground state as well as on
work where this attitude is also taken for the charmonium ground state.
The relative size between the US scale and $\lQ$ remains to be fixed.  

\begin{table}[ht]
\addtolength{\arraycolsep}{0.2cm}
\begin{tabular}{|c|c|c|}
\hline
reference & order & ${\overline m}_b({\overline m}_b)$ (GeV)
\\
\hline
\cite{Beneke:1999fe} & NNLO&
$4.24 \pm 0.09$ 
\\
\cite{Hoang:1998uv}  & NNLO&
$4.21 \pm 0.09$
\\
\cite{Pineda:2001zq}  & NNLO &
$4.210 \pm 0.090 \pm 0.025$
\\
\cite{Brambilla:2001qk}  & NNLO&
$4.190 \pm 0.020 \pm 0.025$
\\
\cite{Penin:2002zv}  & NNNLO &
$4.349 \pm 0.070$
\\
\cite{Lee:2003hh} & NNNLO &
$4.20 \pm 0.04$
\\
\cite{Contreras:2003zb} & NNNLO &
$4.241 \pm 0.070$ 
\\
\hline
\hline
reference & order & ${\overline m}_c({\overline m}_c)$ (GeV)
\\
\hline
\cite{Brambilla:2001fw}  & NNLO &
$1.24 \pm 0.020$ \\
\hline
\end{tabular}
\caption{\it \label{tableMSmasses} Recent determinations of 
 ${\overline m}_b$ and 
${\overline m}_c$ in the $\MS$ scheme from the
$\Upsilon$(1S) and $J/\psi$(1S) masses. }
\end{table} 

Let us now consider recent determinations available in the literature, which
we quote in Table \ref{tableMSmasses}. In the first three references, as well as in
\cite{Penin:2002zv}, no finite charm mass effects due to the potential and
self-energy, calculated in
\cite{Gray:1990yh,Eiras:2000rh,Hoang:2000fm,Melles:2000dq}, were included. In
all the references, except in \cite{Beneke:1999fe} (at the moment of that
computation the conversion from the pole to the $\MS$ masses was not known
with the required accuracy), the conversion from the threshold (or pole)
masses to the $\MS$ has been performed to three loops.  The NNNLO analyses
should be understood as only almost complete, since the three loop static potential
coefficient was only estimated.  In \cite{Beneke:1999fe}, a NNLO analysis was
done in the PS scheme.  In \cite{Hoang:1998uv}, a NNLO analysis was done in the
1S scheme. In \cite{Pineda:2001zq}, a NNLO analysis was done in the RS scheme
as well as an analysis at NNNLO including the logarithms at this order and
the large $\beta_0$ result. In \cite{Brambilla:2001qk}, a NNLO analysis was
done in the $\MS$ scheme using the upsilon expansion.  In \cite{Penin:2002zv},
a NNNLO analysis was done in the on-shell scheme. We believe that the
difference with respect to  the other results is due to the presence of the renormalon, as
well as the way US and non-perturbative effects were implemented, since the
authors assume $mv^2\gg \lQ$.  In \cite{Lee:2003hh}, a NNNLO analysis was done
in a scheme similar to the RS one. In this reference, the US contribution was
included within perturbation theory. Also in \cite{Contreras:2003zb}, a NNNLO
analysis was done in a scheme similar to the RS one. In this case, the US
contribution was also treated perturbatively but in a different way from the
soft one. It would be extremely interesting to repeat these analyses without
the US contribution. Actually, in \cite{Contreras:2003zb}, it is easy to
separate the US contribution (although it is not fully clear in which
scheme). If one eliminates the US contribution in this case, the bottom mass
goes down by around 50 MeV leading to  good agreement with previous
analyses. Nevertheless, it remains to be seen what would happen in
\cite{Lee:2003hh} if a similar approach were applied.

We would also like to mention the determination of the charm mass from the
$J/\psi$(1S) mass \cite{Brambilla:2001fw}. The authors perform a complete NNLO
analysis in the 1S scheme. It would be very interesting to perform a similar
analysis with a different threshold mass, as well as to do the NNNLO
analysis, in order to see whether the result remains stable.

In the above analyses, with the exception of \cite{Penin:2002zv}, the
non-perturbative effects have been left unevaluated.  In some cases the
non-perturbative results obtained in the limit $mv^2 \gg \lQ$ has been used to
estimate their size.

The main sources of errors and possible improvements are the following. 
None of the above analyses has yet incorporated the resummation of logarithms
available at NNLL. It would be interesting to see its effect on the mass of
the heavy quarkonium.
So far all the (almost complete) NNNLO evaluations have been done assuming
that the US contribution can be computed within perturbation theory. It would
be most interesting to perform the NNNLO analysis without the US piece.
Two of the (potentially) major sources of errors in this kind of evaluations
of the heavy quarkonium mass are the non-perturbative contribution
(\ref{energyUS}) and possible effects due to subleading renormalons (see the
discussion in \cite{Pineda:2001zq}).  Any reliable determination of
Eq.~(\ref{energyUS}) will have an immediate impact on our understanding of the
theoretical errors. On the one hand, it would put on more solid ground our
implicit assumption that the LO solution corresponds to a
Coulomb-type bound state. Once this is achieved, it would bring the error
estimates of the non-perturbative effects from a qualitative level to a
quantitative one, (hopefully) decreasing their size significantly. On the other
hand, one may think of cross-checking these results with other
determinations. The fact that the difference happens to be relatively small
supports the belief that (perturbative and non-perturbative) higher-order
effects are indeed not very large.  In order to have an independent handle on
the size of the non-perturbative corrections, one may consider the difference
between the lattice simulation of the static potential and the perturbative
prediction \cite{Pineda:2002se}.  If one neglects possible effects of
unquenching, one gets (in the static limit) non-perturbative contributions,
which are not larger than $\sim 100$ MeV. A precise determination would 
require an accurate determination of the chromoelectric correlator which
appears in Eq.~(\ref{energyUS}) from  the lattice.  In this respect we note that,
by using the data on the gluelump masses reported in Table
\ref{tablegluelumps}, one obtains $\Lambda_E \simeq 1.25$ GeV, which is much
larger than the US scale. This may indicate that the actual situation, even
for the ground-state bottomonium, is $mv \gg \lQ \gg mv^2$, at least as far as
the computation of Eq.~(\ref{energyUS}) is concerned. Then the  results of
sec.~\ref{sectwostep} would  apply.

\subsection{Spectroscopy in the weak-coupling regime}
\label{spectroscopy}
Along the lines of the previous section, once it is assumed that the
$\Upsilon(1S)$ can be described by the weak-coupling version of pNRQCD, it
should be possible to give a prediction for the $\eta_b(1S)$ mass. If this
belief is extended to the $J/\psi(1S)$, it should also be possible to give
predictions for the $B_c(1S)$\footnote{Although its mass has been measured to be
$6.40\pm 0.39\pm 0.13$ GeV \cite{Abe:1998wi}, 
the precision is not good enough to really test the theory.} 
and $B_c(1S)^*$ masses, as well as to
check the theory by comparing with the experimental value of the $\eta_c(1S)$
mass.

Working in the 1S scheme at NNLO \cite{Brambilla:2000db} obtained a 
prediction for the $B_c(1S)$ mass. It reads
\be
\label{Bcmass}
M(B_c)  = 6326^{+29}_{-9} \, {\rm MeV},
\ee
where the error accounts only for higher-order perturbative corrections and 
uncertainties in $\als(M_Z)$. The error due to non-perturbative contributions 
has been estimated to be in the range of $ (40$ - $100) \, {\rm MeV}$.
It is argued there that the non-perturbative contributions to the $B_c$ mass 
in the $1S$-mass scheme come out as the following combination 
of non-perturbative contributions in the pole-mass scheme:
$ - \delta E(J/\psi)^{\rm np}/2$ 
$- \delta E(\Upsilon(1S))^{\rm np}/2$  $+ \delta E(B_c)^{\rm np}$. 
Therefore, cancellations 
may occur if all three corrections are of the same type and size. This may
substantially reduce the total size of the 
non-perturbative corrections to the $B_c$ in the $1S$-mass scheme.
In \cite{Brambilla:2001fw} a similar determination has been done, using the 
${\overline {\rm MS}}$ $c$ and $b$ masses . The result is very similar: $M(B_c) = 6324 \pm 23 \, {\rm MeV}$.
Again the error only accounts for higher-order perturbative corrections and 
uncertainties in $\als(M_Z)$. The error due to non-perturbative
contributions  has not been estimated there.
In \cite{Brambilla:2001qk} charm-mass effects were also included in the analysis. 
They lower a little bit the central value:  $M(B_c) = 6309 \pm 17 \, {\rm MeV}$.
The error is as above. 

In the case of bottomonium, \cite{Kniehl:2003ap} calculated the hyperfine splitting of 
the ground state at NLL in the on-shell scheme (the effects 
due to the pole mass renormalon are subleading). 
For this observable, the resummation of the logarithms along the lines 
discussed in sec.~\ref{RGreview} seems important. 
The authors have given a rather precise prediction for the mass 
of the $\eta_b(1S)$, which uses the experimental value of $M_{\Upsilon(1S)}$,
\begin{equation}
\label{etabmass}
M(\eta_b(1S))=9421\pm 11\,{(\rm th)} \,{}^{+9}_{-8}\, 
(\delta\als)~{\rm MeV}\,,
\end{equation}
where the errors due to the higher-order perturbative corrections and the
non-perturbative effects are added up in quadrature in ``th'', whereas
``$\delta\als$'' stands for the uncertainty in
$\als(M_Z)=0.118\pm0.003$. They also obtained a value for the charmonium 
ground state hyperfine splitting, $M(J/\psi(1S)-M(\eta_c(1S) \simeq 104$ MeV, 
to be compared with 
the experimental figure $117.7$ MeV. \cite{Recksiegel:2003fm} have performed 
a numerical NLO analysis of these hyperfine splittings. For bottomonium 
they get $\simeq 44$ MeV, which 
compares well with the above number, and for charmonium $\simeq 88$ MeV, 
which is somewhat lower.

\cite{Penin:2004xi} have also calculated the hyperfine splitting of the $B_c$  ground state 
at NLL in a way similar to \cite{Kniehl:2003ap}. They obtain 
\be
M(B^*_c)-M(B_c)=65 \pm
24\,{(\rm th)}\,{}^{+19}_{-16}\,(\delta\als)~{\rm MeV}
\;,
\ee
where the errors read as in Eq.~(\ref{etabmass}).
This result, combined with Eq.~(\ref{Bcmass}), or, eventually, with a more 
accurate experimental determination of the $B_c$ mass, provides 
a prediction for the $B_c^*$ mass.

\cite{Brambilla:2001fw,Brambilla:2001qk} have considered higher excitations of
the bottomonium system at NNLO in the $\MS$ mass scheme and using the
upsilon expansion (the latter reference has also included finite charm mass 
effects).  It is not obvious a priori that these can be described under the
kinematical assumption $mv \gg \lQ$, however, it is worth investigating this
possibility.  The results for the levels that turn out to be stable in this
analysis is shown in table \protect\ref{table:spectraBSV1}.  We note that at
least a part of the higher bottomonium levels seems to be reasonably well
described in perturbation theory. In particular, the equal level spacing,
characteristic for the quarkonium spectrum, is reasonably well reproduced without
making use of a confining potential. This behaviour seems to originate from self-energy
contribution remnants of the renormalon cancellation and may reflect, from the
point of view of the spectrum, the numerical agreement mentioned in
ch.~\ref{sec:static} that is found in some situations between the perturbative
static potential and the lattice data up to very large distances. Indeed, it
was this phenomenological analysis that triggered part of the subsequent
analysis of the static potential measured on the lattice in terms of
perturbative QCD.  Moreover, since for higher levels the experimental data
agree with the theoretical results within the uncertainties, we may expect
this to be the case also for the bottomonium ground state, suggesting very
small non-perturbative corrections to it.  Along similar lines, but within a
numerical analysis, fine splittings of bottomonium and charmonium levels have
been considered at NLO in \cite{Recksiegel:2003fm}.

\begin{table}
\begin{center}
\begin{tabular}{|c|c|c|}
\hline
$\Upsilon$ &$M(\Upsilon)^{\rm exp}$(MeV) &$M(\Upsilon)$(MeV)\\ 
\hline
$\Upsilon(1 ^3P_0)$ &  9860~~ & $9995(83)$ \\
$\Upsilon(1 ^3P_1)$ &  9893~~ & $10004(86)$ \\
$\Upsilon(1 ^3P_2)$ &  9913~~ & $10012(89)$ \\
$\Upsilon(2 ^3S_1)$ & 10023~~ & $10084(102)$ \\
$\Upsilon(1 ^3P_0)$ & 10232~~ & $10548(239)$ \\
$\Upsilon(1 ^3P_1)$ & 10255~~ & $10564(247)$ \\
$\Upsilon(1 ^3P_2)$ & 10269~~ & $10578(258)$ \\
$\Upsilon(3 ^3S_1)$ & 10355~~ & $10645(298)$ \\
\hline
\end{tabular}
\end{center}
\caption{\it \label{table:spectraBSV1}
Comparison of the theoretical predictions of some of the bottomonium levels
obtained in \protect\cite{Brambilla:2001qk} with the experimental data.
The errors come from summing quadratically uncertainties in $\als$, 
higher-order corrections and finite charm mass corrections \protect\cite{Brambilla:2004wf}.
}
\end{table}

\subsection{Electromagnetic inclusive decay widths in the weak-coupling regime}
The electromagnetic inclusive decay widths are known at NNLO (see
sec.~\ref{pNRweakObservables}). 
Nevertheless, they suffer from large scale uncertainties, 
which have so far prevented their use for phenomenological analysis. 
This also affects the accuracy of sum rules (see the discussion in sec.~\ref{sec:sumrules}). 

\begin{figure}[th]
\makebox[-16cm]{\phantom b}
\put(0,0){\epsfxsize=8truecm \epsfbox{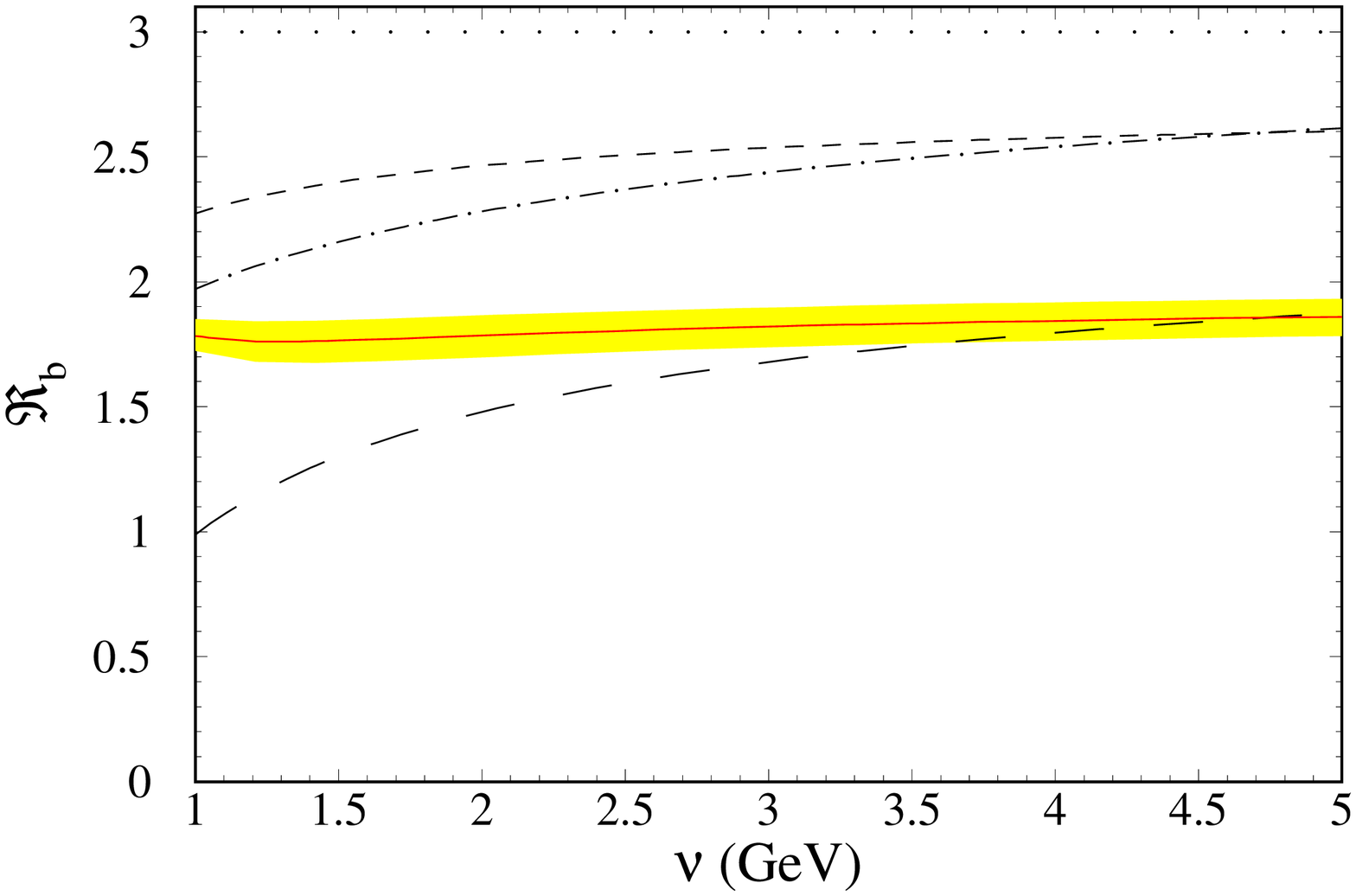}}
\put(240,0){\epsfxsize=8truecm \epsfbox{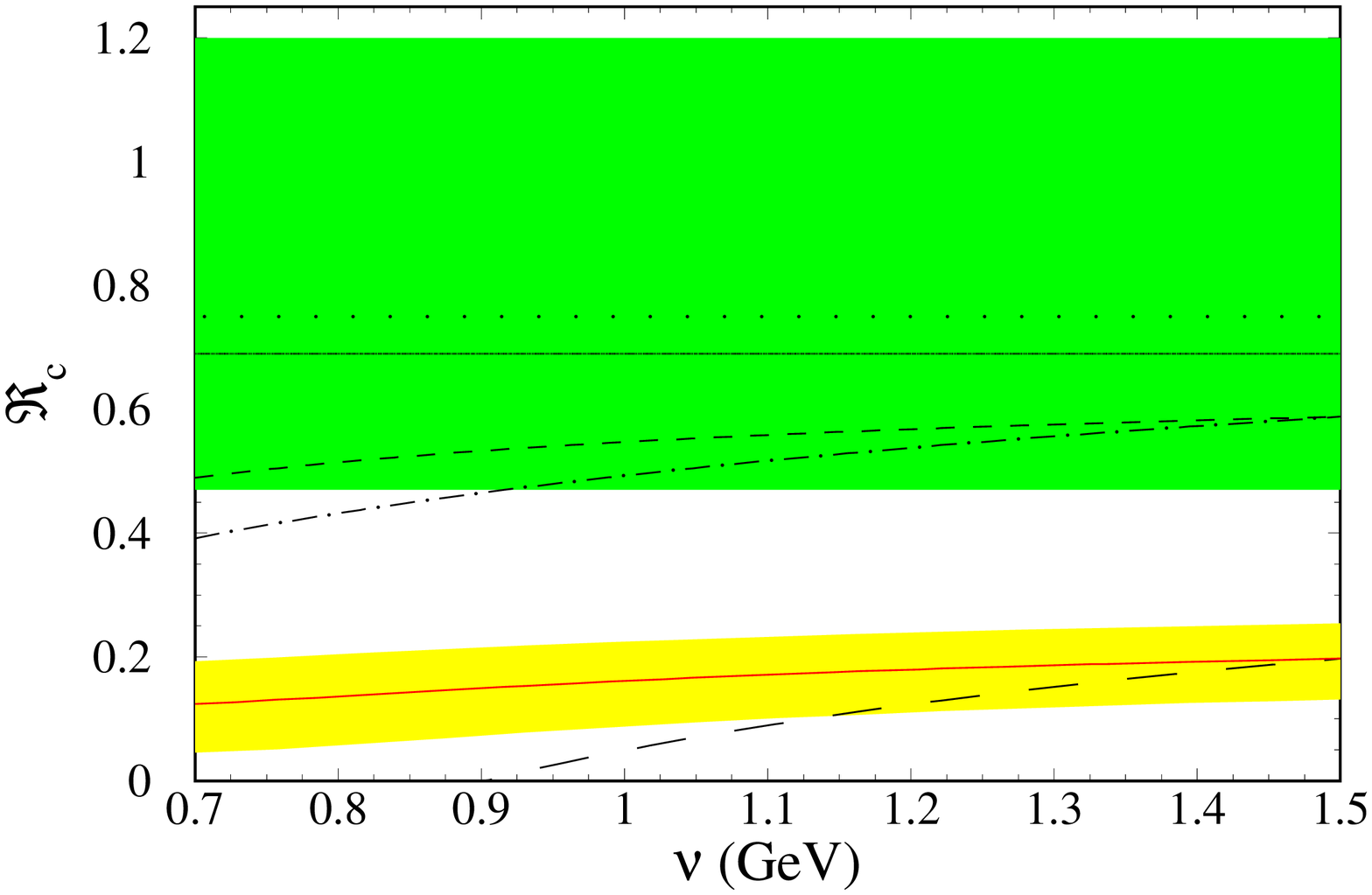}}
\put(-5,130){$(a)$}
\put(235,130){$(b)$}
\caption{\label{figbc} {\it The spin  ratio  as a function of the
renormalization scale $\nu$ in LO$\equiv$LL (dotted line), NLO (short-dashed
line), NNLO (long-dashed line), NLL (dot-dashed
line), and NNLL (solid line) approximation.
For the NNLL result the band reflects the errors 
due to $\als(M_Z)=0.118\pm 0.003$. 
Panel $(a)$ shows the bottomonium ground state case for which $\nu_h=m_b$.
Panel $(b)$ shows the charmonium ground state case for which $\nu_h=m_c$.
In the charmonium case, the upper  band
represents the experimental error of the ratio \cite{Eidelman:2004wy} where the central 
value is given by the horizontal solid line. From \cite{Penin:2004ay}.}}
\end{figure}

Recently, there have been a few phenomenological analyses including the resummation 
of logarithms (see sec.~\ref{RGreview}). The impact of these logarithms appears to be large and 
the overall convergence of the series seems to improve.
For bottomonium, \cite{Penin:2004ay} considered the complete result with 
NNLL accuracy for the ratio of the spin one and spin zero production in the on-shell 
scheme (at this order the effects due to the pole mass renormalon are subleading).
The logarithmic expansion shows nice convergence and
stability (see Fig.~\ref{figbc}a)
despite the presence of US contributions with
$\als$ evaluated at a rather low scale $\nu^2/m_b$. 
At the same time, the perturbative corrections are important
and reduce the LO result by approximately $40\%$.  
For illustration, at the
scale of minimal sensitivity, $\nu=1.295$~GeV, one has the following series:
\be
{\cal R}_b \equiv {\Gamma(\Upsilon(1S) \rightarrow
  e^+e^-)\over\Gamma(\eta_b(1S)\rightarrow\gamma\gamma)} 
={1\over 3e_b^2}\left(1-0.302-0.111\right)\,.
\ee 
In contrast, the fixed-order expansion blows up at the scale
of the inverse Bohr radius. Non-perturbative effects
contribute in the N$^4$LL approximation, that is
far beyond the precision of this computation. 
Note that the non-perturbative 
contribution to the ratio of decay rates
is suppressed by a factor $v^2$ in comparison to the
binding energy and decay rates, where the leading non-perturbative effect
is due to chromoelectric dipole interaction. Thus, by using the 
available experimental data on the
$\Upsilon$ meson as input, one can predict the production and
annihilation rates of the yet undiscovered $\eta_b$ meson. In particular, 
one can predict the $\eta_b(1S)$ decay rate using
the experimental value for the $\Upsilon(1S)$ decay rate 
\cite{Penin:2004ay}:
\be
\label{etabgg}
\Gamma(\eta_b(1S) \rightarrow \gamma\gamma)=0.659\pm 0.089 ({\rm th.}) 
{}^{+0.019}_{-0.018} (\delta\alpha_{\rm s})
\pm 0.015 ({\rm exp.})\; {\rm keV}
\,,
\ee
where $\nu=1.295$ GeV was taken as the central value, the difference between the NLL and NNLL result
for the theoretical error, and $\alpha_{\rm s}(M_Z)=0.118 \pm 0.003$. The
last error in Eq.~(\ref{etabgg}) reflects the experimental error of
$\Gamma(\Upsilon(1S) \rightarrow e^+e^-)=1.314\pm 0.029$ keV \cite{Eidelman:2004wy}.
This value considerably exceeds the result for the absolute value of the
decay width obtained by \cite{Pineda:2003be} on the basis of a full NLL
analysis including the spin-independent part:
$
\Gamma(\eta_b (1S) \rightarrow \gamma\gamma)=0.35 \pm 0.1 ({\rm th.})
\pm 0.05 (\delta\alpha_{\rm s}) \;{\rm keV}.
$ 
This can be a signal of
slow convergence of the logarithmic expansion for the spin-independent
contribution, which is more sensitive to the dynamics of the bound state 
and in particular to the US contribution, as  has been 
discussed above. On the other hand, renormalon effects \cite{Braaten:1998au,
Bodwin:1998mn} could produce some systematic errors in the purely  perturbative
evaluations of the production/annihilation rates. The problem is
expected to be more severe for the charmonium case discussed below.

We would like to point out that the one-loop result for $\nu=m_b$
overshoots the NNLL result by approximately $30\%$.  This casts some doubts
on the accuracy of the existing $\als$ determination from the
$\Gamma(\Upsilon\to{\rm light~hadrons})/\Gamma(\Upsilon\to e^+e^-)$
decay rates ratio, which gives $\als(m_b)=0.177\pm 0.01$, well below
the ``world average'' value \cite{Eidelman:2004wy}.  The theoretical uncertainty in
the analysis is estimated through the scale dependence of the one-loop
result. The analysis of the photon mediated annihilation rates
indicates that the actual magnitude of the higher-order corrections is
most likely quite far beyond such an estimate and the theoretical
uncertainty given in \cite{Eidelman:2004wy} should be increased by a factor of
two. This brings the result for $\als$ into a  $1\sigma$ distance from
the ``world average'' value.

For charmonium, the same analysis was performed in \cite{Penin:2004ay}.
The NNLO approximation becomes negative at an
intermediate scale between $\als m_c$ and $m_c$ (see Fig.~\ref{figbc}b)
and the use of the RG is mandatory in order to get a sensible perturbative approximation. The NNLL
approximation has good stability against the scale variation but the
logarithmic expansion does not converge well.  This is the main factor
that limits the theoretical accuracy, since the non-perturbative
contribution is expected to be under control. For illustration, at the
scale of minimal sensitivity, $\nu=0.645$ GeV, one obtains 
\be
{\cal R}_c \equiv {\Gamma(J/\psi(1S)\rightarrow
  e^+e^-)\over\Gamma(\eta_c(1S)\rightarrow\gamma\gamma)} 
={1\over 3e_c^2}\left(1-0.513-0.326\right)\,.
\ee 
The central value is $2\sigma$ below the experimental
one. The discrepancy may be explained by  large higher-order
contributions. This should not be surprising because of the rather large value
of $\als$ at the inverse Bohr radius of charmonium.  For the charmonium
hyperfine splitting, however, the logarithmic expansion converges well and the
prediction of the RG is in agreement with the experimental data.
One can try to improve the convergence of the series for the
production/annihilation rates by accurately taking into account the
renormalon-related contributions. One point to note is that with a potential
model evaluation of the wavefunction correction  the sign of the NNLO term is
reversed in the charmonium case \cite{Czarnecki:2001zc}. At the same time the
subtraction of the pole mass renormalon from the perturbative static potential
makes explicit that the potential is steeper and closer to lattice results  and
to phenomenological potential models as we have seen in ch.~\ref{sec:static}. 
Therefore, the incorporation of
higher-order effects from the static potential may improve the agreement with
experiment. Finally, we mention that a NLL evaluation for the 
$\eta_c(1S)\rightarrow\gamma\gamma$ decay 
reproduces in the minimal sensitivity region 
the experimental value \cite{Pineda:2003be}.

\subsection{Inclusive decay widths in the strong-coupling regime}
\label{secphendecstr}
At the end of ch.~\ref{sec:NRQCD}, we pointed out that the application
of the NRQCD factorization formulas to inclusive annihilation widths 
of quarkonium was somehow limited by the large number and
poor knowledge of the NRQCD 4-fermion matrix elements. The pNRQCD
factorization formulas presented in sec.~\ref{secinclusivepnrqcd} 
make both problems less severe by reducing the number of non-perturbative 
parameters and by factorizing the wavefunction dependence. 
As a consequence, for systems to which it may be applied,  
pNRQCD in the strong-coupling regime has more predictive power 
than NRQCD. In the following, we will present some of the predictions 
that are specific to pNRQCD. We remark that the problem of the poor 
convergence of the perturbative series for the NRQCD matching coefficients, 
also pointed out at the end of ch.~\ref{sec:NRQCD}, is specific to the hard-scale
factorization and will persist at the level of pNRQCD. 

Let us consider the following ratios of hadronic and electromagnetic annihilation widths
for states with the same principal quantum number ($J=0,2$):
\be
R_n^V = {\Gamma(V_Q (nS) \rightarrow LH) \over
\Gamma(V_Q (nS) \rightarrow e^+e^-)},
\quad 
R_n^P = {\Gamma(P_Q (nS) \rightarrow LH) \over
\Gamma(P_Q (nS) \rightarrow \gamma\gamma)},
\quad
R_n^\chi = {\Gamma(\chi_Q (nJ1) \rightarrow LH) \over
\Gamma(\chi_Q (nJ1) \rightarrow \gamma\gamma)}. 
\ee
It is a specific prediction of pNRQCD that, for states
for which the assumption $\lQ \gg E$ holds, the wavefunction
dependence drops out of the right-hand side of the above equations. 
The residual flavor dependence is encoded in the
powers of $1/m$, in $E_{n0}^{(0)}$ and in the Wilson coefficients, while
the residual dependence on the principal quantum number is encoded in
the LO binding energy $E_{n0}^{(0)}$. The Wilson coefficients 
may be calculated in perturbation theory and the binding energy may be 
derived from the quarkonium mass $M(nS)$: $M(nS) - 2 m \simeq E_{n0}^{(0)}$.
The only unknown quantities are the gluon-field correlators. The crucial point
is that these do not depend on the flavor and the quarkonium quantum
numbers. Therefore, on the whole set of quarkonium states for which the 
pNRQCD formulas apply the number of non-perturbative parameters 
has decreased with respect to NRQCD. As discussed 
in sec.~\ref{secinclusivepnrqcd}, the gluon-field correlators 
may be extracted either from lattice simulations or specific models of the QCD
vacuum or from  experimental data. We will come back to this 
last possibility at the end of the section. 

Here we would first like to consider combinations of ratios in which even the dependence on 
the correlators drops out and predictions  based purely on perturbative QCD 
are possible. Let us consider the ratios between
$R_n^V$ and $R_n^P$ with different principal quantum numbers 
at order $E/m$. Contributions coming from the non-analytic scale 
$\sqrt{m\,\lQ}$ have not been calculated to that order, however, 
they appear to be suppressed in the ratio. We obtain
\bea
& &{R_n^V\over R_m^V} = 1
+ \left({{\rm Im\,}g_1(^3 S_1) \over {\rm Im\,}f_1(^3 S_1)}
- {{\rm Im\,}g_{ee}(^3 S_1) \over {\rm Im\,}f_{ee}(^3 S_1)} \right)
{M(nS) - M(mS) \over m},
\label{rrv}
\\
& &{R_n^P\over R_m^P} = 1
+ \left({{\rm Im\,}g_1(^1 S_0) \over {\rm Im\,}f_1(^1 S_0)}
- {{\rm Im\,}g_{\gamma\gamma}(^1 S_0) \over {\rm Im\,}f_{\gamma\gamma}(^1
S_0)} \right)
{M(nS) - M(mS) \over m}.
\label{rrp}
\eea
Due to the pNRQCD factorization, the octet-type contributions cancel in the
ratio, differently from what is predicted in NRQCD within the standard power
counting \cite{Gremm:1997dq}. In the vector case we get for the $\Upsilon(2S)$ and $\Upsilon(3S)$ state
($m_b \simeq 5$ GeV) $R_2^\Upsilon/R_3^\Upsilon \simeq 1.3$, which is close
to the experimental central value of about $1.4$ that one can obtain from \cite{Eidelman:2004wy}. 
In the pseudoscalar case, since
${{\rm Im\,}g_1(^1 S_0) /{\rm Im\,}f_1(^1 S_0)}
- {{\rm Im\,}g_{\gamma\gamma}(^1 S_0) /{\rm Im\,}f_{\gamma\gamma}(^1 S_0)}$
is of  ${\cal O}(\als)$, we find that, at order $E/m$, $R_n^P$ is the same for all radial excitations.

\begin{figure}
\makebox[-16cm]{\phantom b}
\put(120,0){\epsfxsize=6.5truecm \epsfbox{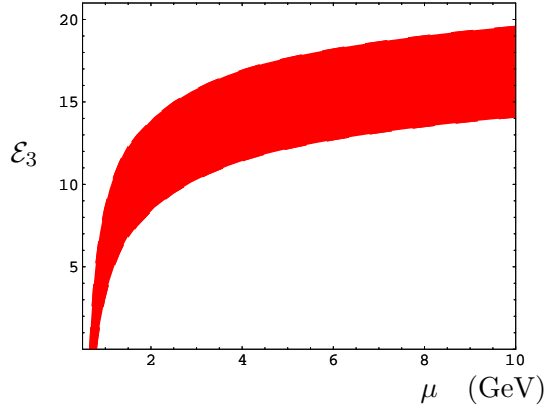}}
\put(265,-2){\small $\mu$ ~ (GeV)}
\put(110,87){\small ${\mathcal E}_3$}
\caption{\it Plot of the 1-loop RG-improved expression for 
$\mathcal E$ vs. $\mu$: $ {\mathcal{E}} (\mu)= {\mathcal{E}}(m) +
\displaystyle {24\, N_c \, C_F \over \beta_0}\ln {\als(m)\over \als(\mu)}$. 
${\mathcal{E}}(m)$ has been extracted from charmonium $P$-wave data. 
The error band accounts only for the uncertainties inherited from the
charmonium data. From \cite{Vairo:2002nh}.
}
\label{figE3}
\end{figure}

As mentioned above, it is possible to fix the gluon-field
correlators on some experimental set of data and use them on some other.
For instance, one may extract them from charmonium data and  
calculate bottomonium widths. This is particularly useful since, at the
moment, bottomonium data are less abundant than charmonium ones. 
The program has been carried out for $P$-wave decays in \cite{Brambilla:2001xy}.
These depend on just one correlator, ${\cal E}_3$, which may be extracted from 
$P$-wave charmonium decay data. The result is shown in Fig.~\protect\ref{figE3}.
At the scale of 1 GeV one finds
\bea
{\mathcal{E}}_3 (1\,{\rm GeV}) = 5.3^{+3.5}_{-2.2}\hbox{(exp)},
\label{numEE}
\eea
where the errors only refer to the experimental uncertainties on the
charmonium decay widths (in particular, uncertainties related 
to higher orders in the perturbative series, which may be potentially large, 
have not been included). In any case, the given figure is compatible with the
values that are usually assigned to the NRQCD octet and 
singlet matrix elements (e.g. from the fit of \cite{Maltoni:2000km} 
one obtains  ${\mathcal{E}}_3 (1\, {\rm GeV}) = 3.6^{+3.6}_{-2.9} \hbox{(exp)}$),  
while the bottomonium lattice data  of \cite{Bodwin:1996tg,Bodwin:2001mk}  
appear to give a lower value. Once ${\cal E}_3$ is known it may be inserted 
in Eqs.~(\ref{hadrchi}) and (\ref{electrchi}) to get the ratios of
annihilation widths of bottomonium $P$-waves. In practice, 
in pNRQCD at the order at which Eqs.~(\ref{hadrchi}) and
(\ref{electrchi}) are valid, the twelve $P$-wave bottomonium and
charmonium states that lie below threshold depend 
on 4 non-perturbative parameters (3 wavefunctions $+$ 
1 chromoelectric correlator ${\mathcal E}_3$).
The reduction of the number of unknown non-perturbative parameters by two 
with respect to NRQCD, allows one to formulate two specific new predictions of pNRQCD:
\be
{ \Gamma(\chi_{b0}(1P)  \rightarrow {\rm{LH}}) 
\over \Gamma(\chi_{b1}(1P)  \rightarrow {\rm{LH}})} 
=
{ \Gamma(\chi_{b0}(2P)  \rightarrow {\rm{LH}}) 
\over \Gamma(\chi_{b1}(2P)  \rightarrow {\rm{LH}})} 
= 8.0 \pm 1.3,  
\label{U01}
\ee
or alternatively 
\be
{ \Gamma(\chi_{b1}(1P)  \rightarrow LH) 
\over  \Gamma(\chi_{b2}(1P)  \rightarrow LH)} 
=
{ \Gamma(\chi_{b1}(2P)  \rightarrow LH) 
\over  \Gamma(\chi_{b2}(2P)  \rightarrow LH)} 
= 0.50^{+0.06}_{-0.04},
\label{U12}
\ee
where ${\mathcal E}_3$ is taken from Fig.~\ref{figE3} and the NRQCD matching 
coefficients are taken at NLO. The errors refer only to the uncertainty in  ${\mathcal E}_3$.
In Fig.~\ref{figPbottomonium} we plot the above ratios as functions of the
factorization scale $\mu$. We note that the scale dependence of ${\cal E}_3$
(see Fig.~\ref{figE3}) has been smoothed out in the plots of Fig.~\ref{figPbottomonium}, as
expected in a physical quantity (compare the cancellation of the 
LO IR divergences between the singlet matching coefficients and the 
octet matrix elements discussed in the paragraph after Eq.~(\ref{gchiem})). The large NLO
corrections are reflected by the extension of the non-overlapping regions in the two bands in
Fig.~\ref{figPbottomonium}. Recent CLEO measurements give (see 
\cite{Cinabro:2002ji} corrected in \cite{Brambilla:2004wf}) 
$\Gamma(\chi_{b0}(2P)  \rightarrow LH) / \Gamma(\chi_{b2}(2P)  \rightarrow LH) 
= 6.1 \pm 2.8$, which agree  inside the large errors with the above
predictions, and $\Gamma(\chi_{b1}(2P)  \rightarrow LH) / \Gamma(\chi_{b2}(2P)  \rightarrow LH) 
= 0.25 \pm 0.09$, which is somewhat lower than above.

The above approach may  eventually be extended to a global fit of all the correlators appearing in
$S$- and $P$-wave annihilation widths. The obtained values could then be used
to predict annihilation ratios of quarkonium states that are 
unknown or to improve present determinations.
This program still requires the calculation of the contribution coming
from the non-analytic scale $\sqrt{m\,\lQ}$ at least at relative order $E/m$ and
$\lQ^2/m^2$ for $S$ waves (that is with the same accuracy as the contributions coming from
the analytic scales listed in sec.~\ref{secinclusivepnrqcd}) and the resummation of large contributions in the
perturbative series of the 4-fermion matching coefficients.

\begin{figure}
\makebox[-16cm]{\phantom b}
\put(25,0){\epsfxsize=6.5truecm \epsfbox{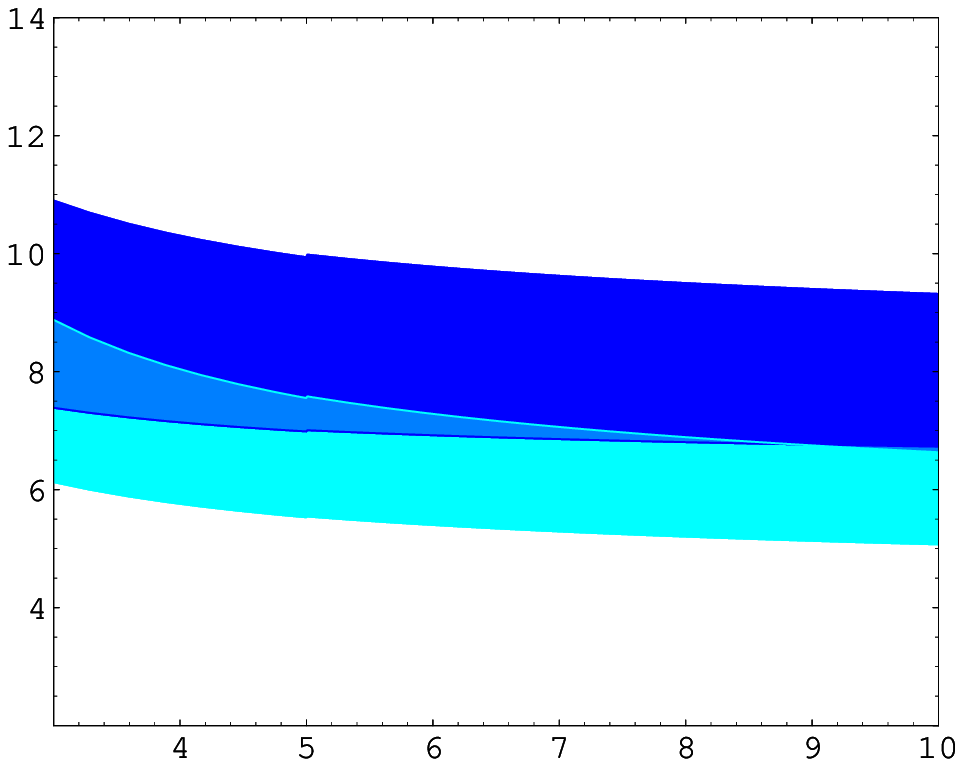}}
\put(170,-2){\small $\mu$  ~ (GeV)}
\put(0,87){\small ${\Gamma_{\chi_{b0}\rightarrow {\rm{LH}}}
               \over\Gamma_{\chi_{b1}\rightarrow {\rm{LH}}}}$}
\put(150,80){\small NLO}
\put(150,55){\small LO}
\put(265,0){\epsfxsize=6.5truecm \epsfbox{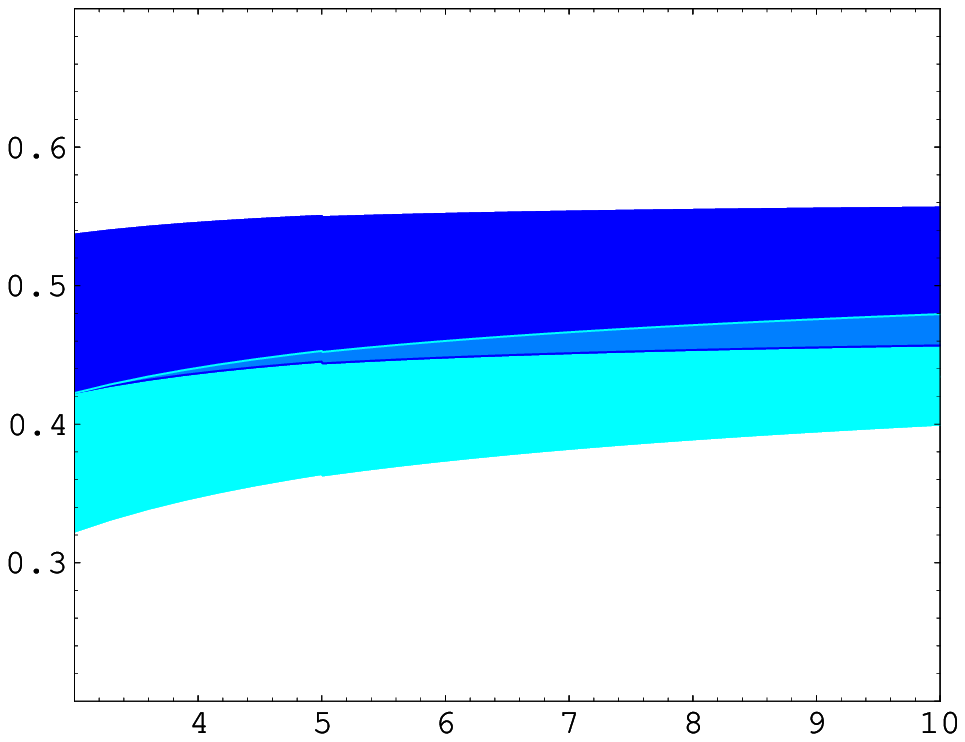}}
\put(410,-2){\small $\mu$  ~ (GeV)}
\put(237,87){\small ${\Gamma_{\chi_{b1}\rightarrow {\rm{LH}}}
                 \over\Gamma_{\chi_{b2}\rightarrow {\rm{LH}}}}$}
\put(390,95){\small NLO}
\put(390,70){\small LO }
\caption{\it The left-hand side of Eqs. (\protect\ref{U01}) and (\protect\ref{U12}) plotted 
vs. $\mu$. We have taken ${\mathcal E}_3$ from Fig.~\protect\ref{figE3}. 
The LO and NLO bands refer to the Wilson coefficients at LO and NLO respectively.
From \cite{Vairo:2002nh}.
}
\label{figPbottomonium}
\end{figure}

\subsection{Non-relativistic sum rules}
\label{sec:sumrules}
NR sum rules are a classical example for the application of
NR EFTs and the determination of the heavy 
quark masses like  charm and bottom. The key point is the 
relation between $\Pi(q^2)$ at 
$q^2=0$ to moments of the 
total cross section $\sigma(e^+e^- \rightarrow Q\bar Q)$. $\Pi(q^2)$
is defined in terms of  the correlator of two electromagnetic heavy quark currents in the 
following way
\be
(q_\mu q_\nu-g_{\mu\nu}q^2)\Pi(q^2)=i\int d^4x e^{iq\cdot x}\langle 
0|T\{j_\mu^v(x)j_\nu^v(0)\}|0\rangle
\,,
\ee
where $j_\mu^v(x)\equiv \bar Q \gamma_\mu Q(x)$. Using causality and the optical 
theorem one obtains
\be
\label{moments}
P_n={12\pi^2e_Q^2 \over n!}\left({d \over dq^2}\right)^n\Pi(q^2)|_{q^2=0} =
\int_{\sqrt{s_{min}}}^\infty {ds \over s^{n+1}}R_{Q\bar Q}(s) \,, 
\ee 
where $R_{Q\bar Q}\equiv \sigma(e^+e^- \rightarrow Q\bar Q)/\sigma(e^+e^-
\rightarrow \mu^+\mu^-)$ and $e_Q$ is the quark electric charge. For low
values of $n$, the left-hand side of Eq.~(\ref{moments}) can be computed using
perturbation theory due to the fact that the energy necessary to reach the
threshold for heavy quarks production  is 
much larger than $\lQ$,\footnote{One should not forget, however, that potential
problems may appear beyond NNLO due to the appearance of physical
decay channels of the heavy quarkonium \cite{Groote:2001py,Portoles:2002rt}.} 
whereas the right-hand
side can be obtained from the experimental data. However, we are 
concerned here with the NR sum rules. These are defined by taking $n$ large. This
implies the existence of new scales in the problem besides $m$ and $\lQ$,
like $m/\sqrt{n}$, $m/{n}$ and so on. Therefore, it is not so clear that one
can actually perform computations within perturbation theory. For $n$ large
enough, one will have $\sqrt{n}\als \sim 1$ and a complete resummation of
these terms should be achieved. The quantity $\sqrt{n}\als$ appears in the
computation through the ratio of two different scales: $(m\als)/(m/\sqrt{n})$.
Hence, we  see the following analogy with the NR situation: $1/\sqrt{n}$ plays
the same role as $v$, the velocity of the heavy quark, and by taking
$\sqrt{n}\als \sim 1$  we are considering the NR limit.

There is also another problem. For sufficiently large  $n$, we can no longer claim
that the induced scales are much larger than $\lQ$ and non-perturbative
effects need to be considered.  How to handle them is a delicate issue. Here,
we will only consider the situation when $m/\sqrt{n} \gg \lQ$.  This seems to
be a safe requirement (at least for bottomonium). It is not clear however that
we can also assume $m/{n} \gg \lQ$. In practical applications the boundary
for doing so is usually taken around $n \sim 10$. We will discuss this issue
further below.

In spite of the above remarks, 
the NR sum rules are ideal from the experimental point of view. 
By taking $n$ large on the right-hand side of Eq.~(\ref{moments})  the contribution from 
high momenta (the continuum region) is suppressed. Actually, this is the region which is
less well known on the experimental side. Therefore, by using NR sum rules, the experimental errors are 
significantly reduced. In practice, the following parameterization is used
\be
P_n^{ex}=\sum_{k=1}^6{9\pi \over 
\alpha^2(2m)}{\Gamma_{\Upsilon(k)} \over 
M_{\Upsilon(k)}^{(2n+1)}}+\int_{\sqrt{s_{B\bar B}}}{ds \over s^{n+1}}r_{cont}(s)
\,.
\ee

The theoretical expressions for the moments $P_n^{th}$ can be computed order 
by order in the NR expansion in $1/\sqrt{n}$ and $\als$, which at each order
resums all the terms proportional to $\als \sqrt{n}$ to any power. Nowadays
they are known in the on-shell scheme at NNLO in the NR expansion, which
includes all corrections up to order $1/n$, $\als/\sqrt{n}$ and $\als^2$
\cite{Kuhn:1998uy,Penin:1998zh,Hoang:1998uv,Melnikov:1998ug,Beneke:1999fe}.
With this accuracy, the dispersion integration for the moments
$P_n$ takes the form
\bea
P_n \, = \,
\frac{18C_A}{4^{n}\,m^{2n+2}\al^2(2m)}\,\int\limits_{E_1}^\infty 
\frac{d E}{m} \,\exp\bigg (\,
-\frac{E}{m}\,n
\,\bigg )\,\bigg(\,
1 - \frac{E}{2\,m} + \frac{E^2}{4\,m^2}\,n
\,\bigg)
\\
\times 
{\rm Im} 
\left[\langle {\bf r} =0 |G_s(E)|  {\bf r} =0 \rangle
\right]
\left[
{\rm Im} f_{\rm EM}^{\rm pNR}({}^3S_1)
+{\rm Im} g_{\rm EM}^{\rm pNR}({}^3S_1){E \over m}
\right]
\,,
\label{Pnexpression1}
\eea
where $E\equiv\sqrt{s}-2m$ and $E_1$ is the binding
energy of the lowest lying resonance. The exponential form of the LO
NR contribution to the energy integration has to be
chosen because $E$ scales like $v^2\sim 1/n$.
For explicit expressions, we refer to \cite{Hoang:1998uv}.

\begin{table}[ht]
\addtolength{\arraycolsep}{0.2cm}
\begin{tabular}{|c|c|c|}
\hline
reference & order & ${\overline m}_b({\overline m}_b)$ (GeV)
\\
\hline
\cite{Melnikov:1998ug} & NNLO (kinetic mass)&
$4.20 \pm 0.10$ 
\\
\cite{Penin:1998kx} & NNLO (pole mass)&
$4.21 \pm 0.11$ 
\\
\cite{Beneke:1999fe} & NNLO (PS mass)&
$4.26 \pm 0.09$ 
\\
\cite{Hoang:2000fm}  & NNLO (1S mass)&
$4.17 \pm 0.05$
\\
\cite{Eidemuller:2002wk}  & NNLO (PS mass)&
$4.24 \pm 0.10$
\\
\hline
\hline
reference & order & ${\overline m}_c({\overline m}_c)$ (GeV)
\\
\hline
\cite{Eidemuller:2002wk}  & NNLO (PS mass)&
$1.19 \pm 0.11$ \\
\hline
\end{tabular}
\caption{\it \label{tablesumrules}  Recent determinations of 
 ${\overline m}_b$ and 
${\overline m}_c$ in the $\MS$ scheme 
from NR sum rules.}
\end{table} 

As we have pointed out  before, working in the on-shell scheme introduces large
errors. Therefore  most of the analyses nowadays use threshold masses, where
the cancellation of the pole mass renormalon is explicit (see Table
\ref{tablesumrules}).  In practical terms this amounts to re-expressing the results
obtained in the on-shell scheme in terms of the threshold masses.
Nevertheless, even if some improvement is obtained, large uncertainties remain
due to a rather strong scale dependence. This scale dependence can be traced
back to the fact that the decay width of the heavy quarkonium to $e^+e^-$ is
strongly scale dependent. For a more detailed discussion of this point see
\cite{Beneke:1999fe}. In this respect, RG techniques have not yet been applied
to these computations. It would be most interesting to do that and to see whether a
more stable result is obtained.

Non-perturbative effects in sum rules are parametrically of the same size as 
in the $\Upsilon(1S)$ mass in the standard counting $1 / \sqrt{n} \sim
\als$.  Nevertheless, it may happen that they are numerically suppressed. This
is indeed the case considering that one can describe the non-perturbative
effects by local condensates \cite{Voloshin:1995sf,Onishchenko:2000yy}.
However, one can use the expression in terms of local condensates only
 in the situation  ${m / n} \gg \lQ$ (although one can use that result
as an order of magnitude estimate of the non-perturbative effects).  This
would be analogous to the assumption $m\als^2 \gg \lQ$, which
may be difficult
to fulfill. Therefore, it is more likely that the non-perturbative corrections
will also depend on a non-local condensate of the same type
(chromoelectric correlator) as the $\Upsilon(1S)$ mass
does. Thus, in order to estimate the non-perturbative errors in sum rules
evaluations, it would be most welcome to have at least the explicit
expression of the non-perturbative effects in the situation ${m / n} \sim
\lQ$, which is still lacking. In that way one could relate the
non-perturbative effects for different moments in the sum rules to each others
 or to the
non-perturbative effects in the $\Upsilon(1S)$ mass.

\subsection{$t$-$\bar t$ production near threshold}
\label{tt}
Future linear electron-positron colliders will produce large samples of
$t$-$\bar t$ pairs near threshold
\cite{Bagger:2000iu,Aguilar-Saavedra:2001rg,Abe:2001nq,Abe:2001gc}.  In this
regime, the top and the  antitop will move slowly with respect each other
and pNRQCD becomes applicable. Since the top quark mass $m_t\sim$ 175 GeV and
the expected (electroweak) decay width $\Gamma_t\sim$ 1.5 GeV are large in
comparison with $\lQ$, non-perturbative effects due to $\lQ$ are expected to
be small in the whole threshold region and hence a weak-coupling analysis is
very reliable. In addition since $\Gamma_t\sim m_t\als^2$ which is the US scale, a
remnant of the would-be toponium $1S$ state is expected to show up as a bump
in the total cross section. This will serve to obtain the top quark mass with
a high accuracy.

The $t$-$\bar t$ pair will be dominantly produced via $e^+ e^- \rightarrow
\gamma^\ast \, ,\, Z^\ast \rightarrow t\bar t$.  The total production cross
section may be written as \cite{Hoang:2000yr}
\begin{eqnarray}
  \sigma_{\rm tot}^{\gamma,Z}(s) = \frac{4\pi\alpha^2}{3 s} \Big[\,
  F^v(s)\,R^v(s) + F^a(s) R^a(s) \Big] \,,
\label{totalcross}
\end{eqnarray}
where $F^v(s)$ and $F^a(s)$ contain electroweak parameters \cite{Hoang:2001mm}
and
\begin{eqnarray} \label{fullR}
 R^v(s) \, = \,\frac{4 \pi }{s}\,\mbox{Im}\,\left[-i\int d^4x\: e^{i q\cdot x}
  \left\langle\,0\,\left|\, T\, j^v_{\mu}(x) \, {j^v}^{\mu} (0)\,
  \right|\,0\,\right\rangle\,\right] \,, \nn\\[2pt] R^a(s) \, = \,\frac{4 \pi
  }{s}\,\mbox{Im}\,\left[-i\int d^4x\: e^{i q\cdot x}
  \left\langle\,0\,\left|\, T\, j^a_{\mu}(x) \, {j^a}^{\mu} (0)\,
  \right|\,0\,\right\rangle\,\right] \,,
\end{eqnarray}
where $q=(\sqrt{s},0)$ and $j^{v}_\mu$ ($j^{a}_\mu$) is the vector
(axial-vector) current that produces a quark-antiquark pair defined by
Eq.~(\ref{NRcurrent}) (Eq.~(\ref{NRcurrenta})).  Hence, the full QCD
calculation can be split into (i) calculating the matching coefficients $b^v_1
(m_t,\nu )$, $b^v_2 (m_t,\nu )$ and $b^a_1 (m_t,\nu )$, and (ii) calculating
current correlators in pNRQCD. Up to NNLO (${\cal O}(\als^ 2)$ corrections), 
the latter reduces to a purely quantum-mechanical calculation along the lines
of sec.~\ref{pNRweakObservables} (US gluons do not play any role). The
potential is only needed at the order displayed in
Eqs.~(\ref{newpot0})-(\ref{Dsten2}).  This calculation has been carried out by
several groups\footnote{For an analytical expression  for the
$\gamma\gamma \rightarrow t\bar t$ cross section  at NNLO see \cite{Penin:1998mx}.} and
the final outcome is summarized in \cite{Hoang:2000yr} (previous computations
at LO \cite{Fadin:1988fn} and NLO \cite{Strassler:1990nw} relied on potential
models which needed phenomenological input). Several comments are in order.

{\bf 1}. At NNLO, the scale dependence which appears in the matching coefficient
$b^v_1(m_t,\nu )$ ($b^v_2= b^a_1=1$ at this order) is compensated by the
scale dependence introduced by regulating and renormalizing the UV divergences
of the quantum-mechanical perturbation theory ({\it potential} loops).

{\bf 2}. The top quark width is introduced by replacing $m_t \rightarrow m_t
-i\Gamma_t /2$.  A consistent inclusion of electroweak effects is still
lacking.

{\bf 3}. In order to obtain stable results for the top quark mass in going from LO
to NLO to NNLO, it is very important to use the so-called "threshold masses"
rather than the pole mass. These are discussed in ch.~\ref{secrenormalons}.

{\bf 4}. The large logarithms arising due to the various scales in the problem can be
resummed using RG techniques as described in sec.~\ref{RGreview}. This problem is
non-trivial because all scales (hard, soft, potential and US) play a role.  It
was first addressed within the vNRQCD framework \cite{Hoang:2000ib}.  However,
the correct result for $b_1^v (m_t,\mu )$ at NLL was first given within pNRQCD
in \cite{Pineda:2001et} and later  reproduced within vNRQCD
\cite{Hoang:2002yy}. \cite{Hoang:2001mm,Hoang:2003ns}
have computed some partial results for the NNLL contribution.
The resulting series \cite{Hoang:2003ns} does not show a
very good convergence (even if the absolute value of the corrections is small).
This, however, may be due to the scheme dependence of the result.
\cite{Penin:2004ay} have obtained a
complete (and therefore scheme independent) result with NNLL accuracy for
the ratio of the spin-one and spin-zero production. In this case a good convergence
is found, but one should keep in mind that this ratio is less sensitive to the
US scale than the full current. Therefore, it is premature to
draw any definite conclusion about the convergence of the series before getting
the complete NNLL evaluation, which, even if difficult, is
within reach.  This is of utmost importance for future determinations of
the top mass and the Higgs-top coupling at a future Linear Collider
\cite{Martinez:2002st}.

{\bf 5}. At NNNLO, as well as for the resummations above, US gluons start to  play a
role.  The double logarithmic contributions were calculated in
\cite{Kniehl:1999mx} and the single logarithmic ones in
\cite{Kniehl:2002yv}. The finite pieces are still missing.  These can in
principle be calculated with the potentials given in \cite{Kniehl:2002br}
together with the three loop static potential (which is still missing), and
the LO terms for the US gluons given by (\ref{Lpnrqcd}).  The
matching coefficients $b^v_1 (m_t,\nu )$, $b^v_2 (m_t,\nu )$ and $b^a_1
(m_t,\nu )$ also need to be calculated to one higher order in $\als$.

Figure \ref{figtopplots} exemplifies the current status of theoretical results
for the total cross section for $e^+e^- \to \gamma^{*} \to t^+ t^-$.

\begin{figure} 
\makebox[-19.5cm]{\phantom b}
\put(0,0){\epsfxsize=11truecm \epsfbox{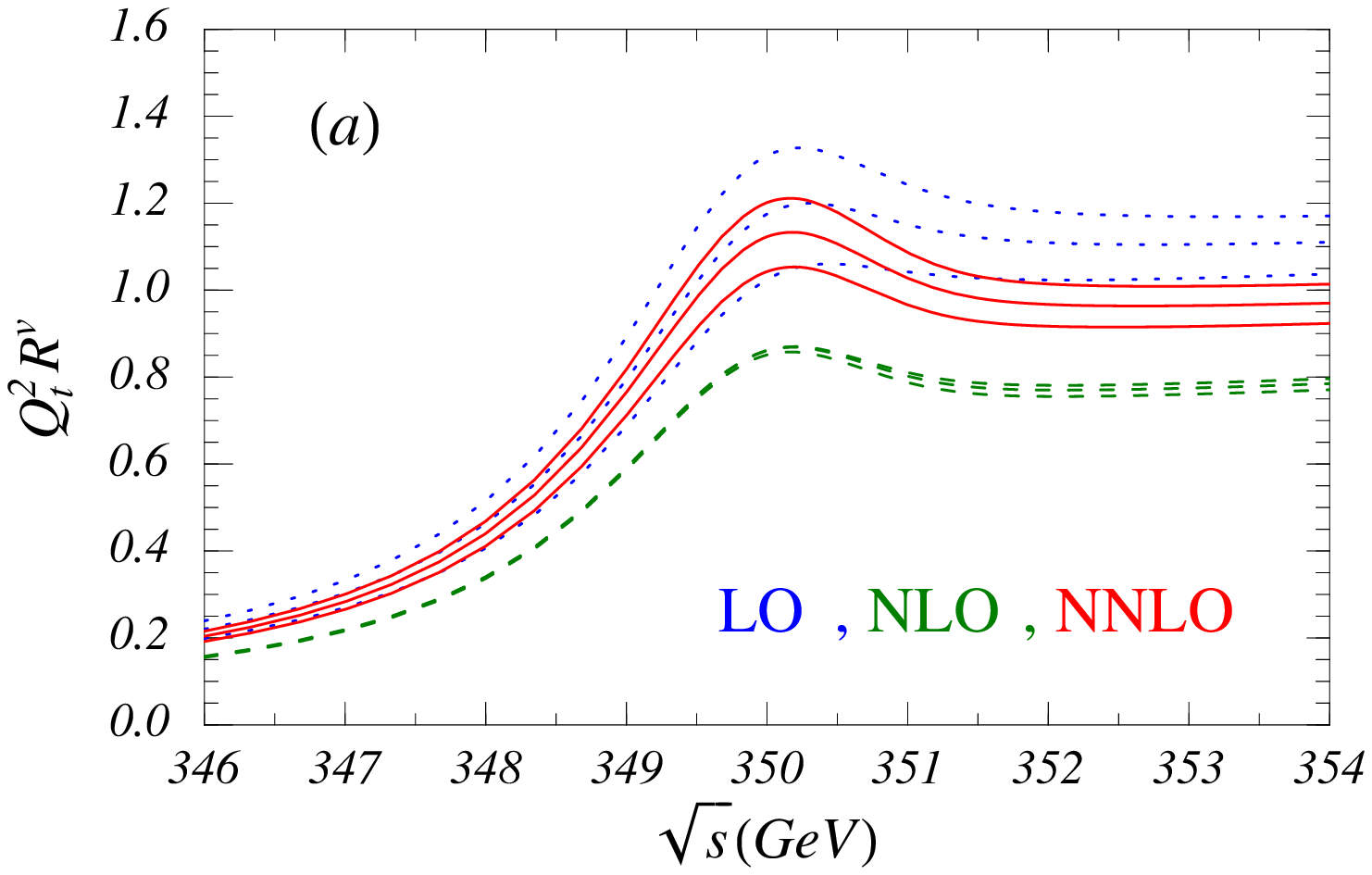}}
\put(240,0){\epsfxsize=11truecm \epsfbox{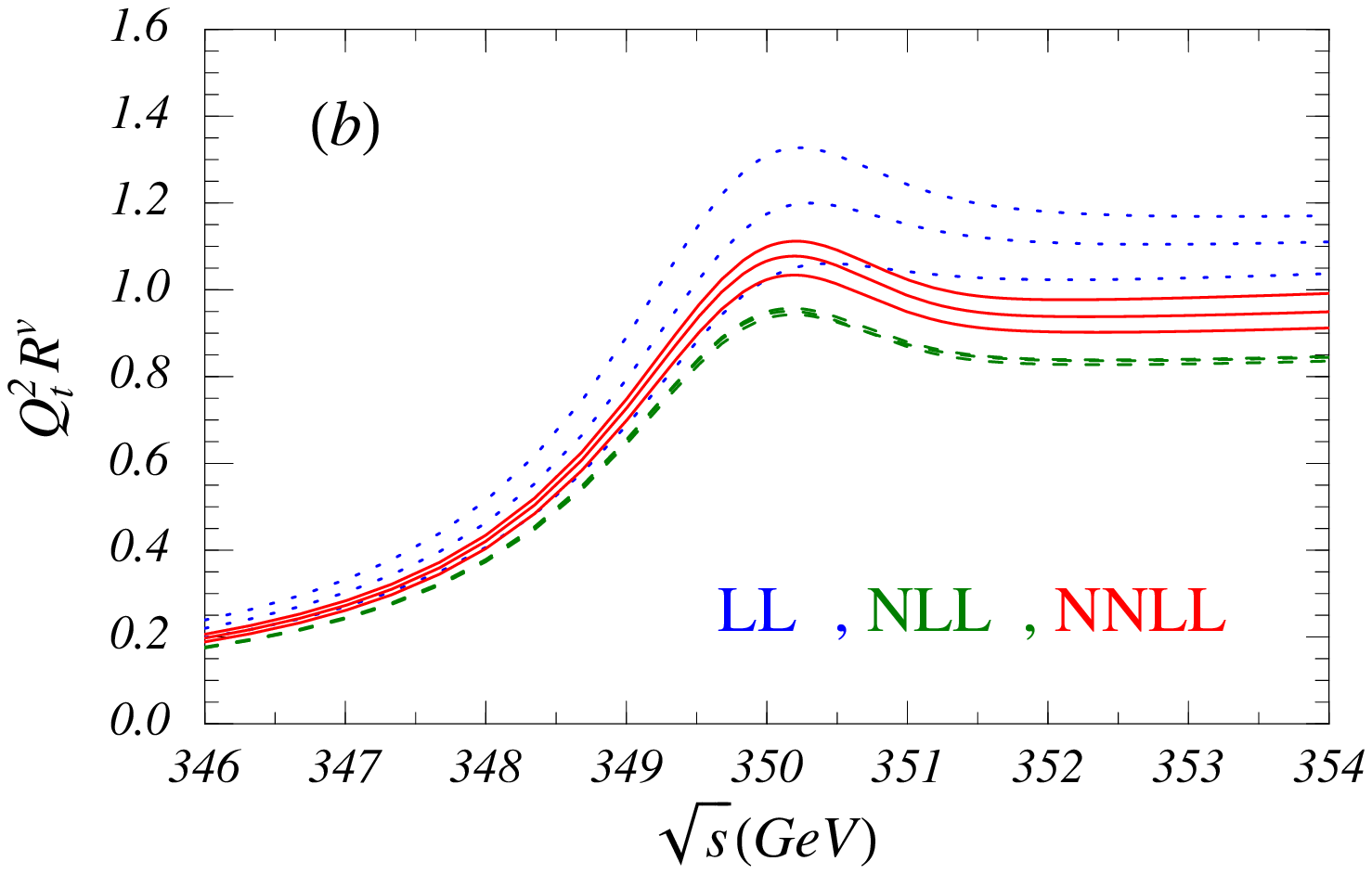}}
\vspace{-8cm}\\
\caption{\it Panel (a) shows the results for $e_t^2 R^v$ ($e_t=2/3$ is the top quark electric charge) 
 with $m_t=175$\,GeV (the $1S$ threshold mass is used, see Section \ref{secrenormalons}) 
 and $\Gamma_t=1.43$\,GeV in fixed-order perturbation theory at
 LO (dotted lines), NLO (dashed lines) and NNLO (solid lines). 
 Panel (b) shows the results for $e_t^2 R^v$ with the same parameters 
 in RG improved perturbation theory at
 LL (dotted lines), NLL (dashed lines) and (partial) NNLL (solid lines) order. 
 For each order, curves are plotted for $\nu_p/m_t=0.15$, $0.20$, and $0.3$. 
 From \cite{Hoang:2003xg}.
\label{figtopplots} 
}
\end{figure}

\subsection{Semi-inclusive radiative decays}
\label{semi} 
We have seen that NRQCD and pNRQCD are particularly suitable for describing inclusive
decays of heavy quarkonia to light particles. Semi-inclusive and
fully exclusive decays can also be addressed but they require
additional theoretical considerations. Similarly to what happens for
inclusive decays, pNRQCD is expected to provide supplementary
information here as well. Semi-inclusive radiative decays to light
hadrons, in which only the photon energy is measured, are the simplest
of them and will be briefly discussed in the following.

We shall restrict our discussion to the so-called direct contributions, 
for which the photon is emitted from a heavy quark electromagnetic current. 
Fragmentation contributions also play an important role \cite{Catani:1994iz}. 
The starting point is the QCD formula \cite{Rothstein:1997ac}
\begin{equation}
{d \Gamma\over dz}=z{M\over 16\pi^2} \, {\rm Im} \, T(z)\,,
\qquad T(z)=-i\int d^4 x e^{-iq\cdot x}\left< 
V_Q (nS)
 \vert T\{ j_{\mu}^v (x) j_{\nu}^v (0)\} \vert
V_Q (nS)
 \right> g^{\mu\nu}_{\perp},
\label{gdz}
\end{equation}
where $M$ is the heavy quarkonium mass,
and we have restricted ourselves to $^3S_1$ states.
$q$ is the photon momentum, which in the rest frame 
of the heavy quarkonium is $q=\left(q_{+},q_{-}, q_{\perp}\right)=(zM,0,0)$. 
We have used light-cone coordinates $q_\pm=q^0\pm q^ 3$.  $z\in [0,1]$ is defined as $z=2E_\gamma /M$, 
namely the fraction of the maximum energy that the photon may have in the heavy quarkonium rest frame.

For $z$ away from the lower and upper end-points ($0$ and $1$ respectively), 
no further scale is introduced beyond those inherent in the NR system.  
The integration of the scale $m$ in the time-ordered 
product of currents in (\ref{gdz}) leads to local NRQCD operators with matching 
coefficients which depend on $m$ and $z$. At LO one obtains
\begin{equation}
\label{LOrate}
\frac1{\Gamma_0} \frac{d\Gamma
_{\rm LO}}{dz} =  
\frac{2-z}{z} + \frac{z(1-z)}{(2-z)^2} +
2\frac{1-z}{z^2}\ln(1-z) - 2\frac{(1-z)^2}{(2-z)^3} \ln(1-z),
\end{equation}
where 
\begin{equation}
\Gamma_0 = \frac{32}{27}\,\alpha\,\als^2\,e_Q^2
\frac{\langle  V_Q (nS)\vert {\cal O}_1(^3S_1)\vert V_Q (nS)\rangle}{m^2},
\label{gamma0}
\end{equation}
and $e_Q$ is the charge of the heavy quark. The $\als$ correction to this rate was
calculated numerically in \cite{Kramer:1999bf}. The contribution of color-octet operators 
turns out to be strongly suppressed away from the upper end-point region (the lowest order color-octet 
contribution identically vanishes) \cite{Maltoni:1998nh}. The expression corresponding to (\ref{gamma0}) 
in pNRQCD is obtained at lowest order in any of the possible regimes by just making the substitution 
\begin{eqnarray}
\label{singletWF}
\langle  V_Q (nS) \vert {\cal O}_1(^3S_1) \vert  V_Q (nS) \rangle &=&
 \frac{N_c}{2\pi} |R_{n0}(0)|^2,
\end{eqnarray}
The final result coincides with the one of early QCD calculations \cite{Brodsky:1977du,Koller:1978qg}. 

For $z\rightarrow 0$, the emitted low energy photon can only produce transitions within the 
NR bound state without destroying it. Hence the direct low-energy photon emission 
takes place in two steps: (i) the photon is emitted (dominantly by electric dipole and magnetic 
dipole transitions) and (ii) the remaining (off-shell) bound state is annihilated 
into light hadrons. It has a suppression  $\sim z^3$ with respect to
$\Gamma_0$ (see \cite{Manohar:2003xv,Voloshin:2003hh} for recent analyses of this region in QED). 
 Hence, at some point the direct photon emission is overtaken by the so-called 
fragmentation contributions $\bar Q Q \rightarrow ggg \rightarrow gg\bar q q \gamma $ 
\cite{Catani:1994iz,Maltoni:1998nh}.

\begin{figure}
\centering
\epsfxsize=8.5truecm \epsfbox{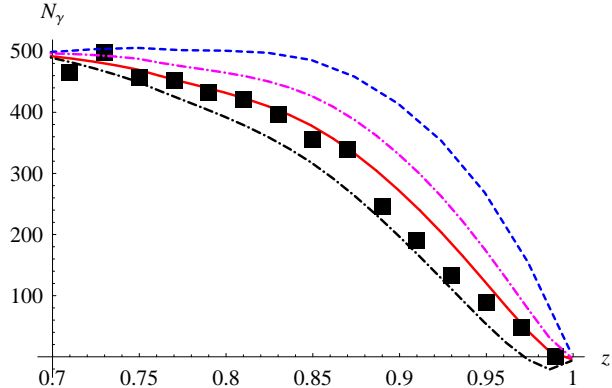}
\caption{\it End-point region of the photon spectrum in semi-inclusive $\Upsilon$ decay. 
The points are  CLEO data \cite{Nemati:1996xy},
the dashed line is the (best) curve obtained in \cite{Fleming:2002sr} 
and the solid and dot-dashed lines are the results of  \cite{GarciaiTormo:2004jw} 
(the solid line is 
the central value
and the dot-dashed lines are obtained by a $2^{\pm 1}$ variation of the 
relevant scale). From \cite{GarciaiTormo:2004jw}.}
\label{figgrafic}
\end{figure}

For $z\rightarrow 1$, momentum conservation implies that the gluons emitted in
the short-distance annihilation process must have a direction roughly opposite
to that of the photon. They produce a jet-like event with momentum
$p_X=((1-z)M, M, 0)$ (in light-cone coordinates). This implies that two more
scales become relevant, ${p_X}_+= (1-z)M$ and $p_X^2=(1-z)M^2$, producing an
additional hierarchy $M \gg M\sqrt{1-z} \gg M(1-z)$. In recent years an EFT
named Soft-Collinear Effective Theory (SCET) has been introduced in order to
efficiently exploit this hierarchy of scales. The main ideas which led to SCET
were outlined in \cite{Bauer:2000ew}. Nowadays, SCET is being developed by
several groups \cite{Bauer:2000yr,Chay:2002vy,Beneke:2002ph,Hill:2002vw}. It
has been applied to $\Upsilon (1S)$ radiative decays by
\cite{Bauer:2001rh,Fleming:2002rv,Fleming:2002sr,Fleming:2004rk,GarciaiTormo:2004jw}.
We shall not review SCET here (a complete analysis connecting pNRQCD and SCET
is still lacking), but only mention its relevant features for the case we are
concerned with. For $z\rightarrow 1$, upon integrating out the scale $m$, the
time-ordered product of currents in (\ref{gdz}) does not reduce to local NRQCD
operators anymore.  Additional degrees of freedom are needed. These are
collinear gluons (and collinear light quarks), which are defined as those
having a typical momentum (in light-cone coordinates) $p\sim ((1-z)M, M,
\sqrt{1-z}M)$. They are incorporated in SCET together with the remaining
degrees of freedom in NRQCD. Then one matches the QCD electromagnetic currents
$j_{\mu}^v (x)$ (rather than the full time-ordered product) to SCET
currents. Next, the scale $\sqrt{1-z}M$ is integrated out (assuming that it is
large enough to use perturbation theory) by matching the time-ordered product
of currents in SCET (now renamed SCET$_{\rm I}$) to (non-local) operators of
the so-called SCET$_{\rm II}$, which does not contain collinear modes of
virtuality $\sim (1-z)M^ 2$ anymore (see \cite{Beneke:2003pa} for a detailed
description of the modes involved in SCET$_{\rm I}$ and in SCET$_{\rm II}$).
By calculating the anomalous dimensions of the various operators appearing in
both matchings and using standard RG techniques, one can resum large (Sudakov)
 logarithms $\ln (1-z)$. For the color-octet currents, this was done by
\cite{Bauer:2001rh} and for the color-singlet ones by
\cite{Fleming:2002sr,Fleming:2004rk}. For the color-singlet sector, the final
outcome corrects the old results of \cite{Photiadis:1985hn}. For
the color-octet sector, the final result may be given in terms of so-called
shape functions \cite{Rothstein:1997ac}, which involve expectation values in
the heavy quarkonium state of two color-octet NRQCD currents separated along a
light-cone direction, for instance
\begin{equation}\label{softfun}
S(\ell^+) = 
 \int \frac{dx^-}{4 \pi} \, e^{\frac{-i}{2} \ell^+ x^-} 
\langle V_Q (nS) | 
  \big[ \psi^\dagger  T^b  \chi\big] (x^-) 
  \, \phi_{bc}^{\rm adj}(0,x^-)
  \big[ \chi^\dagger  T^c  \psi \big] (0)| 
 V_Q (nS) \rangle \,.
\end{equation}
If the heavy quarkonium state is in the weak-coupling regime, as it is
likely to be in the case of the  $\Upsilon (1S)$ system, one can use
pNRQCD in that regime to calculate the shape functions. This was done
by \cite{GarciaiTormo:2004jw} (see also \cite{Beneke:1999gq}). When
these results are combined with those of the singlet sector, an
excellent description of data \cite{Nemati:1996xy} is obtained (see
Fig.~\ref{figgrafic}) for the end-point region.  Although, as discussed in
\cite{GarciaiTormo:2004jw}, there are still some calculations missing
in order to have a totally unambiguous theoretical result, the
agreement with data is very encouraging. Indeed, the end-point region
of the photon spectrum has been very elusive to theoretical
descriptions. The color-singlet contribution (sometimes referred to
as the color-singlet model) lies well above the data. For the 
color-octet contributions,   different models were used in the past to
estimate the shape functions, generically producing results
incompatible with data \cite{Wolf:2000pm,Bauer:2001rh}.  These facts
were used to argue that the introduction of a non-vanishing gluon mass
was necessary in order to fit the experimental data
\cite{Field:2001iu}. This is no longer the case, at least as far as 
the $\Upsilon (1S)$ system is concerned.

\section{Conclusions}
\label{conclusions}
The application of EFTs  of QCD to heavy quarkonia has considerably 
increased our understanding of those systems from a fundamental point of view. 
This has occurred at several levels.

\begin{itemize}
\item
Long standing puzzles have been resolved. For instance, the fate of IR
divergences in the decay widths to light particles has been resolved in NRQCD
by  introducing  color-octet operators, and the fate of the IR
divergences in the QCD static potential has been resolved in pNRQCD by the
explicit use of US gluons.
\item
Heavy quarkonium physics in the strong-coupling regime has been brought into
the realm of systematic calculations in QCD. This has led to the discovery of 
new terms in the potential which were missed in the past, both analytic and
non-analytic in $1/m$, and to express the color-octet NRQCD matrix elements
in terms of wavefunctions at the origin and additional
bound-state-independent non-perturbative parameters.  This puts NR
phenomenological potential models in a QCD context in the kinematic regime
where this EFT description applies.
\item
In the weak-coupling regime, it has allowed higher-order calculations 
to be carried out in a systematic and much simpler manner. Errors are under
parametric control. Moreover, it has made possible the application of RG
techniques, which have been used to resum infinite series of IR QCD logarithms, being
so far the only known way to carry out such resummations. This has opened the
possibility of having precision determinations of the Standard Model parameters
to which the heavy quarkonium is sensitive: $\als$ and the heavy quark masses.
\end{itemize}

Although the virtues by far exceed the drawbacks, the latter are not absent in EFTs of
heavy quarkonium. They are all related to the fact that the actual bottom and,
especially, charm masses are not so much larger than $\lQ$ (for
toponium-like states the EFT should work very well).  This means that the
scales $m\gg p \gg E$, which are assumed to be 
well separated, may actually not be that well separated,
and  hence expansions in various ratios may show a slow
convergence. In the strong-coupling regime  it is still too early to judge. 
For $\Upsilon (1S)$, in the weak-coupling regime, the convergence seems to be
good.  In addition, it is a fact that most of the matching coefficients of
NRQCD show a poor convergence in $\als (m)$ both for bottom and charm masses,
which is jeopardizing many practical applications of NRQCD. We may expect,
however, that once the renormalon singularities in each series have been
identified and properly subtracted, the situation will improve considerably, as
it has occurred with the introduction of {\it threshold masses}.

\section{Prospects}
\label{secPRO}
Non-relativistic EFTs for heavy quarkonium still have an enormous potential
and may evolve in many different directions. Some of
them are more or less obvious improvements or extensions of what has been
presented here.  Others require the introduction of new concepts and
techniques.

Among the obvious improvements are those which consist in calculating 
matching coefficients and observables to a
higher accuracy in both NRQCD and
pNRQCD.  In NRQCD, it would be important to have the NNLO calculation of the
imaginary parts of the matching coefficients of the four fermion operators, at
least for $S$ and $P$ waves. This would allow one to see
whether the poor convergence observed at NLO is corrected or remains, and in 
either case it would facilitate renormalon-based improvements.  It would also be
important to have further and more accurate lattice calculations of the NRQCD
matrix elements (see sec.~\ref{NRdecay}).  In the weak-coupling regime of
pNRQCD, some perturbative calculations are missing, which seem to be in reach
of the current computational power. Let us only mention the complete
three-loop static potential, which is necessary for the complete NNNLO
spectrum and for electromagnetic production processes (for instance in $t$-$\bar t$);
the complete NNLL resummation of the creation and annihilation currents; and
the NNNLO calculation of  electromagnetic production. These would allow us  an 
increase in the precision of the determinations of $m_b$, $m_c$, $\als$, and,
eventually, $m_t$ (see sec.~\ref{Groundstatemass}).  For the case of $\Upsilon
(1S)$, the accuracy is limited by the poor knowledge of the non-perturbative
contributions, which are precisely given in terms of chromoelectric field
correlators. A proper lattice evaluation of the latter would be most
welcome. For the $t$-$\bar t$ system, the level of accuracy calls for the
consistent inclusion of electroweak effects, which is also missing (see
sec.~\ref{tt}).

In the strong-coupling regime of pNRQCD, on the one hand it is necessary to
update the early lattice evaluations of the potentials including the more
recently found $1/m$ and $1/m^2$ potentials.  On the other hand a systematic
matching procedure of the potentials to the continuum limit and a rigorous
lattice renormalization scheme should be developed (see sec.~\ref{sec:latmod}).  
This will lead to
 a fully consistent lattice version of pNRQCD.

In the same regime, the inclusion of pseudo-Goldstone bosons (pions) and low
energy photons is still lacking. This would allow the description of 
electromagnetic and hadronic transitions in that situation.

Let us next mention some applications of EFTs to heavy quarkonium that
require further theoretical elaborations.

The systematic study of semi-inclusive (see sec.~\ref{semi}) and exclusive
decays may need the introduction of further degrees of freedom in addition to those of
NRQCD or pNRQCD.

NRQCD production matrix elements should also have definite expressions in
pNRQCD both in the weak and in the strong coupling regime, which have not been
worked out yet. It is expected that, as happens for the decay matrix
elements, new relations may appear and the number of nonperturbative parameters 
is consequently reduced.

States close to or above the heavy-light meson pair threshold cannot be treated
using pNRQCD, at least in its current formulations. Hence one has to stay at
the NRQCD level. A hadronic version of NRQCD, including heavy quarkonium
states, heavy-light mesons, and pseudo-Goldstone bosons, in the spirit of
\cite{Burdman:1992gh,Mannel:1995jt,Casalbuoni:1996pg,Voloshin:2003kn} might
prove useful  and will eventually help to understand the nature of present
\cite{Choi:2003ue} and possibly future potential states in that region.

Including finite temperature in NRQCD and pNRQCD would make it possible to
address important questions  like $J/\psi$ suppression as a sign of
deconfinement \cite{Matsui:1986dk} in current and future heavy ion collision experiments.

Finally, by slightly changing the fundamental degrees of freedom, EFTs may be
built which are similar to pNRQCD, but also suitable to describe bound states made
of two heavy particles other than heavy quarkonium.  An example are heavy
baryons made of two heavy quarks, like those recently discovered at SELEX
\cite{Mattson:2002vu}. These systems are theoretically quite interesting due
to the interplay of HQET and NRQCD \cite{Soto:2003ft,Rosch:2003aa}.  Also
quarkonia-quarkonia scattering may be studied along the lines of
\cite{Peskin:1979va,Bhanot:1979vb,Fujii:1999xn,Vairo:2000ia}.

We feel that we are at the beginning of a time where most aspects of the
physics of heavy quarkonium, and of similar systems, will be addressed in
terms of EFTs of QCD. This is more than a change in language.  It is moving
this physics from being a battleground of competing models  to being a source
of some of the fundamental parameters of the Standard Model, a reliable test
of its validity in the strong interaction sector and a unique laboratory for
the study of QCD properties.

\bigskip
\noindent
{\bf Acknowledgments.} \par 
We are thankful to A. G. Grozin for comments on the manuscript 
and to C. Ewerz for a careful reading of the manuscript and many useful comments and suggestions.
We gratefully acknowledge financial support from 
"Azioni Integrate Italia-Spagna 2004 (IT1824)/Acciones Integradas 
Espa\~na-Italia (HI2003-0362)'', and from the cooperation agreement INFN04-16 (CICYT-INFN). 
AP and JS are also funded by the MCyT and Feder (Spain) grant FPA2001-3598, 
the CIRIT (Catalonia) grant 2001SGR-00065 and the network EURIDICE (EU) 
HPRN-CT2002-00311. JS has been partially funded by the National Science 
Foundation under Grant No. PHY99-0794. NB, JS and AV would like to thank the  
{\it Benasque Center of Physics} and JS the {\it Kvali Institute of Theoretical Physic} for hospitality
while part of this work was written up.

\bibliographystyle{apsrmp}
\bibliography{rmp-pin}

\end{document}